\documentclass[twoside,12pt,a4paper]{article}

\usepackage{epsfig}
\usepackage{graphicx}
\usepackage{afterpage}
\usepackage{setspace}

\def\Journal#1#2#3#4{{#1} {#2} (#4) #3}

\def\NCS{{\em Nuovo Cimento Suppl.}}

\def\NPA{{\em Nucl. Phys.} A}

\def\NPB{{\em Nucl. Phys.} B}

\def\PLB{{\em Phys. Lett.} B}

\def\NPBPS{{\em Nucl. Phys.} B {\em{(Proc. Suppl.)}}}

\def\PRL{\em Phys. Rev. Lett.}
\def\PREV{\em Phys. Rev.}
\def\PREP{\em Phys. Rep.}

\def\PRD{{\em Phys. Rev.} D}
\def\PRC{{\em Phys. Rev.} C}

\def\JPG{{\em J. Phys.} G}
\def\ZPC{{\em Z. Phys.} C}

\def\EPJA{{\em Eur. Phys. J.} A}
\def\EPJC{{\em Eur. Phys. J.} C}

\def\APPB{{\em Acta Phys. Pol.} B}
\def\ARNPS{{\em Annu. Rev. Nucl. Part. Sci.}}
\def\CPC{{\em Comp. Phys. Comm.}}
\def\BJP{{\em Braz. J. Phys.}}
\def\PPNP{{\em Prog. Part. Nucl. Phys.}}
\def\JHEP{{\em JHEP}}
\def\AP{{\em Ann. Phys.}}
\def\NIMA{{\em Nucl. Instrum. Meth. Phys. Sect.} A}

\newcommand{\be}{\begin{equation}}
\newcommand{\ee}{\end{equation}}
\newcommand{\bea}{\begin{eqnarray}}
\newcommand{\eea}{\end{eqnarray}}
\newcommand{\nn}{\nonumber}

\newcommand{\Eq}[1]   {Eq.~(\ref{#1})}

\newcommand{\Fi}[1]   {Fig.~\ref{#1}}
\newcommand{\Fis}[2]  {Figs.~\ref{#1} and~\ref{#2}}

\newcommand{\Ta}[1]   {Table~\ref{#1}}

\newcommand{\Tar}[2]  {Tables~\ref{#1}--\ref{#2}}

\newcommand{\Se}[1]   {Section~\ref{#1}}

\newcommand{\agev}    {\mbox{$A$~GeV}}               
\newcommand{\gevc}    {\mbox{GeV$/c$}}

\newcommand{\mevcc}   {\mbox{MeV$/c^2$}}
\newcommand{\rb}[1]   {\mbox{\textrm{\scriptsize #1}}}
\newcommand{\rbt}[1]  {\mbox{\textrm{\tiny #1}}}

\newcommand{\pbar}    {\ensuremath{\bar{\textrm{p}}}}
\newcommand{\vzero}   {\ensuremath{\textrm{V}^{0}}}
\newcommand{\lam}     {\ensuremath{\Lambda}}
\newcommand{\lab}     {\ensuremath{\bar{\Lambda}}}
\newcommand{\sig}     {\ensuremath{\Sigma^{0}}}

\newcommand{\xim}     {\ensuremath{\Xi^{-}}}

\newcommand{\xip}     {\ensuremath{\bar{\Xi}^{+}}}

\newcommand{\om}      {\ensuremath{\Omega}}
\newcommand{\omm}     {\ensuremath{\Omega^-}}
\newcommand{\omp}     {\ensuremath{\bar{\Omega}^+}}
\newcommand{\omomp}   {\ensuremath{\Omega^{-}+\bar{\Omega}^{+}}}

\newcommand{\pim}     {\ensuremath{\pi^{-}}}
\newcommand{\pip}     {\ensuremath{\pi^{+}}}
\newcommand{\pipm}    {\ensuremath{\pi^{\pm}}}
\newcommand{\kmin}    {\ensuremath{\textrm{K}^{-}}}
\newcommand{\kplus}   {\ensuremath{\textrm{K}^{+}}}
\newcommand{\kpm}     {\ensuremath{\textrm{K}^{\pm}}}
\newcommand{\kzero}   {\ensuremath{\textrm{K}^{0}_{\rb{S}}}}
\newcommand{\jpsi}    {\ensuremath{\textrm{J}/\psi}}
\newcommand{\psip}    {\ensuremath{\psi^{\prime}}}

\newcommand{\omavg}   {\ensuremath{\langle \Omega \rangle}}

\newcommand{\piavg}   {\ensuremath{\langle \pi \rangle}}
\newcommand{\pipavg}  {\ensuremath{\langle \pi^{+} \rangle}}
\newcommand{\pimavg}  {\ensuremath{\langle \pi^{-} \rangle}}
\newcommand{\kpavg}   {\ensuremath{\langle \textrm{K}^{+} \rangle}}
\newcommand{\kmavg}   {\ensuremath{\langle \textrm{K}^{-} \rangle}}

\newcommand{\pt}      {\ensuremath{p_{\rb{t}}}}
\newcommand{\mt}      {\ensuremath{m_{\rb{t}}}}

\newcommand{\mtavg}   {\ensuremath{\langle m_{\rb{t}} \rangle - m}}
\newcommand{\ptavg}   {\ensuremath{\langle p_{\rb{t}} \rangle}}

\newcommand{\sqrts}   {\ensuremath{\sqrt{s_{_{\rbt{NN}}}}}}
\newcommand{\dedx}    {\ensuremath{\textrm{d}E/\textrm{d}x}}
\newcommand{\dndy}    {\ensuremath{\textrm{d}N/\textrm{d}y}}

\newcommand{\npart}   {\ensuremath{\langle N_{\rb{part}} \rangle}}
\newcommand{\npartch} {\ensuremath{N_{\rb{part}}}}
\newcommand{\npc}     {\ensuremath{N_{\rb{PC}}}}
\newcommand{\np}      {\ensuremath{N_{\rb{P}}}}
\newcommand{\nq}      {\ensuremath{N_{\rb{q}}}}
\newcommand{\navg}    {\ensuremath{\langle N \rangle}}

\newcommand{\tch}     {\ensuremath{T_{\rb{ch}}}}
\newcommand{\tth}     {\ensuremath{T_{\rb{fo}}}}
\newcommand{\tc}      {\ensuremath{T_{\rb{C}}}}
\newcommand{\mub}     {\ensuremath{\mu_{\rbt{B}}}}
\newcommand{\gams}    {\ensuremath{\gamma_{\rb{s}}}}
\newcommand{\gamq}    {\ensuremath{\gamma_{\rb{q}}}}
\newcommand{\lams}    {\ensuremath{\lambda_{\rb{S}}}}

\newcommand{\betap}   {\ensuremath{\beta_{\perp}}}

\newcommand{\bpavg}   {\ensuremath{\langle \beta_{\perp} \rangle}}
\newcommand{\betas}   {\ensuremath{\beta_{\rb{s}}}}
\newcommand{\der}     {\ensuremath{\textrm{d}}}

\newcommand{\ybeam}   {\ensuremath{y_{\rb{beam}}}}

\newcommand{\raa}     {\ensuremath{R_{\rb{AA}}}}
\newcommand{\rcp}     {\ensuremath{R_{\rb{CP}}}}
\newcommand{\nbin}    {\ensuremath{N_{\rb{bin}}}}
\newcommand{\vtwo}    {\ensuremath{v_{\rb{2}}}}
\newcommand{\epspart} {\ensuremath{\epsilon_{\rb{part}}}}

\begin{document}


\title{ \vspace{1cm} Strange hadron production in heavy ion collisions from
SPS to RHIC}

\author{C.\ Blume$^{1}$ and C.\ Markert$^{2}$ \\
$^{1}$Institut f\"{u}r Kernphysik, \\ Johann Wolfgang Goethe-Universit\"{a}t
Frankfurt, Frankfurt, Germany. \\
$^{2}$University of Texas at Austin, Austin, Texas 78712, USA.}
\maketitle


\begin{abstract}
Strange particles have been a very important observable in the search
for a deconfined state of strongly interacting matter, the quark-gluon
plasma (QGP), which is expected to be formed in ultra-relativistic
heavy ion collisions.  We review the main experimental observations
made at the Super Proton Synchrotron (SPS) at CERN, Geneva, and at the
Relativistic Heavy Ion Collider (RHIC) at Brookhaven National
Laboratory (BNL).  The large amount of recently collected data allows
for a comprehensive study of strangeness production as a function of
energy and system size.  We review results on yields, transverse mass
and rapidity spectra, as well as elliptic flow.  The measurements are
interpreted in the context of various theoretical concepts and their
implications are discussed.  Of particular interest is the question
whether strange particles are in any way sensitive to a partonic
phase.  Finally, a compilation of experimental data is provided.
\end{abstract}


\tableofcontents


\clearpage

%
\begin{table*}[th]
\begin{center}
\caption{Properties of the strange particles discussed in this review
\cite{PDG10}.  Listed are the valence quark content, the strangeness
$S$, the isospin, spin and parity $I (J^{P})$, the mass and the decay
channels which are most important for experiments, together with
their branching ratio (B.R.) and decay length $c \tau$.}
\vspace{5pt}
\begin{tabular}{llllllll} \hline\hline
\label{tab:strangeparts}
Particle        &
Quarks          &
$S$             &
$I(J^{P})$       &
Mass            &
Decay particles &
B.R.            &
$c \tau$        \\ 
                &
                &
                &
                &
(\mevcc)        &
                &
(\%)            &
(cm)            \\ \hline
\kplus\ (\kmin) & $u\bar{s}$ ($\bar{u}s$)         
                & +1 (-1)
                & $\frac{1}{2} (0^{-})$
                &  493.677 
                & $\mu^{+} \nu_{\mu}$ ($\mu^{-} \bar{\nu}_{\mu}$) 
                & 63.55 
                & 371.2      \\ 
\kzero          & $d\bar{s}$, $s\bar{d}$
                & ---                      
                & $\frac{1}{2} (0^{-})$
                &  497.614 
                & \pip\ \pim                  
                & 69.2   
                & 2.68       \\
$\phi$          & $\bar{s}s$
                & 0
                & $0 (1^{-})$
                & 1019.455
                & \kplus\ \kmin
                & 48.9
                & $4.63 \times 10^{-12}$ \\
                &
                &
                &
                &
                & e$^{+}$ e$^{-}$
                & $2.95 \times 10^{-2}$
                &                        \\
                &
                &
                &
                &
                & $\mu^{+} \mu^{-}$
                & $2.87 \times 10^{-2}$
                &                        \\ \hline
\lam\ (\lab)    & $uds$ ($\bar{u}\bar{d}\bar{s}$)
                & -1 (+1) 
                & $0 (\frac{1}{2}^{+})$
                & 1115.683 
                & p \pim\ (\pbar\ \pip)       
                & 63.9   
                & 7.89       \\
\xim\ (\xip)    & $dss$ ($\bar{d}\bar{s}\bar{s}$) 
                & -2 (+2)
                & $\frac{1}{2} (\frac{1}{2}^{+})$
                & 1321.71  
                & \lam\ \pim\ (\lab\ \pip)    
                & 99.887 
                & 4.91       \\
\omm\ (\omp)    & $sss$ ($\bar{s}\bar{s}\bar{s}$) 
                & -3 (+3)
                & $0 (\frac{3}{2}^{+})$
                & 1672.45  
                & \lam\ \kmin\ (\lab\ \kplus) 
                & 67.8   
                & 2.46       \\
\hline\hline
\end{tabular}
\end{center}
\end{table*}
%

\section{Introduction}
\label{introduction}


This review article attempts to summarize the main ideas and
observations concerning strangeness in heavy ion reactions that have
emerged during the SPS and RHIC program.  Heavy ion collisions at
these energies are believed to create energy densities that allow to
explore states of matter beyond the deconfinement phase transition
predicted by quantum chromodynamics (QCD).  It separates matter
composed of interacting hadrons from a new state of matter, the
so-called quark-gluon plasma (QGP), where the confinement of quarks
and gluons inside hadrons is removed.  The order of this phase
transition is not finally established.  While for a vanishing baryonic
chemical potential $\mub = 0$ different lattice QCD studies agree on
a cross over type, the situation for $\mub > 0$ is still unclear.  The
exact value of the critical temperature \tc\ for $\mub = 0$ is also
still under debate.  While according to \cite{AOKI1} \tc\ lies in the
range 146~--~170~MeV, depending on the investigated order parameter,
the authors of \cite{BAZAVOV1} find critical temperatures between
180~--~200~MeV.  Several calculations predict that the cross over line
will turn into a first order phase transition \cite{FODOR1,ALLTON1,
HATTA1}, thus giving rise to the presence of a critical point in the
QCD phase diagram, while other lattice QCD investigations result in a
cross over for all \mub\ \cite{FORCRAND1}.

The study of strange hadron production always played a special role in
the investigation of QGP matter.  Initially this was motivated by the
early suggestion that an enhanced production of strange particles,
relative to p+p collisions, might provide a signature of a QGP
formation.  Even though strangeness enhancement has been established
experimentally, many new theoretical developments have shed new light
on the experimental facts.  Over the years also many unexpected
experimental findings have altered the viewpoint on strangeness as an
observable.  Strange particles have also been instrumental in the
investigation of new phenomena that appeared in heavy ion physics
(e.g. quark number scaling of elliptic flow).
Table~\ref{tab:strangeparts} summarizes the properties of the
different strange particle species that have been measured in heavy
ion reactions.  After the bulk of the heavy ion program at the SPS has
finished and the RHIC program has passed its 10th year, this is a good
point in time to take a snapshot of the current situation and to
summarize the main observations.  

\subsection{Strangeness enhancement as QGP signal}

%
\begin{figure}[th]
\begin{center}
\begin{minipage}[b]{0.53\linewidth}
\begin{center}
\includegraphics[width=0.9\linewidth]{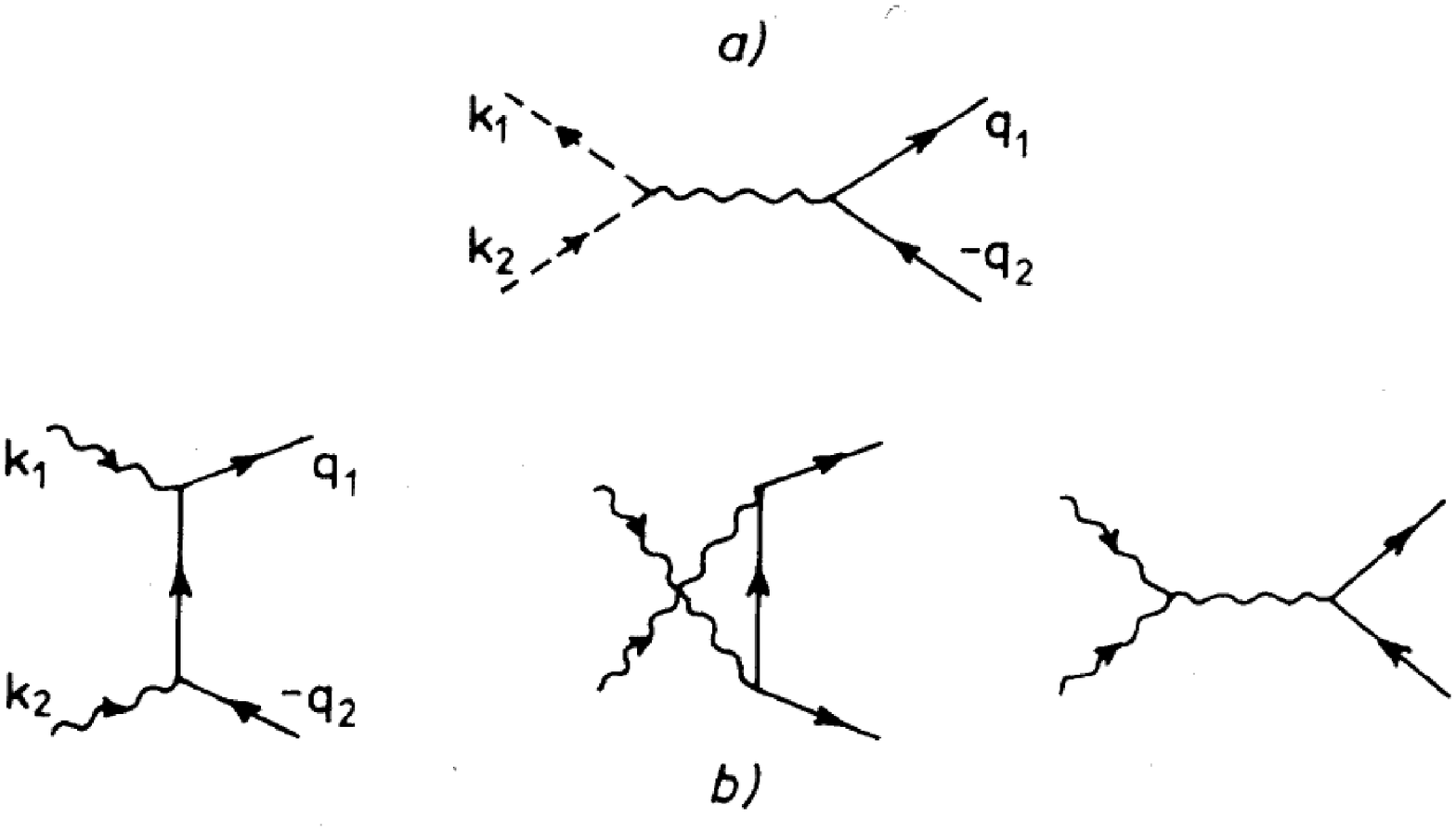}
\end{center}
\end{minipage}
\begin{minipage}[b]{0.45\linewidth}
\begin{center}
\includegraphics[width=0.55\linewidth]{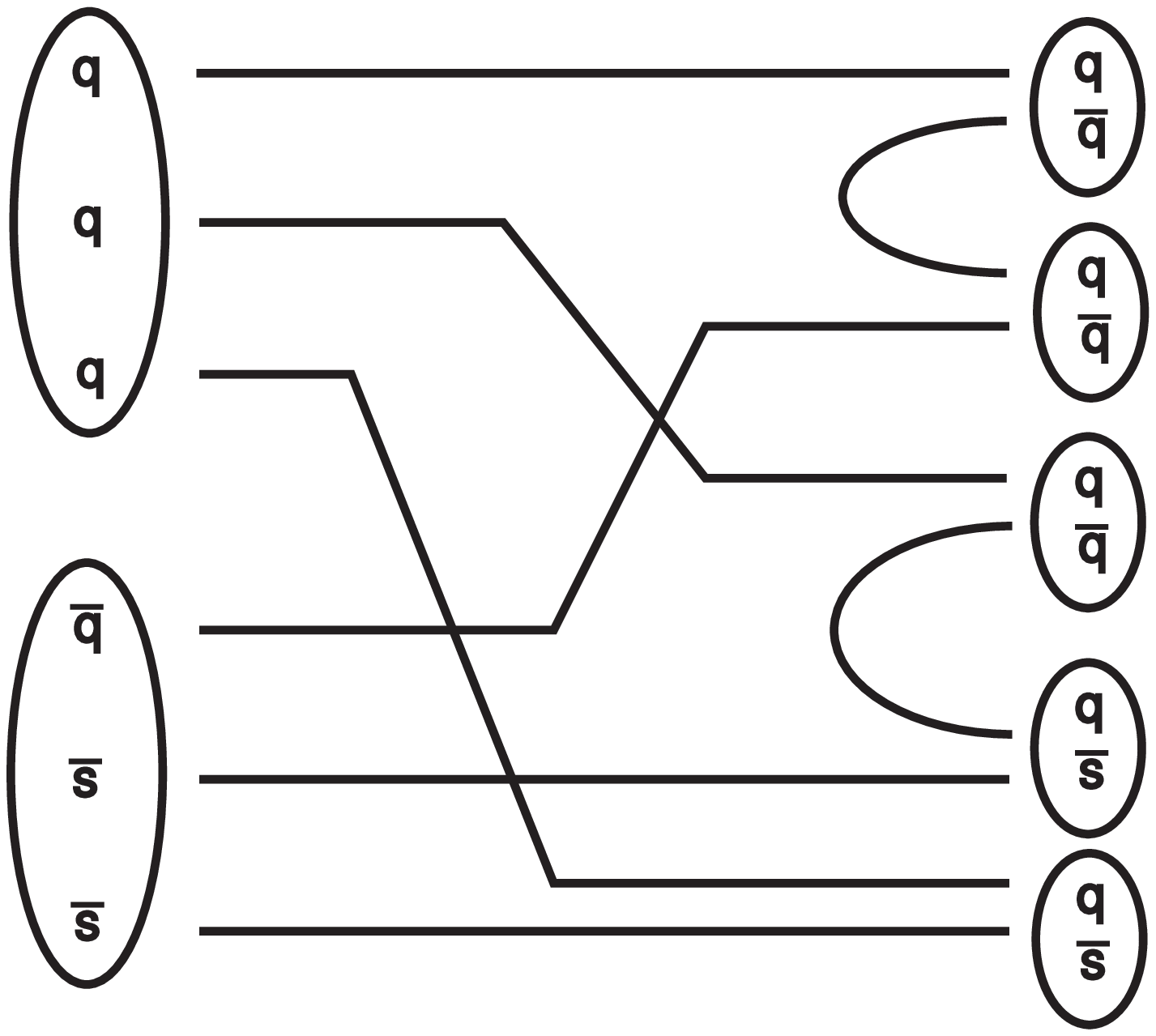}
\vspace{10pt}
\end{center}
\end{minipage}
\end{center}
\caption{Left: The lowest order QCD diagrams for $s\bar{s}$ production
((a) $q\bar{q} \rightarrow s\bar{s}$, (b) $gg \rightarrow s\bar{s}$)
\cite{RAFELSKI1}.
Right: Schematic picture for the reaction $\bar{\Xi} + \textrm{N}
\rightarrow 3 \pi + 2 \textrm{K}$ \cite{CGREINER1}.}
\label{fig:strange_proc}
\end{figure}
%

The main motivation for measuring strange particles in heavy ion
reactions is the expectation that their production rates per
participating nucleon should be enhanced with respect to elementary
nucleon--nucleon collisions, if a quark-gluon plasma is formed.
Strangeness enhancement has been among the first signals suggested for
a QGP state \cite{RAFELSKI1,KOCH1,KOCH2}.  The argumentation by
Rafelski and M\"{u}ller is based on two considerations.  One is that
in a plasma of quarks and gluons strangeness can be easily produced
via pair production of strange-anti-strange quark-pairs.  The basic
processes are the fusion of two gluons or of two light quarks into a
$s\bar{s}$-pair (see also left panel of \Fi{fig:strange_proc}):
\be
g + g \leftrightarrow s + \bar{s} \;\;\;\;\;\;\;\;\;\;\;\;
q + \bar{q} \leftrightarrow s + \bar{s} \;\;\;\;\;\; (q = u, d). \nn
\ee
For these reactions the $Q$-value corresponds to only the current mass
of the produced $s\bar{s}$-pair $Q_{\rb{QGP}} = 2 m_{s} \approx
200$~MeV \cite{PDG10}.  In contrast, the energetically cheapest way of
producing strangeness in a nucleon-nucleon reaction is via associated
production channels:
\be
\textrm{N} + \textrm{N} \rightarrow \textrm{N} + \Lambda + \textrm{K}. \nn
\ee
In this case the mass difference and thus the $Q$-value is already
much higher: $Q_{\rb{ass.}} = m_{\Lambda} + m_{\rb{K}} - m_{\rb{N}} \approx
670$~MeV.  As a consequence it should be much easier to generate
strangeness once a plasma state has been formed.

The second important point is that the equilibration times of partonic
reactions, especially due to the gluon fusion process, are much
shorter than the ones of hadronic reactions. The difference is
especially large, if rare multi-strange (anti-)baryons are considered.
In a partonic scenario with typical temperatures of $T = 200$~MeV
equilibration times of $\tau^{\rb{eq}}_{\rb{QGP}} \approx 10$~fm are
theoretically achievable in an ideal gas of quarks and gluons
\cite{RAFELSKI1}.  This is on the order of the total duration of a
heavy ion reaction, measured from the first collisions to the final
freeze-out of the hadrons.  However, the timespan of the QGP phase
will be still shorter than this, so that these partonic processes
might not be sufficient to drive the system to a complete chemical
equilibrium.  On the other hand, it was found in \cite{KOCH2}, that in
a gas of free hadrons, including resonances, the typical times to
reach an equilibrium state depend strongly on the strange particle
species.  While for particles with strangeness $|S| = 1$, like the
kaon and \lam, chemical equilibrium might be attainable after
$\tau^{\rb{eq}}_{\rb{HG}}(\textrm{K}) \approx 30$~fm, the timescales
for rare (anti-)hyperons should be an order of magnitude longer.
Following these arguments, it would thus be very difficult to produce
multi-strange particles ($\Xi$, $\bar{\Xi}$, \omm, \omp) in large
abundances in a hadron resonance gas, while the presence of a QGP
would be reflected in much higher production rates of these particles.
In the latter case, the multi-strange particles would just form via
quark coalescence at freeze-out and the yields should be close to the
chemical equilibrium expectation.  A review of these theoretical
considerations can be found in \cite{RAFELSKI2,RAFELSKI4}.


\section{Theoretical aspects}

In the following we review the basic theoretical ideas that motivate
the measurement of strange particles in heavy ion reactions.
Beginning with the early suggestion of strangeness enhancement as a
signature for a QGP formation, we discuss the further evolution of
this idea and alternative approaches.  Since transport and statistical
models are widely employed for the interpretation of strangeness data,
we discuss their main features in two extra sections.

\subsection{Recent developments}

Since these early calculations mentioned above, the theoretical
understanding of strangeness production has evolved further.  An
overview over more recent theoretical developments is given in
\cite{SCHAFFNER1}.  One important point is the fact that the assumption
of an ideal gas, as used in the calculations in
\cite{RAFELSKI1,KOCH2}, is not valid for heavy ion reactions.  As the
data on elliptic flow at RHIC have shown, the matter produced in heavy
ion collisions at high energies corresponds in its properties rather
to a liquid with very low viscosity \cite{BRMSWP,PHBSWP,STARWP,
PHNXWP}.  Results from lattice QCD indicate that QGP matter is
characterized by the presence of strong correlations and bound states,
which will prevail for temperatures significantly larger than the
critical temperature \cite{SHURYAK1,DATTA1,ASAKAWA1}.  Therefore, it
is inappropriate to treat partonic scattering processes
non-perturbatively, as done in the original calculations.  Attempts to 
take this into account, e.g. by hard thermal loop resummation
\cite{ANDERSEN1}, cut-off model for gluons \cite{WGREINER1}, and
massive, thermal gluons \cite{BIRO1,ALTHERR1,BILIC1}, seem to indicate
that the partonic equilibration times might rather increase
\cite{SCHAFFNER1}.  However, this issue is still not solved and further
theoretical developments in this direction are needed.  New insights
into the problem of equilibration in a QGP might arise from approaches
such as parton cascades \cite{XU1} or the investigation of plasma
instabilities \cite{MROWCZYNSKI1}.  Both result in a very fast 
equilibration on timescales equal or even below the expected QGP
lifetimes in heavy ion reactions.  Still, quantitative evaluations of
the implications for strangeness production are missing.

Another aspect relevant for strangeness production in heavy ion
collisions is that some features of particle production observed at
RHIC for intermediate transverse momenta \pt, such as baryon-meson
ratios and \vtwo~scaling of elliptic flow, have raised new interest in
quark coalescence as the possibly dominating hadronization process in
this \pt\ region.  Coalescence has been suggested very early on
\cite{HWA-RECO} and is an important ingredient of the original line of
arguments to suggest strangeness enhancement as a signature for QGP
formation \cite{RAFELSKI1,KOCH1,KOCH2}.  Recent implementations of
this idea \cite{BASS1,GRECO,HUANG,BASS2,HWAYANG06} have been
successful in describing the basic features of the measurements (see
discussion in \Se{sec:bmratios}).

In the recent years also new mechanisms to generate strange particles
in a hadron gas have been suggested.  These include multi-meson fusion
processes, such as
\be
\bar{\textrm{Y}} + \textrm{N} \leftrightarrow 
   n \: \pi + n_{\rb{s}}(\bar{\textrm{Y}}) \: \textrm{K},
\ee
which can additionally contribute to the yield of strange anti-baryons
in a dense hadronic medium \cite{CGREINER1}.
$n_{\rb{S}}(\bar{\textrm{Y}})$ denotes the number of anti-strange
quarks in anti-hyperon $\bar{\textrm{Y}}$.  The right panel of
\Fi{fig:strange_proc} shows a scheme for the reaction $\bar{\Xi} +
\textrm{N} \rightarrow 3 \pi + 2 \textrm{K}$ at the quark level, whose
back reaction might give a sizable contribution to the $\bar{\Xi}$
rates, since in a hot and dense fireball the number densities $n$ and
$n_{\rb{S}}(\bar{\textrm{Y}})$ may be high enough.  However, at RHIC
energies this effect seems not to be sufficient for an equilibration
during the lifetime of the hadronic state of the reaction
\cite{KAPUSTA1,HUOVINEN1}.  In \cite{PBM6} it has been pointed out
that the contribution from hadronic many body collisions should be
strongly enhanced close to the phase boundary between hadron gas and
QGP, due to a rapid increase of the particle densities with
temperature in the vicinity of \tc.  The observation of equilibrated
strangeness yields might therefore be interpreted as a direct evidence
for a phase transition.  Another approach that would explain a fast
equilibration inside the lifetime of the hadronic phase is the
inclusion of so-called Hagedorn states \cite{NORONHA1,NORONHA2}.
Hagedorn states (HS) are high mass resonances produced near \tc\ that
follow an exponential mass spectrum \cite{HAGEDORN1,HAGEDORN2} and
provide an efficient way of producing (anti-)baryons and kaons via
reactions such as $n \pi \rightarrow \textrm{HS} \rightarrow
n^{\prime} \pi + \textrm{B}\bar{\textrm{B}}$ and $n \pi \rightarrow
\textrm{HS} \rightarrow n^{\prime} \pi + \textrm{K}\bar{\textrm{K}}$
due to their large decay widths.

Another point of view is based on the argument that particle
production via the strong interaction always leads to a configuration
of maximum entropy, only constrained by energy, baryon number, and
strangeness conservation \cite{HEINZ1,STOCK1,BECATTINI5}.  The
observed particle abundances would thus not result from a dynamical
equilibration, but are a consequence of the hadronization process.
Therefore strangeness yields would always be close to their
equilibrium value and an enhancement in A+A collisions is rather due
to a suppression in p+p collisions due to the much stronger effect of
the conservation laws.

Strangeness production in a hadron gas would also be enhanced, if
hadron masses decrease due to medium modifications inside the hot and
dense fireball and thus lead to reduced $Q$~values for hadronic
reactions.  At the chiral phase transition, which lattice QCD predicts
to happen before \cite{AOKI1} or in coincidence with \cite{BAZAVOV1}
the deconfinement transition, the masses of the strange mesons should
approach to ones of non-strange mesons, leading to a flavor
equilibrated state \cite{SCHAFFNER1}.  In \cite{ZSCHIESCHE1} it has
been investigated within a chiral SU(3) model for the chemical and
thermal equilibrium case to what extend a chiral phase transition is
reflected in the observable particle ratios.  Since this turns out to
have a large effect, the experimentally determined particle ratios can
only be understood, if the chemical freeze-out happens slightly below
the chiral phase transition line.


\subsection{Transport models}
\label{sec:transport}

Hadronic transport models provide an important baseline for
comparisons to the experimentally observed strangeness production.
Since these models are based on hadronic degrees of freedom and string
excitation only, they allow to explore to what extend the observed
features of the data can be understood in a purely hadronic scenario.
They can therefore provide a benchmark that, to a certain extent,
allows to judge, whether there is room for a potential partonic effect
in the interpretation of a given experimental observation.  However,
when interpreting results from transport calculations one should be
aware that these models might reach conditions (e.g. local energy
densities) that cannot easily be reconciled with a purely hadronic
scenario.  But on the other hand, these type of models very easily
allow to generate comparison data for different experimental
situations, like centrality selection, center-of-mass energies, etc.

Examples of hadronic transport models used in heavy ion physics for
studying strangeness production are Ultra-relativistic Quantum
Molecular Dynamics (UrQMD) \cite{URQMD1,URQMD2}, Relativistic Quantum
Molecular Dynamics (RQMD) \cite{SORGERQMD}, Hadron-String Dynamics
(HSD) \cite{HSD1}.  Other models, such as A-Multi-Phase-Transport
(AMPT) model \cite{AMPT1} include both, hadronic and partonic, degrees
of freedom.  The EPOS model (Energy-conserving quantum mechanical
multiple scattering approach, based on Partons, Off-shell remnants,
and Splitting of parton ladders) \cite{EPOS1}, on the other hand,
follows a phenomenological approach that is based on a initial parton
model.

%
\begin{figure}[th]
\begin{center}
\includegraphics[width=0.45\linewidth]{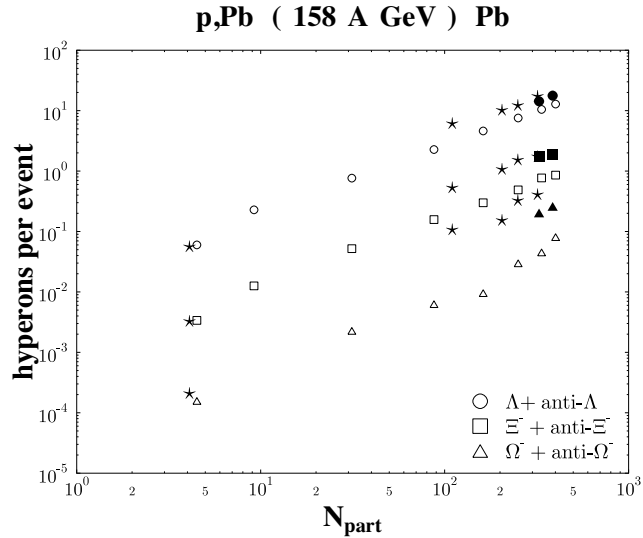}
\end{center}
\caption{The sum of hyperons and anti-hyperons at midrapidity as
a function of the number of participants for Pb+Pb and p+Pb collisions
at 158\agev.  The experimental data by the WA97 collaboration
\cite{WA97ENHANCE} are represented by stars, while the open symbols
correspond to default UrQMD calculations.  The full symbols denote
UrQMD results derived with reduced masses or, equivalently, enhanced
string tension \cite{URQMDENHANCE}.
}
\label{fig:URQMD}
\end{figure}
%

%
\begin{figure}[th]
\begin{center}
\begin{minipage}[b]{0.49\linewidth}
\begin{center}
\includegraphics[width=0.92\linewidth]{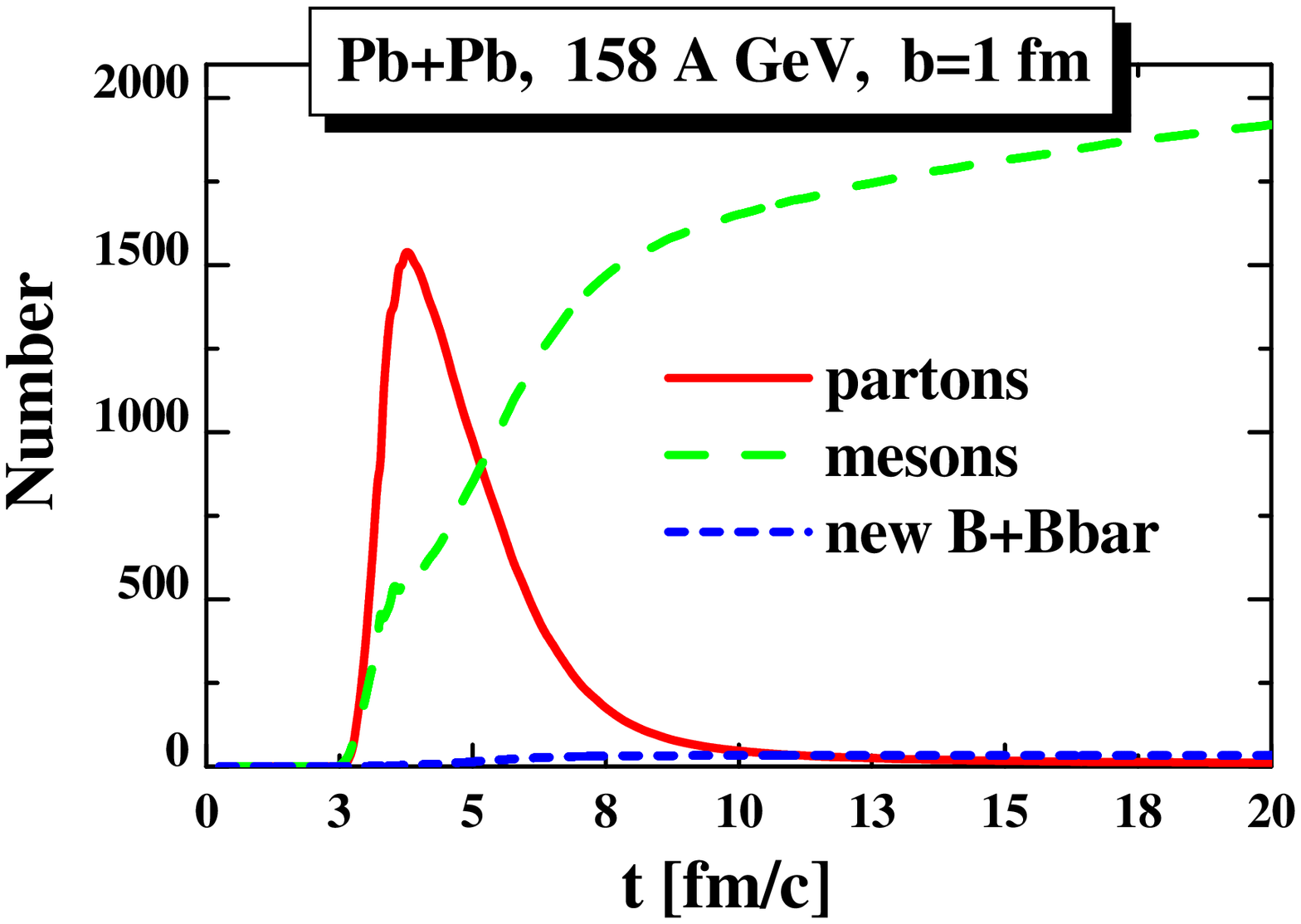}
\end{center}
\end{minipage}
\begin{minipage}[b]{0.49\linewidth}
\begin{center}
\includegraphics[width=\linewidth]{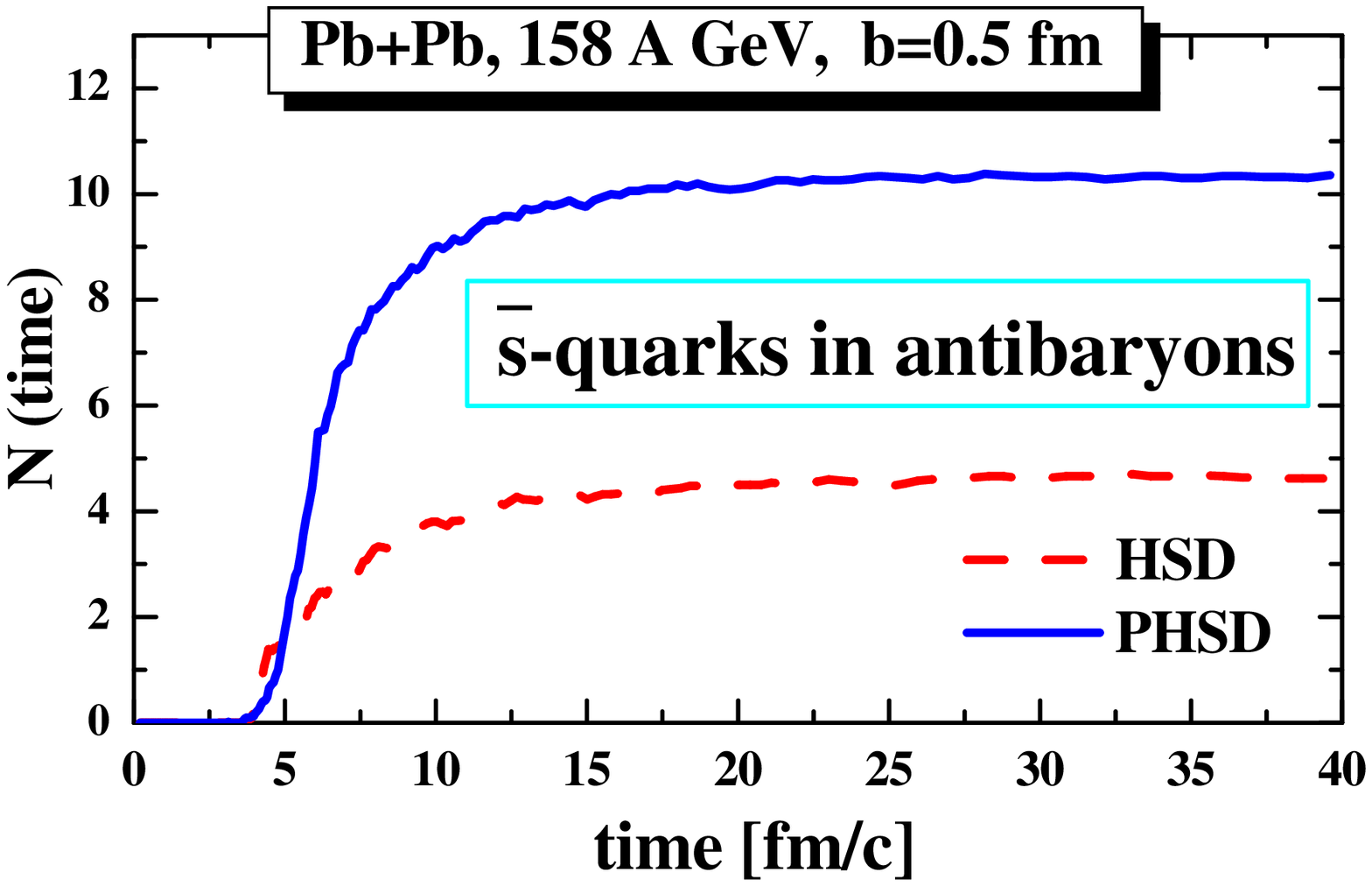}
\end{center}
\end{minipage}
\end{center}
\caption{Left: The number of produced partons (solid red line), mesons
(long dashed green line), and baryon-anti-baryon pairs (short dashed
blue line) as a function of time for central Pb+Pb at 158\agev\
(impact parameter $b$~=~1~fm), as calculated with the PHSD model
\cite{PHSD1,PHSD2}.  
Right: The number of $\bar{s}$ quarks contained in anti-baryons for
central Pb+Pb collisions at 158\agev\ from PHSD (solid blue line) and
HSD (dashed red line) \cite{PHSD1}.}
\label{fig:PHSD}
\end{figure}
%

The UrQMD and the HSD model both implement binary elastic and
inelastic hadronic scattering processes at lower center-of-mass
energies and string excitation at higher $\sqrt{s_{\rm{coll}}}$.  They
differ slightly in the number of hadronic states that are implemented.
UrQMD~1.3, for instance, includes all baryonic resonances up to masses
of 2~GeV and mesonic resonances up to 1.9~GeV, as documented by the
Particle Data Group.  Above these masses hadronic excitations are
treated within a string model with meson formation times on the order
of 1~-~2~fm/$c$.  In the recent version 2.3 the hadron resonance
spectrum has been extended toward a continuous distribution of
resonance states for meson-baryon reactions between 1.67~GeV~$<
\sqrt{s_{\rm{coll}}} <$~3~GeV \cite{URQMD3}.  This version includes
Pythia \cite{PYTHIA6} to model initial hard collisions for high energy
reactions ($\sqrt{s_{\rm{coll}}} > $~10~GeV).  The HSD model on the
other hand includes nucleons, $\Delta$, N$^{*}$(1440) and
N$^{*}$(1535) resonances, strange baryons and their resonances (\lam,
$\Sigma$, $\Sigma^{*}$, $\Xi$, $\Xi^{*}$, and $\Omega$), as well as
their anti-particles, and the $0^{-}$ and $1^{-}$ octet meson states.
Above $\sqrt{s_{\rm{coll}}} \approx 2.6$~GeV inelastic hadronic
collisions are described by the FRITIOF model \cite{FRITIOF}.  Both
model approaches are adjusted to reproduce the known elementary
nucleon-nucleon, meson-nucleon, and meson-meson cross sections over a
wide kinematic range as well as possible.  Beyond the simple
superposition of independent nucleon-nucleon collisions, therefore
additional effects are modeled as they are present in a nuclear
environment.  These include rescattering processes between the
reaction products, which can lead to strangeness exchange reactions
like $\bar{\textrm{K}} + \textrm{Y} \leftrightarrow \pi + \Xi$ (Y =
\lam, $\Xi$), and thus enhance the production of $\Xi$.

Generally, transport models provide a reasonable description of the
production rates of particles with only one strange quark (kaons and
\lam) \cite{HSDURQMD,PHSD1,URQMDHYDRO}, even though there are
remaining discrepancies in various details.  However, it has
frequently been shown that the observed enhancement factors of
multi-strange particles cannot easily be understood in the context of
conventional hadronic transport models without invoking more exotic
features as, e.g., color ropes \cite{NA49EDEPHYP,NA49EDEPOM,
URQMDENHANCE,WA97MODEL,NA49SDEPHYP}.  This is illustrated in
\Fi{fig:URQMD} which shows a comparison of measured hyperons yields at
158\agev\ to UrQMD results obtained with standard parameters.  While
the data for p+Pb collisions are reasonably described by the model,
substantial differences between measurement and model are visible for
central nucleus-nucleus reactions.  In case of the $\Omega$ there is
an order of magnitude between data and the transport model result.
The rates of multi-strange baryons can be matched by the model by
either lowering the masses of the constituent quarks or increasing the
string tension ($\kappa = 3$~GeV/fm, compared to the standard value of
$\kappa = 1$~GeV/fm) in an artificial manner (see filled symbols in
\Fi{fig:URQMD}) \cite{URQMDENHANCE}.  So far it has not been possible
to describe the enhancement of multi-strange baryons in a conventional
hadronic scenario.  However, multi-meson fusion processes
\cite{CGREINER1}, which can contribute to the yield of strange
anti-baryons in a dense hadronic medium and might thus reduce the
discrepancy between hadronic models and the data (see discussion in
the previous section), should also be taken into account in a
transport approach.  First steps into this direction have been done
within the HSD model \cite{HSD1,HSD2}, but a systematic and
quantitative evaluation of this effect has not yet been performed,
mainly due to the large amount of required computing resources.

An extension of the purely hadronic transport models is the 
Parton-Hadron-String Dynamics (PHSD) transport approach, which
includes also interacting partons as dynamic quasiparticles matched to
reproduce recent lattice QCD results \cite{PHSD1,PHSD2}.  The
transition from partonic to hadronic degrees of freedom is described
by fusion of quark-antiquark pairs or three quarks (antiquarks),
obeying color neutrality, flavor and energy-momentum conservation.  An
analysis of SPS data with this model shows that up to 40~\% of the
collision energy is stored in partonic degrees of freedom.  The left
panel of \Fi{fig:PHSD} illustrates how at top SPS energy (\sqrts~=
17.3~GeV) already around 1500 partons are created in the early phase
of the collision.  This partonic phase can have a significant effect
on the transverse mass distributions of kaons, due to repulsive
partonic mean fields and initial parton scatterings, as well as on the
production rates of multi-strange anti-baryons.  As shown in the right
panel of \Fi{fig:PHSD} the number of anti-strange quarks contained in
anti-baryons increases by roughly a factor two in comparison to the
purely hadronic HSD model.  This approach therefore allows to estimate
to which extent the presence of a partonic phase is needed in order to
describe (multi-)strange particle production in heavy ion collisions.


\newpage

\subsection{Statistical models}
\label{sec:statmodels}

Statistical models represent another theoretical approach that plays
an important role for the understanding of strangeness production in
heavy ion collisions.  Several implementations of this model are
frequently applied in this context, e.g. Becattini et al. 
\cite{BECATTINI3}, Braun-Munzinger et al. \cite{PBM3}, THERMUS
\cite{WHEATON1}, SHARE \cite{TORRIERI1}, and THERMINATOR
\cite{KISIEL1}.  The codes for latter three models are publicly
available.  Based on the assumption that particle yields at the end of
all inelastic interactions are corresponding to the chemical
equilibrium expectation, these type of models have generally been
quite successful in describing measured multiplicities with a small
set of parameters.  The basic ingredient of these models are the
partition functions, either grand-canonical, canonical, or
micro-canonical.  While the canonical or even micro-canonical
ensembles have to be used to describe small systems (e.g. p+p)
\cite{BECATTINI5}, the grand-canonical approximation provides a good
description in the case of large systems such as central Au+Au or
Pb+Pb collisions.  In the grand-canonical formulation of the
statistical model, the mean hadron multiplicities $\langle N_{i}
\rangle$ are defined as \cite{BECATTINI4}:
\be
\label{eq:statmodel}
\langle N_{i} \rangle = (2 J_{i} + 1) \frac{V}{(2 \pi)^{3}}
                        \int \der^{3} p \: \frac{1}{\gams^{-s_{i}}
                        \exp [(E_{i} - {\bf \mu} \cdot {\bf q_{i}}) / \tch] \pm 1}
\ee
The parameters are the chemical freeze-out temperature \tch, the
chemical potentials {\bf $\mu$}, the volume $V$, the strange quark
fugacity \gams, the spin $J_{i}$ and the number of strange quarks
$s_{i}$ of a given particle type $i$.  To account for the different
statistical behavior, the $+$~sign in the denominator is valid for
fermions, while the $-$~sign is used for bosons.  The strange quark
fugacity \gams\ allows for the possibility that strangeness might not
be fully equilibrated ($\gams < 1$) and is used in some
implementations of the statistical model \cite{BECATTINI3,SOLLFRANK1},
while other versions assume a full equilibrium also for strange
particles ($\gams = 1$) \cite{PBM3,CLEYMANS1}.  Other implementations
allow the non-strange quarks to be out of equilibrium as well and
therefore introduce yet an additional factor \gamq\ \cite{LETESSIER1,
LETESSIER2}.  The chemical potentials $\mu_{\rb{S}}$ and
$\mu_{\rb{Q}}$ are determined by the constraint of global strangeness
and charge conservation, while the baryonic chemical potential \mub\
is a free fit parameter.  Some implementations of the statistical
model (e.g. \cite{BECATTINI3}) directly calculate multiplicities
according to \Eq{eq:statmodel} and therefore also have to adjust the
volume $V$.  If instead particle ratios are fitted, as it is done in
\cite{PBM3}, the volume parameter drops out and only \tch\ and $\mub$
remain as free parameter (for $\gams = 1$).  Another aspect that has
been frequently debated is whether the statistical model analysis
should be restricted to midrapidity yields or if rather 4$\pi$
integrated multiplicities should be used \cite{BECATTINI3,ANDRONIC1}.
This issue plays mainly a role for SPS energies, at which it is
difficult to separate the central fireball and the fragmentation
regions, where the physics will be different.  At RHIC energies these
regions will be well separated and for most particle species anyway
only midrapidity \dndy\ values are experimentally available.  The
necessity of including \gams\ in the fits as a free parameter seems to
some extend arise from the choice of 4$\pi$ multiplicities as input.
At least at SPS energies \gams\ comes out closer to one if
midrapidity values are used \cite{ANDRONIC1,BECATTINI1}.

%
\begin{figure}[th]
\begin{center}
\begin{minipage}[b]{0.49\linewidth}
\begin{center}
\includegraphics[width=\linewidth]{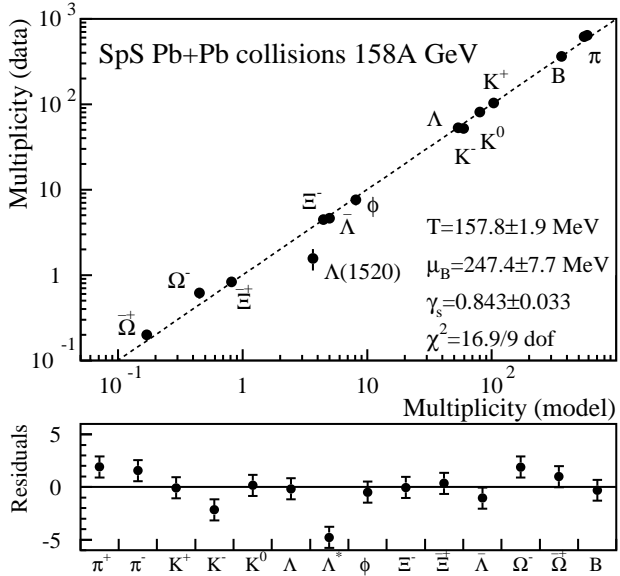}
\end{center}
\end{minipage}
\begin{minipage}[b]{0.49\linewidth}
\begin{center}
\includegraphics[width=0.9\linewidth]{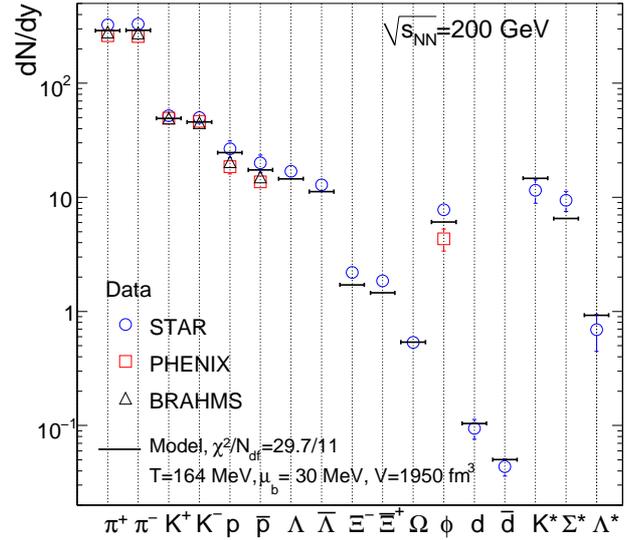}
\end{center}
\vspace{1pt}
\end{minipage}
\end{center}
\caption{Left: The measured particle multiplicities versus the result
of a statistical model fit for Pb+Pb collisions at \sqrts~= 17.3~GeV
\cite{BECATTINI1}.  The lower panel shows the fit residuals.
Right: Measured hadron yields around midrapidity for Au+Au collisions
at \sqrts~= 200~GeV, together with a statistical model fit, resulting
in the shown parameters \cite{ANDRONIC3}.}
\label{fig:StatModels}
\end{figure}
%

Figure~\ref{fig:StatModels} shows two different fits with statistical
models to heavy ion data, measured at \sqrts~= 17.3~GeV
\cite{BECATTINI1} and \sqrts~= 200~GeV \cite{ANDRONIC3}.  These
examples show that the yields of all strange particles, including the
rare $\Omega$ and \xip, agree quite well with the chemical equilibrium
expectation, both at the SPS and at RHIC.  In models that include
\gams\ as a free parameter it is found that \gams\ is usually larger
than 0.6~-~0.7 \cite{BECATTINI7}, thus indicating that strange
particle production seems to be at least not far away from the
equilibrium value.  One notable exception to this behavior is the
\lam(1520) resonance, whose yield has been found to be below the
prediction of the statistical model by almost five standard deviations
(see left panel of \Fi{fig:StatModels}).  Due to its short lifetime
($c \tau =$~12.6~fm) the resonance decays to a large part still inside
the fireball.  As a consequence its decay particles will be subject to
elastic scattering processes in the medium which in turn reduce the
yield of \lam(1520) that is reconstructible via an invariant mass
analysis.  In addition the regeneration cross section in the hadronic
phase is smaller than the one for scattering processes of the decay
particles, which results in overall suppression of the measurable
resonances \cite{URQMDRESO}.  This is the case for the \lam(1520) and
the K(892) at SPS and RHIC energies \cite{STARRESO}.  Depending on the
suppression a lower limit for the lifetime of the hadronic phase can
be derived \cite{RAFELSKIRESO}.  Due to this effect it is suggested to
exclude the resonances from statistical model fits in order to
determine freeze-out parameters.

%
\begin{figure}[t]
\begin{center}
\begin{minipage}[b]{0.50\linewidth}
\begin{center}
\includegraphics[width=0.95\linewidth]{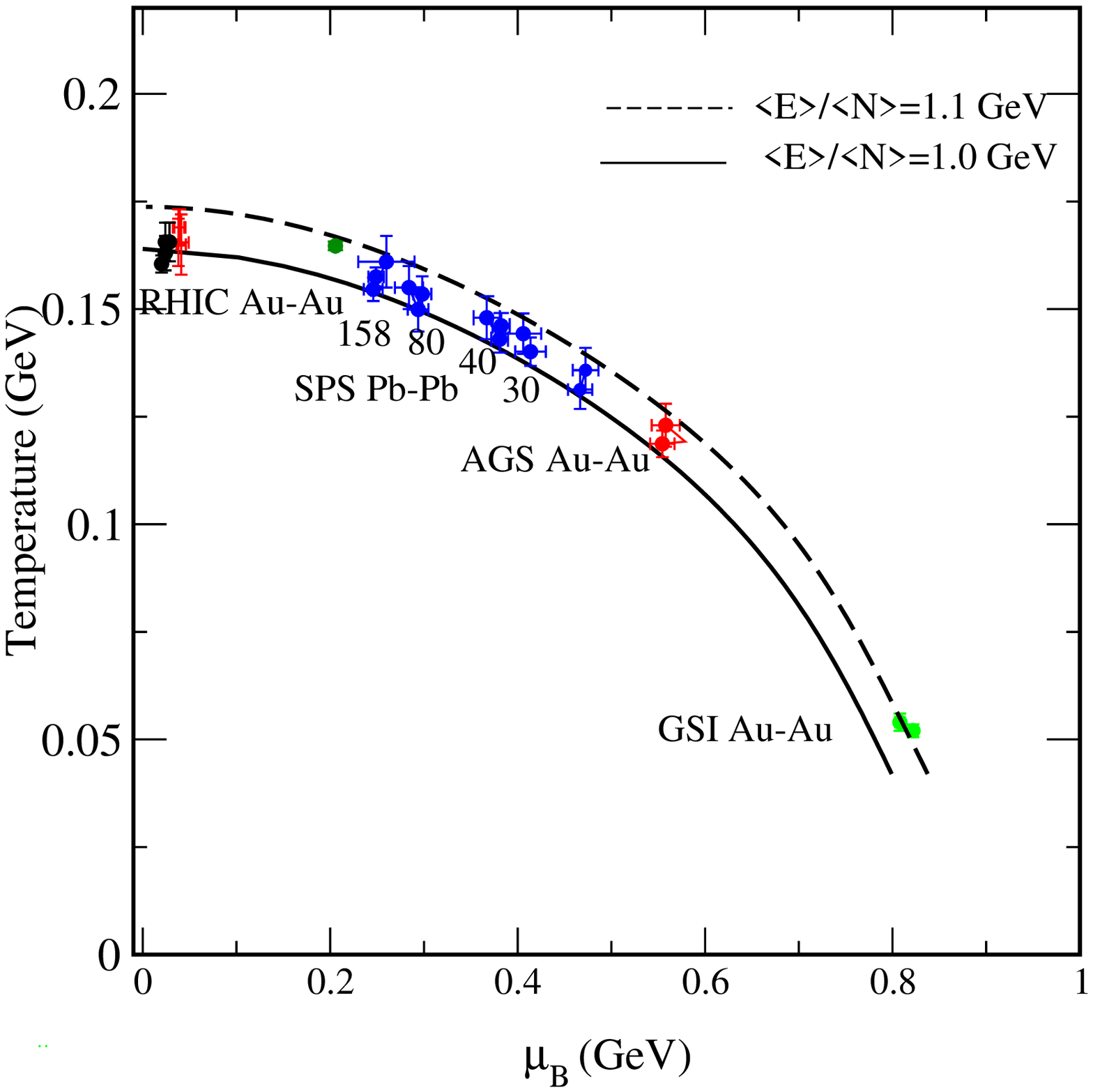}
\end{center}
\vspace{55.0pt}
\end{minipage}
\begin{minipage}[b]{0.45\linewidth}
\begin{center}
\includegraphics[width=\linewidth]{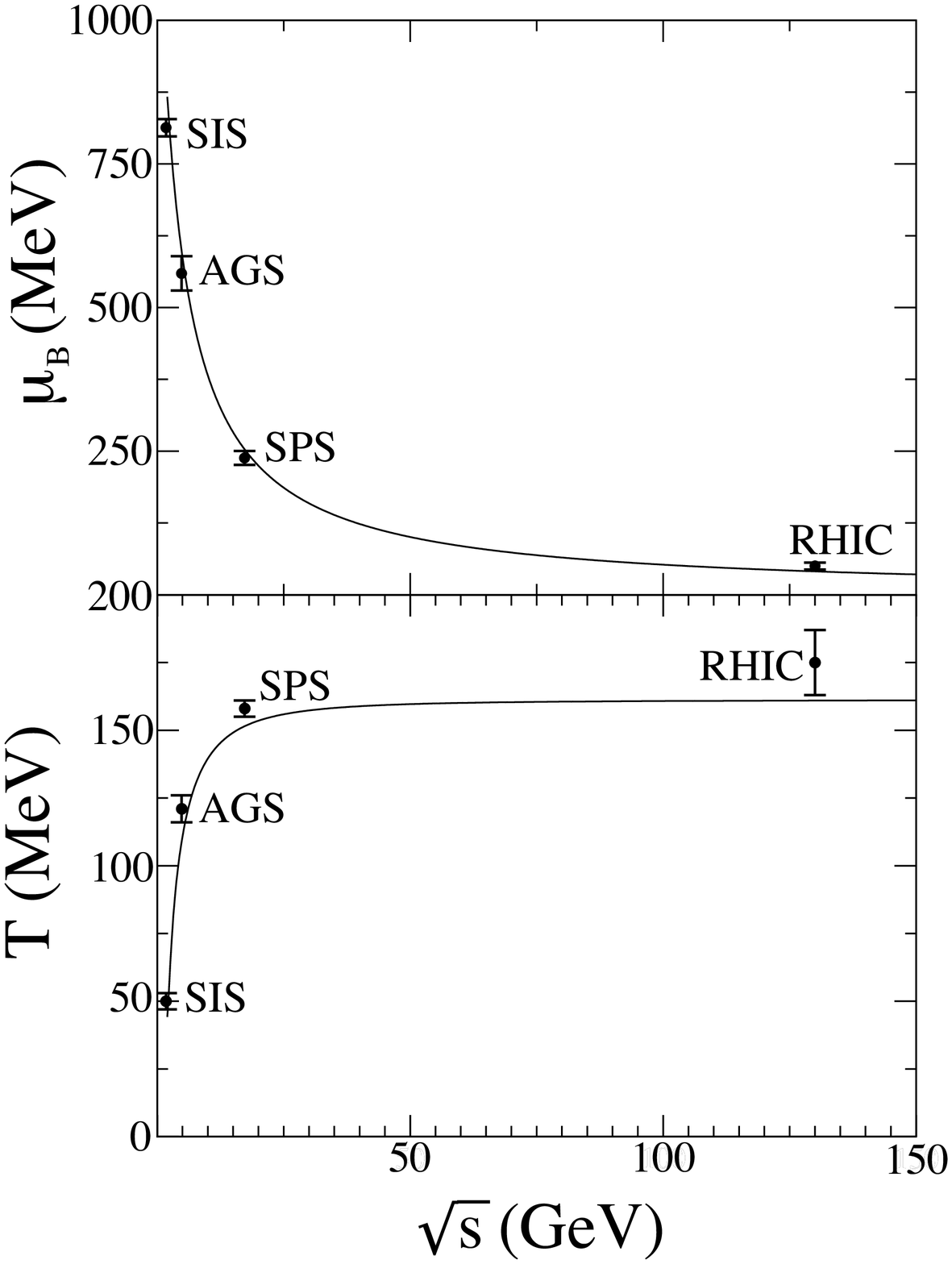}
\end{center}
\end{minipage}
\end{center}
\caption{Left: The chemical freeze-out parameters \tch\ and \mub\ for
different energies, as obtained from fits to data from RHIC, SPS, AGS,
and SIS.  The lines correspond to the freeze-out conditions $\langle E
\rangle/\langle N \rangle = 1$~GeV and 1.1~GeV \cite{CLEYMANS3}.
Right: The variation of \tch\ and \mub\ as a function of energy.  The
lines are the parametrization according to \Eq{eq:mub_param}, using
the freeze-out condition $\langle E \rangle/\langle N \rangle = 1$~GeV
(Fig. is adapted from \cite{PBM2}).}
\label{fig:freeze_out}
\end{figure}
%

It has been realized quite early that the resulting parameters \tch\ and
\mub\ lie on a single freeze-out curve, which follows a constant
energy density over particle density of $\langle E \rangle/\langle N
\rangle \approx 1$~GeV \cite{CLEYMANS1} (see left panel of
\Fi{fig:freeze_out}).  This has been interpreted as a consequence of
the changing composition of the hadron gas at chemical freeze-out,
moving from a nucleon gas at low temperatures to a mainly meson
dominated gas at high temperatures.  As a result the average rest mass
per particle $\langle M \rangle$ of the hadron gas decreases with
increasing temperature, thus compensating the increasing thermal
contribution to the energy density such that, in a non-relativistic
approximation, $\langle E \rangle/\langle N \rangle \approx \langle M
\rangle + 3/2 \, \tch \approx 1$~GeV results \cite{CLEYMANS1}.
Therefore, the two chemical freeze-out parameters \tch\ and \mub\ do not
vary independently of each other, when the center-of-mass energy of
the reaction is changed.  If furthermore the \sqrts\ dependence of
\mub\ is parametrized, particle ratios at different energies can be
predicted within the statistical model approach.  In \cite{PBM2} the
following phenomenological parametrization was suggested:
\be
\label{eq:mub_param}
\mub(s_{\rbt{NN}}) \simeq \frac{a}{(1 + \sqrts/b)};
\:\:\: a \simeq 1.27 \: \textrm{GeV};
\:\:\: b \simeq 4.3  \: \textrm{GeV}.
\ee
The resulting \sqrts~dependence of \tch\ and \mub\ is shown in the right
panel of \Fi{fig:freeze_out}.  Based on this parametrization it was
shown in \cite{PBM2} that a maximum of the relative strangeness
production is to be expected around $\sqrts = 7 - 8$~GeV (see
\Se{sec:edepstruct} for a detailed discussion).  Alternative
suggestions to parametrize the chemical freeze-out curve have been
made (e.g. by requiring a fixed total (anti-)baryon densities of
$n_{\rm{B}}  \approx 0.12\:\textrm{fm}^{-3}$ \cite{PBM4}, or a fixed
value of the entropy density, $s / T^{3}$, of approximately 7
\cite{CLEYMANS4,TAWFIK1}).  Other parameterizations of the energy
dependence of the chemical freeze-out parameters can be found in
\cite{BECATTINI7}.  Even though these parameterizations result in
similar \tch~--~\mub\ curves, they are motivated by different physical
interpretations.  For instance, fixed baryon density criterion is
based on the picture that the freeze-out line is determined by a
critical $n_{\rm{B}}$ below which inelastic baryon-baryon and
baryon-meson interactions cease to change the chemical composition.

At low energies or for small systems, such as p+p, the description of
particle yields with a statistical model has to be based on the
canonical, or micro-canonical, ensemble \cite{BECATTINI5,BECATTINI6,
KRAUS1}.  In these ensembles strangeness production will be
suppressed due to the small correlation volume in which strangeness
conservation has to be fulfilled \cite{RAFELSKI3}.  Therefore, it has
been suggested to model the system size dependence of strangeness
production by a transition from a canonical ensemble to a
grand-canonical one \cite{HAMIEH1}.


\newpage

\section{Experimental overview}

This sections summarizes the history of the main experimental
observations that have been made in order to establish strangeness
enhancement in high energy heavy ion reactions.  Also, we shortly
review the experiments at the SPS and RHIC and the experimental
techniques that have been employed to measure strange particles.

\subsection{Basic observations}
\label{sec:basicobservations}

The prediction outlined above triggered numerous experimental efforts
in order to establish whether strangeness enhancement is realized
in high energy heavy ion reactions.  Early attempts at the AGS and SPS
were limited to the measurement of particles with $|S| = 1$
(i.e. kaons, \lam, and \lab), and to smaller reaction systems (Si+Au,
O+Au, S+S, S+Au) \cite{E802KPMSIAU,E802KPMPA,E802KPPIP,E810V0A,E810V0B,
WA85LAM,NA35POAU,NA35PSS,NA36PSPB}.  But already with these data it
was possible to establish that there is indeed an enhancement of
strangeness production for kaons and \lam\ relative to pion production
compared to p+p and p+A collisions \cite{E802KPMPA,NA35PSS,
NA35SSAGAU}.  However, since the yields of $|S| = 1$ particles
measured in A+A can as well be understood in terms of a hadron
resonance gas, including rescattering processes \cite{HSDURQMD,WANG1},
this observation cannot unambiguously be interpreted as a signature
for QGP formation.  Therefore, dedicated experiments focused in the
following on the measurement of multi-strange (anti-)baryons in S+A and
p+A collisions \cite{WA85XIM,WA85LAMXIM,WA85OM,WA85XIOM,WA85LAMXIMB,
WA85PWHYP,WA85ENHANCE,WA94SS,WA94PS}.  One of the important results of
the first heavy ion period at the SPS was that also \xim\ and \xip\
production is enhanced in S+W collisions with respect to p+W
interactions \cite{WA85ENHANCE}.  With the availability of Pb beams at
the SPS these studies were extended in the years after 1994 toward a
systematic study of the system size dependence of this effect,
including the rare $\Omega$ hyperon \cite{WA97ENHANCE,NA57ENHANCE,
NA49ENHANCE}.  A review of the experimental results available after
the first round of CERN experiments with Pb beams at 158\agev\ can be
found in \cite{MARGETIS1}.

\begin{figure}[th]
\begin{center}
\includegraphics[width=0.8\linewidth]{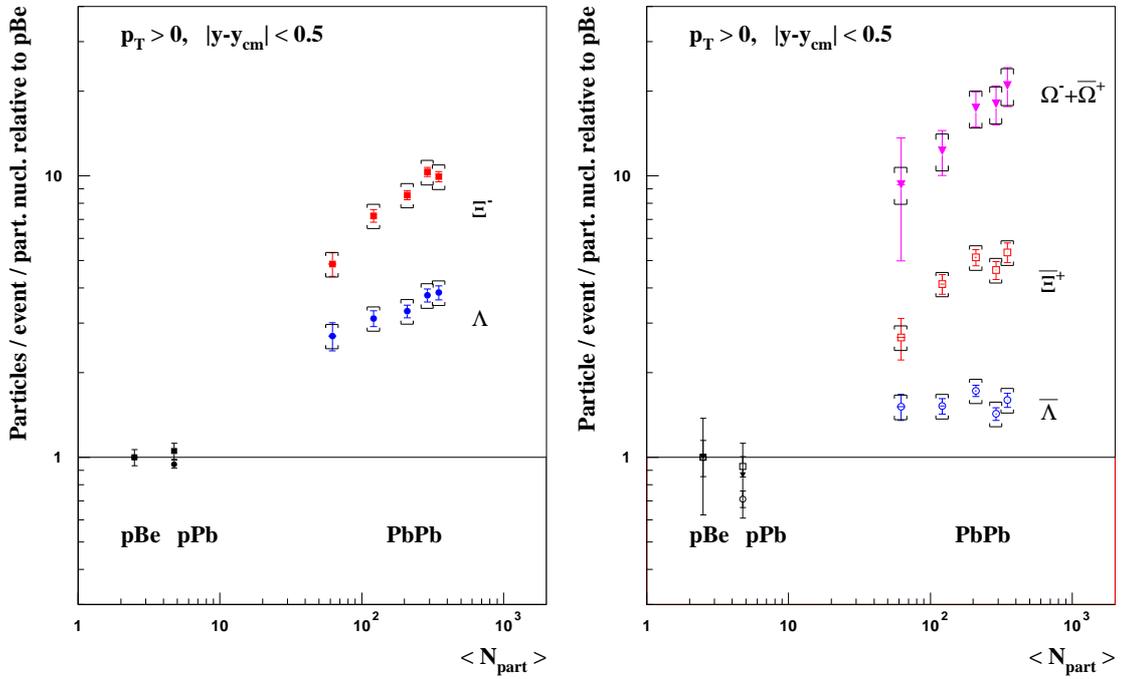}
\end{center}
\caption{The enhancement of strange particle production with respect
to p+Be collisions as a function of the number of participating
nucleons, measured by the NA57 collaboration for Pb+Pb collisions at
158\agev\ (Fig. adapted from \cite{NA57ENHANCE}).}
\label{fig:na57_enh}
\end{figure}

Figure~\ref{fig:na57_enh} shows the strange particle enhancement in
Pb+Pb collisions relative to p+Be interactions as observed by the NA57
experiment \cite{NA57ENHANCE}.  The strangeness enhancement factor
$E_{\rb{S}}$ is defined as:
\be
E_{\rb{S}} = \left. \left( \frac{1}{\npart}
             \left. \frac{\textrm{d}N(\textrm{Pb+Pb})}{\textrm{d}y}\right|_{y=0}
     \right) \right/
     \left( \frac{1}{2}
             \left. \frac{\textrm{d}N(\textrm{p+p(Be)})} {\textrm{d}y}\right|_{y=0}
     \right)
\label{eq:enhancefactor}
\ee
In order to account for the larger reaction volume of the
nucleus-nucleus system, the measured yields are normalized by the
averaged number of participating nucleons \npart.  This quantity is
usually derived from a Glauber model calculation \cite{GLAUBER1,
MILLER1}.  A clear hierarchy of the enhancement factors for the
different particle species is observed: While for the \lam\ ($S = -1$)
the enhancement reaches values of 3~--~4, $E_{\rb{S}}$ is found to be
close to 10 for the \xim\ ($S = -2$) and 20 for the $\Omega$ ($S =
-3$) in central Pb+Pb collisions.  Such an enhancement pattern is
expected in the case of a full chemical equilibrium for a large system
\cite{HAMIEH1}, which, according to the early publications
\cite{RAFELSKI1,KOCH1,KOCH2}, might be interpreted as a sign for fast
partonic equilibration in a QGP.  It should be pointed out that
equilibration in a partonic phase at high temperatures can lead to an
over-saturation of strangeness production relative to the expectation
of an equilibrated hadron gas at typical chemical freeze-out
temperatures, if the transition happens very suddenly
\cite{TORRIERI2}.  In any case it demonstrates that in high energy
nucleus-nucleus collisions additional mechanisms for strange particle
production are at work.  The strong enhancement of multi-strange
baryons is the most dramatic effect observed when comparing p+p to A+A
collisions (other effects like J/$\psi$ suppression or
high-\pt~suppression are on the order of a factor 5, while the
$\Omega$~enhancement reaches values up to 15~-~20 for central
reactions).

%
%

The NA57 results \cite{NA57ENHANCE} are a refinement of the previous
WA97 measurements \cite{WA97ENHANCE} and thus supposed to supercede
these data.  While NA57 and WA97 use p+Be collisions as a baseline
measurement, other experiments compare the A+A data to p+p collisions
\cite{NA49ENHANCE,NA49ENHANCE2,STARENHANCE}.  There are indications
that strangeness production is already enhanced in p+A collisions
relative to p+p scaled by \npartch\ \cite{NA49PA,E910PALAM} and that a
simple participant scaling is therefore violated.  A projectile
nucleon that collides more than once behaves different than one that
is participating in a single binary collision (see also the discussion
on the core corona model in \Se{sec:corecorona}).  Even though this
effect cannot be explained by a QGP formation, since the created
fireball is too small, and it also not sufficient to explain the
enhancement in A+A \cite{NA49PA}, it shows that an understanding of
p+A reactions is an important prerequisite for the interpretation of
heavy ion collisions.  A consequence of this effect is that the
measured enhancement in A+A will be less if p+A is used as a reference
instead of p+p.


\subsection{Experimental techniques}

%
\begin{figure}[th]
\begin{center}
\includegraphics[width=0.6\linewidth]{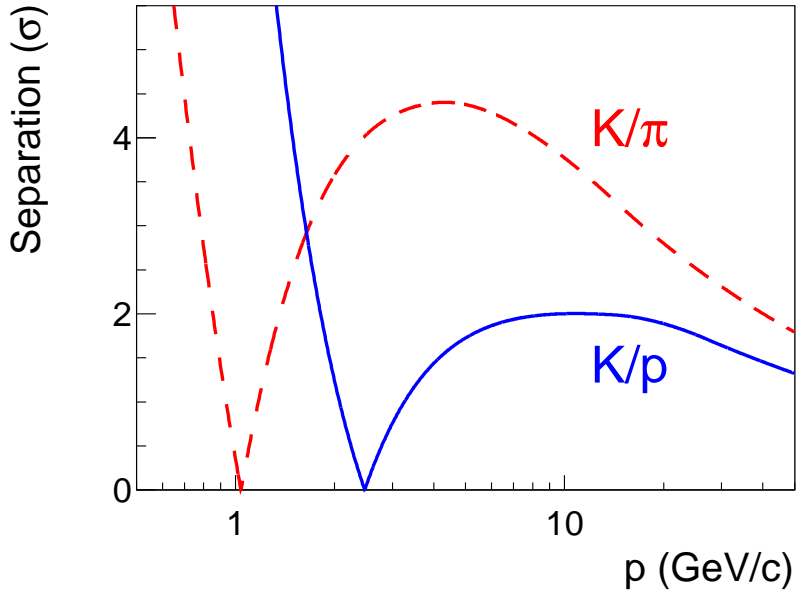}
\end{center}
\caption{The separation of kaons to pions and protons as achievable by
specific energy loss (\dedx) measurements for a resolution of 4\% as
obtained by the NA49 experiment \cite{NA49NIM}.}
\label{fig:na49_dedx}
\end{figure}
%

The strange particles listed in \Ta{tab:strangeparts} have been
measured by several experiments, both at the SPS and at RHIC.  In
Pb+Pb era at the SPS, these are the WA97, NA44, NA45, NA49, NA50,
NA57, and NA60 experiments, while at RHIC data on strange particle
production were gathered mainly by the STAR, PHENIX and BRAHMS
collaborations (see \Ta{tab:experiments}).  A detailed summary of the
available data measured at different center-of-mass energies can be
found in section \ref{sec:compedep}.

\begin{table*}[h]
\begin{center}
\caption{Overview on experiments measuring strange particles at the
SPS and at RHIC.}
\vspace{5pt}
\begin{tabular}{lllll} \hline\hline
\label{tab:experiments}
Experiment       &
Particles        &
Reaction systems &
\sqrts (GeV)     &
4$\pi$           \\ \hline
NA44             & \kpm                                & Pb+Pb                   & 17.3                      & --- \\
NA57 (WA97)      & \kzero, \lam, $\Xi$, $\Omega$       & p+A, Pb+Pb              & 8.7, 17.3                 & --- \\
NA45             & $\phi \rightarrow e^{+}+e^{-}$        & Pb+Pb                  & 17.3                      & --- \\
NA49             & \kpm, $\phi$, \lam, $\Xi$, $\Omega$ & p+p, C+C, Si+Si, Pb+Pb  & 6.3, 7.6, 8.7, 12.3, 17.3 & yes \\
NA50             & $\phi \rightarrow \mu^{+}+\mu^{-}$    & Pb+Pb                  & 17.3                      & --- \\
NA60             & $\phi \rightarrow \mu^{+}+\mu^{-}$    & In+In                  & 17.3                      & --- \\ \hline
STAR             & \kpm, $\phi$, \lam, $\Xi$, $\Omega$ & p+p, d+Au, Cu+Cu, Au+Au & 9.2, 62.4, 130, 200       & --- \\
PHENIX           & \kpm, $\phi$, \lam\                 & Au+Au                   & 130, 200                  & --- \\
BRAHMS           & \kpm\                               & p+p, Au+Au              & 62.4, 200                 & yes \\
\hline\hline
\end{tabular}
\end{center}
\end{table*}

The reconstruction of strange particles is done via two basic
experimental methods, which are employed by the experiments listed
above.  On one side there is the direct reconstruction as a charged
particle track (\kpm) and the indirect reconstruction via their weak
decay topology (\kzero, hyperons, and \kpm\ via kinks).  In the
following, a short description of the basic methods is given and their
strengths and associated difficulties (e.g. required corrections) are
discussed.

\subsubsection{Charged kaons}

In the case of charged kaons, a direct identification of the particles
can be performed.  Their trajectory can be reconstructed and their
identity derived from either their specific energy loss (\dedx) in
the detector material, their time-of-flight (TOF), the kink decay
topology, or their \v{C}erenkov signal.

The main advantage of using the specific energy loss for particle
identification is that it is usually measured with the same device
that is used for tracking (e.g. time projection chambers (TPC) or
silicon detectors) and that therefore the same acceptance can be
covered.  This allows to identify particles in large regions of phase
space.  The charged particles traveling through the active volume of
the detectors lose energy by atomic collisions in the gas (TPC, drift
chamber) or silicon (Si-detectors).  The released charge depends only
on the velocity of the particle, as described by the Bethe-Bloch
formula
\be
- \frac{\der E}{\der x} = K z^{2} \frac{Z}{A} \frac{1}{\beta^{2}}
                          \left[ \frac{1}{2}\ln
                          \left(
                          \frac{2 m_{\rb{e}} c^{2} \beta^{2}
                                \gamma^{2} T_{\rb{max}}}
                               {I^{2}}
                          \right)
                          - \beta^{2} - \frac{\delta(\beta \gamma)}{2}
                          \right] \:,
\ee
and thus allows to separate different particle species experimentally,
if their momentum is known.  $K$ is a constant, $z$ the charge of the
incident particle, $m_{\rb{e}}$ the electron mass, $Z$ the atomic
number and $A$ the atomic weight of the detector medium, $\beta$ the
velocity of the incident particle in units of $c$, $T_{\rb{max}}$ the
maximum kinetic energy which can be imparted to a free electron in a
single collision, $I$ the mean excitation energy of the detector
medium in eV, $\gamma = (1 - \beta^{2})^{-1/2}$, and $\delta(\beta
\gamma)$ the density effect correction \cite{PDG10}.  At lower
momenta ($p < 3 - 4$~GeV for pions), where the specific energy loss
follows the $1/\beta^{2}$ dependence of the Bethe-Bloch curve, the
kaons can be easily separated from protons and pions.  Therefore the
\dedx\ method is very well suited for kaon identification in collider
experiments up to momenta of $p \le 0.7$~\gevc\ \cite{STARSYST}.
Above this momentum the kaon band crosses the pion band and a clear
separation is difficult.  In fixed target experiments at higher
energies, where due to the additional Lorentz-boost particles have to
be measured at higher laboratory momenta, the \dedx\ method becomes
more involved.  Because in the high momentum region the specific
energy loss is described by the relativistic rise region of the
Bethe-Bloch curve, where the bands of pions, kaons, and protons are
relatively close together.  But also in these cases the \dedx\
signal has been successfully employed for kaon identification.  For
instance the NA49 experiment was specifically designed to reach a
\dedx\ resolution of 4\% \cite{NA49NIM} (the resolution for STAR is
better than 8\% \cite{STARNIM}), which is sufficient to
de-convolute the measured \dedx\ spectra and to extract charged kaon
yields on a statistical basis \cite{NA49KPI40158,NA49KPI2030}.
Figure~\ref{fig:na49_dedx} shows the separation power that can be
achieved with this resolution.

The time-of-flight method, in many cases combined with a \dedx\
measurement, can provide a clean identification of charged kaons in
the momentum range $p \le 2.0$~\gevc\ \cite{NA49KPI40158,NA49KPI2030,
PHNXKPI130B,PHNXKPI200,BRMSCENT200}.  To apply this method, the path
length and the momentum of a given particle has to be known, as well
as the precise time-of-flight.  The required time resolution typically
has to be below 100~ps, which can easily be achieved with scintillator
based TOF detectors in combination with a fast start detector.  Since
the TOF method requires a separate detector, the acceptance region of
the TOF measurement is in most cases, due to cost reasons, smaller
than the one available for tracking.  Modern techniques, such as
resistive plate chambers (RPC), allow to build TOF detectors with a
large acceptance.  These detectors can cover, e.g., the full 2$\pi$
azimuth of the TPC in the STAR and ALICE experiments.

%
\begin{figure}[t]
\begin{center}
\begin{minipage}[b]{0.49\linewidth}
\vspace{40.0pt}
\begin{center}
\includegraphics[width=\linewidth]{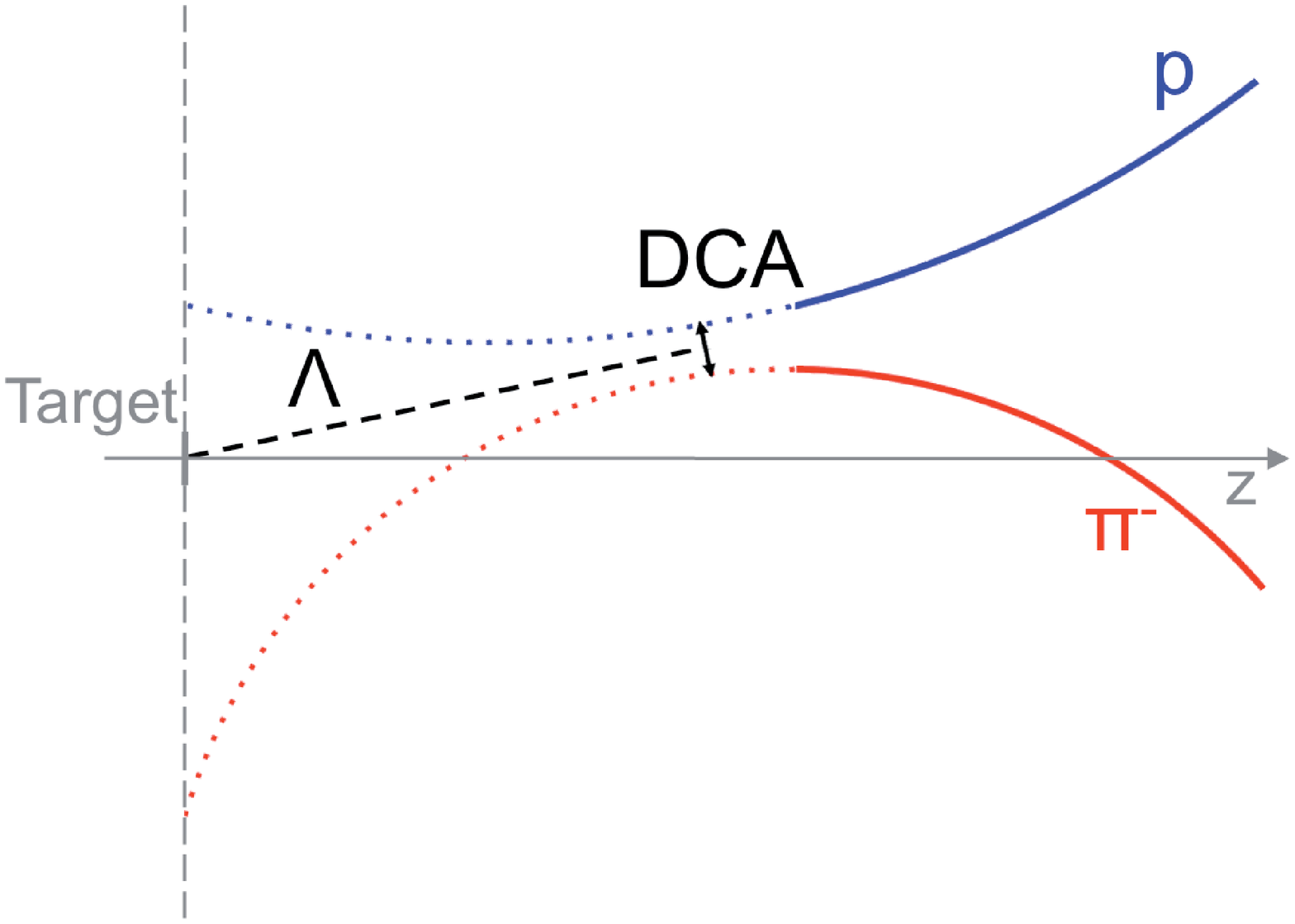}
\end{center}
\vspace{2pt}
\end{minipage}
\begin{minipage}[b]{0.49\linewidth}
\begin{center}
\includegraphics[width=\linewidth]{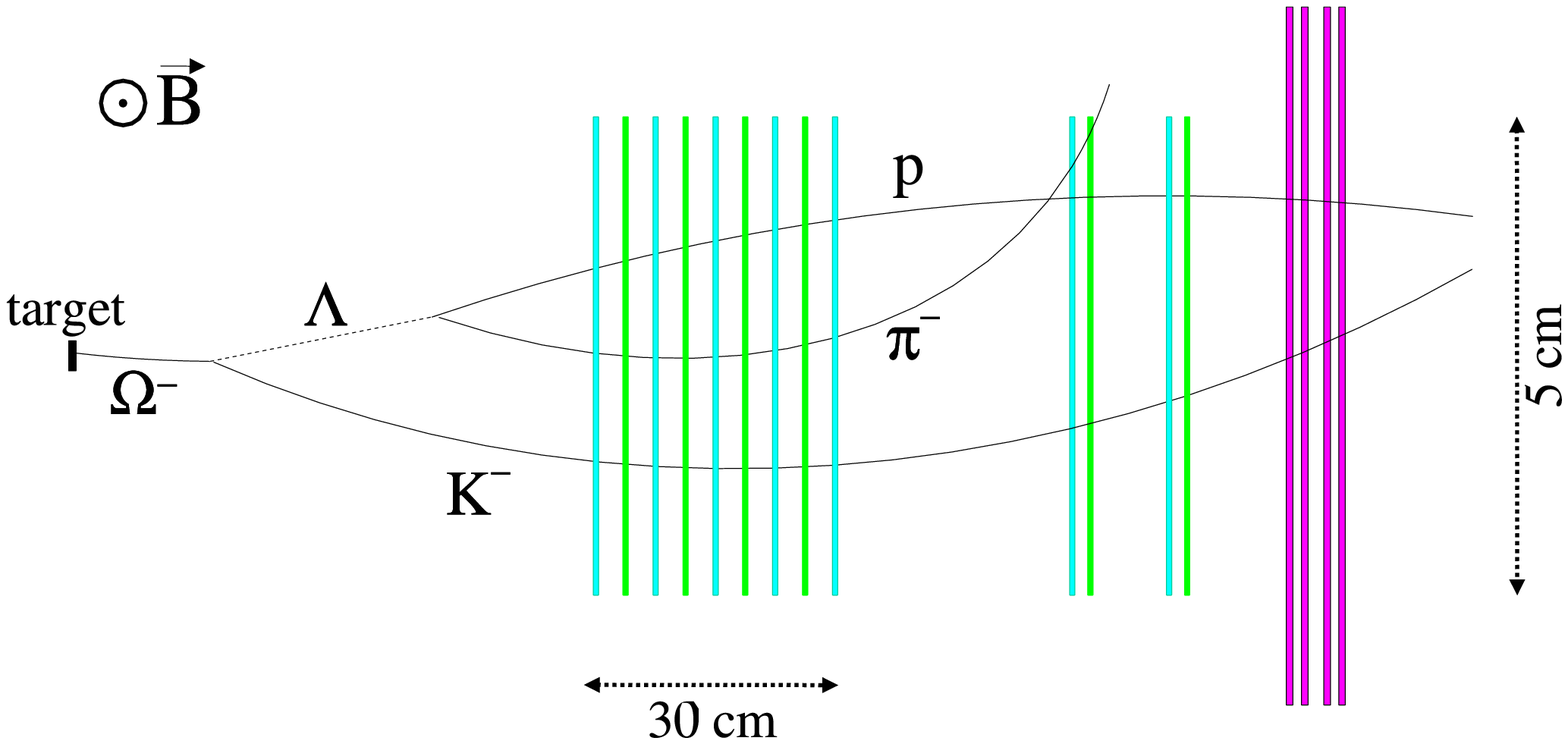}
\end{center}
\vspace{34.0pt}
\end{minipage}
\end{center}
\caption{Left: The topology of the decay $\lam\ \rightarrow \textrm{p}
  + \pim$.  The solid lines correspond to the reconstructed particle
tracks of the proton and the negative pion, while the dotted ones
represent their extrapolations through the magnetic fields toward the
target plane.  Shown is the distance of closest approach (DCA) between
the two charged tracks.
Right: Illustration of the reconstruction of the decay $\omm
\rightarrow \lam + \kmin$ within the NA57 setup (not to scale)
\cite{NA57ENHANCE}.}
\label{fig:topology}
\end{figure}
%

Sometimes an alternative way of reconstructing charged kaons is
followed, which exploits the topology of a weak decay, namely
$\kpm \rightarrow \mu^{\pm} + \nu_{\mu}(\bar{\nu}_{\mu})$ or $\kpm
\rightarrow \pipm + \pi^{0}$.  Since the neutral particles are not
detected, the reconstruction algorithms try to identify this decay by
looking for kinks in the tracks of charged particles.  This method is
complementary to \dedx\ or TOF measurement and can provide an
independent way of particle identification with different systematic
effects \cite{STARKPI130}.  However, there are substantial sources of
background, like charged hyperon and pion decays, as well as multiple
scattering in the detector material \cite{MARGETIS1}.  The method is
limited to particles that decay inside the sensitive area of the
tracking devices which reduces the available statistics.

\subsubsection{Hyperons and \kzero}

The other method, which is used for the majority of strange particles,
is based on the analysis of their decay topology.  This method
requires the simultaneous reconstruction of 2 (\kzero, \lam, \lab) or
3 (\xim, \xip, \omm, \omp) charged tracks. Figure~\ref{fig:topology}
illustrates the topologies of a \lam, respectively \omm, decay.  In
order to search for \vzero\ candidates (i.e. \kzero, \lam, and \lab),
all oppositely charged tracks are combined into pairs and the distance
of closest approach (DCA) between them is searched for.  The position
of the DCA gives the possible secondary vertex position.  Since weak
decays are characterized by long lifetimes of the mother particles, a
large fraction of their decay vertices has a substantial spatial
distance to the interaction vertex.  Therefore, the requirement of a
minimal distance between both allows to reduce the combinatorial
background, which to a large extend results from combining primary
tracks.  As the corresponding decay lengths are on the order of
centimeters (see \Ta{tab:strangeparts}), the secondary vertex
resolution has to be in the same order or better for collider
experiments.  The requirements are less stringent in the case of fixed
target experiments, since the average decay lengths are larger due to
the additional Lorentz boost.  In many cases (e.g. E895
\cite{E895LAMXI}, NA49 \cite{NA49EDEPHYP}, STAR
\cite{STARHYP130,STARHYP200}, and PHENIX \cite{PHNXLAM130}) TPCs or
drift chambers provide sufficient spatial resolution for this purpose.
Other experiments (e.g. NA57 \cite{NA57ENHANCE}) use silicon detectors
as a tracking device, which provide much better vertex resolution in
the sub-millimeter region and a higher rate capability at the expense
of a reduced acceptance.  In large collider experiments both
techniques, TPC and silicon, are combined (e.g. ALICE
\cite{ALICEPPRII} and STAR).  By applying additional topological cuts
(e.g. on the pointing angle of the mother momentum vector relative to
the main vertex position) or using additional particle identification
on the daughter tracks, the combinatorial background can be further
reduced.  For the particle pair assigned to a given \vzero\ candidate
the invariant mass is calculated based on a mass hypothesis by
assigning either the pion mass to both tracks (\kzero) or the pion and
the proton mass (\lam) to the positively and negatively charged track,
respectively.  From the resulting invariant mass distributions the
yields of the corresponding mothers particles can be extracted.
Typically a mass resolution around 4~MeV is achieved.

\lam\ and \kzero\ can also be reconstructed without measuring the
decay vertex position, as has been demonstrated by PHENIX
\cite{PHNXLAM130} and NA45 \cite{NA45KZERO158}.  The drawbacks of this
approach are an increased combinatorial background and a poorer mass
resolution, while the benefit is a somewhat reduced sensitivity to the
accuracy of the detector simulation.

\lam\ (\lab) candidates from this first step can then be
combined with a third charged track to form \xim\ (\xip) and \omm\
(\omp) candidates.  Since the \lam\ are in these cases daughter
particles, it is usually required that their momentum vector does not
point back to the interaction vertex, in contrast to the case of the
primary \lam.  This method has been successfully applied to heavy ion
reactions to measure systematically the production of the rare \omm\
and \omp\ \cite{NA49EDEPOM,NA57ENHANCE,STARHYP200}, although usually
the sum $\omm + \omp$ is shown in order to increase the statistical
significance.

\subsubsection{Corrections}

The different detection techniques for strange particles differ in
their necessary corrections.  The ones for the direct measurement of
charged kaons are mainly determined by the single track acceptance and
reconstruction efficiency, which is usually quite high (typically $>
90$~\% at higher momenta), thus resulting in relatively small
correction factors.  In contrast, the reconstruction of the decay
topology requires the measurement of two charged tracks.  Therefore
any acceptance losses and inefficiencies enter quadratically.
Additionally, the algorithms for the secondary vertex reconstruction
usually cause losses which have to be corrected, mainly due to the
cuts applied to reduce the combinatoric background.  As a consequence,
the corrections for this method are higher than for the single track
measurements and have to be carefully determined by a Monte Carlo
simulation (e.g. for \lam\ the efficiencies vary typically between
5~\% and 50~\%, depending on momentum).  On the other side, the
particle identification of the topological method is clean and
unambiguous.

Another correction that is necessary for some particles is the
correction for feed-down from weak decays.  This is important for
\lam\ and \lab, where part of the measured yield is originating from
the decays $\xim(\xip) \rightarrow \lam(\lab) + \pim(\pip)$
\footnote{The weak decays $\omm(\omp) \rightarrow \lam(\lab) +
\kmin(\kplus)$ are usually ignored, since their contribution is
marginal.}.  Due to the long decay lengths of the weak decays, the
fraction of accepted feed-down particles strongly depends on the
detector geometry, the reconstruction algorithm, and the applied cuts
(typical values for the size of the feed-down correction are 5~\% to
20~\% for \lam, for \lab\ it can be even higher).  Therefore, this
correction has to be applied at the analysis stage and cannot be
calculated properly afterward without exact knowledge of the
experimental conditions.  This argument does not apply to
strong and electro-magnetic decays, which all happen close to the
reaction zone.  There, all decay particles will be included in the
measured yields and can thus easily be subtracted, if necessary.
For instance, the measured \lam\ multiplicities usually refer to the
sum $\lam + \sig$, since \lam\ from the decay $\sig \rightarrow \lam +
\gamma$ cannot be separated experimentally from the primary ones
without performing an additional photon measurement.


\clearpage

%
\begin{figure}[th]
\begin{center}
\includegraphics[width=0.55\linewidth]{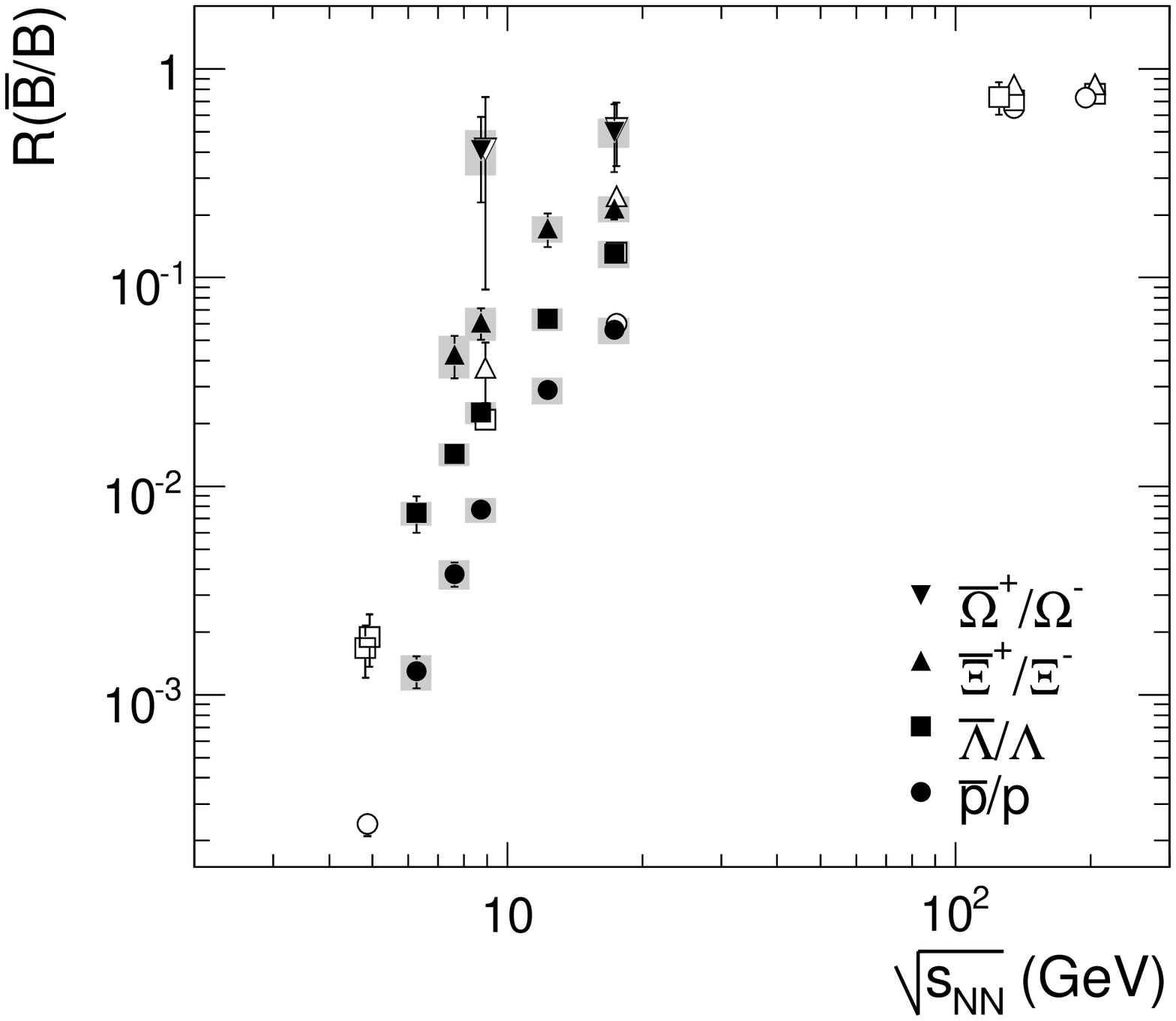}
\end{center}
\caption{ The \pbar/p, \lab/\lam, \xip/\xim, and \omp/\omm\ ratios
around midrapidity in central Pb+Pb and Au+Au collisions as a function
of \sqrts.  Shown are data by the SPS experiments NA49
\cite{NA49EDEPHYP,NA49EDEPOM,NA49PPBAR}, NA44 \cite{NA44PPBAR} and NA57
\cite{NA57ENHANCE,NA57EDEPHYP}, from  AGS \cite{E802PPBAR,E896LAM,
E917LAB,E891LAM} and RHIC experiments \cite{PHNXKPI200,STARHYP130,
STARHYP200,PHNXLAM130,STARPR130,STARLAM130,STARPR200}.}
\label{fig:bbbarratios}
\end{figure}
%

\section{Energy dependence}

The large amount of data at different center-of-mass energies that
have been accumulated at the SPS and at RHIC over the recent years
allows, together with AGS measurements from low energies, to compile
a relatively complete picture of the energy dependence for most
strangeness related observables.  Of special interest is the \sqrts\
evolution of the observed strangeness enhancement.  Also, several
intriguing structures in the energy dependences have emerged, for
instance in the \kplus/\pip~ratio.  The $\phi$ meson, which might
be directly sensitive to the partonic phase of the collision, deserves
a special attention and is treated in a separate subsection.  We also
discuss the \sqrts\ dependence of transverse and longitudinal spectra.


\subsection{Strangeness enhancement at different energies}

%
\begin{figure}[th]
\begin{center}
\begin{minipage}[b]{0.43\linewidth}
\begin{center}
\includegraphics[width=\linewidth]{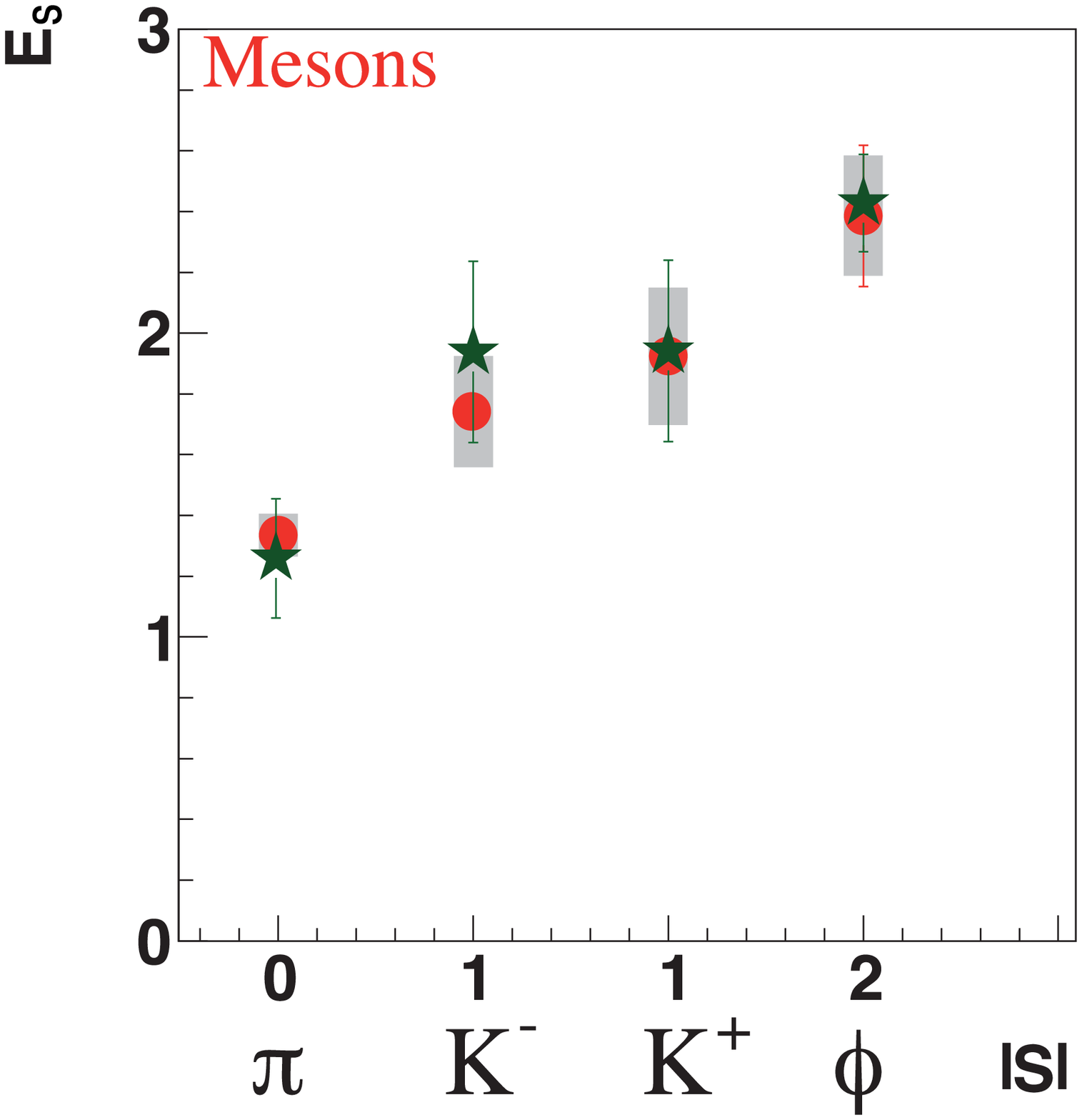}
\end{center}
\end{minipage}
\begin{minipage}[b]{0.43\linewidth}
\begin{center}
\includegraphics[width=\linewidth]{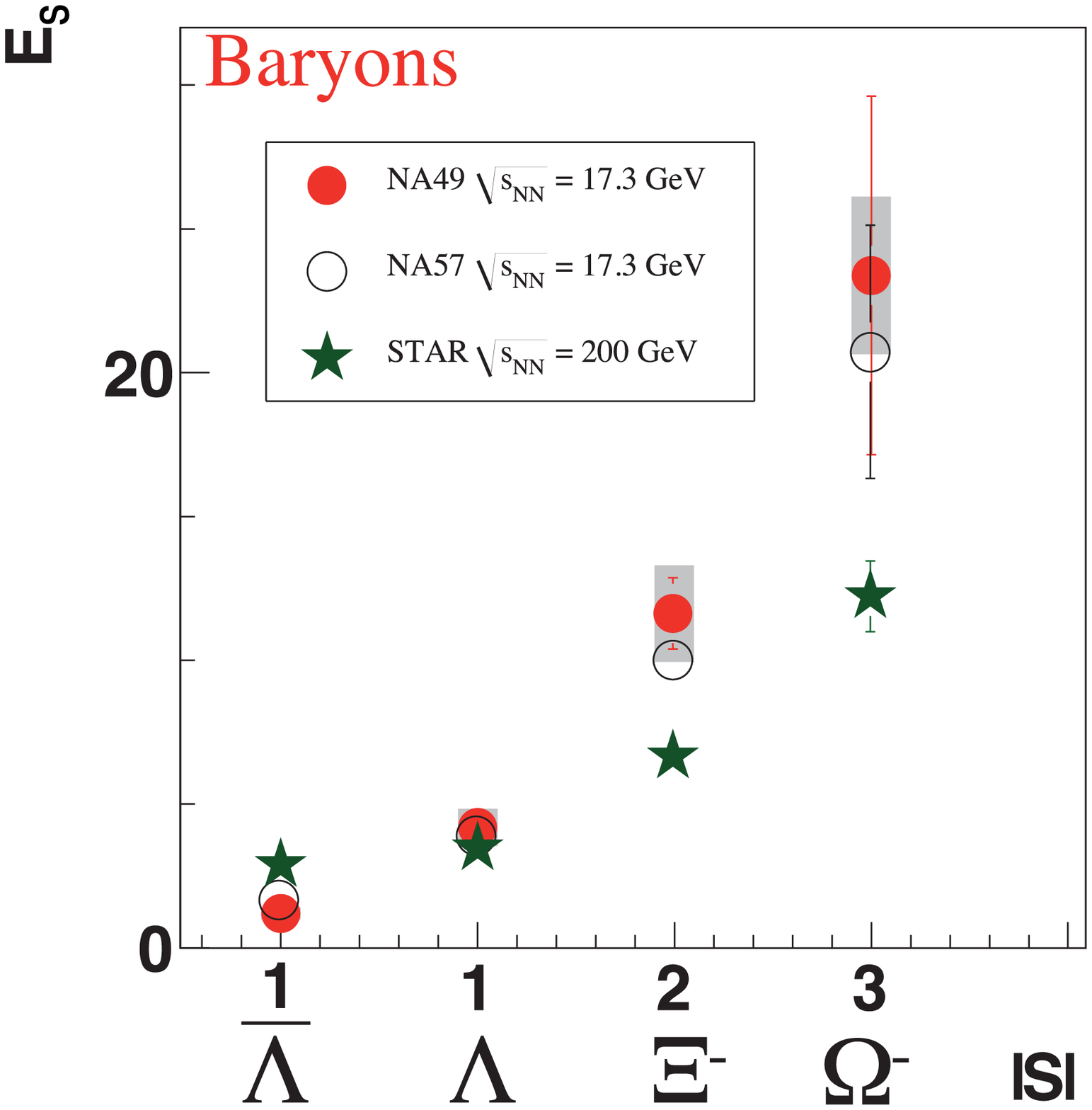}
\end{center}
\end{minipage}
\end{center}
\caption{The enhancement factor $E_{\rb{S}}$ for midrapidity yields
measured in central Pb+Pb (Au+Au) collisions at \sqrts~=~17.3~GeV
(200~GeV) versus the number of strange valence quarks for mesons
(left) and baryons (right) \cite{NA49ENHANCE2}.  Shown are data by the
NA49 collaboration (filled circles) \cite{NA49EDEPHYP,NA49EDEPOM,
NA49KPI40158,NA49PPPION,NA49SDEPSTRNG,NA49EDEPPHI}, by the NA57
collaboration (open circles) \cite{NA57ENHANCE}, and by the STAR
collaboration (filled stars) \cite{STARENHANCE,STARSYST,STARPHIESDEP}.
NA49 and STAR use p+p collisions as baseline, while in the case of
NA57 $E_{\rb{S}}$ is calculated relative to p+Be collisions.}
\label{fig:NA49_enh}
\end{figure}
%

%
\begin{figure}[th]
\begin{center}
\begin{minipage}[b]{0.49\linewidth}
\begin{center}
\includegraphics[width=0.80\linewidth]{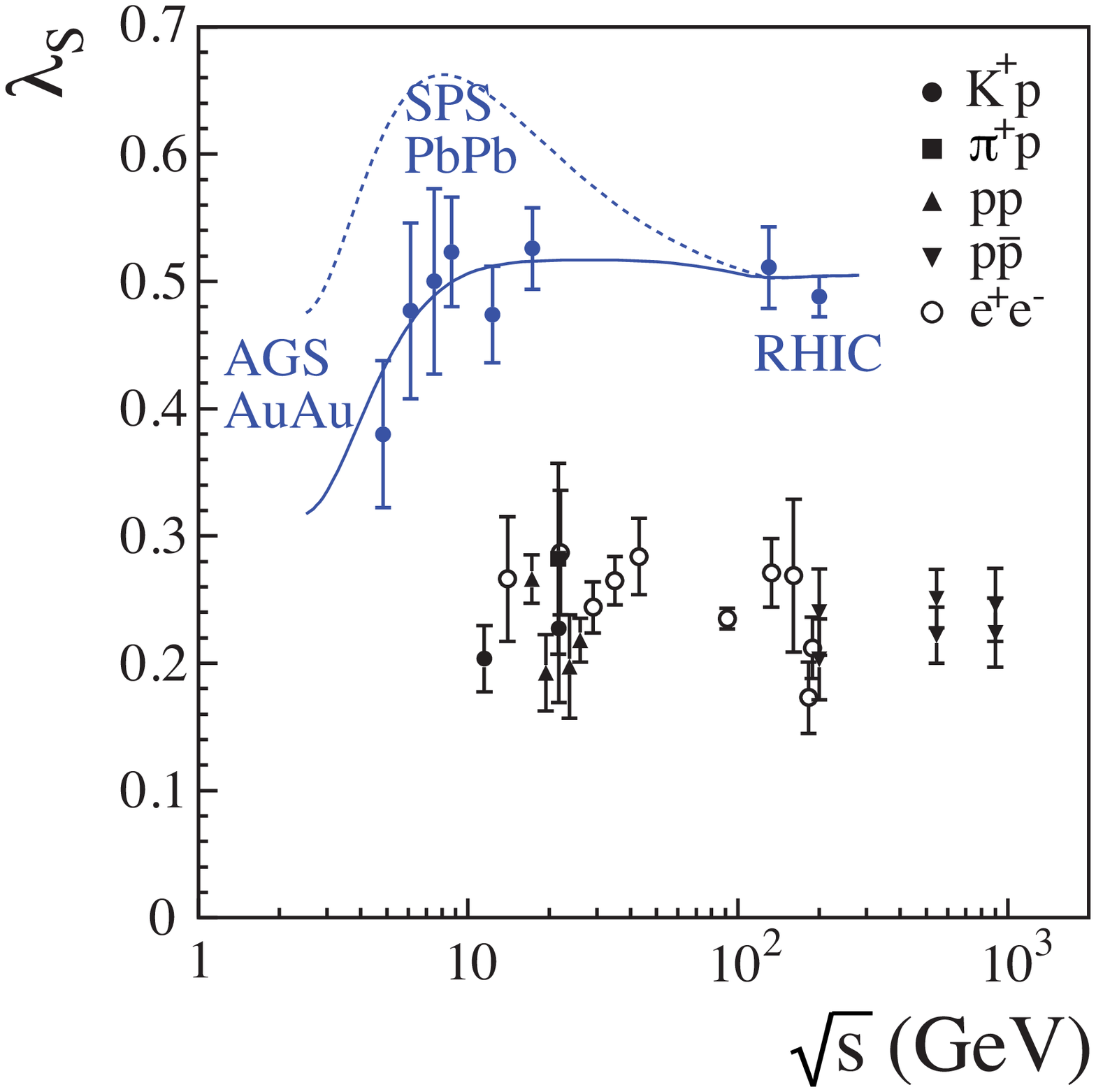}
\end{center}
\end{minipage}
\begin{minipage}[b]{0.49\linewidth}
\begin{center}
\includegraphics[width=1.18\linewidth]{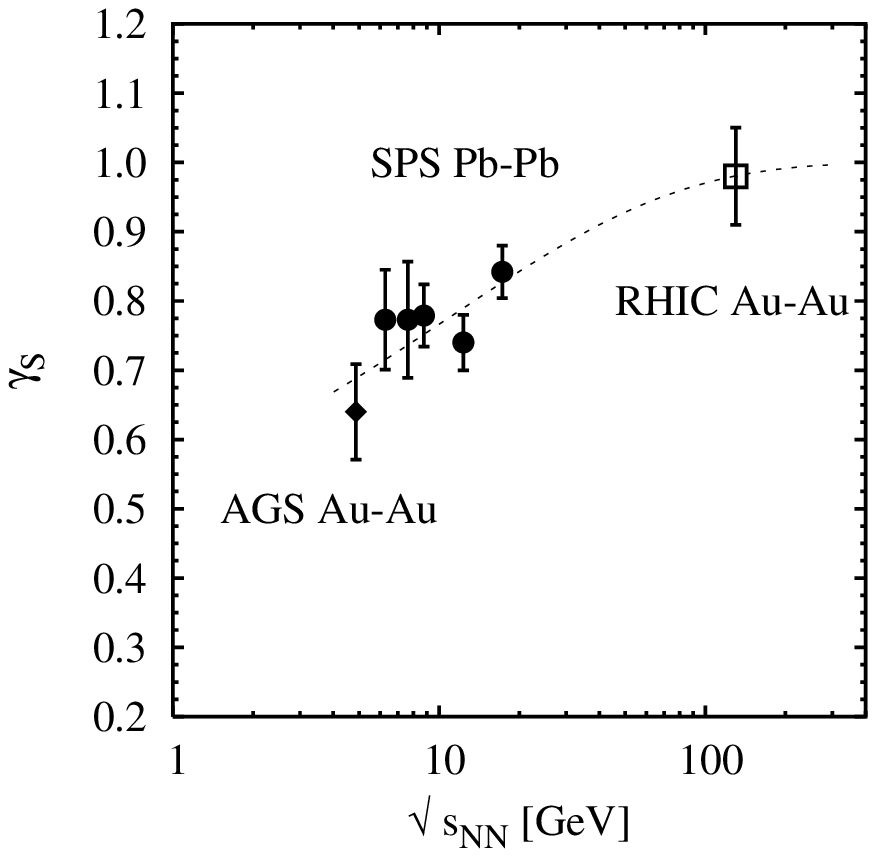}
\end{center}
\end{minipage}
\end{center}
\caption{Left: The energy dependence of the Wroblewski factor \lams\
as determined via fits of a statistical hadron gas model to
multiplicities measured in elementary (e$^{+}$e$^{-}$, \kplus p, \pip
p, pp, p\pbar) as well as central heavy ion collisions
\cite{BECATTINI2}.  The dashed line is a prediction for a fully
equilibrated hadron gas, the solid line an interpolation of fits
including an additional strangeness suppression factor \gams\
(Fig. is adapted from \cite{BECATTINI2}).
Right: The energy dependence of the strangeness undersaturation factor
\gams\ at chemical freeze-out \cite{BECATTINI7}.  The dashed line is a
phenomenological parametrization in the form $\gams = 1 - 0.606 \exp
\left\{ -0.0209 \,\sqrt{A\,\sqrts} \, \right\}$, where $A$ is the
atomic mass number, fulfilling $\gams \rightarrow 1$ for $\sqrts
\rightarrow \infty$ \cite{BECATTINI7}.}
\label{fig:Wroblewski_gams_free}
\end{figure}
%

%
\begin{figure}[th]
\begin{center}
\begin{minipage}[b]{0.49\linewidth}
\begin{center}
\includegraphics[width=\linewidth]{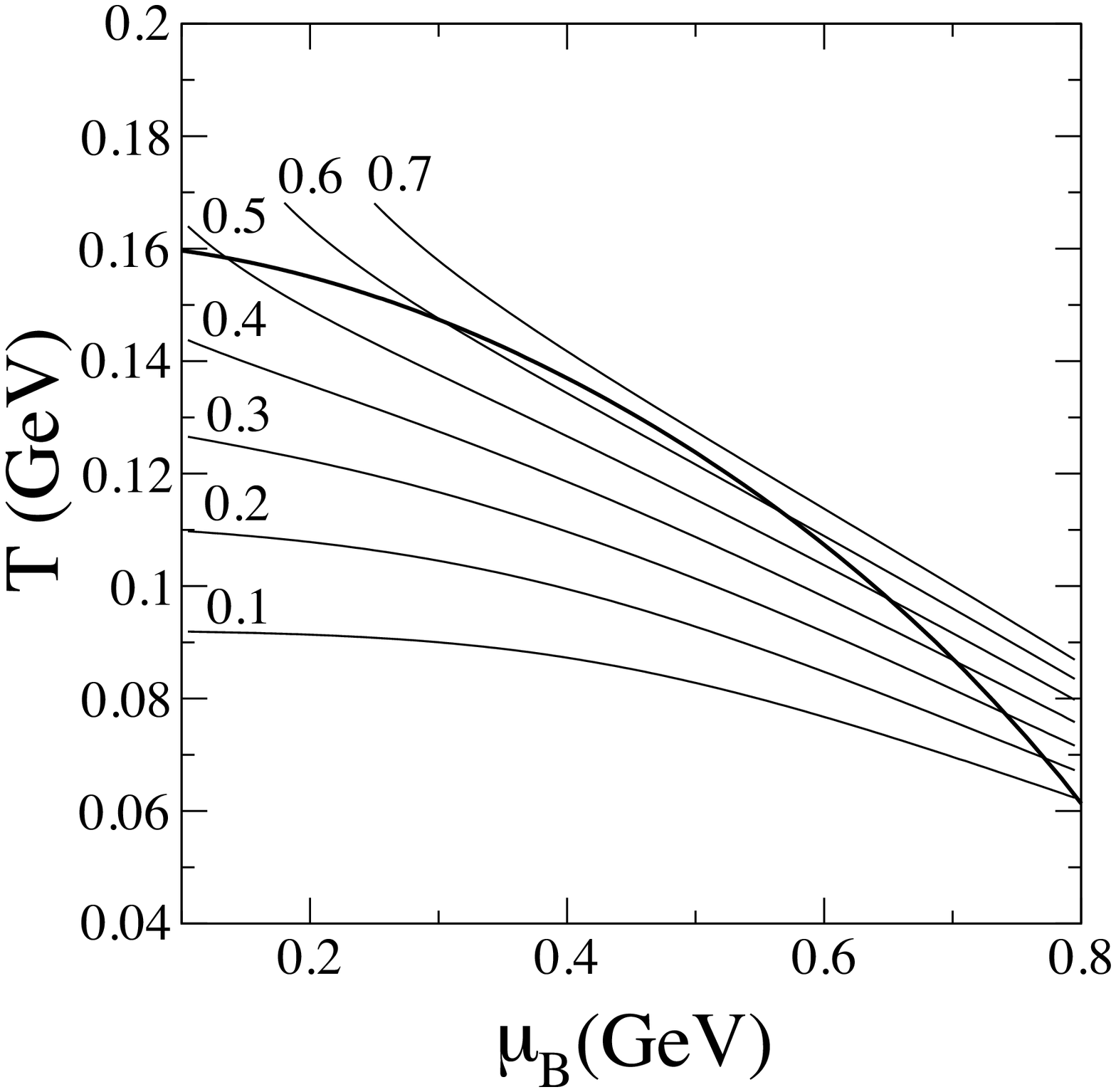}
\end{center}
\end{minipage}
\begin{minipage}[b]{0.49\linewidth}
\begin{center}
\includegraphics[width=1.14\linewidth]{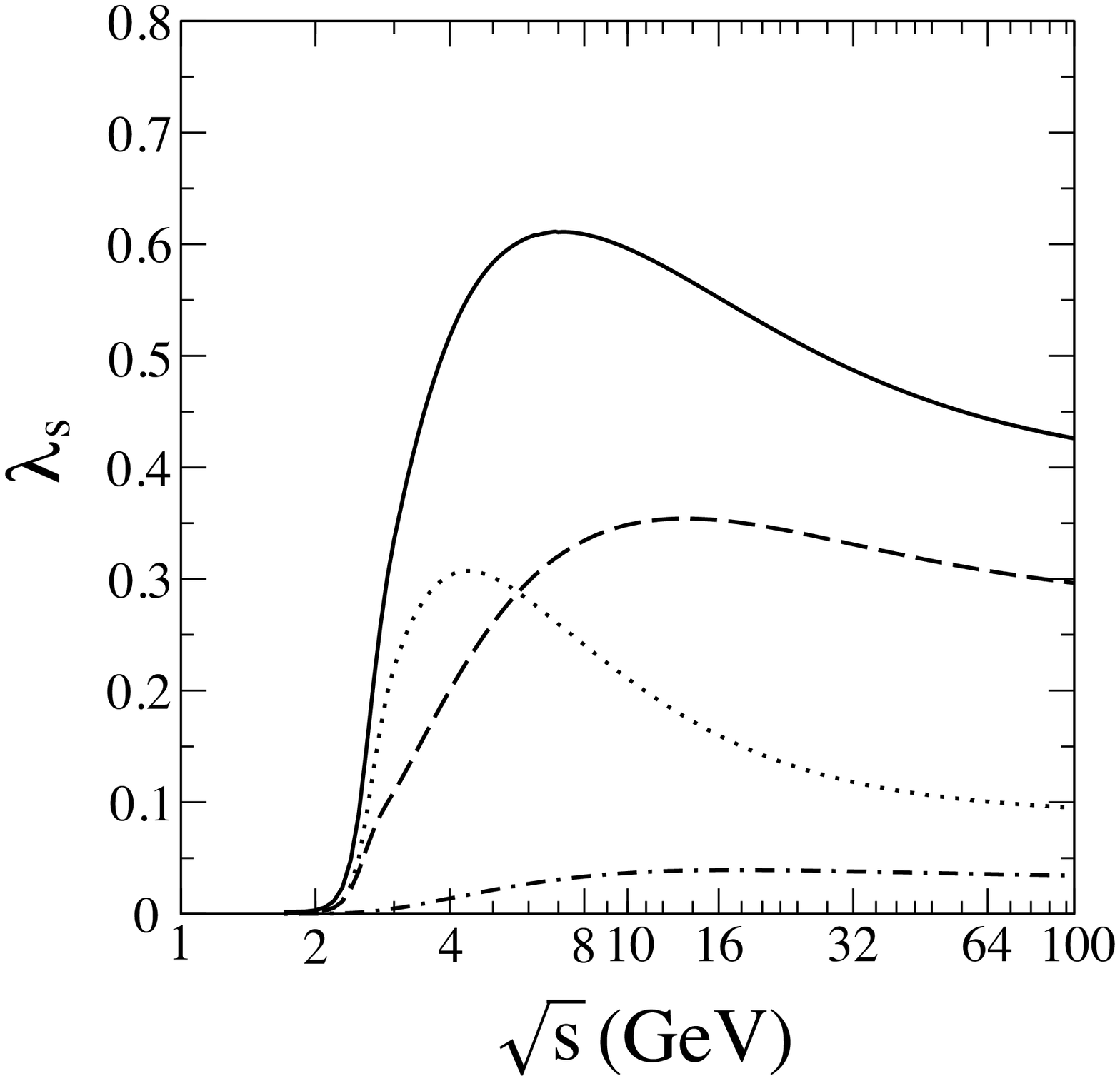}
\end{center}
\end{minipage}
\end{center}
\caption{Left: Lines of constant Wroblewski factor \lams\ in the $\tch
  - \mub$ plane.  The thick solid line without labeling represents the
chemical freeze-out curve defined by $\langle E \rangle/\langle N
\rangle \simeq 1$~GeV \cite{PBM2}.
Right: The contributions to the Wroblewski factor from strange baryons
(dotted line), strange mesons (dashed line), and with hidden
strangeness (dash-dotted line) in the full equilibrium case
(i.e. \gams~=~1) \cite{PBM2}.  The sum of all contributions is given
by the solid line. (Figs. are adapted from \cite{PBM2}.)}
\label{fig:Wroblewski_gams_fixed}
\end{figure}
%

In order to answer the question whether the observed strangeness
enhancement can really be caused by a fast equilibration in a partonic
phase, the investigation of its energy dependence provides important
information.  Following this argument one should observe an onset of
strangeness enhancement when going from low center-of-mass energies
toward higher ones.  At lower energies, where the energy densities
should be too low to form a deconfined state, any enhancement relative
to elementary collisions can only be due to secondary hadronic
interactions, which are, at least in the case of multi-strange
baryons, unlikely to drive the system toward chemical equilibrium.
It has been pointed out quite early that the enhancement for kaons is,
in contrast to naive expectations, rising toward lower energies
\cite{GAZDZICKIKAON,DUNLOP}.  A similar observation has been made for
\lam\ \cite{NA49EDEPHYP}, while for multi-strange particles the
current experimental situation still is unclear.  Data on
multi-strange particles for heavy ion collisions at lower energies are
scarce and have large errors, while p+p reference data are essentially
not existent.  While first attempts into this direction have been done
for heavy ion collisions \cite{E895LAMXI,HADESXI}, there is still a
lack of reference data from p+p or p+A collisions at the same
energies.  Threshold effects play an important role at very low
energies and complicate any interpretation.  Nevertheless, it is
interesting to observe that \xim\ measurements at very low energies
are close  (Au+Au at 6\agev, E895 collaboration \cite{E895LAMXI}), or
even above (Ar+KCl at 1.76\agev, HADES collaboration \cite{HADESXI}),
the expectation for an equilibrated hadron gas \cite{PBM2,ANDRONIC2}.

At higher energies, a clear trend in the energy dependence of
strangeness enhancement emerges.  The comparison of the hyperon
enhancement factors measured at top SPS energy (\sqrts~= 17.3~GeV) and
top RHIC energy (\sqrts =~200~GeV) shows that the enhancement for
baryons is actually higher at SPS.  As shown in the right panel of
\Fi{fig:NA49_enh}, the enhancement factor $E_{\rb{S}}$
(\Eq{eq:enhancefactor}) is larger for \xim\ and \omm\ at \sqrts~=
17.3~GeV than at \sqrts~= 200~GeV.  However, for \lam\ and mesons
(\kplus, \kmin, $\phi$) the enhancement is essentially the same at
both energies (left panel of \Fi{fig:NA49_enh}).  Preliminary data
by the NA57 collaboration indicate that in central Pb+Pb collisions at
\sqrts~= 8.7~GeV the \xim\ enhancement is again larger as compared to
p+Be collisions than at \sqrts~= 17.3~GeV \cite{NA57SANDOR}.  The main
reason for this behavior is that the strangeness production in p+p
(p+Be) collisions is rising faster with energy than in nucleus-nucleus
reactions.  For instance, the \xip/\pip\ ratio changes by a factor of
$\sim$~3.8 in p+p and of only $\sim$~2.9 in central A+A between
\sqrts~= 17.3~GeV and 200~GeV \cite{NA49EDEPHYP,NA49PA,STARSYST,
STARHYP200,NA49PPPION,STARPPHYP}.  This is even more pronounced for
the \xim/\pim\ ratio, since for central nucleus-nucleus collisions
this ratio is decreasing between the two energies by roughly 20\% (see
\Fi{fig:NA49_hyp_edep}), while at the same time it is rising in p+p
collisions by 25\%.  In the context of statistical models the
suppression of strangeness production is understood as a consequence
of the small available volume, requiring a canonical ensemble, which
results in a strong reduction of the available phase space for strange
particles.  Along with the center-of-mass energy the overall
multiplicity in p+p reaction increases, and thus the effective volume,
with the result of a less effective phase space suppression.
Proton-proton collisions therefore slowly approach the chemical
equilibrium values for large systems (which then can be described by a
grand-canonical ensemble), such as heavy ion collisions.

The \lab\ deviates from the general trend, because its enhancement
significantly increases with center-of-mass energy.  But in this
case the strong change of the net-baryon density  in A+A collisions
with energy and the consequently strongly changing anti-baryon/baryon
ratio is dominating the energy evolution of the \lab\ enhancement.
The sensitivity of a given anti-baryon to this effect depends on its
strangeness content.  As a consequence a clear hierarchy of the
anti-baryon/baryon ratio $R$ is visible, $R(\omp/\omm) > R(\xip/\xim)
> R(\lab/\lam) > R(\pbar/\textrm{p})$, and the energy dependence of
the ratios gets significantly weaker with increasing strangeness
content (see right panel of \Fi{fig:bbbarratios}).

Another way of looking at the energy dependence of strangeness
enhancement is studying the Wroblewski factor \lams\
\cite{BECATTINI3}.  It is defined as the ratio of the mean
multiplicities of the newly produced valence quark-antiquark
pairs \cite{WROBLEWSKI}:
\be
  \lams = \frac{2 \: \langle\textrm{s}\bar{\textrm{s}}\rangle}
               {     \langle\textrm{u}\bar{\textrm{u}}\rangle
                   + \langle\textrm{d}\bar{\textrm{d}}\rangle}
\ee
Since usually not all hadron multiplicities for a given reaction
system are measured, the determination of \lams\ has to rely on some
input from models.  In \Fi{fig:Wroblewski_gams_free} \lams\ has been
extracted from fits with a statistical hadron gas model for elementary
collisions, as well as for central heavy ion reactions
\cite{BECATTINI2}.  The \lams\ values determined for nucleus-nucleus
collisions are found to be a factor 2 higher than for elementary
collisions, even at the lowest SPS energy (\sqrts~= 6.3~GeV).  There
is an indication for a rise of \lams\ at lower energies (\sqrts~$<$
6~GeV), but from then on \lams\ stays at a constant value of $\approx
0.5$, well above the value of $\approx 0.25$ found for elementary
collisions.  This demonstrates again that strangeness production is
globally enhanced in heavy ion reactions, already at very low energies.
However, it remains to be seen whether this is not only true for the
global strangeness production, but as well for the rare multi-strange
baryons.

In the statistical model approach by Becattini et al. \cite{BECATTINI7}
the strangeness undersaturation factor \gams\ is a free parameter in
the fits to the particle yields.  The right panel of
\Fi{fig:Wroblewski_gams_free} shows the energy dependence of \gams\
for nucleus-nucleus collisions between AGS and RHIC energies.  A slow
increase with \sqrts\ from 0.7~-~0.8 at SPS toward a value of $\gams
= 1$ at RHIC is observed.  This would indicate that strangeness
production is not yet fully equilibrated at the SPS, even though it
is already very close to the full equilibrium case.

Under the full equilibrium condition (i.e. \gams~=~1) the statistical
model results in a well defined connection between a given chemical
freeze-out position in the $\tch - \mub$ plane and the Wroblewski
factor.  The left panel of \Fi{fig:Wroblewski_gams_fixed} shows the
lines of constant \lams\ in the $\tch - \mub$ plane \cite{PBM2}.
Based on the parametrization of the \sqrts\ dependence of \mub, as
e.g. given in \Eq{eq:mub_param}, and the  $\langle E \rangle/\langle N
\rangle \approx 1$~GeV freeze-out criterion (solid line in the left
panel of \Fi{fig:Wroblewski_gams_free}), which connects \tch\ to a
given \mub, a unique curve for the energy dependence of \lams\ can
thus be constructed \cite{PBM2}.  This curve is shown as solid line in
the right panel of \Fi{fig:Wroblewski_gams_fixed}  and is
characterized by a distinct maximum of relative strangeness production
around \sqrts\ of 6~-~8~GeV (corresponding to $\mub \approx 500$~MeV
and $\tch \approx 130$~MeV).  The dotted line in the left panel of
\Fi{fig:Wroblewski_gams_free} is based on a similar approach, but
uses different parameterizations of the energy dependence of the
chemical freeze-out parameters \cite{BECATTINI2}.  It follows from
these considerations that the relative contributions to \lams\ change
with energy.  While at lower energies, where \mub\ is large, the
produced strange quarks are to a large extent contained in strange
baryons (dotted line), the relative strangeness production at higher
energies ($\sqrts > 6$~GeV, corresponding to $\mub < 500$~MeV) is
dominated by strange mesons (dashed line).  The maximum in \lams\ at
lower energies is thus mainly due to the contribution of the strange
baryons.  Also strange mesons contribution exhibits a maximum, which
is, however, less pronounced and shifted toward higher energies.
This expected maximum of relative total strangeness production is
reflected in the energy dependence of various strange to non-strange
particle ratios, as discussed in the following sections.


\clearpage

%
\begin{figure}[th]
\begin{center}
\begin{minipage}[b]{0.47\linewidth}
\begin{center}
\includegraphics[width=\linewidth]{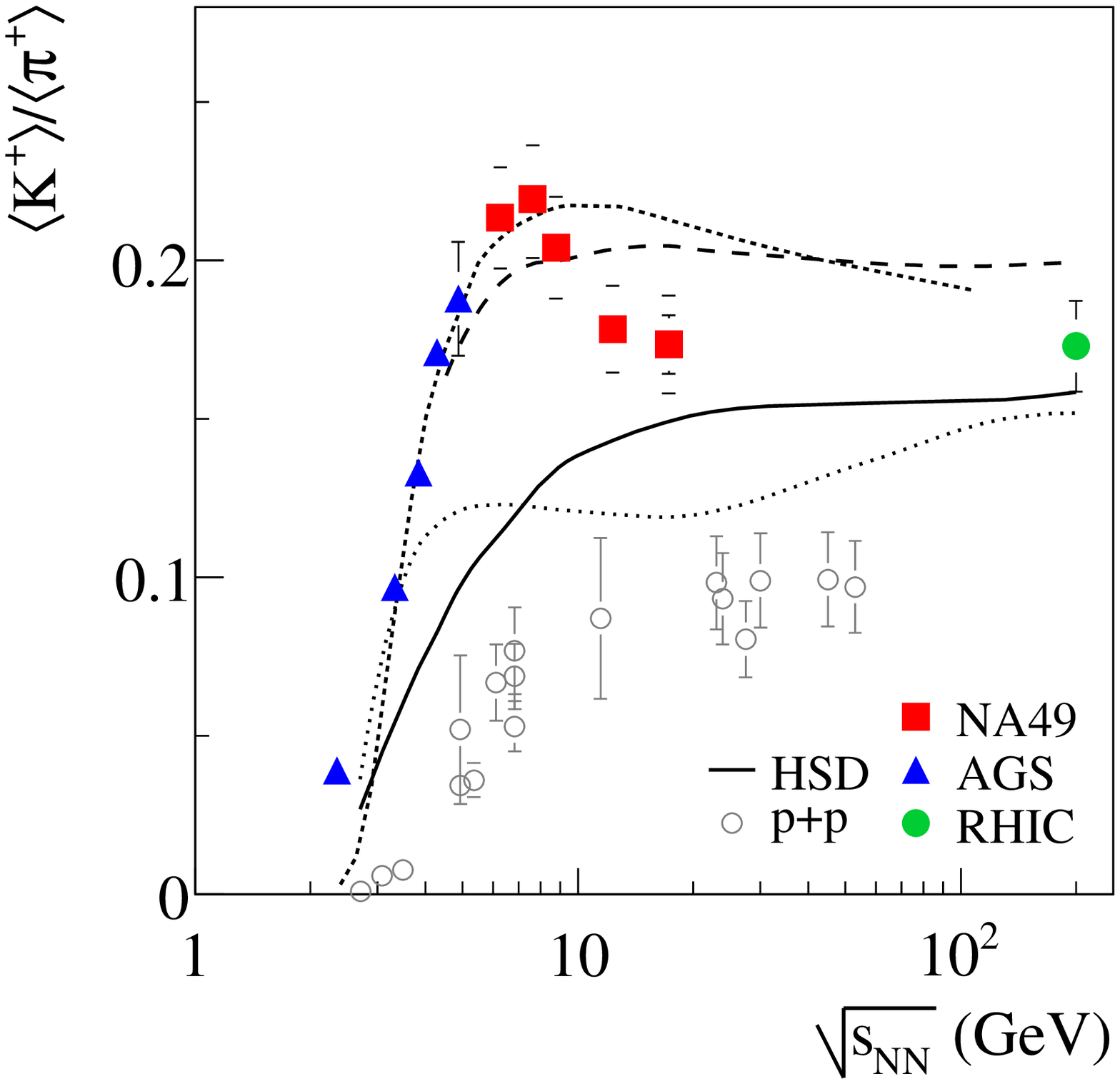}
\end{center}
\end{minipage}
\begin{minipage}[b]{0.47\linewidth}
\begin{center}
\includegraphics[width=\linewidth]{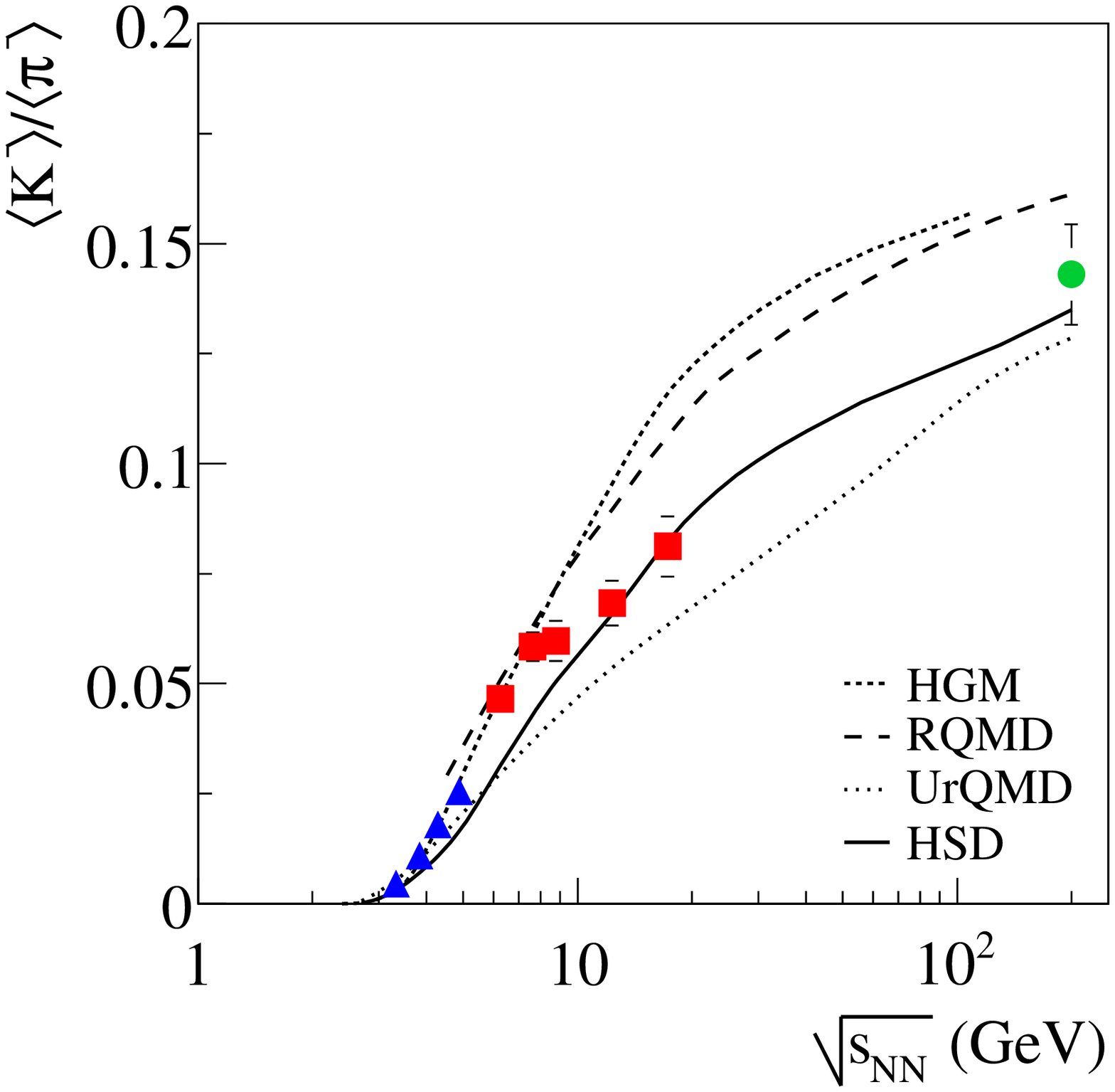}
\end{center}
\end{minipage}
\end{center}
\caption{The energy dependence of the \kpavg/\pipavg~ratio (left
panel) and the \kmavg/\pimavg~ratio (right panel) measured in central
Pb+Pb and Au+Au collisions (solid symbols) \cite{NA49KPI40158,
NA49KPI2030,E802PI116,E802KPM116,E802SDEP111,E866E917KP,E866E917KPM,
E877PI116,E895PION,BRMSRAP200} compared to results from
p+p($\bar{\textrm{p}}$) reactions (open symbols) \cite{GAZDZICKIKAON,
GAZDZICKIPION} and from the transport models HSD (solid line)
\cite{HSD1,HSD2}, UrQMD (dotted line) \cite{URQMD1}, and RQMD (long
dashed line) \cite{SORGERQMD} and from a statistical hadron gas model
(short dashed line) \cite{CLEYMANS1,PBM2}.}
\label{fig:NA49_kp_edep}
\end{figure}
%

\subsection{Structures in the energy dependence of particle yields}
\label{sec:edepstruct}

Apart from the search for an onset of strangeness enhancement, the
measurement of strange particle abundances in heavy ion collisions at
different center-of-mass energies revealed new insights into the
physics of strangeness production.  Data from several experiments at
the AGS, SPS, and RHIC allow to compile a comprehensive overview for
central A+A collisions.  In the following, only some of the most
interesting features seen in the energy dependences shall be
discussed.  A compilation of the \sqrts\ dependences of the
midrapidity yields of all strange particles is given in section
\ref{sec:compedep} (see \Fis{fig:dndy_vs_sqrt_meson}
{fig:dndy_vs_sqrt_baryon} and \Tar{tab:comp_kp}{tab:comp_om}).

\subsubsection{The K/$\pi$ ratios}
\label{sec:kpiratios}

%
\begin{figure}[th]
\begin{center}
\begin{minipage}[b]{0.38\linewidth}
\begin{center}
\includegraphics[width=\linewidth]{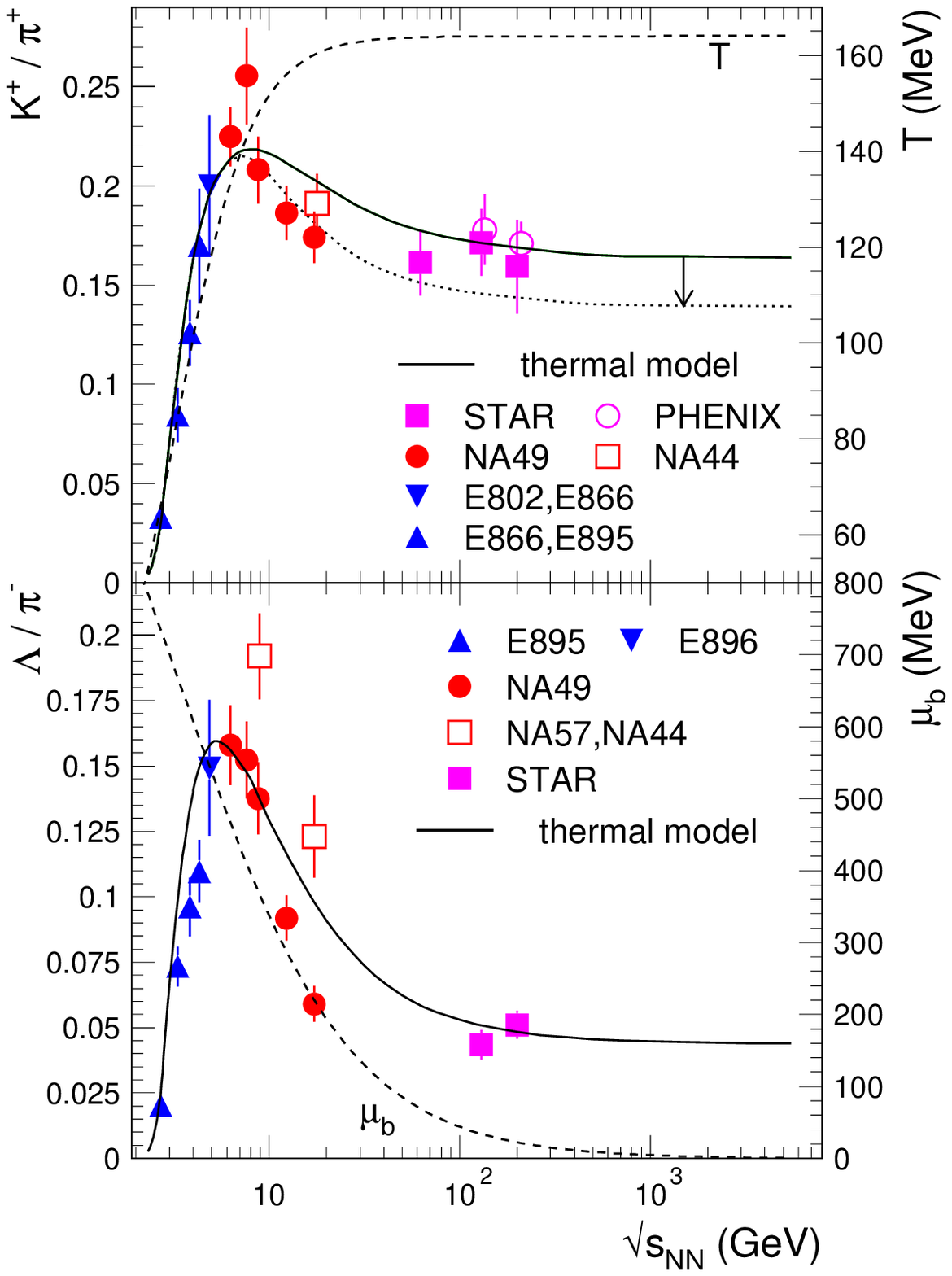}
\end{center}
\end{minipage}
\begin{minipage}[b]{0.60\linewidth}
\begin{center}
\includegraphics[width=\linewidth]{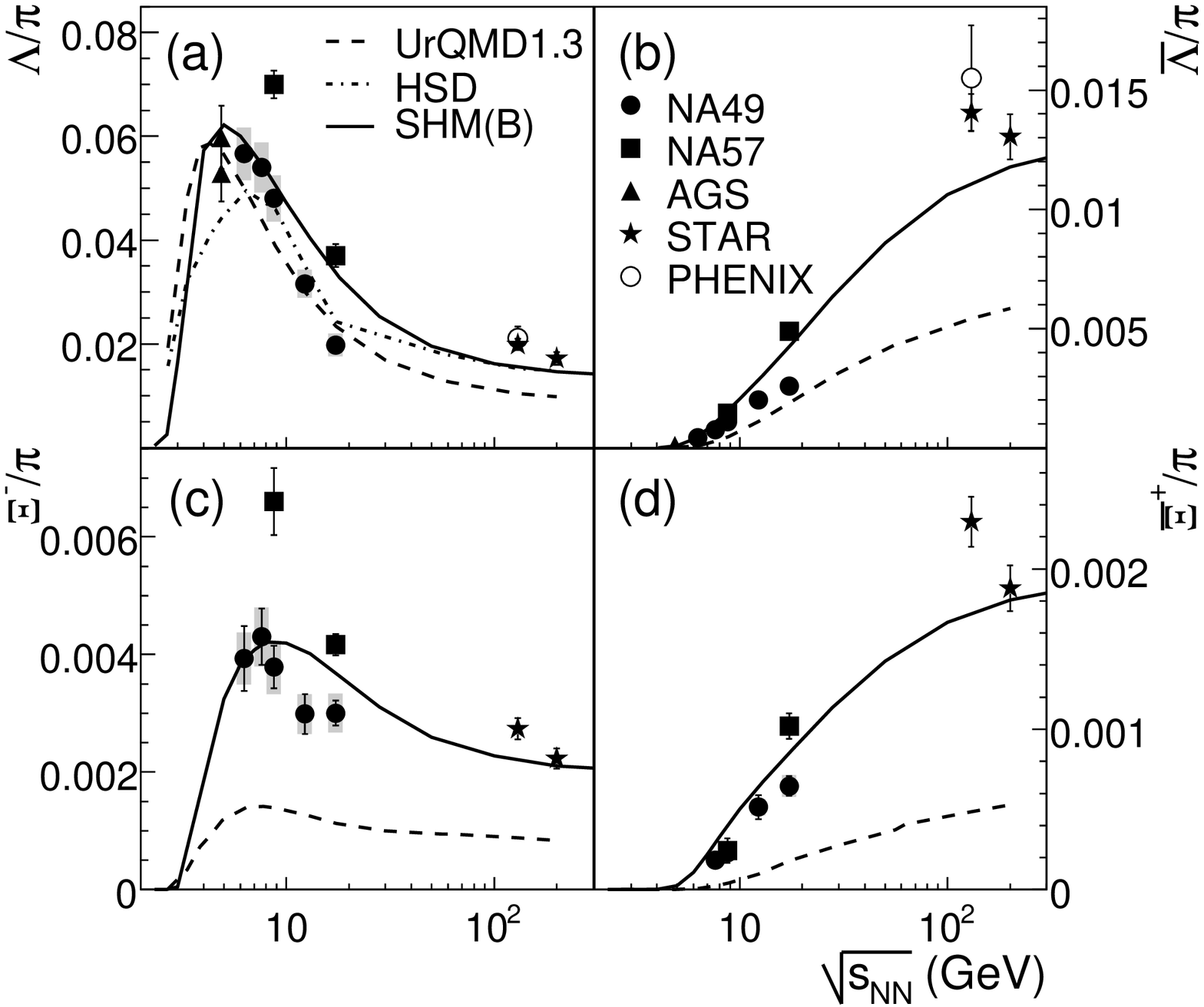}
\end{center}
\end{minipage}
\end{center}
\caption{Left: The energy dependence of the rapidity densities \dndy\
around midrapidity of \kplus\ and \lam, divided by the ones of \pip,
respectively \pim, for central Pb+Pb and Au+Au collisions.  The solid
line is the result of a statistical model calculation.  The dotted
line gives the \kplus/\pip~ratio including the additional effect of
higher mass resonances.  The dashed lines show the energy dependence
of the temperature \tch\ (upper panel) and of the baryonic chemical
potential \mub\ (lower panel) \cite{ANDRONIC3}.  Right: The energy
dependence of the rapidity densities \dndy\ around midrapidity of
\lam, \lab, \xim, and \xip\ divided by the total pion rapidity
densities (\mbox{$\pi = 1.5 \:(\pim + \pip)$}) for central Pb+Pb and
Au+Au collisions \cite{NA49EDEPHYP,NA57ENHANCE,STARKPI130,STARHYP130,
STARHYP200,PHNXLAM130,NA57EDEPHYP,E896LAM,E917LAB,E891LAM,STARLAM130,
STARPR200,E802PI116,PHNXKPI130}.  Please note that the vertical error
bars correspond to the statistical errors only.  Also shown are
results for the transport models UrQMD (dashed line) \cite{URQMD1} and
HSD (dash-dotted line) \cite{HSD1,HSD2}, as well as a statistical
hadron gas model (solid line) \cite{ANDRONIC1}.}
\label{fig:NA49_hyp_edep}
\end{figure}
%

%
\begin{figure}[t]
\begin{center}
\begin{minipage}[b]{0.47\linewidth}
\begin{center}
\includegraphics[width=\linewidth]{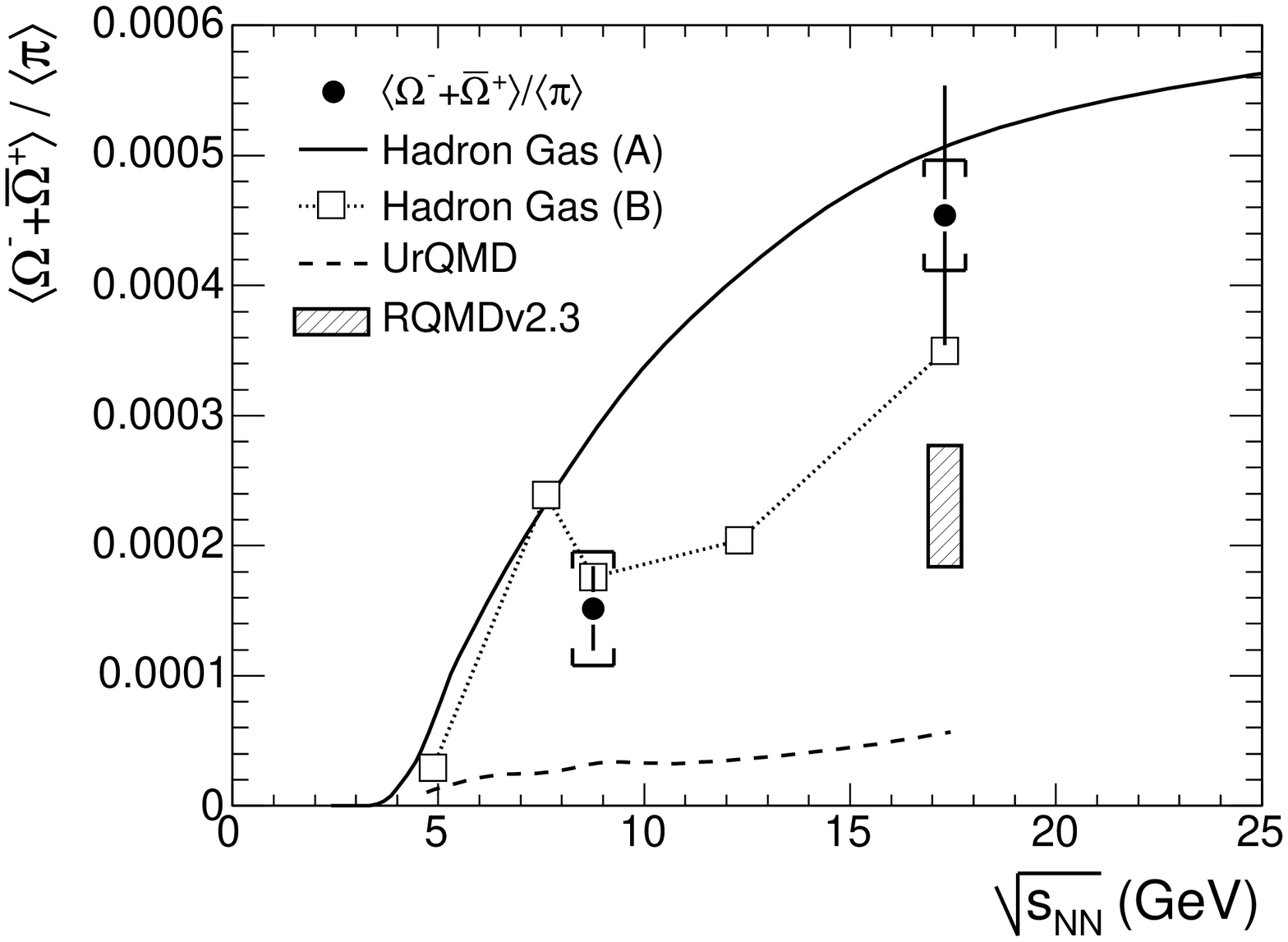}
\end{center}
\vspace{0.25pt}
\end{minipage}
\begin{minipage}[b]{0.43\linewidth}
\begin{center}
\includegraphics[width=\linewidth]{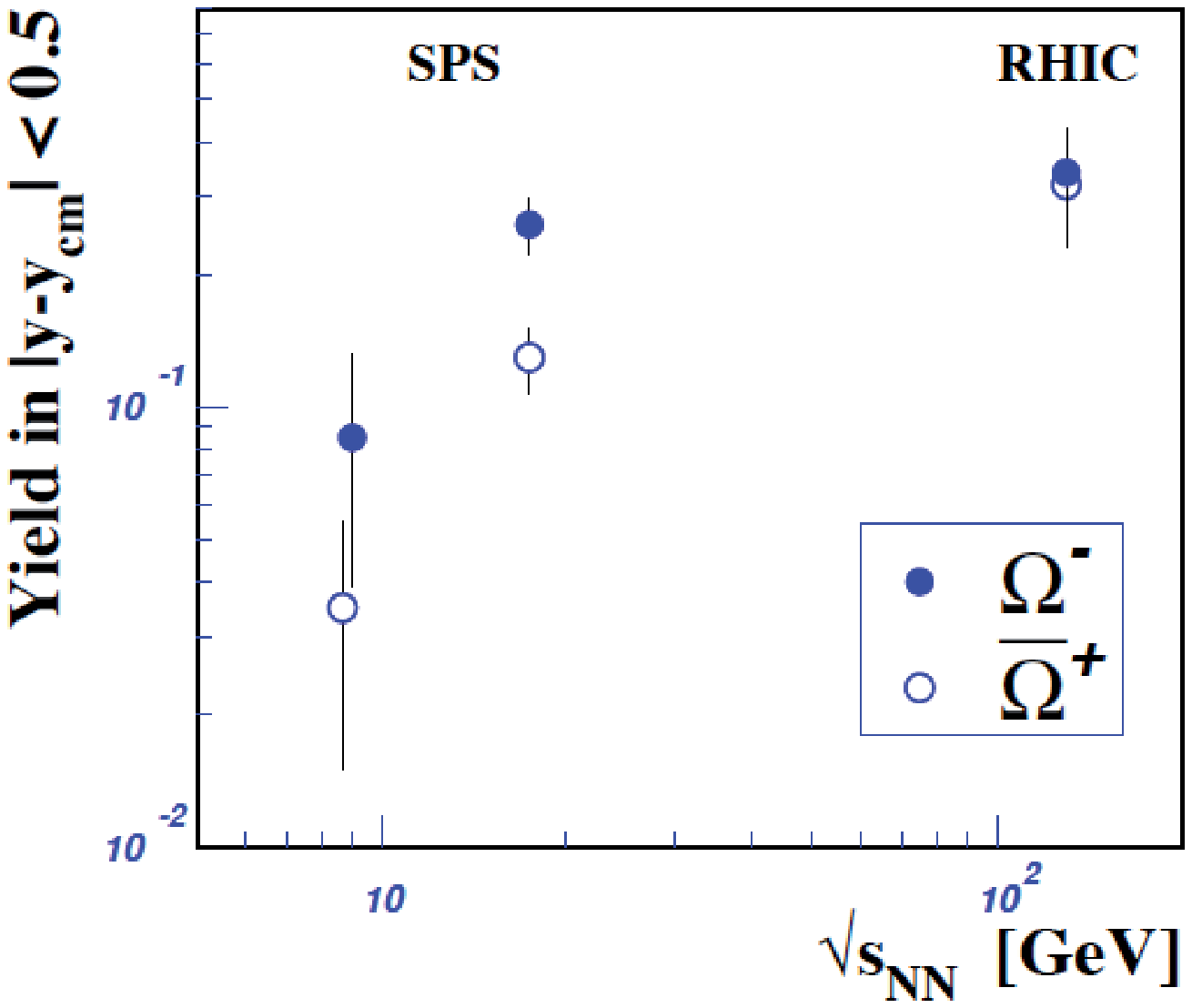}
\end{center}
\end{minipage}
\end{center}
\caption{Left: The energy dependence of the \omavg/\piavg~ratio in
central Pb+Pb collisions \cite{NA49EDEPOM}.  Also shown are results
for the transport models UrQMD (dashed line) \cite{URQMD1} and RQMD
(hashed box) \cite{HEINZRQMD}, as well as of statistical hadron gas
models, one with strangeness undersaturation factor \gams\ (dotted
line with empty boxes) \cite{BECATTINI1} and one without (solid line)
\cite{PBM2}.  Right: the energy dependence of the \omm\ and \omp\
rapidity densities \dndy\ around midrapidity for central Pb+Pb
\cite{NA57EDEPHYP} and Au+Au \cite{STARHYP130} collisions.}
\label{fig:NA49_NA57_om_edep}
\end{figure}
%

%
\begin{figure}[t]
\begin{center}
\includegraphics[width=0.40\linewidth]{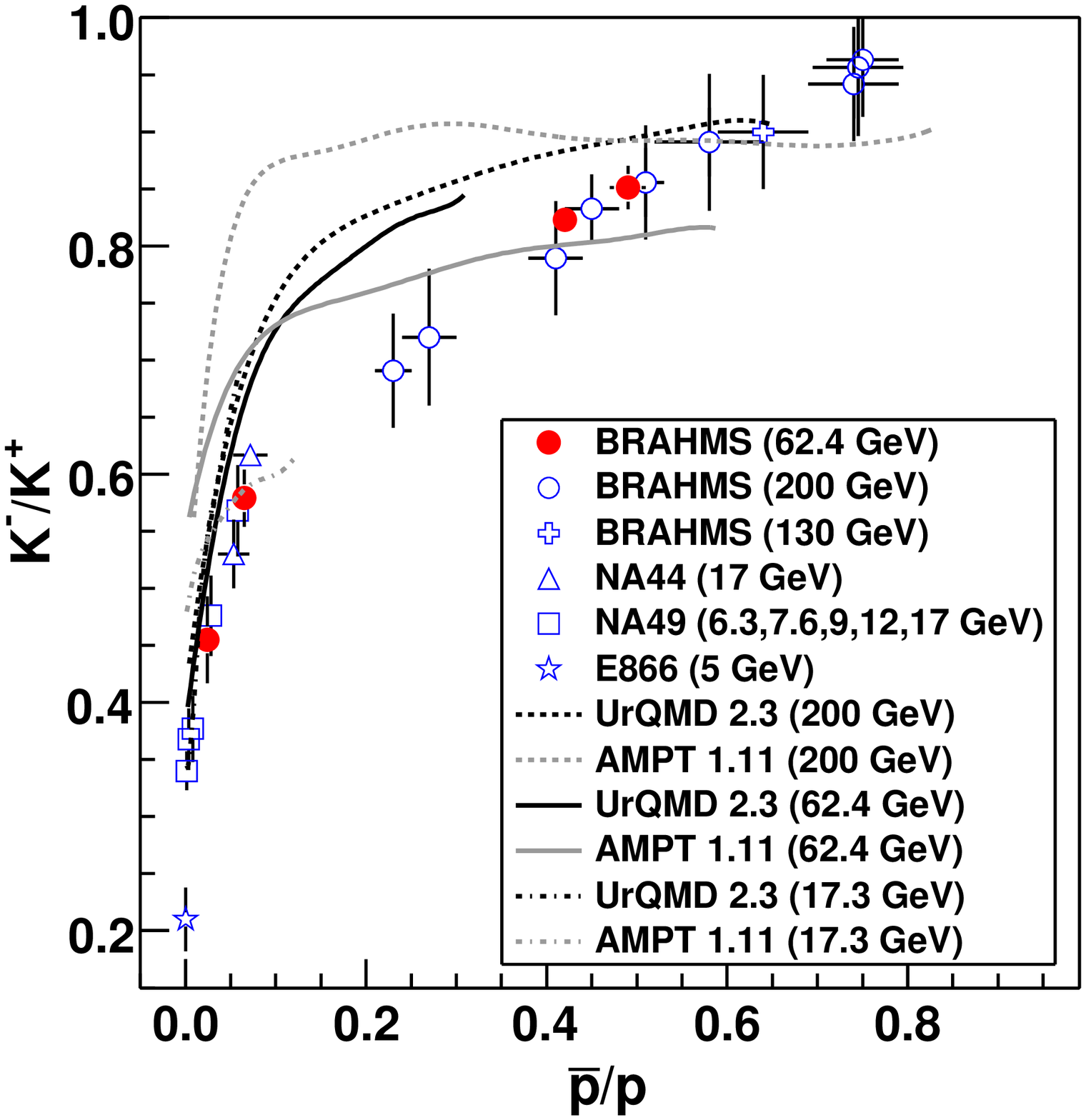}
\end{center}
\caption{The \kmin/\kplus\ ratio as a function of the \pbar/p ratio
\cite{BRMSRAP62}.  The data are taken from several experiments
\cite{NA49KPI40158,NA49KPI2030,NA49PPBAR,E802PPBAR,BRMSRAP200,
BRMSRAP62,NA44B} and are measured for central collisions at different \sqrts\
and rapidities.  Also shown are various model predictions
\cite{URQMD1,URQMD2,AMPT1,AMPT2}.}
\label{fig:BRAHMS_kmkp_ppbar}
\end{figure}
%

One of the most prominent observations in this context is a pronounced
maximum in the energy dependence of the \kplus/\pip~ratio around
\sqrts~=~7~GeV \cite{NA49KPI2030}, as shown in the left panel of
\Fi{fig:NA49_kp_edep}.  The \kmin/\pim~ratio, in contrast, exhibits
a rather smooth energy dependence with only a slight indication for
a kink at the position of the maximum in the \kplus/\pip~ratio (right
panel of \Fi{fig:NA49_kp_edep}).  Transport models like RQMD
\cite{SORGERQMD}, UrQMD \cite{URQMD1}, and HSD \cite{HSD1,HSD2,HSD3}
are generally not able to describe the structures seen in the data.
For instance, they just predict localized turnover in the energy
dependence of the \kplus/\pip~ratio, but not the observed maximum.
Similar discrepancies are present for the \kmin/\pim~ratio, although
they are not so visible in this case, due to the less pronounced
structure in its energy dependence.

On the other hand, a structure like the observed one has been
predicted in \cite{GAZDZICKIREV} for the energy dependence of the
strangeness to entropy ratio as a consequence of the onset of
deconfinement.  This prediction is based on the statistical model of
the early stage \cite{GAZDZICKISMES}, which is an extension of Fermi's
statistical model of particle production that takes into account two
different states of matter, a confined (hadron gas) and a deconfined
(QGP) state.  The initial state is described by Fermi-Landau
conditions, which allows to define the energy density at a given
center-of-mass energy.  At a critical temperature \tc\ a first order
phase transition is assumed.  This temperature corresponds via the
energy density to a certain center-of-mass energy and is adjusted such
that the transition occurs around $\sqrts \approx 7 - 8$~GeV.  The
model assumes full statistical equilibrium in the initial stage of the
reaction.  Strangeness and entropy are conserved throughout the
evolution of the system.  Once the transition energy is passed, a
change in the energy dependence of the strangeness to entropy ratio is
expected due to the change of mass of the strangeness carriers and of
the available number of degrees of freedom.  At low energies
strangeness is confined in hadrons with high mass, leading to a steep
energy dependence of the strangeness to entropy ratio.  At high
energies, above \tc, the mass of the strangeness carriers is assumed
to be reduced to the mass of the strange quarks, resulting in an
almost flat energy dependence of this ratio.  The charged pion
multiplicity is assumed to be proportional to the entropy production.
Since the \kplus\ are carrying almost half of the produced
anti-strange quarks\footnote{The other half is mainly contained in
  K$^{0}$, due to isospin symmetry, while strange anti-baryons do not
  contribute significantly at lower energies.} and are thus directly
sensitive to the total (anti-)strangeness production, the
\kplus/\pip~ratio should exhibit the same behavior as the total
strangeness to entropy ratio.  Even though this interpretation is
highly debated and relies on simplified assumptions (e.g. ideal gas
equations), the structure in the \kplus/\pip~ratio might nevertheless
be indicative for a sudden change in the nature of the reaction system
produced in this energy region.

The statistical hadron gas approach, using a parametrized \sqrts\
dependence of the chemical freeze-out temperature \tch\ and the
baryonic chemical potential \mub\ (see \Eq{eq:mub_param}), as based on
fits to the measured particle ratios \cite{CLEYMANS1,PBM2}, also
results in a maximum of the \kplus/\pip~ratio at approximately the
right position, although its width is clearly wider than the data (see
short dashed curve in left panel of \Fi{fig:NA49_kp_edep}).  However,
a recent extension of this model that includes the scalar $\sigma$
meson and high mass resonances ($m > 2$~GeV) \cite{ANDRONIC3},
following the early ideas of Hagedorn \cite{HAGEDORN1,HAGEDORN2},
leads to an improved description of the data (left panel of
\Fi{fig:NA49_hyp_edep}).  In \cite{ANDRONIC3} it has therefore been
argued that the special form of the energy dependence of the
\kplus/\pip~ratio would be a consequence of reaching the limiting
temperature predicted by Hagedorn \cite{HAGEDORN3} and therefore
indirectly indicate the presence of a quark-gluon plasma phase.  

\subsubsection{Other particle ratios}
\label{sec:baryonratios}

Similar maxima as in the case of the \kplus/\pip~ratio, although not
as sharp, can be observed in the ratios of strange baryons to pions
(panel (a) and (c) of \Fi{fig:NA49_hyp_edep}) \cite{NA49EDEPHYP}.
While the \lam/$\pi$~ratio is relatively well described by transport
models, the \xim/$\pi$~ratio is clearly under-predicted by the HSD
model \cite{HSD1,HSD2}.  Generally, hadronic transport models are not
able to match the measured abundances of multi-strange (anti-)baryons
(see \Fi{fig:NA49_hyp_edep}(d) and left panel of
\Fi{fig:NA49_NA57_om_edep}).  The primary production mechanism of
string fragmentation does not favor the generation of particles with
multiple strangeness without going to unrealistic parameter settings
(see discussion in \Se{sec:transport}) and subsequent rescattering
processes add only little to the final multiplicities.  Multi-meson
fusion processes might help to approach the data, but the currently
available transport codes do not yet include this feature.

Statistical hadron gas models generally provide a relatively good
description of multi-strange particles\footnote{Please note that
  concerning the hyperon multiplicities there are still remaining
  discrepancies between the experiments NA49 and NA57 at the SPS
  (\Fis{fig:NA49_hyp_edep}{fig:dndy_vs_sqrt_baryon}), which exceed the
  quoted systematic errors. These are present for all particles at
  158\agev\ and for \lam\ and \xim\ at 40\agev.  Possible reasons
  might be difficulties in correcting trigger inefficiencies (NA57) or
  an insufficient correction of the multiplicity dependent
  reconstruction efficiency (NA49).  But despite extensive discussion
  between the collaborations the origin of these discrepancies were
  never finally resolved.}
\cite{BECATTINI1,ANDRONIC1}, including the $\Omega$ (left panel of
\Fi{fig:NA49_NA57_om_edep}).  So far, all measurements of strange
particle multiplicities in heavy ion reactions are in reasonable
agreement with the expectation for an equilibrated hadron gas at all
energies, although in some implementations of this model, which use
4$\pi$ integrated multiplicities as input, the additional strangeness
undersaturation parameter \gams\ is needed to provide a proper fit to
the data everywhere \cite{BECATTINI1} (see discussion in
\Se{sec:statmodels}).  There is no indication yet for any significant
departure from the chemical equilibrium values at lower energies
\cite{E895LAMXI,HADESXI}.

In the statistical model approach the maxima of the \lam/$\pi$,
\xim/$\pi$, and \kplus/$\pi$ ratios are due to the specific dependence
of the baryonic chemical potential \mub\ on the center-of-mass energy.
This results in a maximum of relative strangeness production around
\sqrts~= 7~GeV \cite{PBM2}.  The right panel of
\Fi{fig:Wroblewski_gams_fixed} shows how, according to this model, the
Wroblewski factor \lams\ changes with energy.  Especially the
contribution from strange baryons exhibits a distinct maximum around
\sqrts~= 4~GeV, which in the data is most clearly visible in the
\lam/$\pi$~ratio (\Fi{fig:NA49_hyp_edep}(a)).  The position of the
maximum is expected to shift toward higher energies with increasing
strangeness content of the baryon, and the maximum should be less
pronounced for the \xim\ and the \omm\ \cite{CLEYMANS2}.  The
observed peak in the \xim/$\pi$~ratio seems to be smaller in the NA49
measurement \cite{NA49EDEPHYP}, while the NA57 data suggest a stronger
energy dependence \cite{NA57EDEPHYP} (see \Fi{fig:NA49_hyp_edep}(c)).
For the $\Omega$ there is no evidence for a maximum, but the current
data are limited in significance and coverage (see
\Fi{fig:NA49_NA57_om_edep}).

Figure~\ref{fig:BRAHMS_kmkp_ppbar} shows the \kmin/\kplus\ ratio as a
function of the \pbar/p ratio \cite{BRMSRAP62}.  The data points
measured at RHIC energies by the BRAHMS collaboration (\sqrts~=
62.4, 130, and 200~GeV) have been measured at different rapidities,
while the SPS and AGS data correspond to midrapidity.  Generally, a
universal dependence of the \kmin/\kplus\ on the \pbar/p ratio is
observed.  Also included are predictions from two different transport
models (UrQMD \cite{URQMD1,URQMD2} and AMPT \cite{AMPT1,AMPT2}).
Even though they roughly follow the trend, they do not result in the
universality observed in the data.  In the statistical model approach,
on the other side, the \pbar/p ratio is defined by the baryo-chemical
potential \mub\ and the chemical freeze-out temperature \tch\ as
\cite{PBM5}
\be
\frac{\pbar}{\textrm{p}} = \exp\left(\frac{-2 \mub}{\tch}\right).
\ee
Thus, the \pbar/p ratio, measured in the given rapidity range, should
provide a good estimate of the local \mub.  Under the assumption that
the freeze-out temperature \tch\ is constant, the \pbar/p ratio would
therefore fully determine the chemical composition of the system.  A
fireball observed at high rapidities and high energies (e.g. \sqrts~=
200~GeV) would be equivalent to one at midrapidity and lower energies
(e.g. \sqrts~= 17.3~GeV), and the measured particle ratios should be
the same, as shown for the \kmin/\kplus\ ratio in
\Fi{fig:BRAHMS_kmkp_ppbar}.  A detailed analysis \cite{BECATTINI8}
shows that both parameters, \mub\ and \tch, depend on rapidity, but can
be related via the freeze-out condition $\langle E \rangle/\langle N
\rangle \approx 1$~GeV \cite{CLEYMANS1}.  However, the dependence of
\mub\ on rapidity is found to be stronger at SPS energies than at
RHIC.


\subsubsection{The $ \phi$ Meson}
\label{sec:phimeson}

%
\begin{figure}[t]
\begin{center}
\begin{minipage}[b]{0.42\linewidth}
\begin{center}
\includegraphics[width=\linewidth]{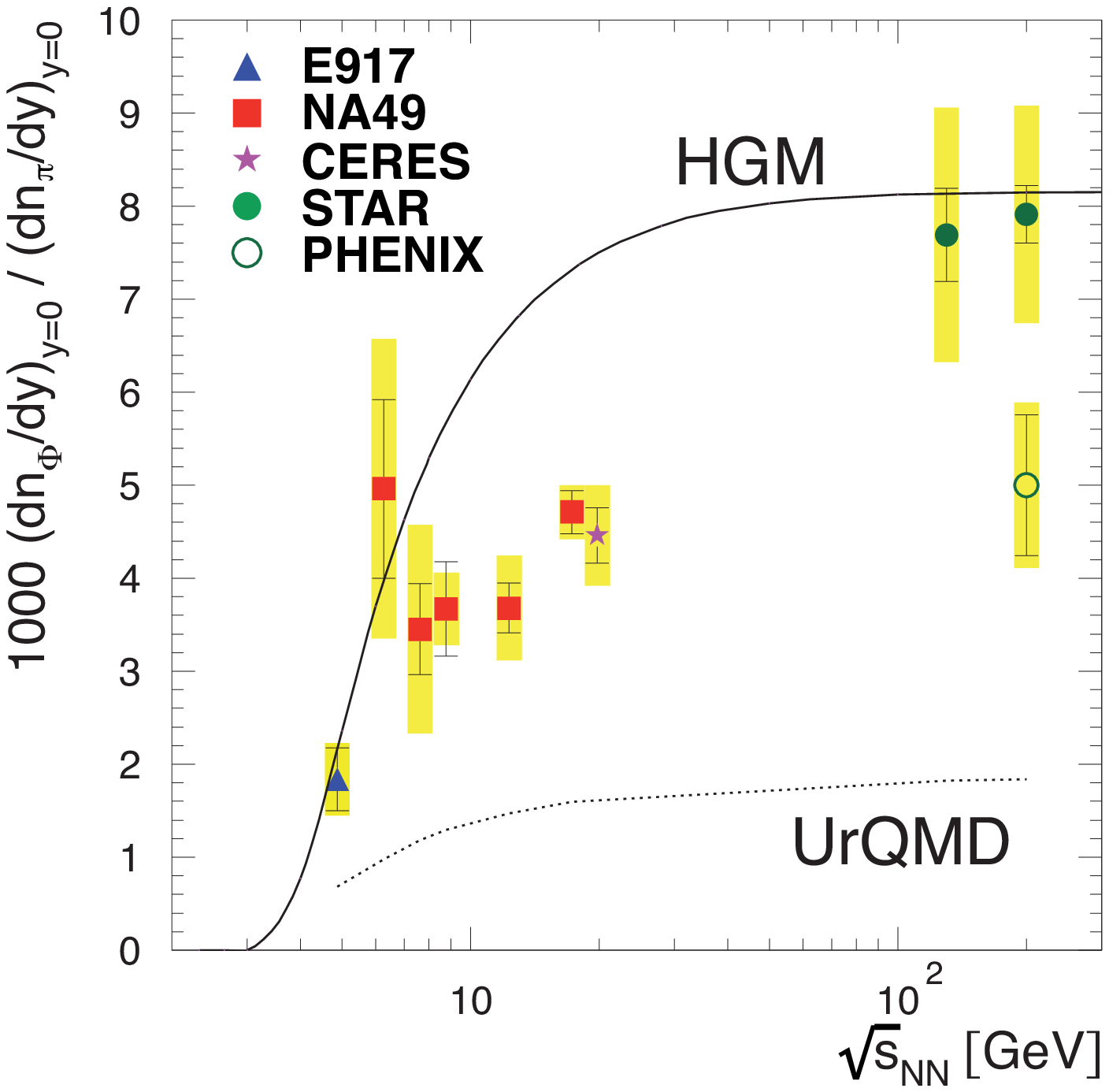}
\end{center}
\end{minipage}
\begin{minipage}[b]{0.44\linewidth}
\begin{center}
\includegraphics[width=\linewidth]{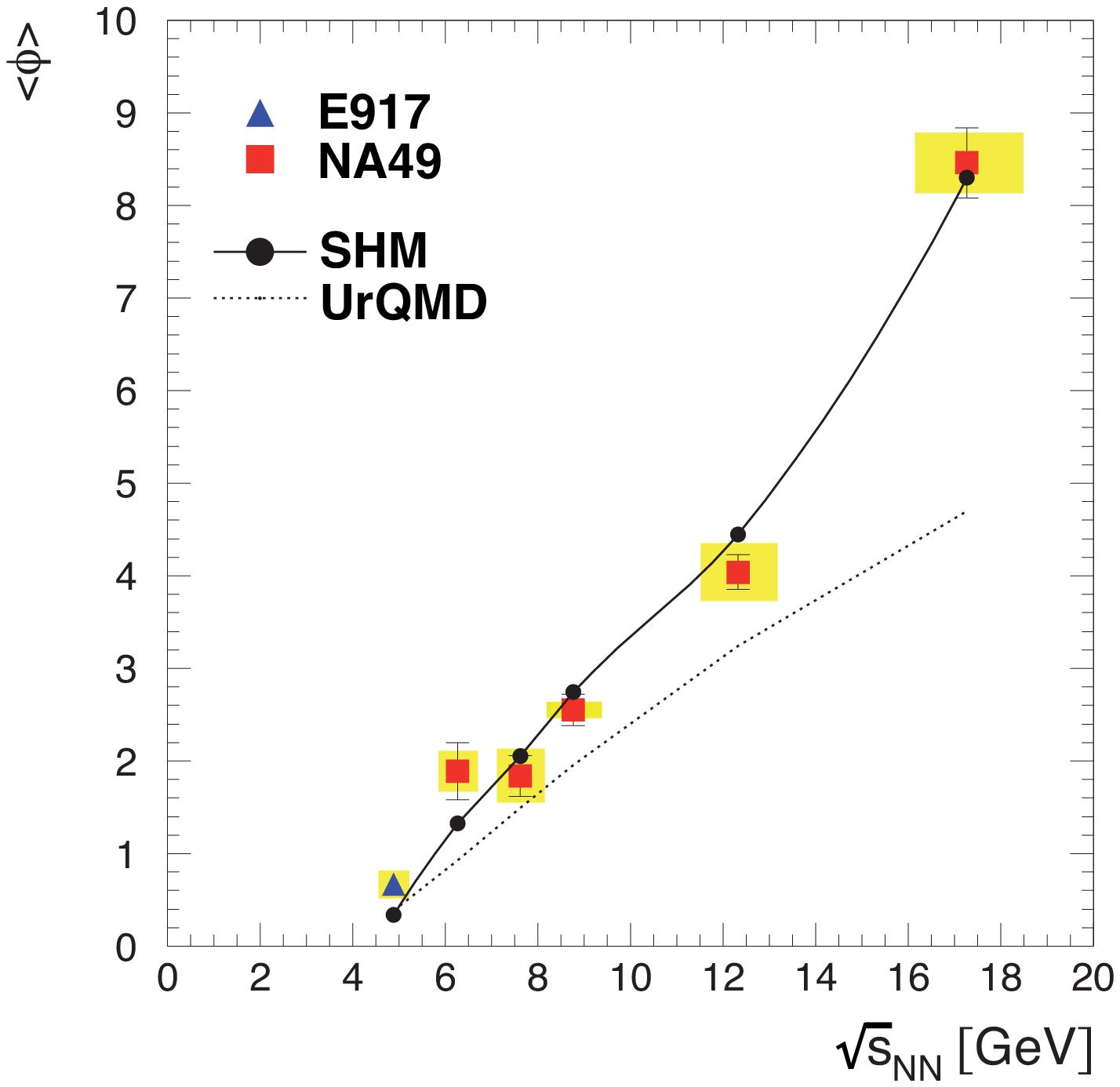}
\end{center}
\end{minipage}
\end{center}
\caption{Left: The $\phi$/$\pi$ ratio at midrapidity in central
nucleus-nucleus collisions as a function of the center-of-mass energy
\cite{NA49EDEPPHI}.  Also shown are the predictions of a statistical
hadron gas model \cite{ANDRONIC1} and of the transport model UrQMD1.3
\cite{URQMD1}.
Right: The total $\phi$ multiplicity in central nucleus-nucleus
collisions as a function of the center-of-mass energy
\cite{NA49EDEPPHI}.  The filled circles, connected by a solid line,
represent results from a fit with a statistical hadron gas model that
includes a strangeness undersaturation factor \gams\
\cite{BECATTINI7}.  The dotted curve shows the prediction of the UrQMD
model \cite{URQMD1}.  (The Figs. are adapted from \cite{NA49EDEPPHI}.)}
\label{fig:NA49_phi_edep}
\end{figure}
%

%
\begin{figure}[t]
\begin{center}
\begin{minipage}[b]{0.39\linewidth}
\begin{center}
\includegraphics[width=\linewidth]{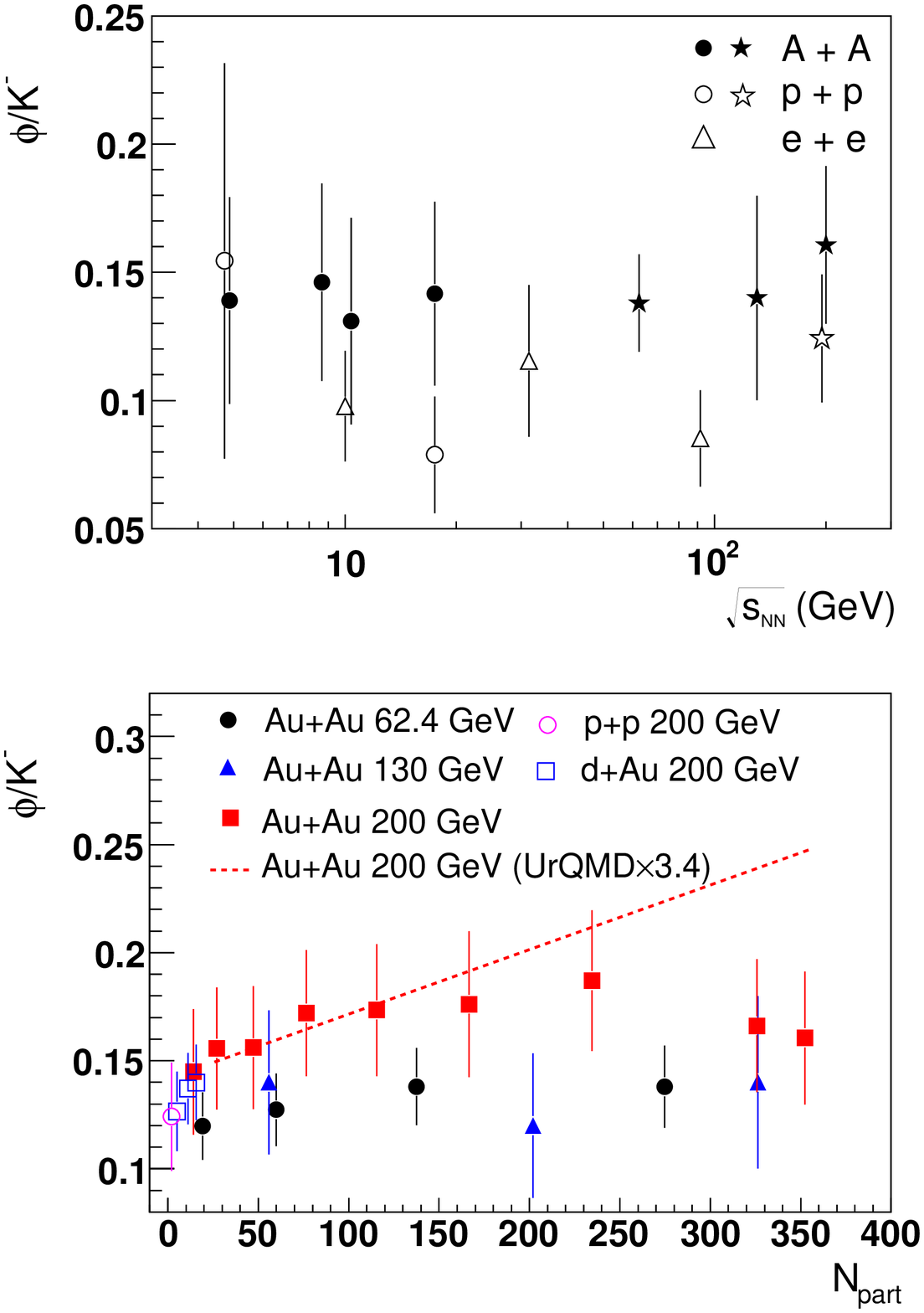}
\end{center}
\end{minipage}
\begin{minipage}[b]{0.48\linewidth}
\begin{center}
\includegraphics[width=\linewidth]{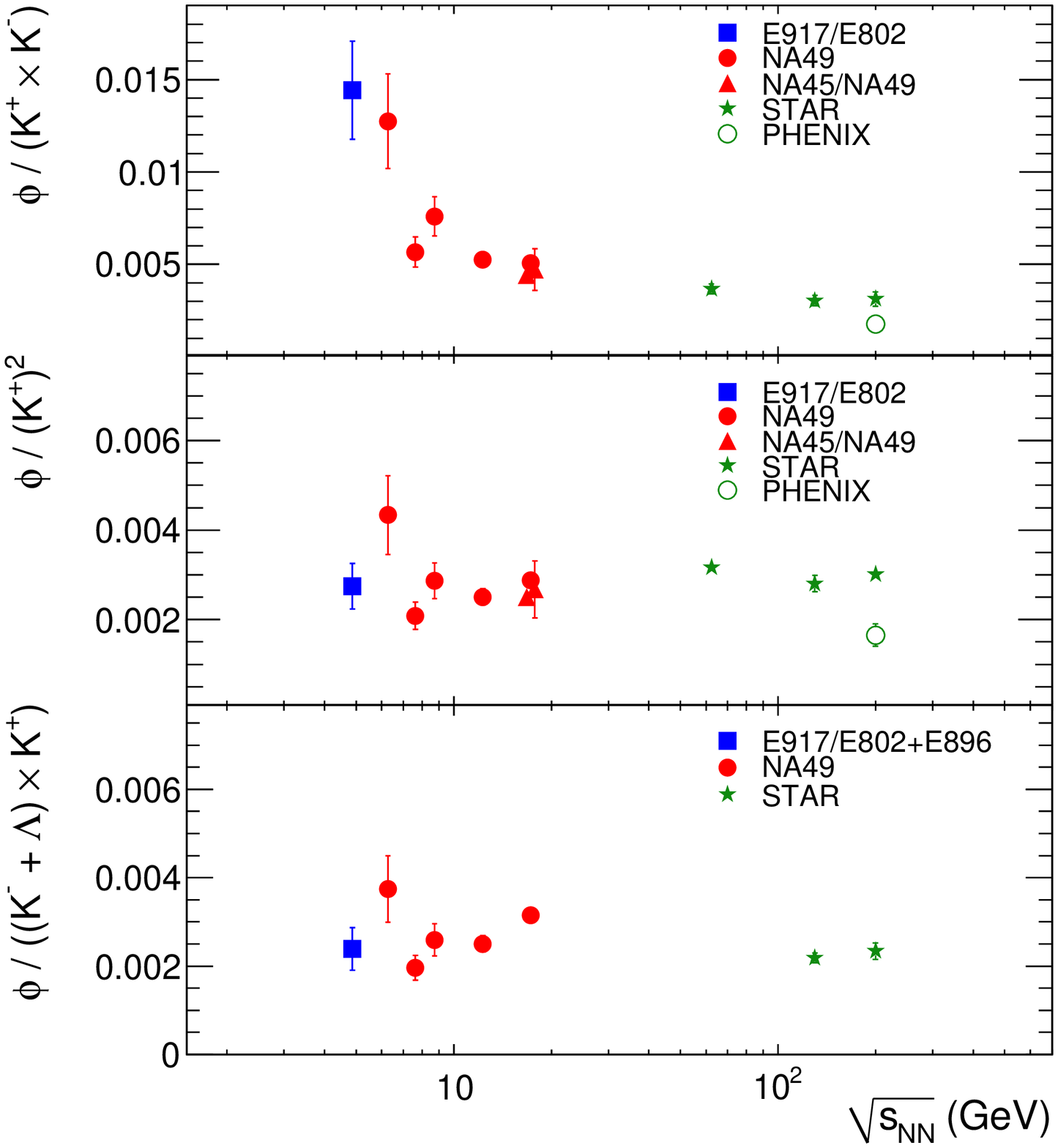}
\end{center}
\end{minipage}
\end{center}
\caption{Upper left: The energy dependence of the $\phi$/\kmin~ratio
in nucleus-nucleus (full symbols), e$^{+}$+e$^{-}$ (open triangles),
and p+p (open circles and star) collisions \cite{STAREDEPPHI}.
Lower left: The system size dependence of the $\phi$/\kmin~ratio in
different collisions systems.  The dashed line shows scaled results
for the transport model UrQMD \cite{URQMD1}.
Right: The energy dependence of the $\phi$ yield, normalized by the
product of the \kplus\ and \kmin\ yield (top), the square of the
\kplus\ yield (middle), and the product of the \kplus\ yield with the
sum of the \kmin\ and \lam\ yields (bottom).  All yields are measured
around midrapidity for central Pb+Pb or Au+Au collisions.  Shown are
only statistical errors.  The data are taken from
\cite{NA49EDEPHYP,STARSYST,NA49KPI40158,NA49KPI2030,PHNXKPI200,
STARHYP200,E896LAM,STARLAM130,NA49EDEPPHI,E802KPM116,STAREDEPPHI,
NA45PHI158,E917PHI,STARPHI130,STARPHI200,PHNXPHI200}.
}
\label{fig:phi_scaling}
\end{figure}
%

Among the strange particles, the $\phi$ meson has been of special
interest.  Due to its $s\bar{s}$ valence quark structure it is
strangeness neutral and should therefore not be sensitive to any
hadro-chemical effects, as they are described in statistical hadron gas
models by a strangeness undersaturation factor \gams\ or the canonical
suppression for small systems.  However, if the $\phi$ production was
dominated by quark coalescence from a partonic phase of the reaction,
its sensitivity would be even more pronounced.  In this case, any
strangeness undersaturation should affect the $\phi$ rather as
$\gams^{2}$.  On the other side, in a hadronic phase $\phi$ mesons can
be produced by kaon coalescence, $\kplus + \kmin \rightarrow \phi$.

The energy dependence of the $\phi$/$\pi$ ratio, measured around
midrapidity in central nucleus-nucleus collisions, is summarized in
the left panel of \Fi{fig:NA49_phi_edep} \cite{NA49EDEPPHI}.  This
ratio increases monotonically with center-of-mass energy without any
significant structure.  The data are compared to a statistical hadron
gas model, assuming full strangeness equilibration (i.e. \gams~=~1)
\cite{ANDRONIC1}.  While this model matches the STAR measurement at
high energies\footnote{Please note that there is an unresolved
  discrepancy between the STAR and PHENIX measurement.  The latter is
  lower by ~40\% than STAR.  The difference is essentially in the low
  \pt\ region, while at higher transverse momenta the measurements are
  roughly in agreement.}, it does not provide a good description at
the intermediate energies.  On the other hand, if \gams\ is included
in the fits as a free parameter, the $\phi$ yields can be described
rather well (see right panel of \Fi{fig:NA49_phi_edep}).  This
sensitivity of the strangeness neutral $\phi$ meson to the \gams\
parameter suggests that the abundances of strange quarks are already
determined in the partonic and not in the subsequent hadronic phase.

In a hadronic scenario the $\phi$ can be produced via kaon
coalescence.  For instance in the hadronic transport model UrQMD
\cite{URQMD1} this is the dominant production mechanism.  The
predictions of this model are compared to the data as well (see
\Fi{fig:NA49_phi_edep}).  At AGS energies the model under-predicts the
$\phi$/$\pi$ ratio at midrapidity by a factor of 2 (left panel) and is
further below the measurements at SPS and RHIC energies.  For the
unnormalized total yields the discrepancy is smaller (right panel),
but still prominent at higher energies.  This leads to the conclusion
that kaon coalescence is not sufficient to explain the amount of
observed $\phi$.  There are two other observations that substantiate
this point of view: the first is that the $\phi$/\kmin~ratio does not
increase with system size, as predicted by UrQMD (see lower left panel
of \Fi{fig:phi_scaling}), but is essentially flat.  The second
experimental indication is based on the comparison of the measured
widths of the $\phi$ rapidity distributions $\sigma_{\phi}$ to the
widths expected in the case of kaon coalescence (see discussion in
\Se{sec:rapidity}).

It is a remarkable observation that the $\phi$/\kmin~ratio seems to be
independent not only of the system size, but also of the
center-of-mass energy of the reaction systems (see left panel of
\Fi{fig:phi_scaling}) \cite{STAREDEPPHI}.  Alternatively, one finds
that the midrapidity $\phi$ yields scale as well with the \kplus\
yield squared (middle right panel of \Fi{fig:phi_scaling}).  Since the
number of \kplus\ is directly proportional to the number of produced
anti-strange quarks (= number of strange quarks), the $\phi$
production thus might only depend on the product of $s$ and $\bar{s}$
quarks.  Since the $s$ quarks are to a large extent contained in the
\kmin\ and \lam, the product $s \cdot \bar{s}$ should be roughly
proportional to $\kplus \cdot (\kmin + \lam)$.  The ratio $\phi /
(\kplus \times (\kmin + \lam))$ is similarly energy independent than
$\phi / (\kplus)^{2}$ (lower right panel of \Fi{fig:phi_scaling}) and
has approximately the same value ($0.2 - 0.3 \cdot 10^{-3}$).  On the
other hand, the ratio $\phi / (\kplus \kmin)$, as it might be
suggested by a naive kaon coalescence scenario, is clearly not energy
independent (upper right panel of \Fi{fig:phi_scaling}).


\clearpage

\subsection{Spectra}

This section discusses the dependence of transverse mass spectra 
($\mt = \sqrt{\pt^{2} + m^{2}}$) on the center-of-mass energy of the
reaction, which gives insight into the evolution of radial flow.  This
is of interest especially for multi-strange baryons, since they could
reveal the onset of a partonic contribution.  The inverse slope
parameters extracted from \mt~spectra of charged kaons exhibit a
step-like feature in their energy dependence, which might be connected
to a first order phase transition.  Further the measured rapidity
spectra at different center-of-mass energies and attempts to connect
thermal parameters determined at different rapidities and at different
energies are discussed.

\subsubsection{Transverse momentum spectra}

%
\begin{figure}[t]
\begin{center}
\begin{minipage}[b]{0.40\linewidth}
\begin{center}
\includegraphics[width=\linewidth]{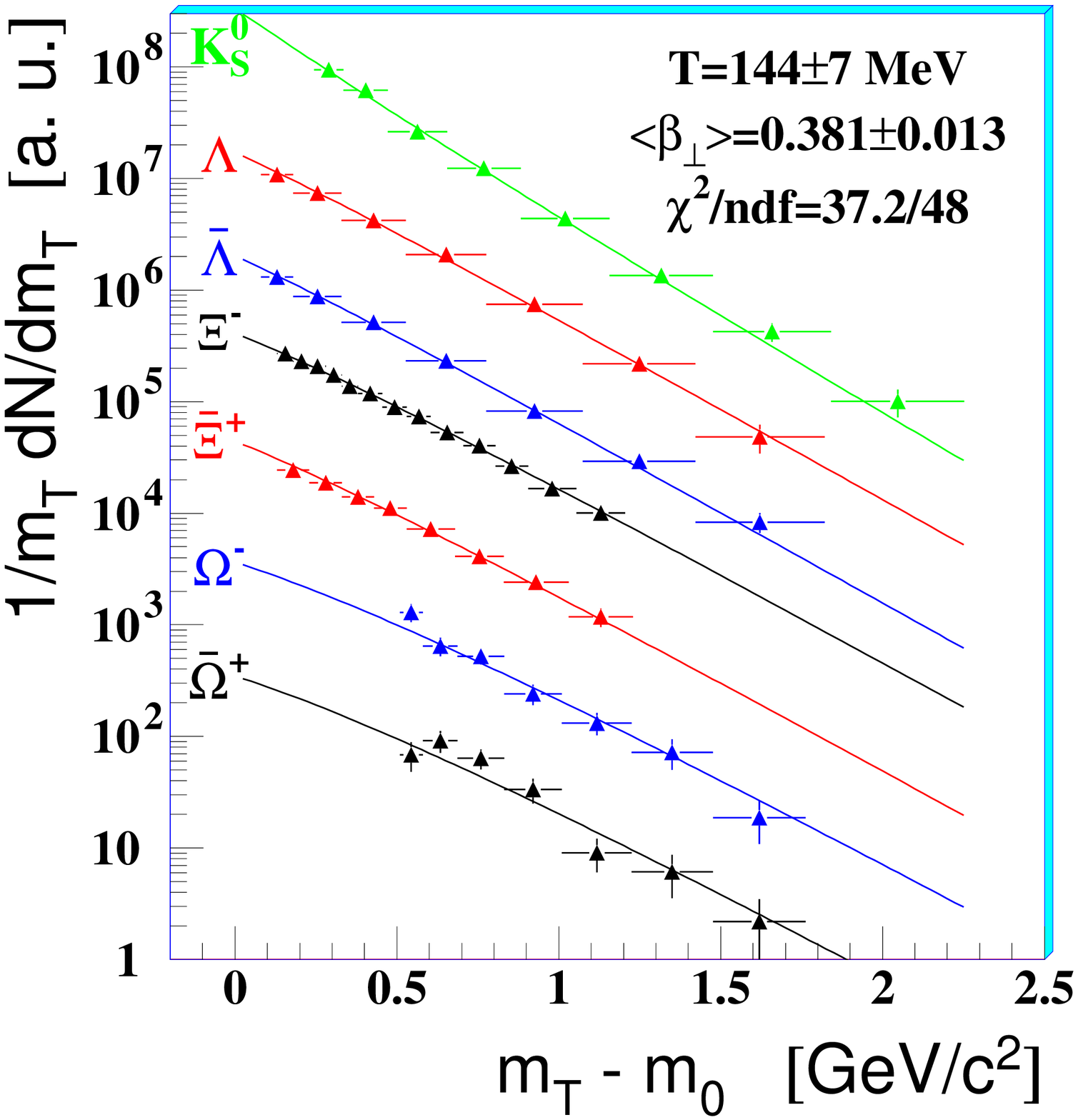}
\end{center}
\end{minipage}
\begin{minipage}[b]{0.40\linewidth}
\begin{center}
\includegraphics[width=\linewidth]{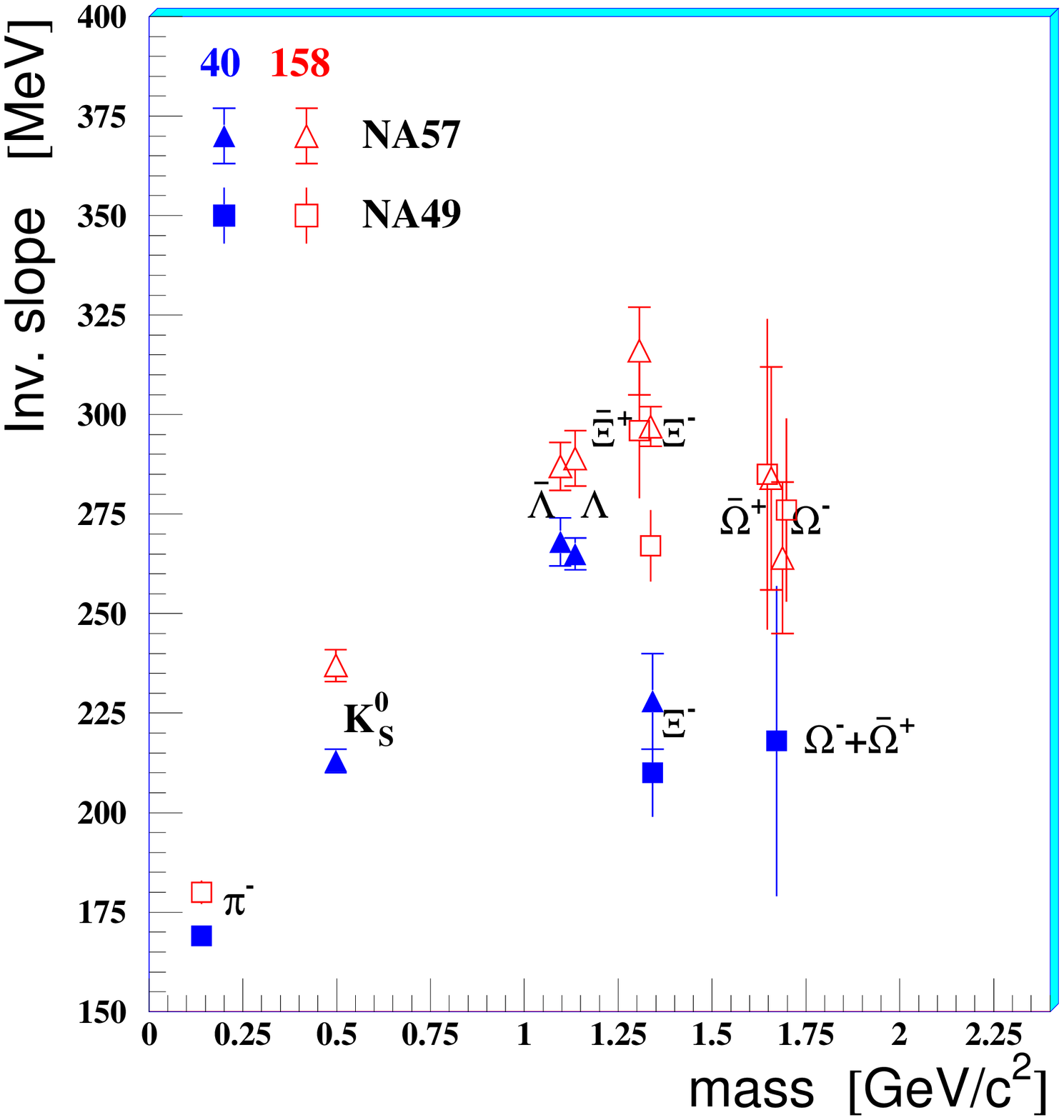}
\end{center}
\end{minipage}
\end{center}
\caption{Left: The \mt~spectra of \kzero, \lam, \lab, \xim, \xip,
\omm, and \omp\ as measured by NA57 for central Pb+Pb collisions at
158\agev\ (Fig. adapted from \cite{NA57MT158}).  This solid lines
represent the result of a fit with a hydrodynamically inspired model
\cite{SCHNEDERMANN}.
Right: The inverse slope parameters as a function of the
particle mass measured in central Pb+Pb collisions at 40\agev\ (closed
symbols) and at 158\agev (open symbols).  The data are from the NA57
\cite{NA57MT40} and the NA49 \cite{NA49EDEPOM,NA49XI158,NA49MEURER}
collaboration.
}
\label{fig:NA57_mt}
\end{figure}
%

%
\begin{figure}[t]
\begin{center}
\includegraphics[width=0.55\linewidth]{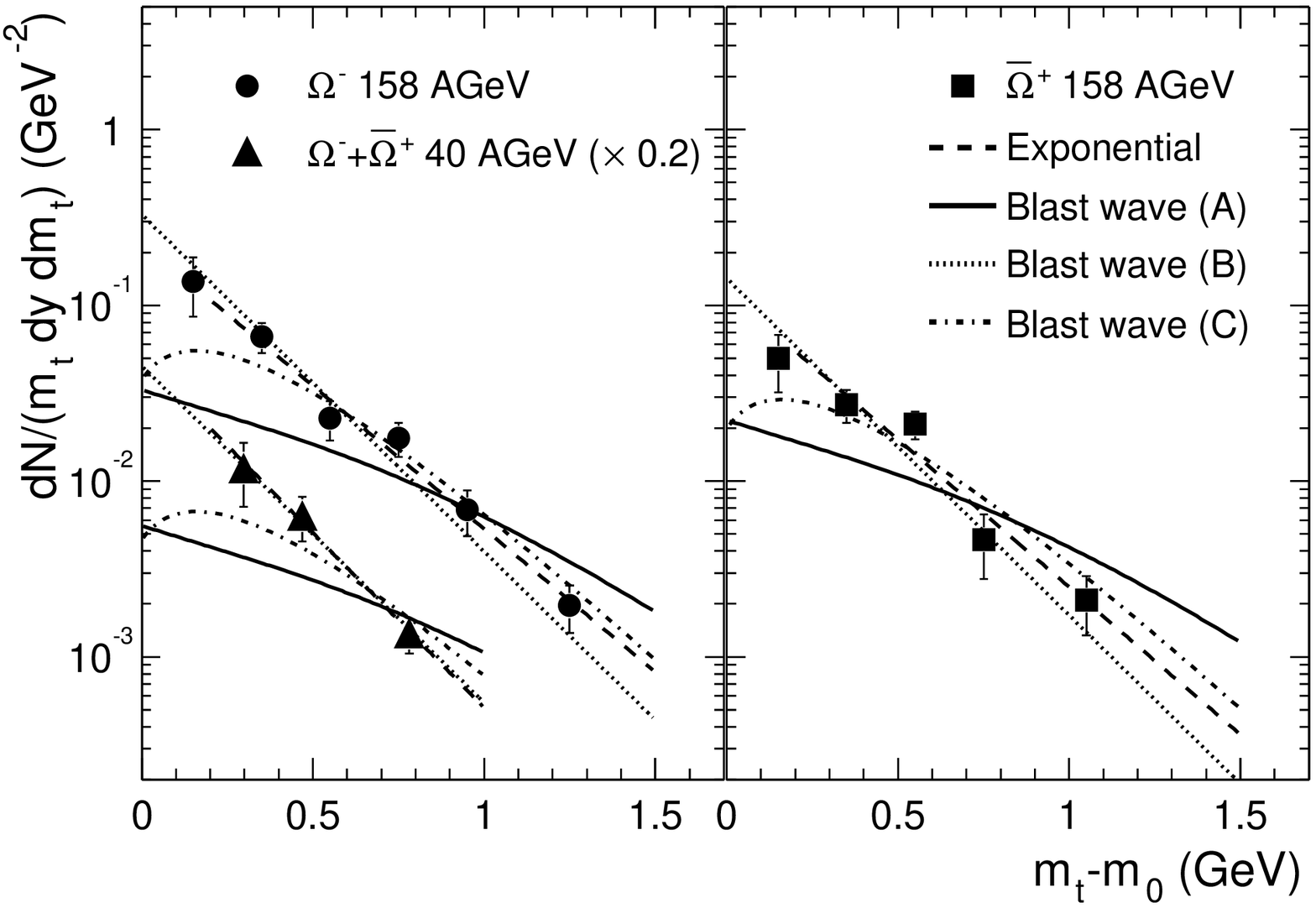}
\end{center}
\caption{The transverse mass spectra of \omm\ and \omp\ in central
Pb+Pb collisions at 158\agev\ and of the sum $\omm + \omp$ at 40\agev\
\cite{NA49EDEPOM}.  The dashed curve shows a fit with an exponential.
The solid, dotted, and dash-dotted curves represent a model including
transverse expansion \cite{SCHNEDERMANN}.  The used parameters are
$\tth = 90$~MeV and $\bpavg = 0.5$ (A), $\tth = 170$~MeV and $\bpavg =
0.2$ (B), and $\tth = 127$~MeV and $\bpavg = 0.5$ (C). In (A) and (B)
a linear velocity profile is used, while (C) was calculated with a
constant expansion velocity.}
\label{fig:NA49_om_slopes}
\end{figure}
%

%
\begin{figure}[t]
\begin{center}
\includegraphics[width=0.85\linewidth]{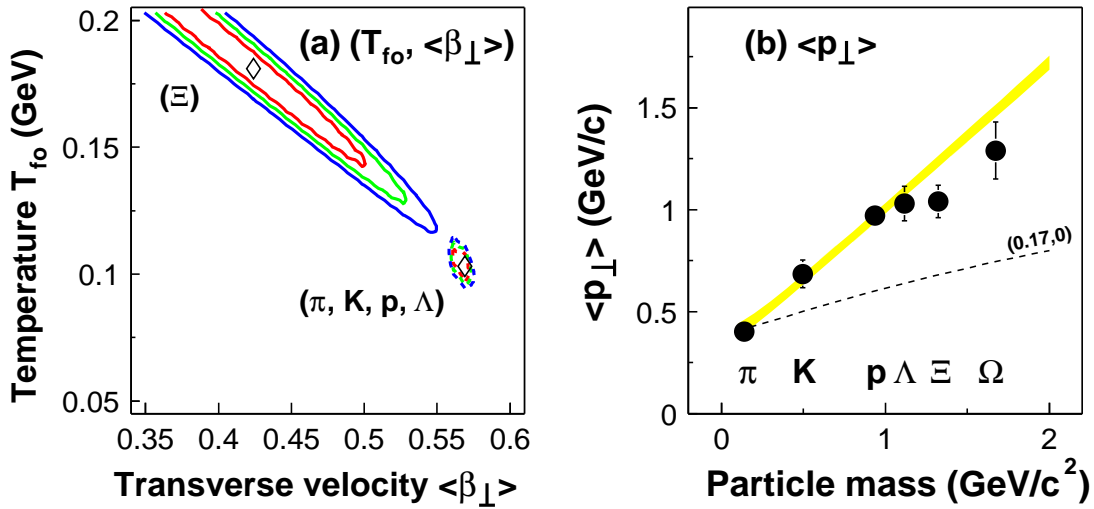}
\end{center}
\caption{Left: The kinetic freeze-out temperature \tth\ versus the
transverse flow velocity $\langle\beta_{\perp}\rangle$ for the
hydrodynamically inspired model fits to the \mt~spectra
\cite{STARHYP130}.  The 1, 2, and 3 sigma contours are shown.  Solid
curves are for a simultaneous fit to the \xim\ and \xip.  Dashed
curves are a separate fit to the $\pi$, K, p, and \lam\ data as
measured by the STAR collaboration.  The diamonds represent the best
fit in both cases (Fig. adapted from \cite{STARHYP130}).
Right: The mean transverse momentum $\langle p_{\perp}\rangle$ for
identified particles as a function of the particle mass
\cite{STARHYP130}.  The band follows from the three sigma contour of
the fit to $\pi$, K, p, and \lam, as shown in the left panel, and the
dashed curve is for $\tth = 170$~MeV and $\langle\beta_{\perp}\rangle
= 0$.}
\label{fig:STAR_bwfit}
\end{figure}
%

%
\begin{figure}[t]
\begin{center}
\begin{minipage}[b]{0.47\linewidth}
\begin{center}
\includegraphics[width=\linewidth]{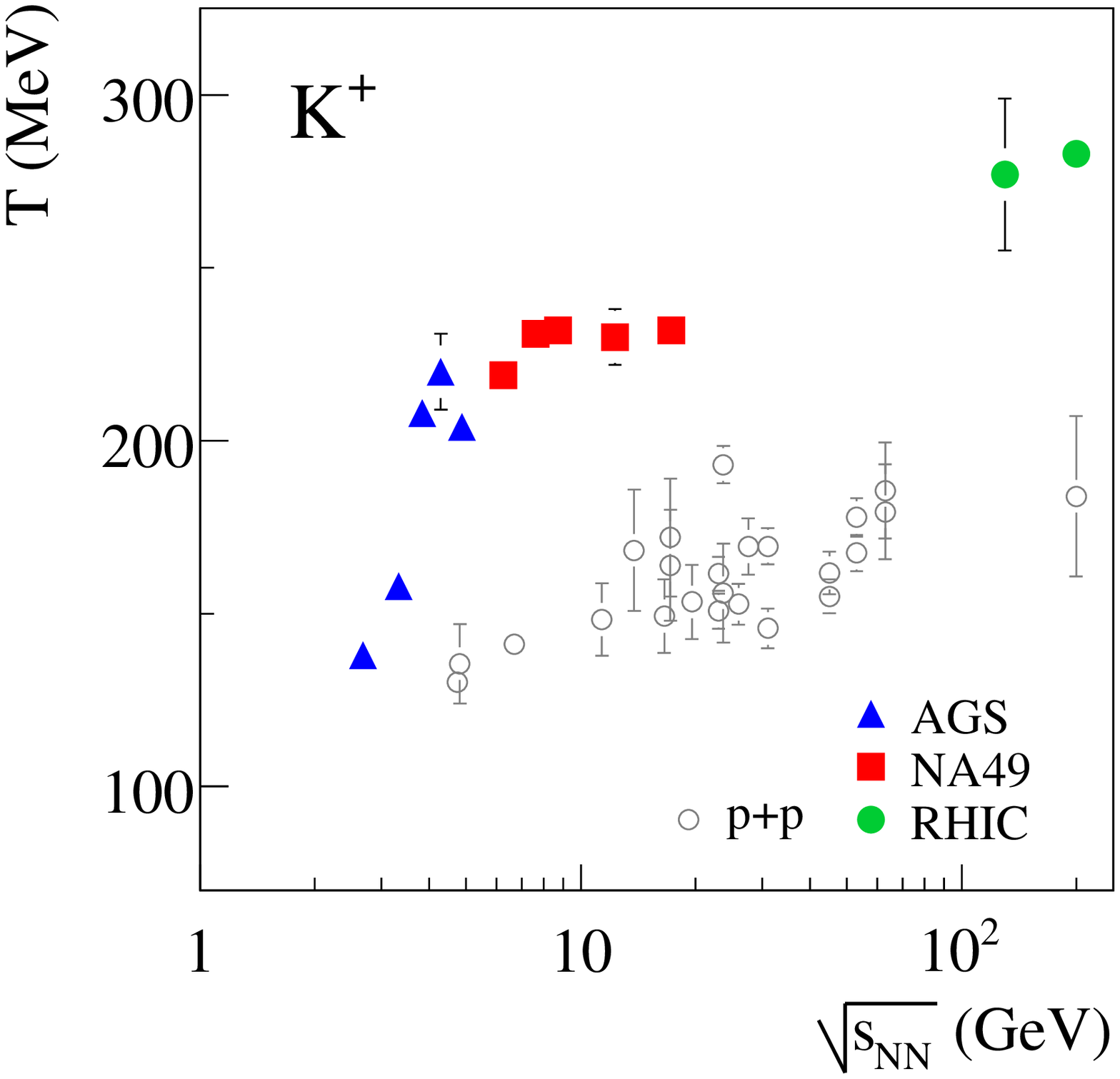}
\end{center}
\end{minipage}
\begin{minipage}[b]{0.47\linewidth}
\begin{center}
\includegraphics[width=\linewidth]{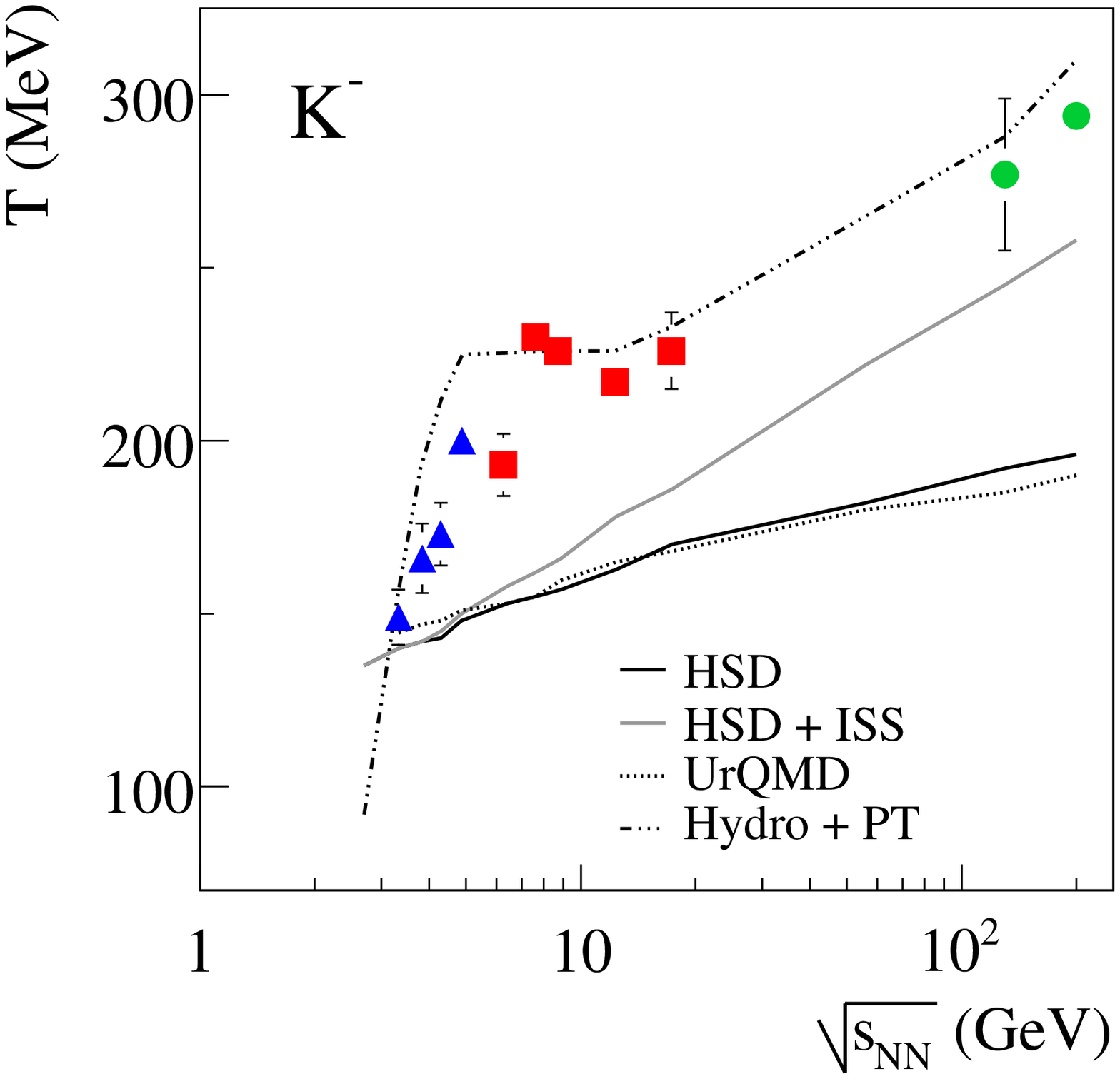}
\end{center}
\end{minipage}
\end{center}
\caption{The energy dependence of the inverse slope parameters $T^{*}$
of \kplus\ (left) and \kmin\ (right) measured at midrapidity in
central Pb+Pb and Au+Au collisions \cite{NA49KPI2030}. The \kplus\
slope parameters are compared to results for p~+~p(\pbar) collisions
\cite{KLIEMANT1} (left).  The curves shown in the right panel
represent various model predictions.}
\label{fig:NA49_kaon_slopes}
\end{figure}
%

Transverse mass spectra have been measured for all strange particle
species at different center-of-mass energies.  As an example, the left
panel of \Fi{fig:NA57_mt} shows the \mt~spectra of \kzero\ and
hyperons, measured by the NA57 collaboration in Pb+Pb collisions at
158\agev.  Since strange particles span a relatively wide range of
rest masses (between $m$(\kpm) = 493.677~\mevcc\ and $m$($\Omega$) =
1672.45~\mevcc), they provide an ideal probe for the effects of
transverse expansion, which is generated by the high energy density in
the fireball.  Especially the question to which extent the radial
flow is caused by the partonic stage of the reactions can be addressed
with multi-strange particles.  In order to quantify the spectral
shape, the invariant \mt~spectra are usually fitted by an exponential
function
\be
  \label{eq:exp}
  \!\! \frac{\der N}{\mt \: \der \mt \der y}
  \propto \exp \left( -\frac{\mt}{T^{*}} \right).
\ee
with $T^{*}$ as the inverse slope parameter.  For kaons the
exponential results in a very good fit over a larger region of \mt.
Heavier particles exhibit deviation from an exponential behavior, due
to the stronger influence of radial flow which introduces a convex
shape in the spectra.  Therefore, the fit result will depend on the
\mt~range in which the fit it performed.  Nevertheless, a comparison
of the inverse slope parameter $T^{*}$ extracted for different
particles at different energies, reveals already interesting features.
In the right panel of \Fi{fig:NA57_mt} such a comparison for 40$A$ and
158\agev\ beam energy is shown \cite{NA57MT40}.  The inverse slope
parameters increase linearly with particle mass up to the \lam.  This
can be understood as a consequence of the radial expansion of the
fireball, which, in a simplified picture, introduces an additional
component to the effective inverse slope parameter $T^{*}$, that
depends on the particle mass $m$ and the expansion velocity
$v_{\perp}$ as
\be
  \label{eq:radflow}
  T^{*} \approx \tth + \frac{1}{2} m v_{\perp}^{2}.
\ee
While the kinetic freeze-out temperature \tth\ is assumed to be the
same for all particles, the second term introduces the observed mass
dependence of $T^{*}$.  The slightly higher $T^{*}$ measured at
158\agev\ would thus indicate that the average radial expansion
velocity still increases between the two energies.  However, the $\Xi$
and $\Omega$ do not fit into this systematics.  At 158\agev\ their
inverse slope parameter is at the same level as the one measured for
the \lam, while at 40\agev\ it is significantly lower.
Apparently, the multi-strange particles do not participate in the
radial flow to the same extent as the particles with one or no strange
constituent quark.  This also follows from the analysis shown in
\Fi{fig:NA49_om_slopes} \cite{NA49EDEPOM}.  Here the \mt~spectra of
the $\Omega$ are compared to a hydrodynamically inspired model, which
assumes a transversely expanding emission source \cite{SCHNEDERMANN}.
Such an analysis avoids the problems connected to the use of an
inverse slope parameter, whose value for heavy particles can depend on
the fit range in \mt.
The parameters of this model are the kinetic freeze-out temperature
\tth\ and the transverse flow velocity $\betas = v_{\rb{s}}/c$ at the
surface. Assuming a linear radial velocity profile $\betap (r) =
\betas\; r/R_{\rb{s}}$, as motivated by hydrodynamical calculations,
the \mt~spectrum can be computed from 
\be
  \label{eq:blast}
  \!\! \frac{\der N}{\mt \: \der \mt \der y}
  \propto \! \int_{0}^{R_{\rb{s}}} \!\!\! r \: \der r\; \mt \:
  I_{0} \!\!\left(\!\frac{\pt \sinh \rho}{\tth} \!\right)\!
  K_{1} \!\!\left(\!\frac{\mt \cosh \rho}{\tth} \!\right),
\ee
where $R_{\rb{s}}$ is the radius of the source and $\rho =
\tanh^{-1}\! \betap$ the boost angle.  The curve labeled (A) is
calculated with the parameters derived from a simultaneous fit to
kaons, (anti-)protons \lam, and $\phi$ ($\tth = 90$~MeV and
$\bpavg = 0.5$).  The clear disagreement to the measured $\Omega$
\mt~spectra illustrates that the freeze-out conditions for the
$\Omega$ are different than for the lighter hadrons.  A much better
agreement can be achieved, if the freeze-out parameters taken from a
fit to \jpsi\ and \psip\ spectra \cite{GORENSTEIN1} are used, as shown
as curve (B) in \Fi{fig:NA49_om_slopes}.  This indicates that the
$\Omega$ has similar kinetic freeze-out conditions than the \jpsi.
The same observation was made at RHIC energies (see
\Fi{fig:STAR_bwfit}).  Using the same model different freeze-out
parameters are extracted for the $\Xi$ than for the lighter hadrons.
The averaged transverse momentum of the $\Xi$ and $\Omega$ departs
from the linear increase with particle mass, similar to the inverse
slope parameters shown in the right panel of \Fi{fig:NA57_mt}.

The interpretation of this behavior follows from the assumption that
rare particles as the $\Xi$ and $\Omega$ have a lower hadronic
scattering cross section than light hadrons \cite{VANHECKE1} and
therefore do not participate in the radial flow that is developing
during the hadronic phase of the fireball evolution.  This leads to
the conclusion that a substantial part of the transverse expansion
probed by these particles has to be generated during the partonic
phase.  Thus, the $\Xi$ and $\Omega$ would be directly sensitive to
the pressure in the early phase of the reaction.  The rapid increase
of the inverse slope parameters measured for these particles in the
SPS energy range (right panel of \Fi{fig:NA57_mt}) would thus indicate
that the partonic part of the flow is beginning to develop at these
energies.

The \mt~spectra of kaons provide important information on the
transverse dynamics of the reaction system.  Calculations within
hybrid hydrodynamics and transport approaches \cite{BASS3,TEANEY1}
indicate that kaons freeze out slightly earlier than nucleons and
\lam\ (e.g. in \cite{TEANEY2} decoupling times of $\tau_{\rb{fo}}
\approx 14$~fm/$c$ are found for mesons and $\tau_{\rb{fo}} \approx
18$~fm/$c$ for nucleons, \lam\ and $\Sigma$), and thus they will be
less sensitive to the late rescattering phase.  Their \mt~spectra can
be very well described by a single exponential (\Eq{eq:exp}) and can
thus be characterized by just one parameter, the inverse slope
parameter $T^{*}$.  The energy dependence of $T^{*}$ for charged kaons
exhibits an interesting feature, as shown in
\Fi{fig:NA49_kaon_slopes}.  While $T^{*}$ is rising rapidly with
center-of-mass energy for $\sqrts < 7 - 8$~GeV, it is rather constant
or only slightly increasing above this energy.  A similar observation
has been made for the averaged transverse mass \mtavg\ of pions and
protons \cite{NA49KPI2030,SQM04}.  In proton-proton collisions such a
behavior is not observed (see left panel of \Fi{fig:NA49_kaon_slopes})
\cite{KLIEMANT1}.

The \dndy\ dependence of the mean transverse momentum has been
suggested as a signature for a first order phase transition to a
deconfined state quite a while ago within the concept of Landau's
hydrodynamical model \cite{VANHOVE}.  Originally, this has been
proposed as an interpretation of the dependence of $\langle \pt
\rangle$ on \dndy\ observed in p~+~\pbar\ collisions at \sqrts~=
540~GeV.  While the fireball volume produced in these reactions
is most likely too small to create a QGP state, attempts have been
made to apply a similar interpretation to the observed structure in
the energy dependence of the inverse slope parameters of kaons
\cite{GAZDZICKIHYD}.  While hadronic transport model calculations are
generally not able to properly reproduce the observed structure (see
right panel of \Fi{fig:NA49_kaon_slopes}) \cite{URQMD1,HSDURQMD,HSD3},
the data can be described by a hydrodynamical model assuming a first
order phase transition between a hadronic and a deconfined phase
(dash-dotted curve in right panel of \Fi{fig:NA49_kaon_slopes})
\cite{GAZDZICKIHYD}.  If initial conditions calculated event-by-event
with the NEXUS event generator \cite{DRESCHER1,DRESCHER2} are used,
the same model is able to simultaneously describe yields, transverse
mass and rapidity spectra.  Another hydro approach, which uses a
dynamical description of the freeze-out conditions \cite{IVANOV1},
also results in a good match to the data, but turns out to be less
sensitive to a specific equation-of-state and thus does not
necessarily require a phase transition.  A similar observation has
been made with a model that combines hadronic transport and
hydrodynamics \cite{PETERSEN1}.  Also here it was found that the way
the freeze-out procedure is implemented can have as much influence on
the final result as the equation-of-state that is used.  Even though
it is quite possible that the measured energy dependence of the
inverse slope parameters might be due to a first order phase
transition, it is currently difficult to establish an unambiguous
theoretical connection.  At the moment it is also still unclear
whether in the center-of-mass energy range discussed here a first
order phase transition is to be expected.  While most lattice QCD
calculation expect a change from a cross over to a first order phase
boundary at higher \mub, corresponding to lower \sqrts\
\cite{FODOR1,ALLTON1,HATTA1}, the results obtained by \cite{FORCRAND1}
rather indicate that the phase transition is of the type of a cross
over also at higher \mub.

\subsubsection{Rapidity spectra}
\label{sec:rapidity}

%
\begin{figure}[t]
\begin{center}
\includegraphics[width=0.60\linewidth]{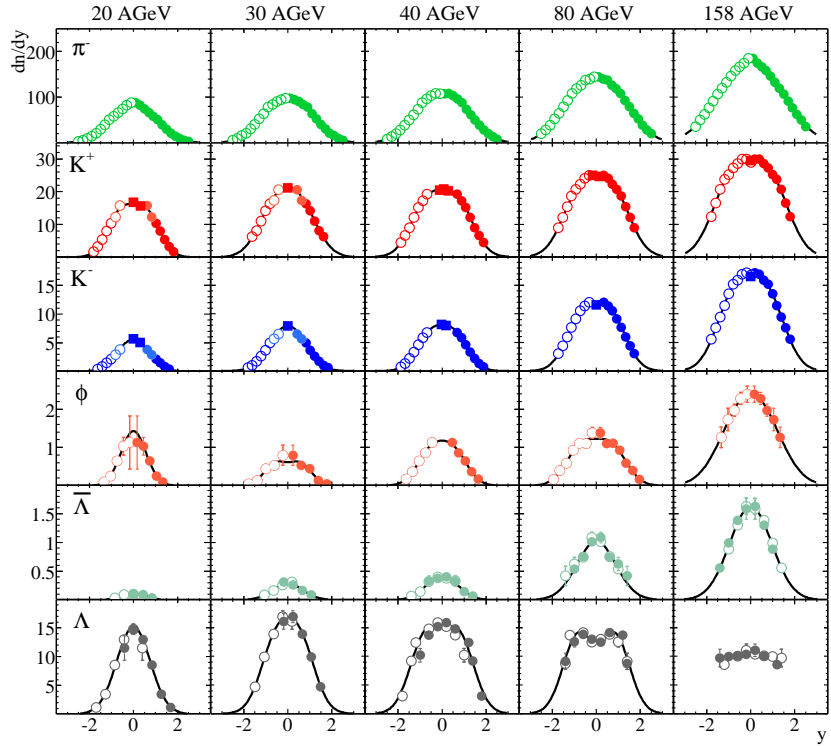}
\end{center}
\caption{The rapidity spectra of \pim, \kplus, \kmin, $\phi$, \lab,
and \lam\ as measured by the NA49 collaboration in central Pb+Pb
collisions at 158\agev\ \cite{SQM04}.  The closed circles indicate
measured data points, while open ones are reflected around
midrapidity.  The solid lines represent fits with a single Gaussian or
the sum of two Gaussians.}
\label{fig:NA49_rapidity}
\end{figure}
%

%
\begin{figure}[t]
\begin{center}
\begin{minipage}[b]{0.37\linewidth}
\begin{center}
\includegraphics[width=\linewidth]{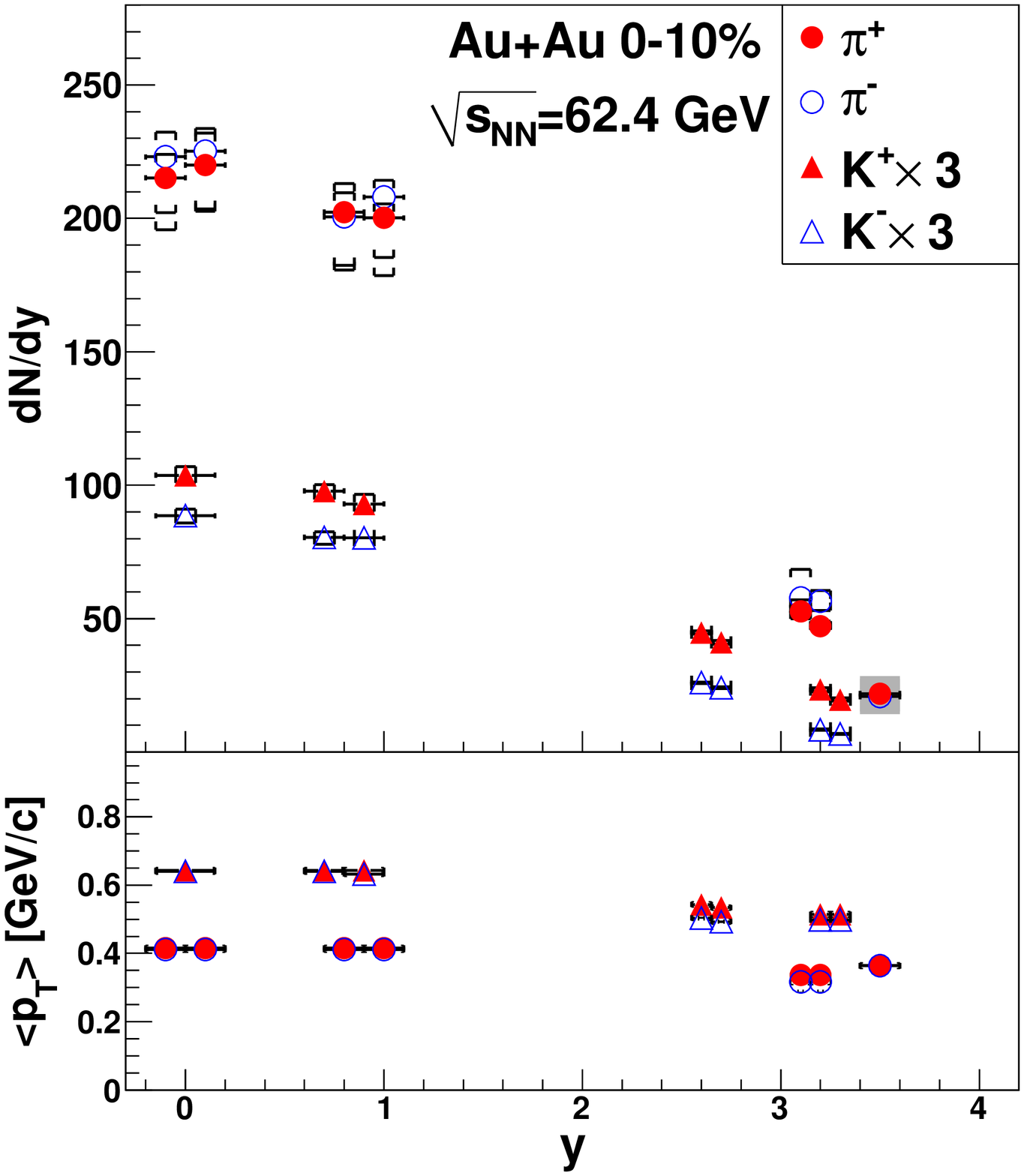}
\end{center}
\end{minipage}
\begin{minipage}[b]{0.47\linewidth}
\begin{center}
\includegraphics[width=0.98\linewidth]{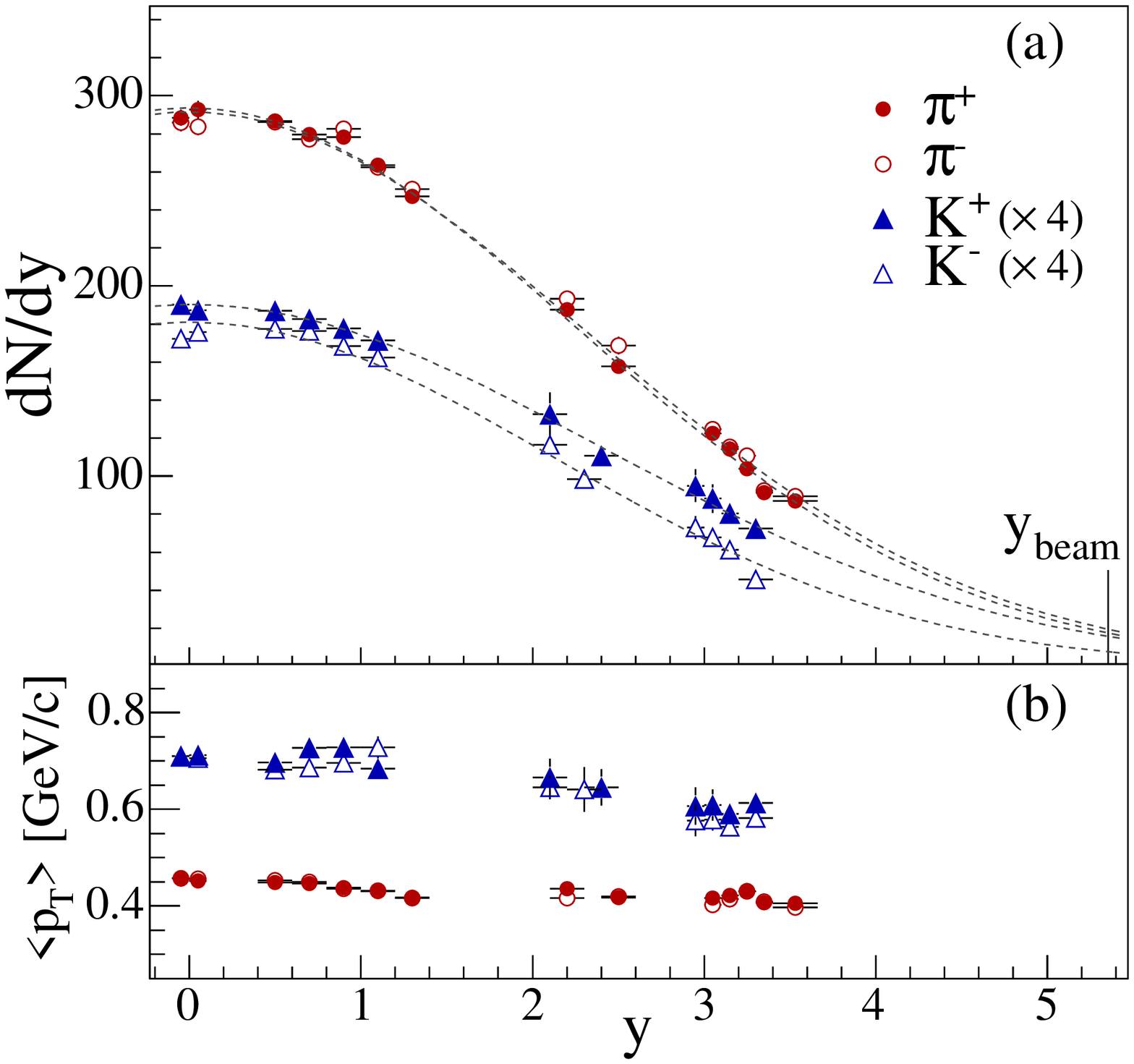}
\vspace{5pt}
\end{center}
\end{minipage}
\end{center}
\caption{\dndy\ (upper panels) and the averaged transverse momentum
\ptavg\ (lower panels) as a function of rapidity for charged pions and
kaons for central Au+Au collisions, as measured by the BRAHMS
collaboration at \sqrts~= 62.4~GeV (left) \cite{BRMSRAP62} and at
\sqrts~= 200~GeV (right) \cite{BRMSRAP200}.}
\label{fig:BRAHMS_rapidity}
\end{figure}
%

%
\begin{figure}[t]
\begin{center}
\begin{minipage}[b]{0.42\linewidth}
\begin{center}
\includegraphics[width=\linewidth]{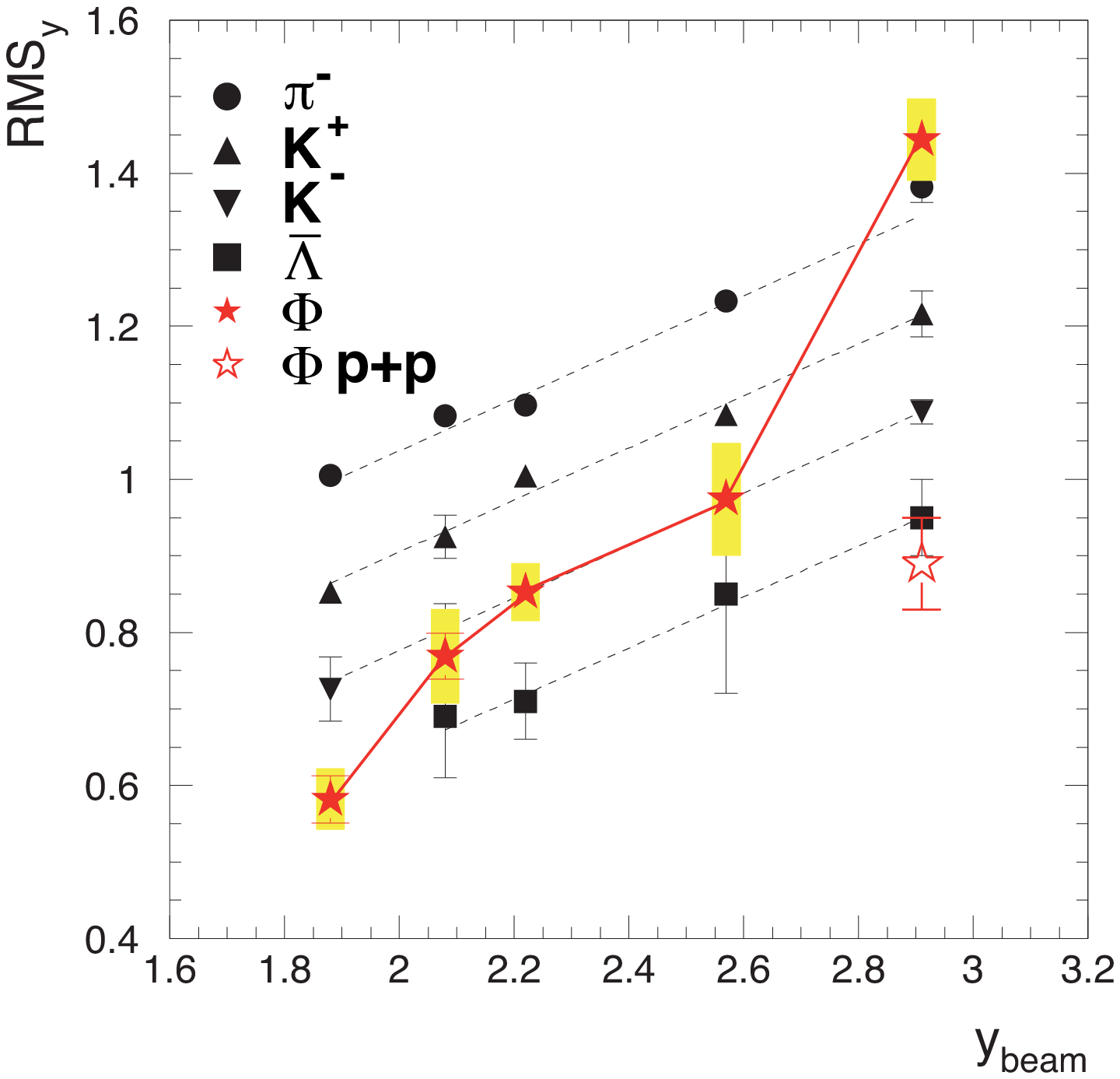}
\end{center}
\end{minipage}
\begin{minipage}[b]{0.42\linewidth}
\begin{center}
\includegraphics[width=\linewidth]{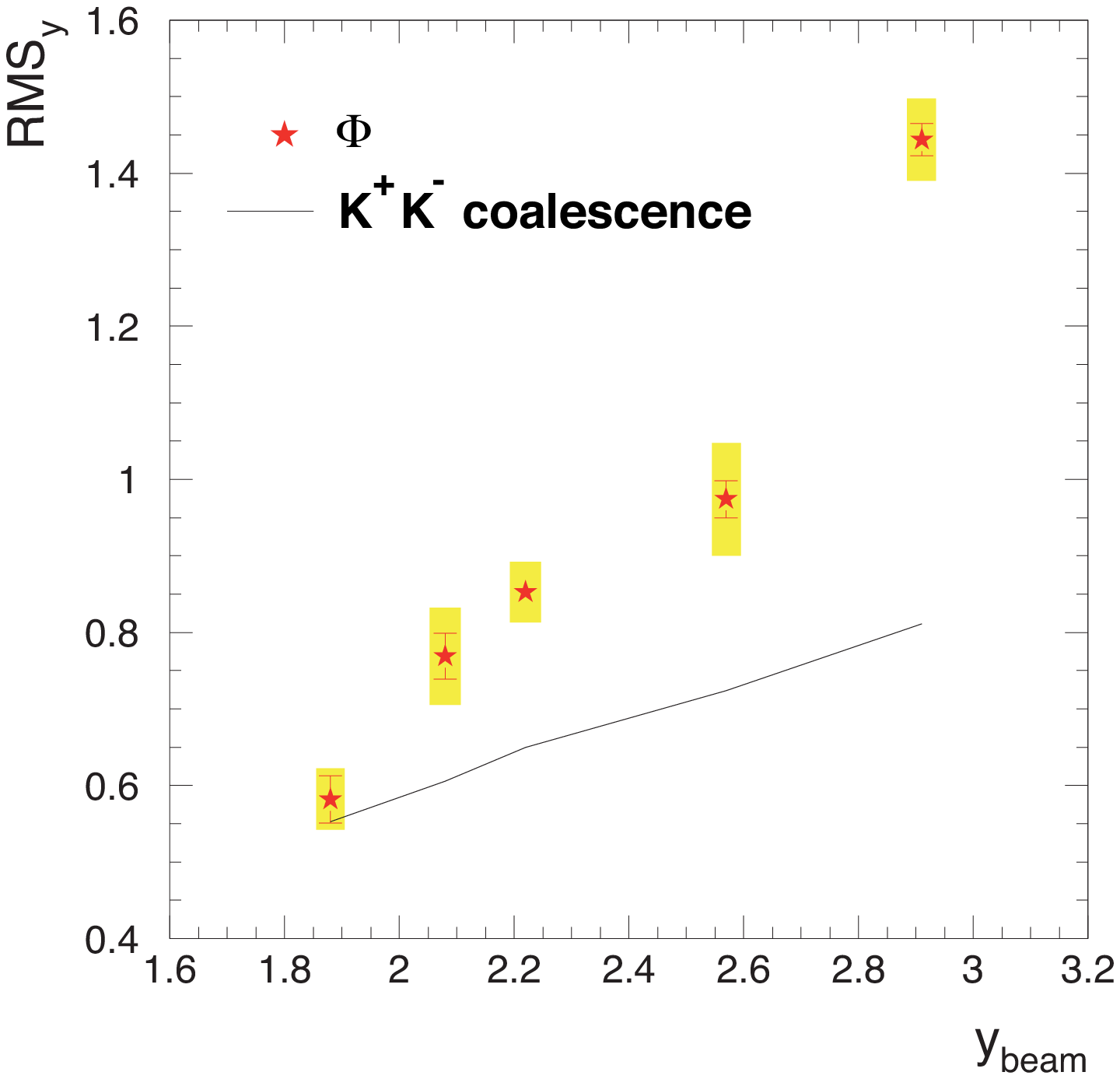}
\end{center}
\end{minipage}
\end{center}
\caption{Left: The RMS widths of the rapidity distributions of \pim,
\kmin, \kplus, and $\phi$ in central Pb+Pb collisions at SPS energies
as a function of the beam rapidity \cite{NA49KPI40158,NA49KPI2030,
NA49EDEPPHI,SQM04}.
Right: The widths of the rapidity distributions of $\phi$ mesons in
central Pb+Pb collisions at SPS energies as a function of the beam
rapidity \cite{NA49EDEPPHI}.  The data are compared to expectations
from kaon coalescence (solid line).  (The Figs. are adapted from 
\cite{NA49EDEPPHI}.)}
\label{fig:NA49_rapwidths}
\end{figure}
%

At the SPS rapidity spectra in the forward hemisphere have been
measured by the NA49 collaboration for charged kaons, $\phi$, and
hyperons at five different beam energies \cite{NA49EDEPHYP,NA49EDEPOM,
NA49KPI40158,NA49KPI2030,NA49EDEPPHI}.  Part of these spectra are
shown for central Pb+Pb collisions in \Fi{fig:NA49_rapidity}.  The
NA57 collaboration performed measurements of rapidity distributions
for \kzero, \lam, \lab, \xim, \xip, and $\omm+\omp$ in the lab
rapidity range $2.5 < y < 3.5$ for the 53\% most central Pb+Pb
collisions at 158\agev\ \cite{NA57RAP158}.  At RHIC no measured
rapidity distributions are available for most particle types, since
most detectors only cover the midrapidity region.  The only exception
is provided by the BRAHMS collaboration, which can cover a large part
of the forward hemisphere.  BRAHMS has measured in central Au+Au
collisions at \sqrts~= 64.2~GeV and 200~GeV rapidity distributions
of \dndy\ and \ptavg\ for charged kaons (see \Fi{fig:BRAHMS_rapidity})
\cite{BRMSRAP200,BRMSRAP62}.  In order to avoid assumptions on
longitudinal phase space distributions, the measurement of \dndy\
distributions over a large range of rapidities is a prerequisite for
the precise determination of 4$\pi$ integrated total yields.
Therefore, measurements of the total multiplicities at RHIC energies
are only available by the BRAHMS collaboration for charged pions and
kaons (see \Tar{tab:comp_kp}{tab:comp_om}).

The rapidity spectra of strange mesons can reasonably be described by
a Gaussian distribution at all energies \cite{NA49EDEPPHI,BRMSRAP200}
(\Fis{fig:NA49_rapidity} {fig:BRAHMS_kpi_rap}), although in some cases
the sum of two Gaussians, displaced relative to midrapidity, can result
in a slightly better fit \cite{NA49KPI40158,NA49KPI2030}.  This
observation is also true for anti-baryons (\lab, \xip)
\cite{NA49EDEPHYP} and for the $\Omega$ \cite{NA49EDEPOM}.  Baryons,
such as the \lam\ and \xim, have a wider distribution which is
changing drastically with beam energy, because their light valence
quark structure is connected to the ones of the originally colliding
nucleons.  This is in particular visible for the \lam, whose
distribution changes from a Gaussian shape at 20\agev\ to a spectrum
that is essentially flat inside the measured region at 158\agev\ (see
lowest row of \Fi{fig:NA49_rapidity}).  No pronounced change with
center-of-mass energy is observed for \xim\ spectral shapes, but the
widths of the rapidity distributions increase significantly faster
with \sqrts\ than the ones measured for \xip\ \cite{NA49EDEPHYP}.  The
reason for this behavior is that \lam\ and \xim\ are sensitive to the
longitudinal redistribution of net-baryon number, which depends
strongly on center-of-mass energies.  While at low SPS energies the
net-baryon number is concentrated around midrapidity, its distribution
changes very rapidly and turns from a maximum around $y = 0$ into a
shallow minimum at RHIC energies \cite{QM06,BRMSSTOPPING}.
Consequently, the midrapidity anti-baryon/baryon ratios exhibit a
strong energy dependence.  With increasing strangeness content of the
baryons this effect becomes weaker, as already shown in the right
panel of \Fi{fig:bbbarratios}.

The systematics of the RMS widths of the rapidity spectra of \pipm,
\kpm, $\phi$, and \lab, i.e. those particles whose longitudinal
distributions can be described by Gaussians, is summarized in the left
panel of \Fi{fig:NA49_rapwidths} \cite{NA49EDEPPHI}.  For charged
pions, kaons, and anti-lambdas the widths $\sigma$ scale linearly with
the beam rapidity $\ybeam = 2 \; \cosh^{-1}(\sqrts/(2 \,m_{\rm{p}}))$
in the SPS energy region \cite{SQM04}.  The widths depend, at least in
the SPS energy region, on the particle type as $\sigma(\pim) >
\sigma(\kplus) > \sigma(\kmin) > \sigma(\lab)$.  However, since at
RHIC the width of the \kplus\ approaches the one of the pions, or is
even larger \cite{BRMSRAP200}, this scaling apparently does not hold
up to top RHIC energy.

A peculiar behavior can be observed for the $\phi$ meson, whose
rapidity widths are increasing much faster with beam energy than the
ones of the other mesons (left panel of \Fi{fig:NA49_rapwidths}).
While at 20\agev\ the $\phi$ has a rapidity distribution that is
narrower than the one of the \kmin, its width is comparable to the one
of the \pip\ at 158\agev.  This behavior is at variance with a simple
expectation based on kaon coalescence (right panel of
\Fi{fig:NA49_rapwidths}).  In this case the $\phi$ rapidity
distributions $\sigma_{\phi}$ should depend on the measured widths of
the kaon rapidity distributions $\sigma_{K^{\pm}}$ in the following
way:
\be
\label{eq:coal}
\frac{1}{\sigma_{\phi}^{2}} = \frac{1}{\sigma_{K^{+}}^{2}}
                            + \frac{1}{\sigma_{K^{-}}^{2}}.
\ee
As shown in the right panel of \Fi{fig:NA49_rapwidths}, this simple
scenario is clearly ruled out for nucleus-nucleus collisions.
However, it is interesting to note that for p+p collisions at top SPS
energy (158 GeV) the coalescence expectation, derived from the
measured kaon rapidity widths according to \Eq{eq:coal}, is relatively
close to the measured $\phi$ rapidity widths
($\sigma_{\phi}^{\rb{meas.}} = 0.89 \pm 0.06$ \cite{NA49PHI158},
compared to $\sigma_{\phi}^{\rb{coal.}} = 0.77$ as based on the kaon
measurements in \cite{NA49PPKAON}).  In addition, it is remarkable
that the rapidity widths measured at top SPS energy are significantly
wider in central Pb+Pb collisions compared to p+p collisions at the
same energy ($RMS_{\rm{y}}(\textrm{Pb+Pb}) = 1.44 \pm 0.021 \pm 0.054$
\cite{NA49EDEPPHI} and $RMS_{\rm{y}}(\textrm{p+p}) = 0.89 \pm 0.06$
\cite{NA49PHI158}, see also left panel of \Fi{fig:NA49_rapwidths}).
For instance, such a significant difference between p+p and A+A is not
seen for charged kaons (\kplus: $RMS_{\rm{y}}(\textrm{p+p}) = 1.20$
and $RMS_{\rm{y}}(\textrm{Pb+Pb}) = 1.22$, \kmin:
$RMS_{\rm{y}}(\textrm{p+p}) = 1.01$ and $RMS_{\rm{y}}(\textrm{Pb+Pb}) =
1.14$, both at \sqrts~= 17.3~GeV \cite{NA49KPI40158,NA49PPKAON}).  

%
\begin{figure}[th]
\begin{center}
\includegraphics[width=0.85\linewidth]{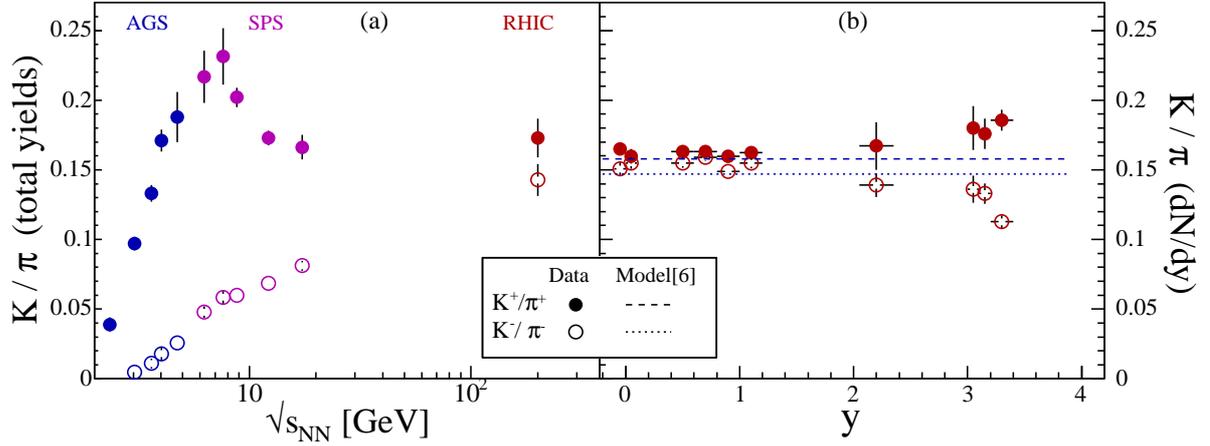}
\end{center}
\caption{The K/$\pi$ ratios, measured in full phase space for central
Pb+Pb/Au+Au collisions, as a function of \sqrts\ (left) and of
rapidity (right) at \sqrts~= 200~GeV \cite{BRMSRAP200}.  The dotted
and dashed lines correspond to predictions from a statistical model
with rapidity independent freeze-out parameters \cite{PBM7}.}
\label{fig:BRAHMS_kpi_rap}
\end{figure}
%

In a publication by the BRAHMS collaboration \cite{BRMSRAP200} the
K/$\pi$ ratios measured at different rapidities for central Au+Au
collisions at \sqrts~= 200~GeV were compared to the corresponding
ratios determined at midrapidity, but at different center-of-mass
energies.  While at top RHIC energy the \kmin/\pim\ ratio drops when
going from midrapidity to forward rapidities, the \kplus/\pip\ ratio
rises.  Since the \kplus\ production is favored by the presence of
baryons via associated production mechanisms (e.g. $\textrm{p} +
\textrm{p} \rightarrow \textrm{p} + \kplus + \lam$), the higher
net-baryon number at forward rapidities at RHIC can thus increase the
number of \kplus\ relative to \kmin\ and explain the rise in the
\kplus/\pip\ ratio.  As the net-baryon number at midrapidity is
increasing when going toward lower center-of-mass energies, the
\kplus/\pip\ ratio should rise to a similar level when the same
net-baryon number is reached.  Figure~\ref{fig:BRAHMS_kpi_rap}
contrasts the two dependences.  The \kplus/\pip\ ratio measured at
very forward rapidities for \sqrts~= 200~GeV roughly corresponds to
the one observed at midrapidity and \sqrts~= 8.7~GeV.  There seems to
be a universal dependence of the \kmin/\kplus ratio on the \pbar/p
ratio, if measurements at different energies and rapidities are
compared (see \Fi{fig:BRAHMS_kmkp_ppbar}).

In the context of statistical models, this dependence on the
net-baryon density is described by the baryo-chemical potential \mub.
If the chemical freeze-out temperature is assumed to be constant,
which is a good approximation for $\sqrts \ge 17.3$~GeV , the change
in \mub\ caused by the changing net-baryon number or the changing
center-of-mass energy would therefore be the only parameter driving
the changes in the other particles ratios.  However, this is not a
universal behavior valid everywhere.  For instance, the \kplus/\pip\
ratio is essentially independent of rapidity at \sqrts~= 62.4~GeV,
even though the net-baryons are changing quite dramatically with
rapidity \cite{BRMSRAP62}.  Assuming that there are no large
differences between the \pip\ and \pim\ rapidity spectra (only \pim\
spectra have been measured), there are indications that for Pb+Pb
collisions at \sqrts~= 17.3~GeV the \kplus/\pip\ ratio drops
instead of rises with increasing rapidity, which is opposite to the
behavior seen at \sqrts~= 200~GeV.  This can be deduced from the fact
that at \sqrts~= 17.3~GeV the \kplus\ rapidity width is smaller that
the one for pions (see left panel of \Fi{fig:NA49_rapwidths})
\cite{NA49KPI40158}, while it is wider at top RHIC energy (see right
panel of \Fi{fig:BRAHMS_rapidity}).  Since at \sqrts~= 17.3~GeV the
net-baryon number is slightly rising toward forward rapidities (at
least up to $y = 1.8$ \cite{QM06}), this would indicate that it is not
the only driving factor for the relative strangeness production.  As
already mentioned in \Se{sec:baryonratios}, a study of the rapidity
dependence of the chemical freeze-out parameters derived from fits
with a statistical model shows that both, \tch\ and \mub, do depend on
rapidity \cite{BECATTINI8}.  In the same analysis, it has been found
that the dependences on $y$ are not the same at different
center-of-mass energies.  The observed dependence of statistical model
fit results on whether midrapidity \dndy\ or 4$\pi$ integrated data
are used as input (see discussion in \Se{sec:statmodels}) seems to be
due to this effect.  Strangeness production in the fragmentation
region is apparently different than in the central fireball and it
starts to affect the measured rapidity distributions for \sqrts\ below
top SPS energies.


\clearpage
\section{System size dependence}

Strange particle production rates in heavy ion collisions, relative to
elementary p+p collisions define the strangeness enhancement
$E_{\rb{S}}$ (\Eq{eq:enhancefactor}) as already discussed in
\Se{sec:basicobservations} of this article. Whether the strangeness
enhancement in larger systems is due only to the phase space
suppression in the smaller systems, such as p+p or p+A, or is indeed
due to an increased strangeness production from a deconfined partonic
medium is still an ongoing discussion.

An rapid increase of the strangeness enhancement factor in small
systems (up to \npartch~=~50) has been suggested by canonical models
due to the restriction in phase space.  In other words the volume is
not sufficiently large to contain the necessary strangeness content
for statistical strange particle production.  Complete statistical
production can be assumed when the system reaches its chemical
equilibrium and is thermalized.  Strangeness equilibration requires a
finite number of binary collisions in an extended volume.  This can
only be achieved if the number of participants is larger than 50. In
comparison, the light quark hadrons equilibrate faster, i.e. in a
smaller system with less binary collisions.  In p+p collisions the
underproduction of strangeness due the limited number of binary
collisions (one binary collision) is largest, which means that
strangeness is suppressed, at least for collision energies up to RHIC
energies.

Any centrality scaling of strangeness production with respect to the
p+p yield therefore includes a strangeness suppression factor.  This
means that any apparent strangeness enhancement factor $E_{\rb{S}}$ as
a function of centrality is foremost dominated by the transition of a
non-equilibrated small heavy ion system into an equilibrated large
system.  Any remaining additional enhancement could be interpreted as
a phase transition signature.  The strangeness enhancement as a
function of system size, \npartch\, and as a function of collision
energy is shown in \Fi{fig:npart1} (left) for the $\Omega$ in Pb+Pb
collisions \cite{TOUNSI1}.

%
\begin{figure}[h]
\begin{center}
\vspace{-0.8cm}
\begin{minipage}[b]{0.45\linewidth}
\begin{center}
\includegraphics[width=\linewidth]{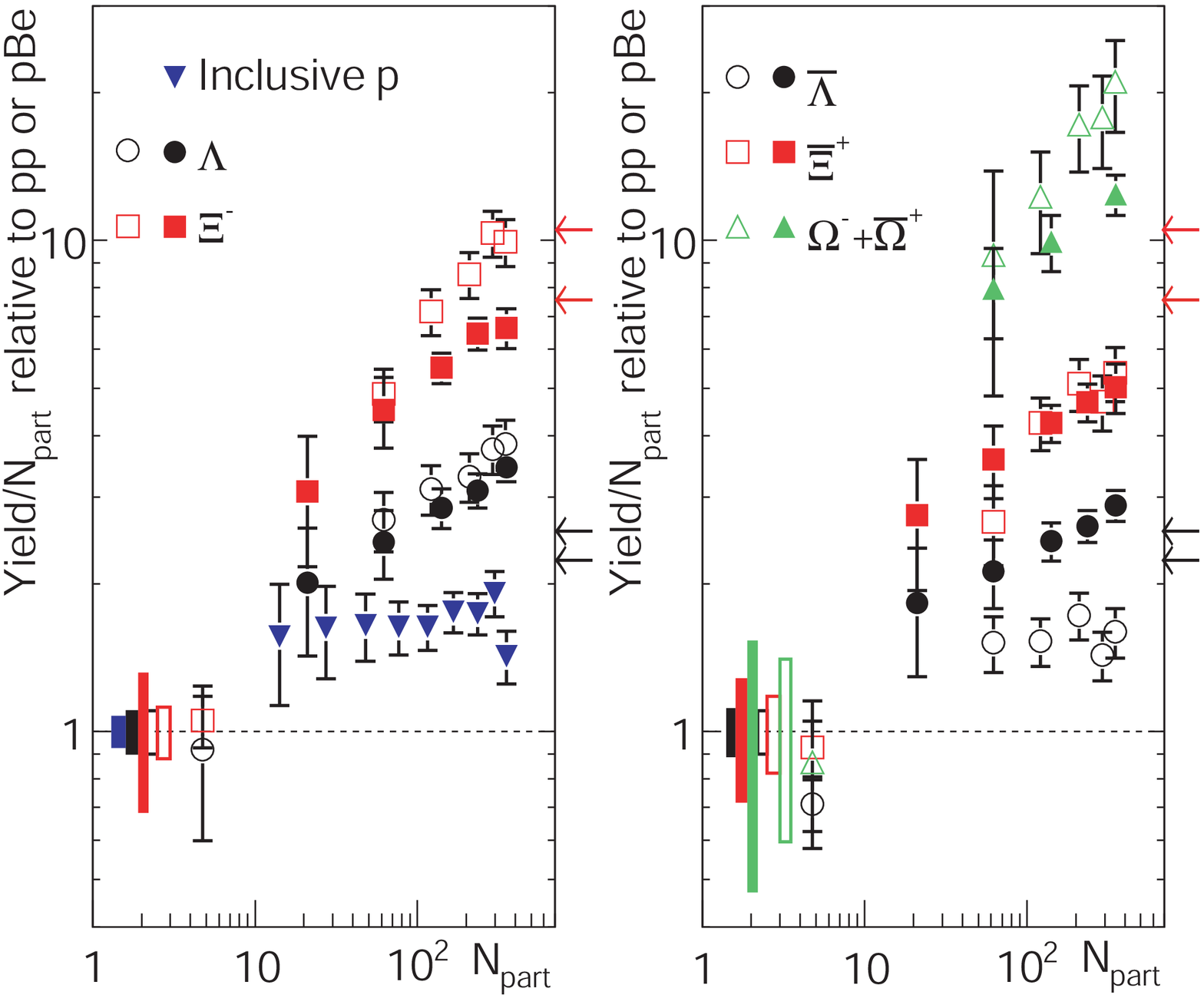}
\end{center}
\vspace{2pt}
\end{minipage}
\begin{minipage}[b]{0.49\linewidth}
\begin{center}
\includegraphics[width=\linewidth]{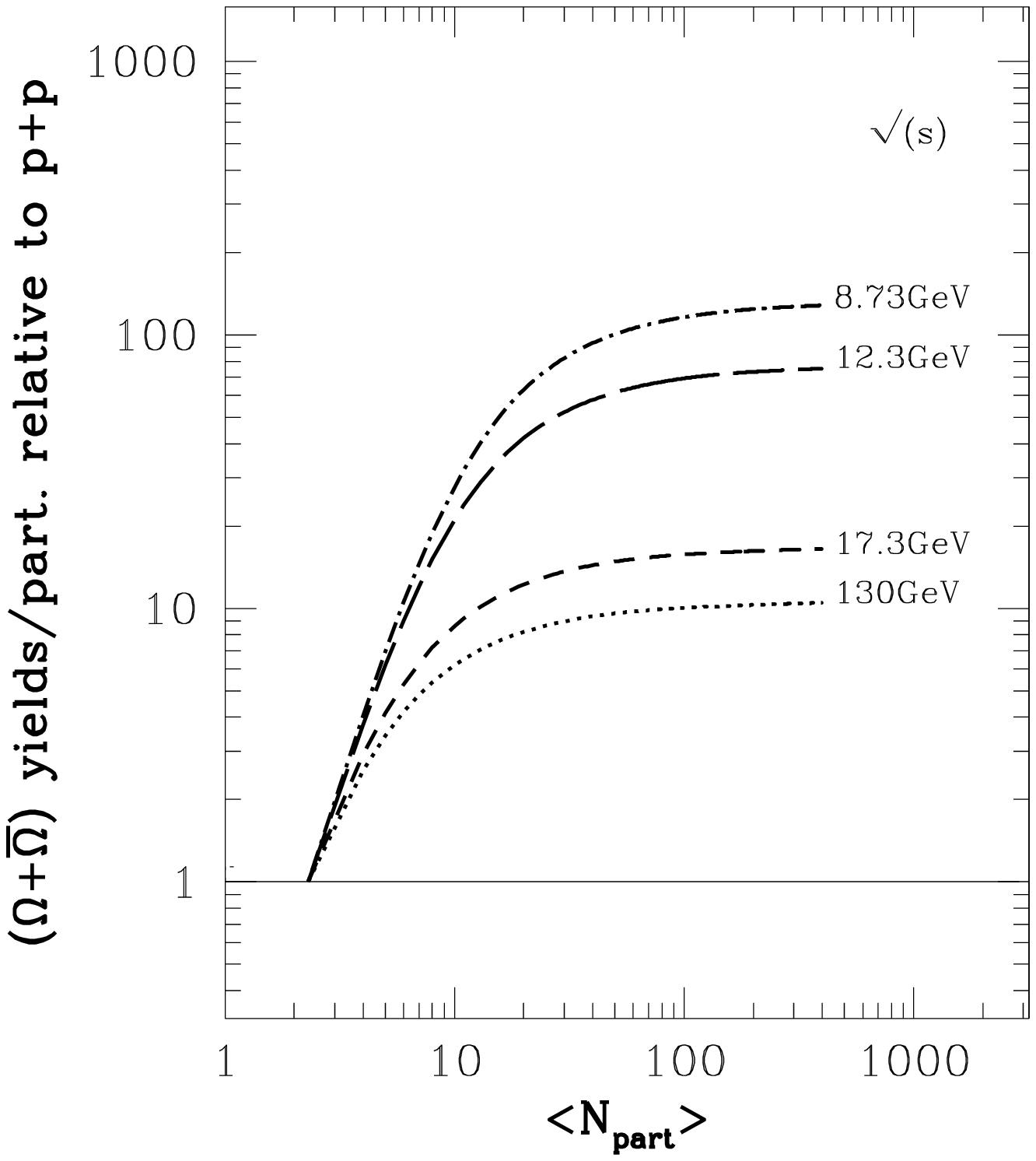}
\end{center}
\end{minipage}
%
\end{center}
\vspace{-1cm}
\caption{Left: The enhancement of inclusive protons, \lam, \lab, \xim,
\xip, and \omomp\ around midrapidity as measured by the STAR
collaboration for Au+Au collisions at \sqrts~=~200~GeV (solid symbols)
\cite{STARENHANCE} and by the NA57 collaboration for Pb+Pb collisions
at \sqrts~=17.3~GeV (empty symbols) \cite{NA57ENHANCE}. The boxes at
unity show the combined statistical and systematical errors in the
p+p, respectively p+Be, reference data.  The arrows mark statistical
model predictions for a grand-canonical ensemble and two different
temperatures ($\tch =$~165~MeV and $\tch =$~170~MeV).  Red arrows
correspond to \xim\ (\xip), black arrows to \lam\ (\lab).
Right: Centrality dependence of relative enhancement of the
yields of \omomp\ normalized by the number of participants in central
Pb+Pb to p+p reactions at different collision energies \cite{TOUNSI1}.
}
\label{fig:npart1}
\end{figure}
%

When the enhancement factor levels out it suggests that the collision
region has reached sufficient size and the small volume size effects
can be neglected.  In statistical models, the predicted equilibration
of the strangeness  production, independent of the incident energy, is
realized for sufficiently large volumes ($\npartch > 50$).  At the
highest SPS energy and the RHIC energies the $\Omega$ yield increases
by about an order of magnitude due to this phase space effect
alone.  Lower energies yield an even larger enhancement factor due to
the more dominant suppression of strangeness production in small
systems.  Any additional enhancement due to deconfinement would have
to be assessed on top of this 'trivial effect', i.e. it should
manifest itself as an enhancement relative to the equilibrium
value.  Since it is assumed that at RHIC energies and below the QGP
will arise only in sufficiently large systems, a systematic study of
the volume dependence of the enhancement factor might give information
about the onset of QGP creation.

Recent results from the LHC in elementary collisions might indicate
that in high multiplicity events a phase transition might occur in p+p
collisions.  At RHIC energy, though, this evidence could not be obtained
due to the low collisions energy \cite{ALICEPP1,ALICEPP2,CMSPP}.  

The dependence of strangeness production on the system size (volume)
of the collisions has been investigated via two methods, either by the
variation of the target and beam nucleus species or by the selection
of the centrality (impact parameter) of the collisions. In order to
characterize the collision geometrical calculations such as Glauber
\cite{GLAUBER1,MILLER1} models are used to determine the interaction
volume, the number of participating nucleons ( = wounded nucleons) and
the number of binary collisions in the heavy ion reaction.

Measurements from STAR and NA57 at \sqrts~= 200~GeV and \sqrts~=
17.3~GeV are shown in left panel of \Fi{fig:npart1}.  They reveal a
steady increase of strange hadron production with increasing system
size (\npartch) and increasing strange quark content. A saturation
(leveling off of the enhancement factor) for large systems as
predicted by statistical models is not detected. In fact, the
enhancement factor continues to increase up to the most central
collision (\npartch~=~350).  This observation can have different
explanations.  For instance, it might suggest that even in the most
central events the full equilibrium value of a grand-canonical
ensemble is not yet reached for multi-strange particles.

%
\begin{figure}[h]
\begin{center}
\begin{minipage}[b]{0.49\linewidth}
\begin{center}
\includegraphics[width=\linewidth]{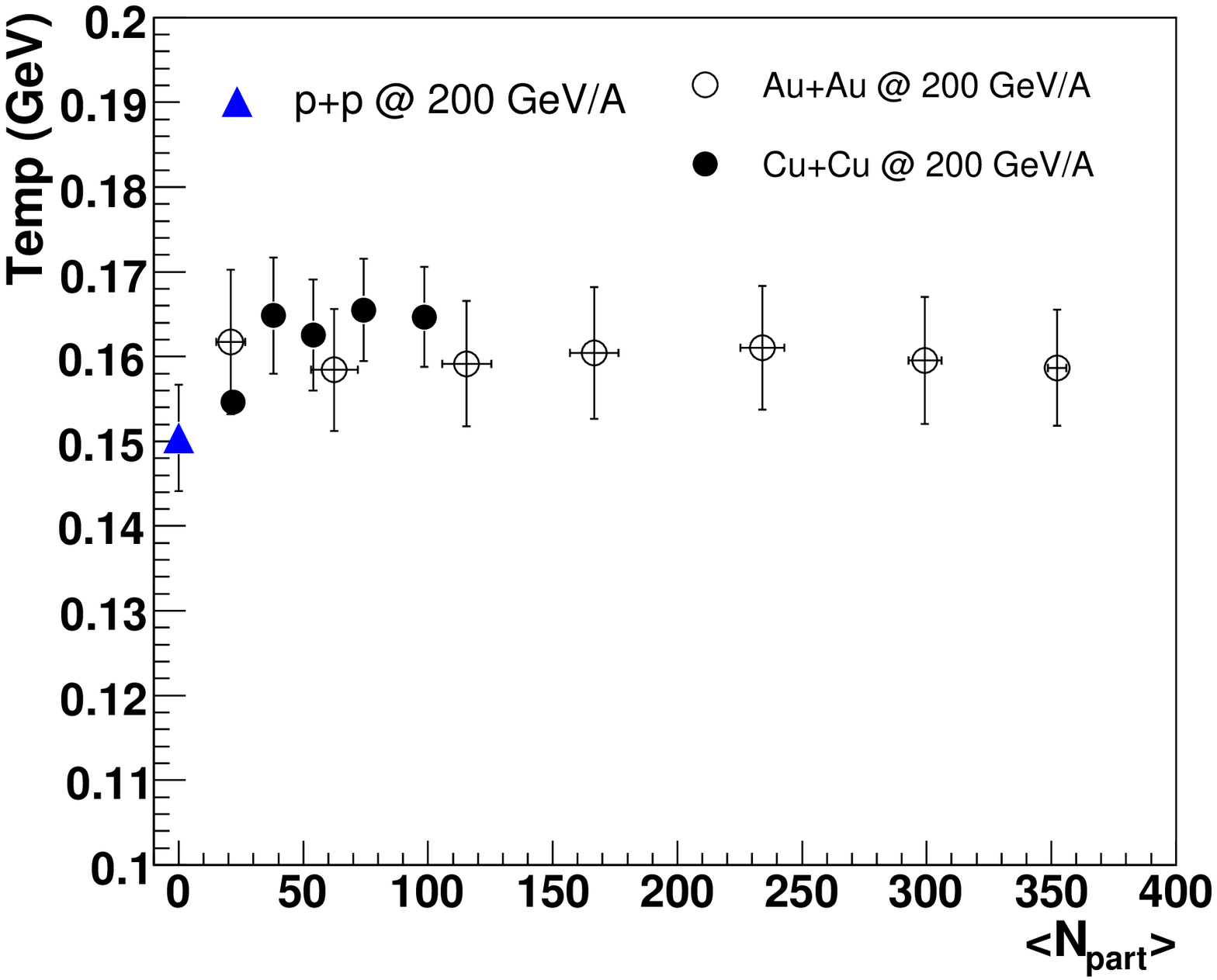}
\end{center}
\end{minipage}
\begin{minipage}[b]{0.49\linewidth}
\begin{center}
\includegraphics[width=\linewidth]{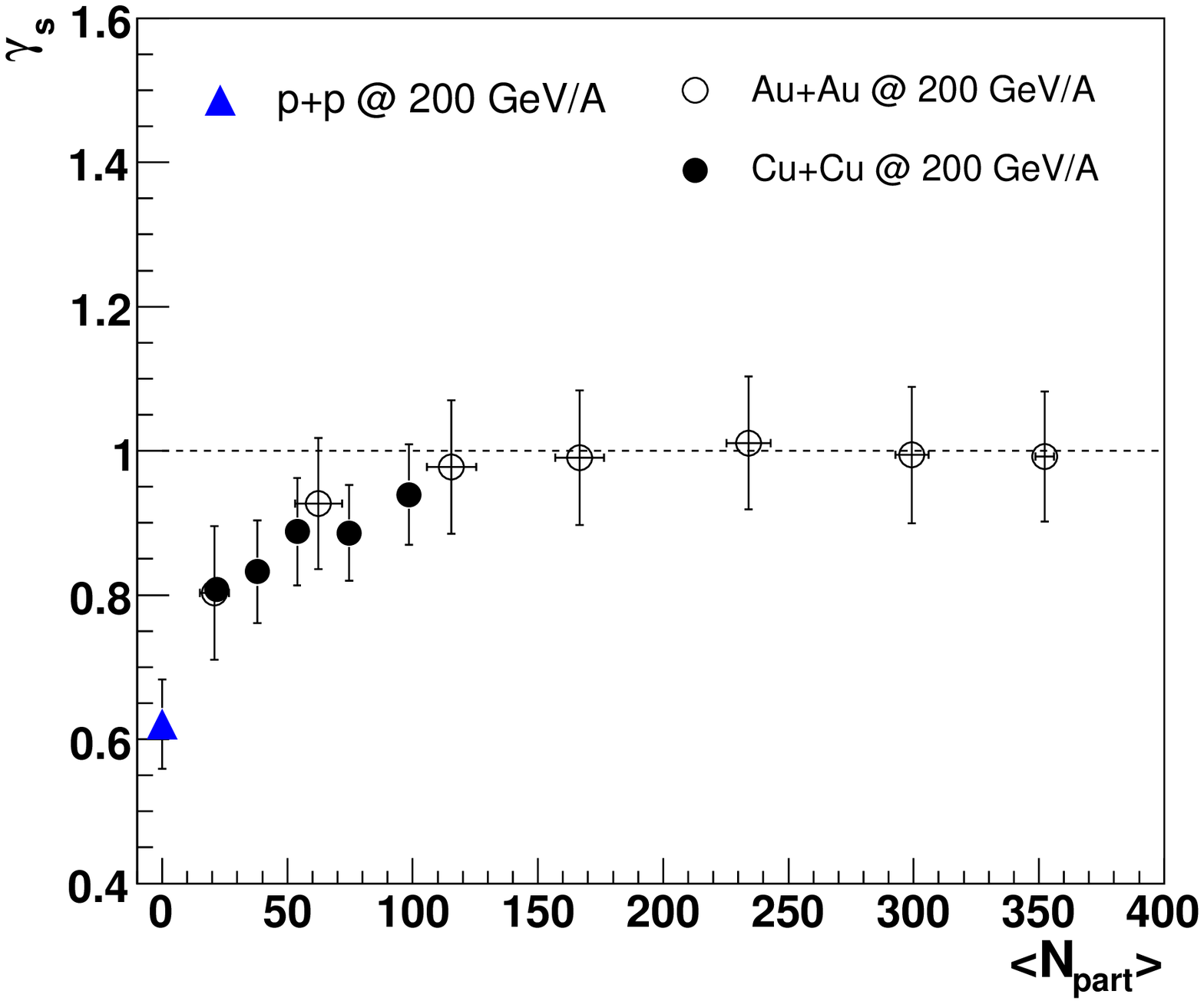}
\end{center}
\end{minipage}
\end{center}
%
\caption{Temperature (left) and strangeness saturation factor (\gams)
(right) as a function of the number of participants (\npartch) for all
particles measured by the STAR experiment \cite{TAKAHASHI1,TAKAHASHI2},
as obtained with the THERMUS code \cite{WHEATON1}.}
\label{fig:gammas}
\end{figure}
%
\newpage

An alternative explanation for the enhancement above \npartch~=~50
would be the observation of strangeness enhancement due to a QGP
phase in a grand-canonical system, i.e. in systems where the volume
effects of the strangeness suppression from p+p is not present
anymore. Calculations of the strangeness saturation factor \gams\
(\Fi{fig:gammas} right), using a thermal model fit to STAR data
\cite{TAKAHASHI1,BARANNIK1} are consistent with this interpretation. 
The saturation factor \gams\ approaches unity for \npartch~=~100,
which can be interpreted as evidence for an equilibrated and
strangeness saturated system at these system sizes.  Therefore, any
further enhancement of strangeness production from \npartch~=~100 on
might be related to an additional contribution from a deconfined
medium.  At RHIC energies this could account for about half of the
enhancement factor for $\Omega$ baryons measured in central Au+Au
collisions and compared to p+p collisions.


\subsection{Core and corona model}
\label{sec:corecorona}

A different conclusion can be reached if one considers a description
of the interaction region in a heavy ion collision, which
distinguishes between independent core and corona volumes in
centrality dependent heavy ion collisions \cite{BECATTINI2,BOZEK,
WERNER1,AICHELIN1}.  Here the dense core behaves like a QGP volume,
whereas the surrounding corona can be understood as a superposition of
independent p+p collisions.  The justification for such an approach is
based on the realistic modeling of the nuclear surface rather than the
very schematic rigid sphere calculations used in earlier models.

The effect of core-corona models is best investigated in the
centrality dependence of particle production since the corona fraction
continues to increase for the more peripheral systems.  In other
words, the core fraction, which represents the QGP volume, increases
with centrality.  The system might be equilibrated in a
grand-canonical regime even at small \npartch\ but due to the volume
to surface ratio increase the strangeness enhancement factor continues
to increase.  

In order to investigate the strangeness enhancement, this volume
effect of the core-corona ratio has to be taken into account and
subtracted from the measured strangeness enhancement factor.  The
remaining enhancement will be the real "strangeness enhancement".

\subsubsection{Centrality dependence}

To investigate the strangeness enhancement as a function of the
collision centrality the additional effects of the change in surface
to volume ratio will be taken into account by unfolding the source
into a corona (surface) and core term.  The total volume is the
superposition of the core and corona volume.  Both areas will be
treated independently, i.e. there are no interactions between core and
corona particles.

As a consequence of this superposition in heavy ion collisions one has
to calculate the relative weight, in terms of particle production, of
core and corona separately as a function of centrality.  The particle
production of the core volume is described by a partonic medium in
thermal equilibrium whereas the corona is described by the
nucleon-nucleon collisions modeled by a statistical hadronization
model \cite{BECATTINI2} (Becattini) or alternatively by the EPOS
model \cite{WERNER1}.

\begin{figure}[h]
\begin{center}
\begin{minipage}[b]{0.47\linewidth}
\begin{center}
\includegraphics[width=\linewidth]{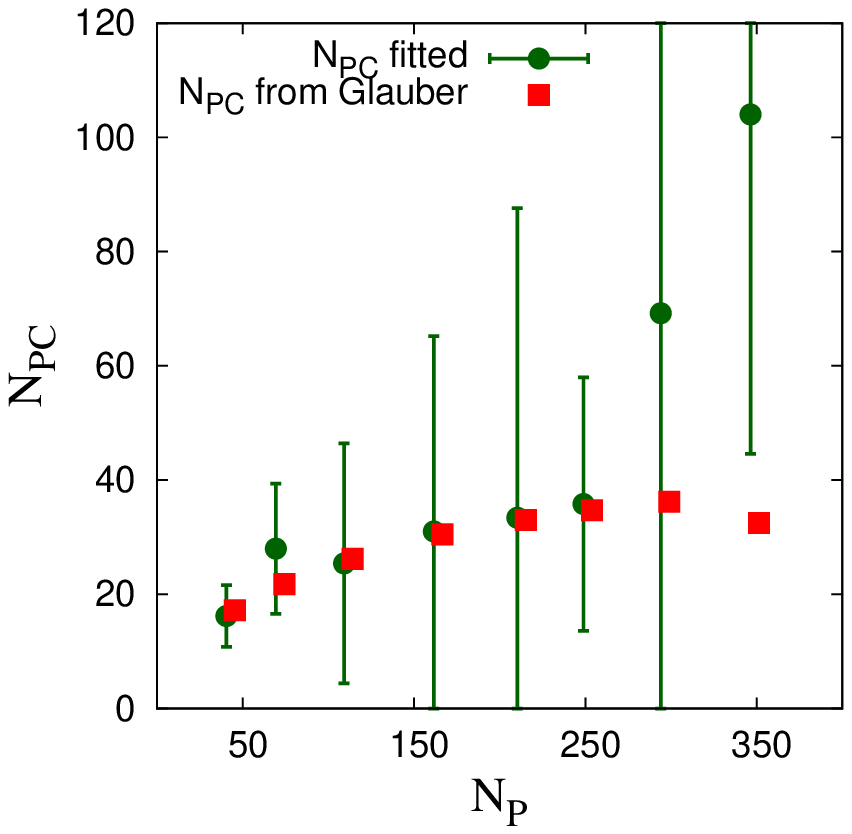}
\end{center}
\end{minipage}
\begin{minipage}[b]{0.47\linewidth}
\begin{center}
\includegraphics[width=\linewidth]{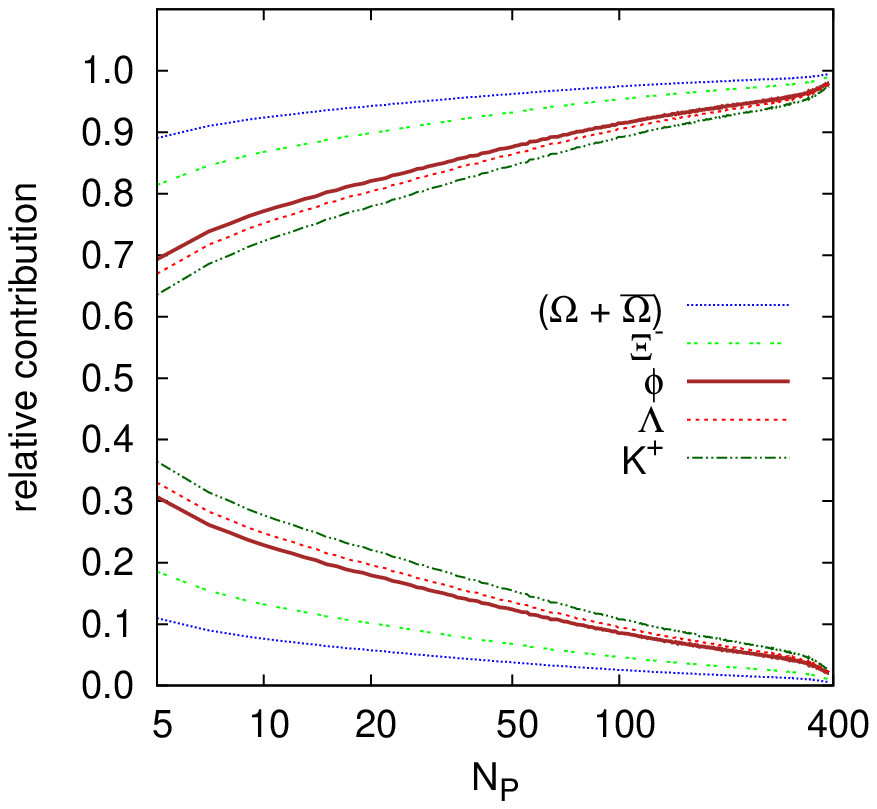}
\end{center}
\end{minipage}
\end{center}
\vspace{-1cm}
\caption{Left: Number of corona participants \npc\ at different
centralities as a function of the total number of participants \np.
The square dots denote the values calculated with a Glauber model
while the round ones denote the values arising from fitting \npc\ as a
free parameter.  The round symbols are shifted 5 units of \np\
leftward for clarity.
Right: Fraction of produced particles coming from core (upper
lines) and corona (lower lines) as a function of centrality
\cite{BECATTINI9} (Becattini).}
\label{fig:core_corona_plot_2}
\end{figure}

Figure~\ref{fig:core_corona_plot_2} (left) shows the corona
participants (\npc) as a function of the number of total
participants \npartch, which increases sightly with centrality.
Although the number of corona participants increases their relative
fraction of (\npc/\npartch) drops dramatically from around 50\% at
\npartch~=~50 to 10\% at \npartch~=~350.  This fraction can be
calculated either with a Glauber model (red squares) or from a fit
(green points) to the strange particle yields of K, $\Lambda$, $\phi$,
$\Xi$ and $\Omega$ (\Fi{fig:core_corona_plot}) when using relative
contributions from corona and core type mechanisms as shown in
\Fi{fig:core_corona_plot_2} (right).  The fitted \npc\ and the \npc\
from Glauber model calculations are in agreement.  However the errors
on the corona participants \npc\ derived from the data are very large
and therefore not sensitive to falsify this approach.

%
%

%
\begin{figure}[h!]
\begin{center}
\includegraphics[angle=270,width=0.65\linewidth]{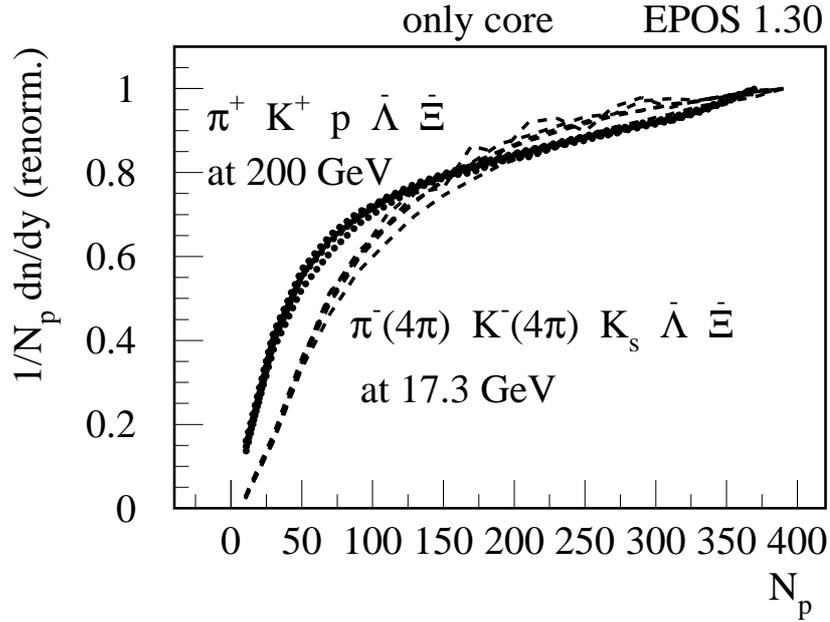}
\end{center}
%
\caption{Multiplicity per participant as a function of \np\
for the core part only. The results for \pim, \kmin, \pbar, \lab,
\xip\ in Au+Au collisions at \sqrts~= 200~GeV (dotted lines), and for
\pim, \kmin, \kzero, \lab, \xip\ in Pb+Pb collisions at \sqrts~=
17.3~GeV (dashed lines) (Fig. is adapted from \cite{WERNER1} (EPOS)).}
\label{fig:core_corona_plot_3}
\end{figure}
%

To quantify the relative contribution of the two components, the
fraction of nucleons that scatter more than once $f(\npartch)$ can be
used.  $f(\npartch)$ can simply be calculated with a Glauber model
\cite{GLAUBER1}.  This quantity allows for a natural interpolation
between the yields $Y$ measured in elementary p+p 
($= Y_{\textrm{corona}}$) and in central nucleus-nucleus collisions 
($= Y_{\textrm{core}}$):
\begin{equation}
Y(\npartch) = \npartch \: [ f(\npartch) \: Y_{\textrm{core}}  \:
                     + \: (1 - f(\npartch)) \: Y_{\textrm{corona}} ]
\end{equation}

The contribution of the core part alone is shown in
\Fi{fig:core_corona_plot_3} for a calculation using the EPOS model,
which can be compared to the result from an alternative model by
Becattini shown in \Fi{fig:core_corona_plot_2} (right).  The overall
trend of increased particle production from the core volume of this
models is the same.  However the particle yields differ, mostly in the
region of small number of participants.  This discrepancy can be
investigated experimentally after statistically more significant data
samples are obtained in particular for either small collisions systems
or peripheral collisions of the heavy systems.

In summary, the derived strangeness enhancement factors from different
core-corona models shown in \Fi{fig:core_corona_plot} (left:
Becattini, right: EPOS) describe the steady increase of the data from
p+p up to the most central heavy ion collisions.  

The \xip\ yields in the more central collisions are better described by
Becattini than the EPOS Model.  However, this model is over-predicting
the \xip\ yield for peripheral collisions in the range of $\npartch <
60$, which might be a hint that the centrality dependence cannot
exclusively be described by a core/corona volume effect.

A final interpretation of this result will depend strongly on the
modeling of the core volume.  If the dynamics of the produced
particles require the system to be partonic in nature, which is also
indicated by measurements of the radial and anisotropic expansion of
the system, then the enhanced strangeness production is not just a
volume effect, but indeed a signature for deconfinement, at least in
the core of the collision volume.  The main contribution of the
core-corona models to this discussion is a more precise understanding
of the centrality dependence of the strange particle production which
goes beyond the simple phase space argument of strangeness suppression
in small systems. 

%
\begin{figure}[h]
\begin{center}
\begin{minipage}[b]{0.44\linewidth}
\begin{center}
\includegraphics[width=\linewidth]{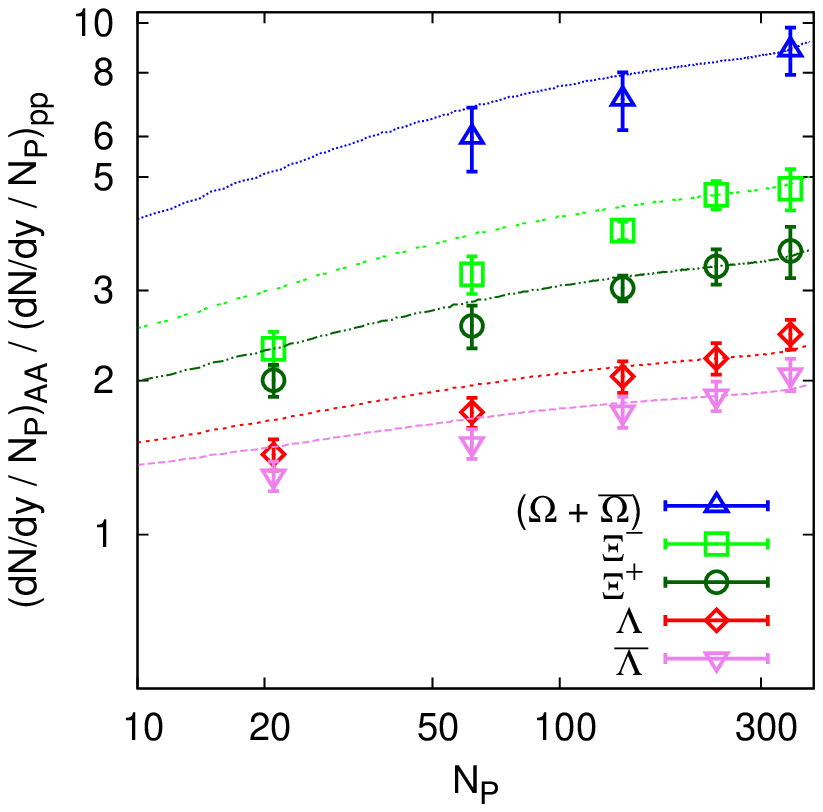}
\end{center}
\end{minipage}
\begin{minipage}[b]{0.55\linewidth}
\begin{center}
\includegraphics[angle=270,width=\linewidth]{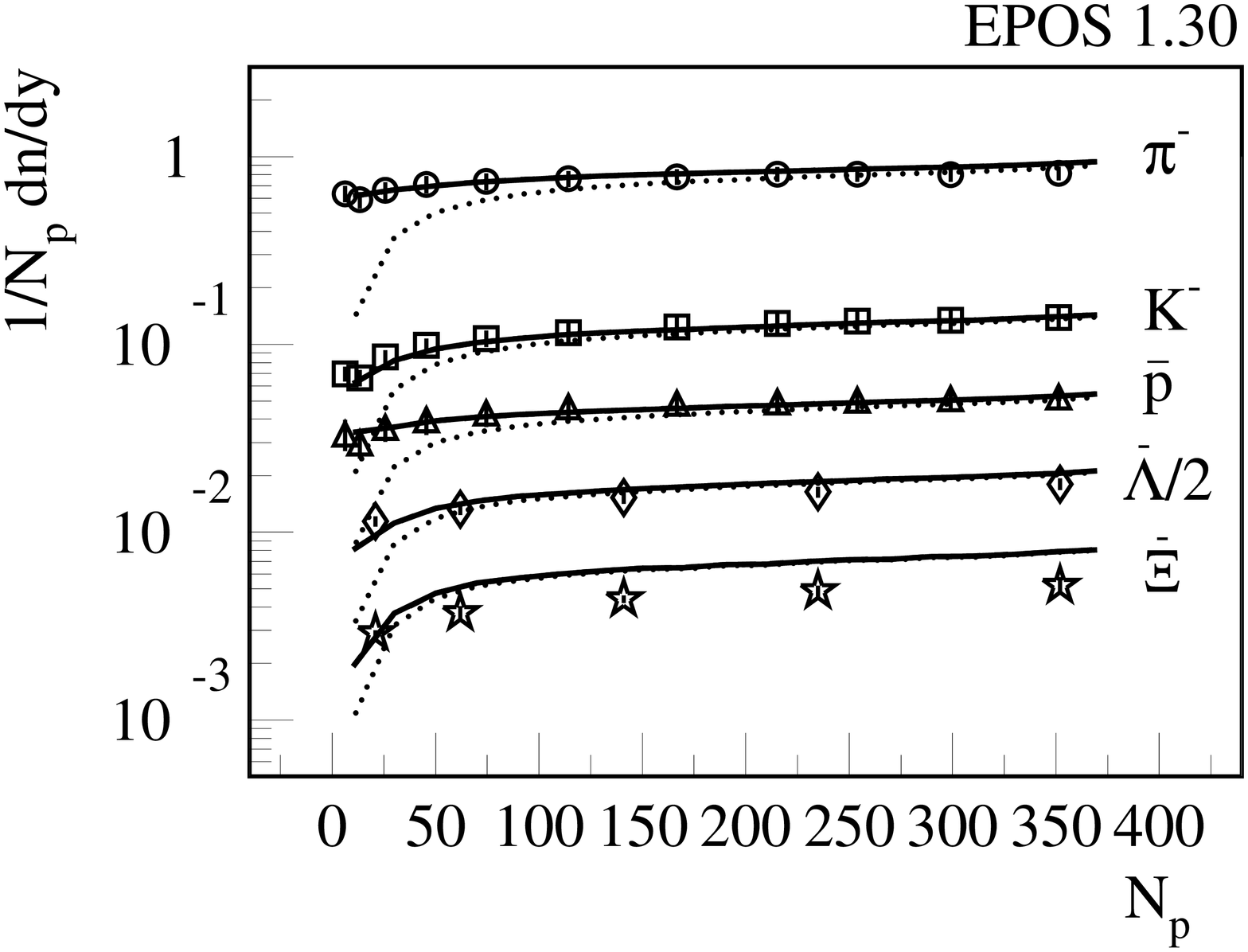}
\vspace{2pt}
\end{center}
\end{minipage}
\end{center}
\vspace{-1.cm}
\caption{Left: Hyperon rapidity density measured in Au+Au collisions
normalized to measurements in p+p collisions and the number of
participants at \sqrts~= 200~GeV as a function of the number of
participants \cite{STARENHANCE}.  The lines are theoretical
calculations from Becattini et al. \cite{BECATTINI2,BECATTINI9}.
Right: Rapidity density \dndy\ per participant as a function of the
number of participants (\np) in Au+Au collisions at \sqrts~= 200~GeV
(RHIC) for \pim, \kmin, \pbar, \lab, \xip\ together with the full
calculation (full lines) and the core part only (dotted lines)
(Fig. is adapted from \cite{WERNER1} (EPOS)).}
\label{fig:core_corona_plot}
\end{figure}
%

\subsubsection{Comparison of different collision systems}

A measurement which might also be explained by the core-corona
approach is the apparent difference of the strangeness enhancement
factors for different collision systems at the same \npartch.  This
was first measured at the SPS \cite{HOEHNE1} and has since then been
corroborated at RHIC energies \cite{TIMMINS1}.

%
\begin{figure}[h!]
\begin{center}
\includegraphics[angle=0,width=0.79\linewidth]{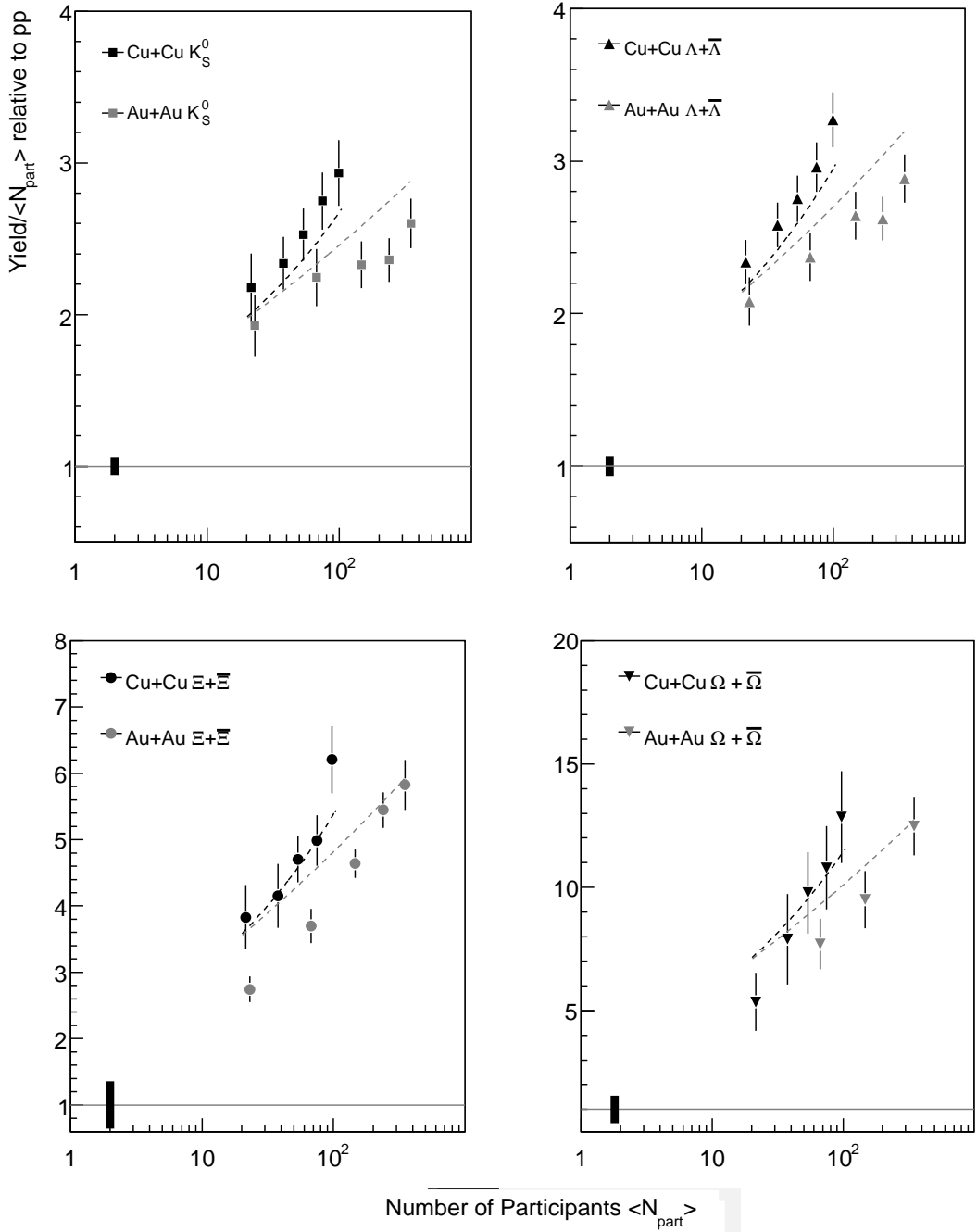}
\end{center}
\caption{The enhancement factor for (multi-)strange particles at
midrapidity in Cu+Cu and Au+Au collisions at \sqrts~= 200~GeV.  
The \lam\ and \lab\ yields have been feed-down subtracted in all
cases.  The Au+Au multi-strange data have been previously published
\cite{STARHYP200}.  The black bars show the normalization
uncertainties, and the uncertainties for the heavy ion points are the
combined statistical and systematic \cite{TIMMINS1}.  The lines
represents calculations with the EPOS model \cite{WERNER1}.}
\label{fig:sys_cucu_auau_3}
\end{figure}
%

Figure~\ref{fig:sys_cucu_auau_3} shows the difference between the
Cu+Cu and Au+Au systems at \sqrts~= 200~GeV collision energy.  For all
strange particle species the smaller system exhibits larger
strangeness enhancement at the same \npartch\ as is summarized in
\Fi{fig:sys_cucu_auau_3}.  A possible explanation is the larger core
in Cu+Cu compared to Au+Au when using a core-corona approach.  
At the same \npartch\ the Cu+Cu collision system features a larger
core fraction and thus a larger number of participants that undergo
multiple collisions according to Glauber calculations.  This is
implemented in EPOS which describes the trend of the data in Cu+Cu
better than in Au+Au collision in  \Fi{fig:sys_cucu_auau_3}.  On the
other hand the effect is also visible in the AMPT model which does not
feature separate core and corona volumes but rather uses string
breaking and mini-jet fragmentation to describe the medium.  In
general though it seems true that at the same \npartch\ the Cu+Cu
system yields a larger contribution to the strangeness yield than the
more peripheral Au+Au system.


\subsection{Scaling with \npartch\ and \nbin}

In the case of a chemically equilibrated bulk system of light quarks
we expect rather \npartch\ scaling than \nbin\ scaling for the
centrality dependence of the particle yields. This is true, for
example, for the protons at RHIC energies (\Fi{fig:npart1}
left).  Their enhancement factor, which presumes \npartch\ scaling, is
constant as a function of the system size.  In other words, the
increase in proton production from peripheral to central heavy ion
collisions scales with the number of participants.  This hints at the
dominant production of up and down quarks through thermal
parton-parton interactions rather than hard scattering.

\npartch\ scaling does not necessarily proof that the system in its
early stage is thermalized.  A particle species, whose initially
produced yield scales with \nbin\ and is thus not chemically
equilibrated, might thermally equilibrate in the hadron phase through
elastic scattering and exhibit flow features and a thermal \pt\
spectrum in central collisions.  On the other hand, species for which
the yield scales with \npartch\ might exhibit non-thermal behavior in
their kinematic spectra. 

\begin{figure}[h]
\begin{center}
\vspace{-0.5cm}
\begin{minipage}[b]{0.48\linewidth}
\begin{center}
\includegraphics[width=\linewidth]{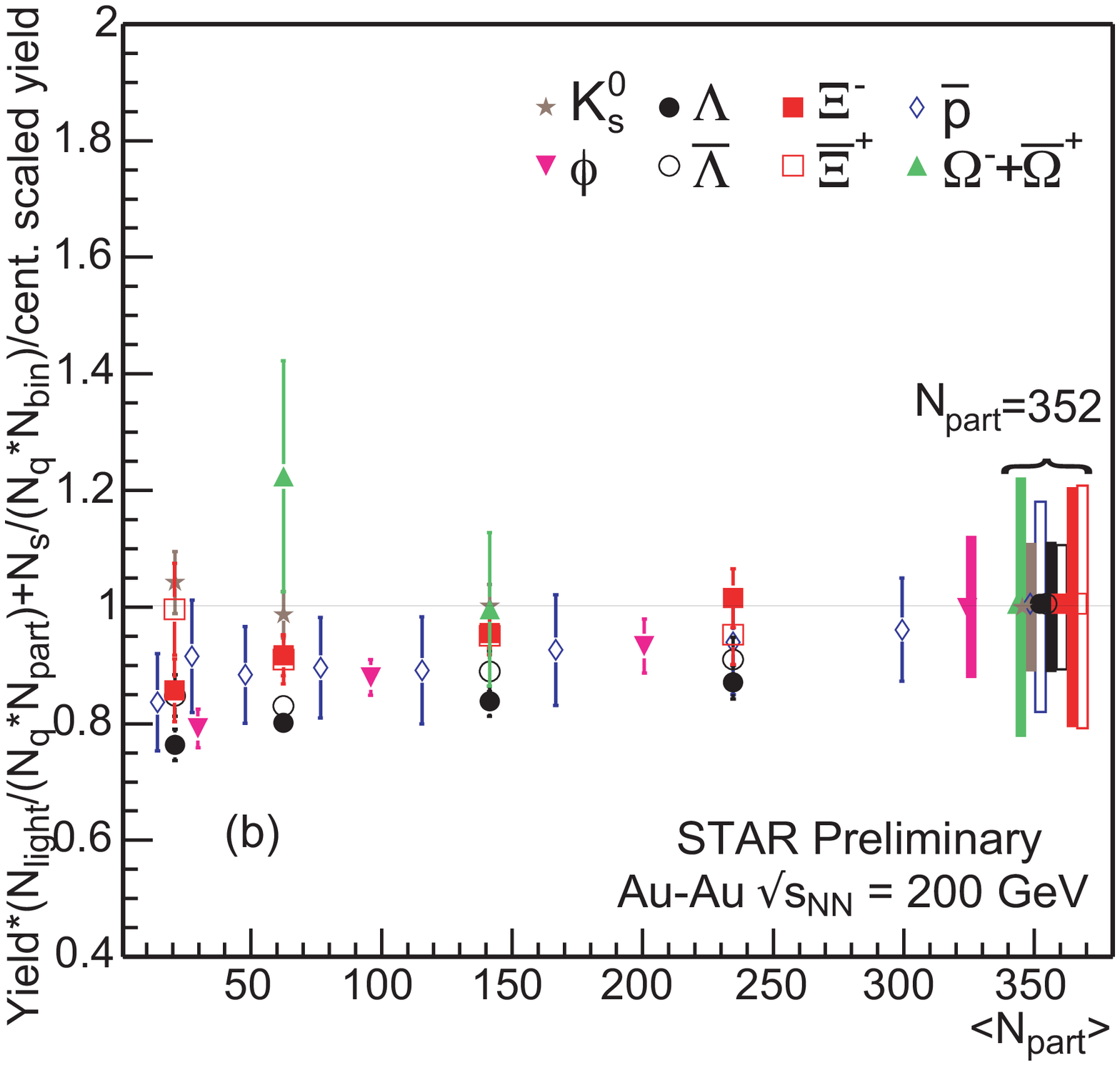}
\vspace{14pt}
\end{center}
\end{minipage}
\begin{minipage}[b]{0.44\linewidth}
\begin{center}
\includegraphics[width=\linewidth]{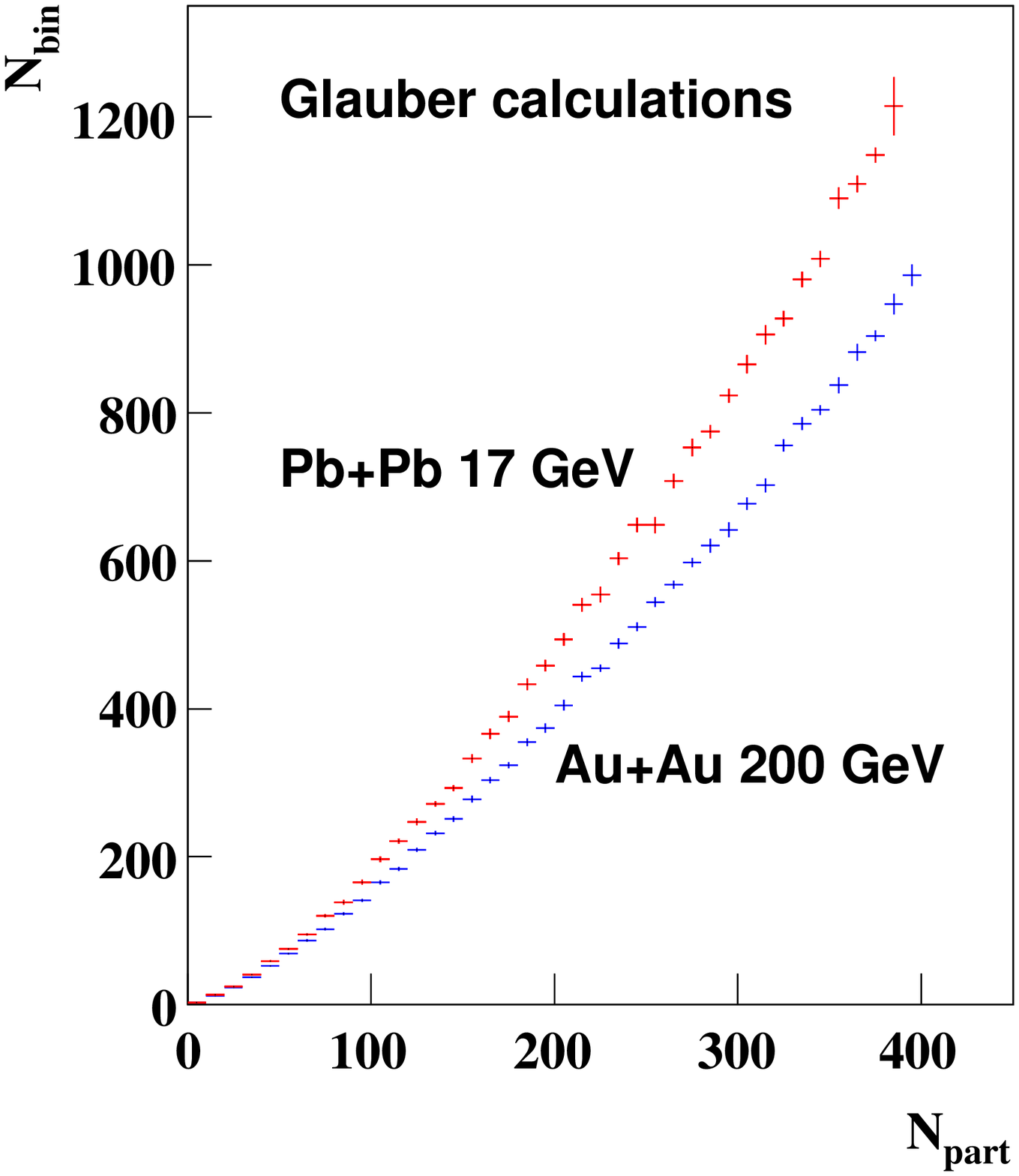}
\end{center}
\end{minipage}
\end{center}
\vspace{-1.2cm}
\caption{Left: Au+Au data scaled as defined by \Eq{eq:sfactor}
except the $\phi$, which is scaled by \npartch.  All data are
normalized to the most central bin \cite{CAINES1}.
Right: Number of participants versus number of binary collisions using
Glauber calculation for Pb+Pb and Au+Au collision systems at \sqrts~=
17.3~GeV and \sqrts~= 200~GeV \cite{GLAUBER1,MILLER1}.}
\label{fig:scaling}
\vspace{-0.5cm}
\end{figure}

Generally heavier quarks (i.e. charm and bottom) are not expected to
equilibrate chemically or thermally, and are mostly produced through
initial parton-parton interactions which scale with the number of
binary collisions, \nbin.  Since the strange particle yield continues
to increase beyond the \npartch\ scaling relative to p+p, this could
be viewed as a non-equilibrium signature for strange quarks.  An
alternative scaling method, based on an independent normalization of
the constituent quarks in a hadron was suggested by H. Caines.  The
method takes into account the specific scaling behavior of the strange
quark, which is the  \nbin\ scaling.  The new factor is defined as
\cite{CAINES1}:
\begin{eqnarray}
C_{\rb{scaling}} = \frac{N_{\rb{light}} \times \npartch}{\nq} +
               \frac{N_{\rb{s}} \times  \nbin}{\nq} \;
\label{eq:sfactor}
\end{eqnarray}
where \nq\ is the number of quarks in the particle, $N_{\rb{light}}$
is the number of light (u and d) quarks, and $N_{\rb{s}}$ is the
number of strange quarks.  Figure~\ref{fig:scaling} right shows the
relation of \npartch\ and \nbin\ for different collision energies
based on a Glauber calculation \cite{GLAUBER1,MILLER1}.  This numbers
are used to calculate the new enhancement factors in \Fi{fig:scaling}
left, which shows a flat distribution for all light and strange quark
particle species. 

This new scaling is successful on the level of 20\%.  Thus one could
conclude that the strange quark indeed scales with \nbin\ and is thus
mostly produced through initial gluon-gluon interactions.  However,
one exception is the scaling of the $\phi$, which according to
\Eq{eq:sfactor} should scale with \nbin, but appears to scale better
with \npartch.  

\newpage

%
%


\section{Strangeness at intermediate and high \pt}

The bulk of the particles ($>$~90\%) produced in heavy ion collisions
carry low momentum ($\pt < 2$~\gevc) and are produced either through
soft multiple parton scattering or statistical production from a
thermal medium.  The intermediate \pt~range is commonly defined as the
region between \pt~=~2 and 6~\gevc\ and can be considered the
transition region between hard processes, described by pQCD, and the
aforementioned soft region.  Hadron production from pure fragmentation
of the hard scattered parton is only dominant at $\pt > 6$~\gevc\
based on the RHIC heavy ion results.  In a medium we expect a modified
parton fragmentation \cite{GLV1,GLV2,ASW} which presumably softens the
fragmentation function, i.e more soft particles are produced in the
emitted hadron jet and pushed into a region which is dominated by the
physics of the bulk.  In this picture the particles in the
intermediate \pt~range can be due to either a large radial expansion
which adds additional kinetic energy to the bulk or an enhanced
quenching of high \pt\ jet fragments traversing the medium.  An
additional alternative arises if one takes into account the
possibility of forming color neutral objects through recombination of
partons in the deconfined medium.  This recombination mechanism for
hadronization has been first introduced for elementary collisions in
the late 70's \cite{HWA-RECO}.  When applied to heavy ion collisions
the initial system is a thermalized partonic medium in which partons
close in phase space can recombine to form a hadron with momentum from
the sum of the quark momentum. 

Since the thermal spectrum of partons is likely to have a slightly
higher mean momentum than the pions from the kinetic freeze-out, and
the total momentum of the final hadrons depends on the number of added
quarks, this mechanism populates predominantly the mid-\pt\ range and
favors baryon over meson production at a fixed intermediate \pt\
\cite{BASS1,GRECO}.  Thus the two main experimental measures to
quantify the interplay of bulk, fragmentation and recombination
processes are the \pt\ dependent baryon to meson ratio and the \pt\
dependent nuclear suppression factors.  Here we will first discuss the
baryon to meson ratios.


\subsection{Baryon to meson ratios}
\label{sec:bmratios}

The enhanced baryon/meson ratio in the transition from $\pt =
2 - 6$~\gevc\ in heavy ion collisions indicates that particle production
from the deconfined medium is neither simply statistical in nature nor
can it be described through fragmentation alone.  The earliest RHIC
measurements of the p/$\pi$ ratio show an increase up to \pt~=
3~\gevc\ to unity \cite{PHNXKPI130B,STARPRL97}, a value much higher
than in elementary p+p collisions \cite{DELPHIEPJC17,ALPERNPB}.

%
\begin{figure}[h]
\vspace{-0.5cm}
\begin{center}
\includegraphics[width=0.55\linewidth]{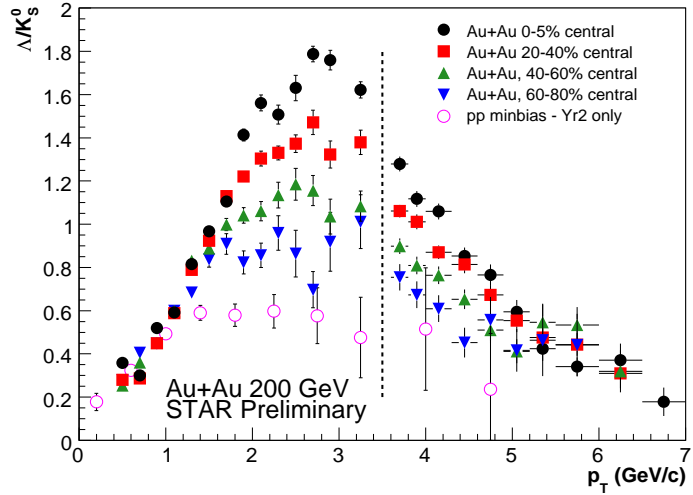}
\end{center}
\vspace{-0.5cm}
\caption{The \lam/\kzero\ ratio as a function of \pt\ for
different collision centralities for Au+Au collisions at \sqrts~=
200~GeV \cite{STARLAMONT}.}
\label{fig:lambda_kaon_ratio}
\end{figure}
%

For the strange sector \Fi{fig:lambda_kaon_ratio} shows the
\lam/\kzero\ ratio in Au+Au collisions at \sqrts~= 200~GeV out to
\pt~= 6~\gevc\ for different centralities \cite{STARLAMONT}.  The
maximum of the distribution is at \pt~= 2.8~\gevc\ for all collision
systems including the elementary p+p collisions. However, the maximum
value of the ratio for p+p collision is below unity, while it
continues to increase for more central Au+Au collisions up to 1.8 in
the 5\% most central collision.  This remarkable value means that in
certain momentum regions baryon production dominates over meson
production by almost a factor of two.  Such an effect is impossible to
model in fragmentation where the baryon production relies on the
formation of di-quark out of the fragmenting proton and is thus
considerably reduced in probability. 

Early attempts to describe this phenomenon tried to merge a radially
expanding bulk system, which pushes higher mass particles (baryons) to
larger momenta than lower mass particles (mesons), with a common
quenching model for high momentum hadrons \cite{VITEV-Q}.  Such a model
qualitatively leads to a 'pile-up' of the baryon to meson ratios at
mid-\pt, although a quantitative comparison has never been attempted.
On the other hand thermal recombination models became more
quantitative over the years and now also take into account unique
strangeness aspects in these particle ratios.  The general
recombination approach is based on the idea that since  the thermal
parton spectrum is exponentially falling as a function of momentum,
the probability to form a high momentum three quark state is higher
than the forming a two quark state of the same final state momentum.
The effect is per se not unique to strange particles as shown by
p/$\pi$ ratio, but certain aspects of the process can be tested when
comparing to non-strange ratios. First, the models require
thermalization of the strange parton in the system prior to hadron
formation \cite{HUANG}.  Second, the relevant degree of freedom in
those models is the constituent quark, i.e. the modeling requires
quarks to have a finite mass which differs between $u$, $d$ and
$s$-quarks by about 100~\mevcc\ \cite{BASS2}.  More recent models even
include recombination between bulk partons and the non-thermal hard
partons \cite{HWAYANG06}.  These models are successful in describing
the pure strange ratio of $\Omega$/$\phi$ up to $\pt \sim 4$~\gevc\ as
shown in \Fi{fig:omega_phi} \cite{STARPHIPRL07}, which includes more
than 95\% of the hadron yield.  This is of particular relevance since
the small hadronic production cross section for $\Omega$ and $\phi$
particles requires the particle generation to occur through partonic
rather than hadronic recombination.  At higher \pt\, the model fails
for the coalescence of thermal quarks and the combination of thermal
and shower (total) quarks.  Only the decreasing trend of
$\Omega$/$\phi$ from the total quark (thermal and shower) combination
is visible.  Since most of the $\phi$-mesons are made via coalescence
of seemingly thermalized $s$-quarks in central Au+Au collisions, the
observations imply that a hot and dense matter with partonic
collectivity has been formed at RHIC energies. 

Although coalescence models assume particle production from a source
of thermal partons their results can be incompatible with thermal
equilibrium of the final hadrons in the low \pt\ (bulk) region.  They
should therefore only be applied to hadrons in a medium \pt\ range
(e.g. 2~-~6~\gevc).  In order to not violate entropy conservation,
another production mechanism is required to produce a thermal
reservoir of bulk particles below 2~\gevc.  Above $\approx 6$~\gevc\
it is likely that parton fragmentation will be the most dominant
mechanism.

%
\begin{figure}[h]
\begin{center}
\includegraphics[angle=0,width=0.45\linewidth]{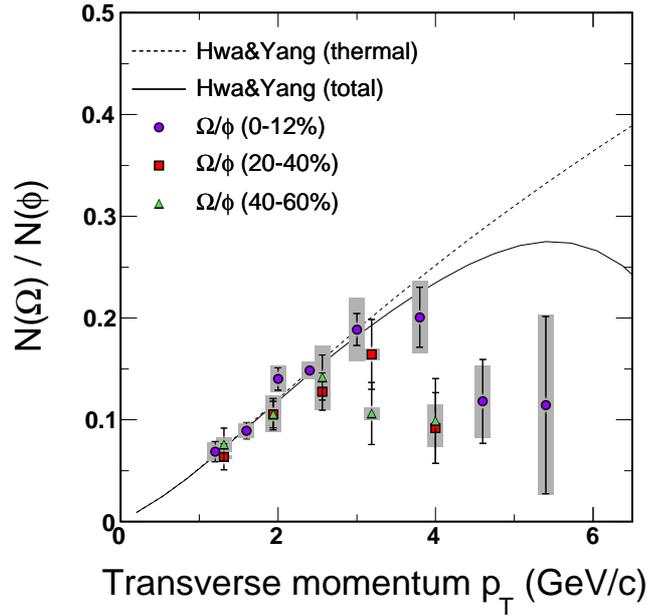}
\end{center}
\vspace{-0.5cm}
\caption{The N(\om)/N($\phi$) ratio vs. \pt\ for three centrality bins
in \sqrts~= 200~GeV Au+Au collisions.  The solid and dashed lines
represent recombination model predictions for central collisions
\cite{HWAYANG06} for total and thermal contributions, respectively
\cite{STARPHIPRL07}.  However the coalescence model should be applied
for $\pt > 2$~\gevc\ only.}
\label{fig:omega_phi}
\end{figure}
%


\subsection{Nuclear suppression factors (\raa\ and \rcp)}

In order to study the medium modification of the hard process one has
to determine the yield as a function of momentum. These \pt\
dependent studies can be quantified via the nuclear modification
factor defined by the yield in p+p ($\der^{2} \sigma^{\rb{NN}} /
\der\pt \der y$) and A+A ($\der^{2} N^{\rb{AA}} / \der\pt \der y$)
collisions, normalized by the binary collision scaling factor \nbin\
and the inelastic cross section for p+p collisions
$(\sigma_{\rb{inel}}^{\rb{NN}})$:
\begin{eqnarray}
\raa = \frac{\sigma_{\rb{inel}}^{\rb{NN}}}{\nbin^{\rb{AA}}} \:
       \frac{\der^{2} N^{\rb{AA}} / \der\pt \der y}
            {\der^{2} \sigma^{\rb{NN}} / \der\pt \der y} \;
\label{eq:raa}
\end{eqnarray}

Since the accumulation of sufficient statistics in the elementary p+p
collisions required several run periods, the early RHIC results on
nuclear suppression were based on \rcp, which is the momentum
dependent yield of central $(\der^{2} N^{\rb{cent}} / \der\pt \der y)$
divided by the peripheral $(\der^{2} N^{\rb{per}} / \der\pt \der y)$
collisions and normalized by their respective number of binary
collisions (\nbin) (see \Fi{fig:raa_helen} right).

\begin{eqnarray}
\rcp = \frac{\nbin^{\rb{per}}}{\nbin^{\rb{cent}}} \:
       \frac{\der^{2} N^{\rb{cent}} / \der\pt \der y}
            {\der^{2} N^{\rb{per}}  / \der\pt \der y} \;
\label{rcp}
\end{eqnarray}

For non-strange light quark particles this observable is roughly
equivalent to \raa\ since the light flavor production scales well from
p+p to peripheral A+A collisions. In the strange sector, though, one
would expect a distinct difference between \rcp\ and \raa\ based on
the strangeness suppression in the elementary collisions which was
discussed in the previous chapter.  \rcp\ attempts to eliminate this
phase space effect by normalizing the central spectrum to a more
peripheral spectrum where one expects the strangeness suppression to
be already minimized.

%
\begin{figure}[h!]
\begin{center}
\vspace{+0.5cm}
\begin{minipage}[b]{0.47\linewidth}
\begin{center}
\includegraphics[width=\linewidth]{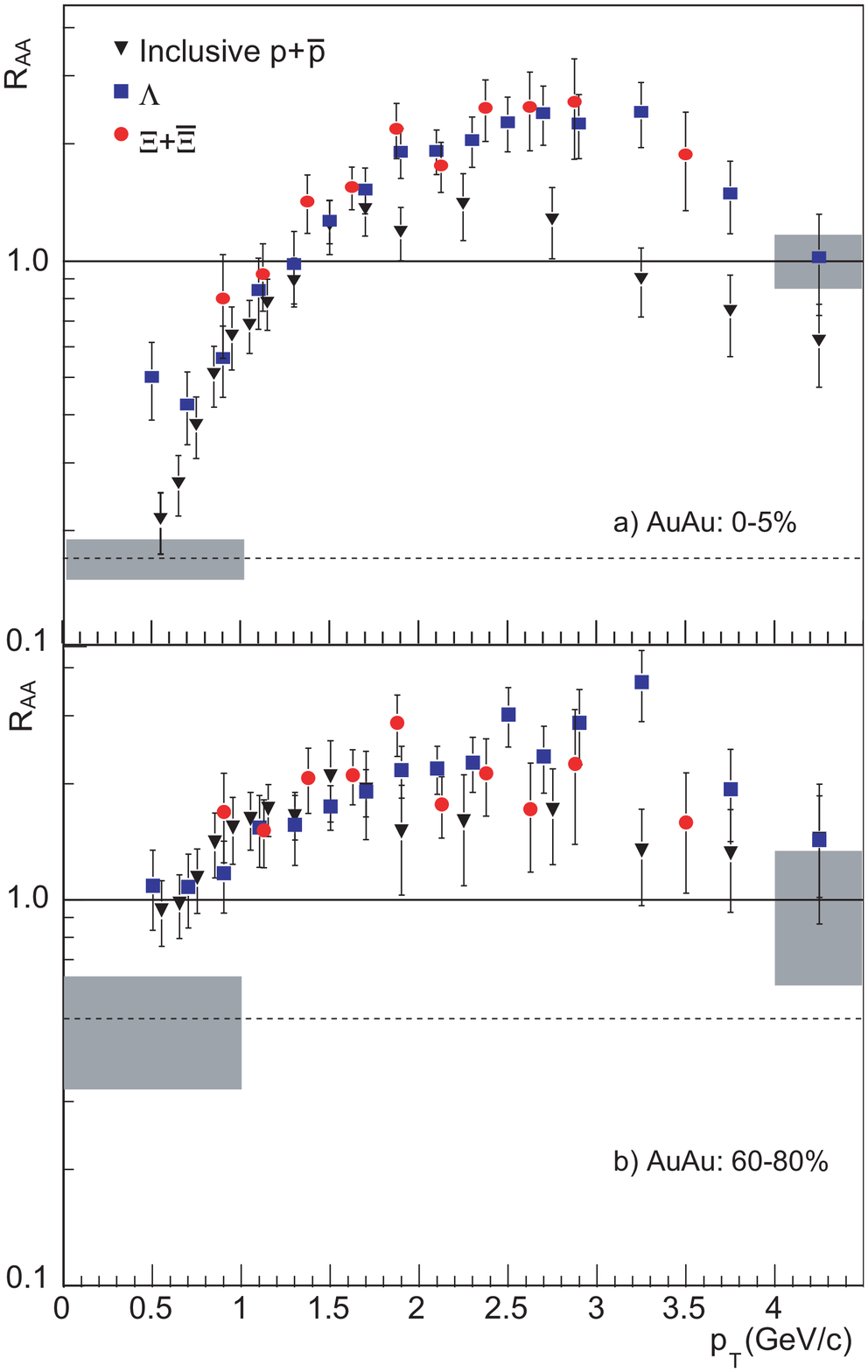}
\end{center}
\end{minipage}
\begin{minipage}[b]{0.47\linewidth}
\begin{center}
\includegraphics[width=\linewidth]{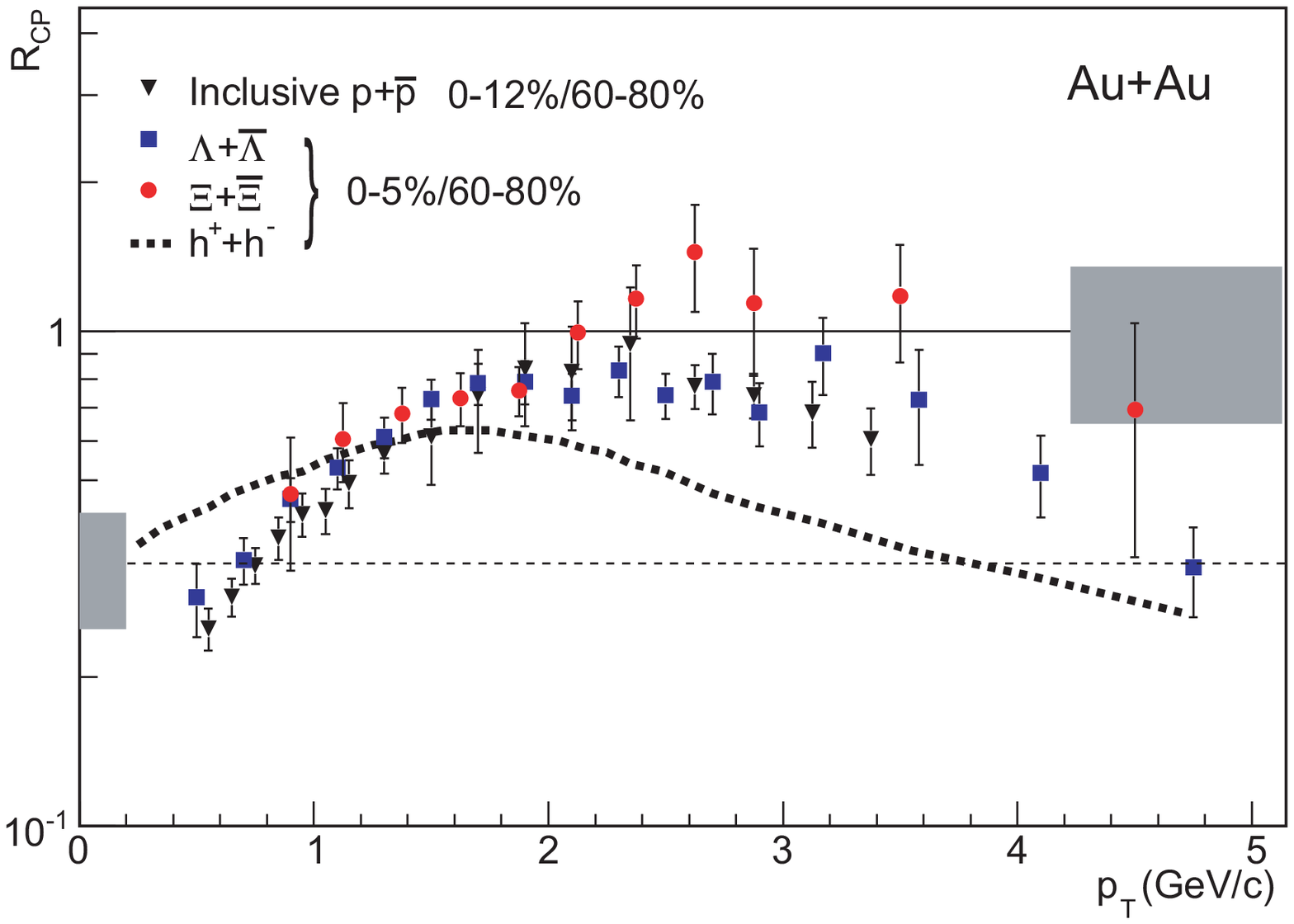}
\end{center}
\end{minipage}
%
\end{center}
\caption{Left: \raa\ for p+\pbar\ \cite{STARPPDAU,STARHIGHPT2},
\lam\ and \xim+\xip\ for a) 0-5\% and b) 60-80\% most central Au+Au
events at \sqrts~= 200~GeV.
Right: \rcp\ measured in Au+Au events for \pbar\ (0-12\%/60-80\%)
\cite{STARHIGHPT2}, \lam+\lab\ and \xim+\xip\ (0-5\%/60-80\%)
\cite{STARHYP200}. The dashed curve shows the results for
h$^{+}$+h$^{-}$ for 0-5\%/60-80\% \cite{STARHIGHPT1}.  In all plots the
statistical and systematical errors are included.  The band at unity
shows the uncertainty on the binary scaling \nbin\ and the dashed line
indicates the participant scaling (\npartch) with its uncertainty in a
gray band.}
\label{fig:raa_helen}
\end{figure}
%

\Fi{fig:raa_helen} right shows that the \rcp\ of \lam\ is in agreement
with the \rcp\ of protons out to \pt~= 3.5~\gevc, while the
\rcp\ of the $\Xi$ shows slight enhancement in the higher \pt\
region.  However, the errors are too large to draw any conclusion about
an enhancement of strange particles with respect to non-strange particles.

\Fi{fig:raa_helen} left shows \raa\ for \lam\ and $\Xi$ compared to
inclusive p+\pbar\ measurements \cite{STARPPDAU,STARHIGHPT2}.  The
\raa\ distributions of strange particles reach a maximum greater
than unity for central (top) and for peripheral (bottom) events, which
would imply a scaling close to binary collisions.  In central Au+Au
collisions the protons are significantly lower than the strange
baryons.  

Measurements of \rcp\ have shown that the Cronin effect cannot account
for the relative difference between \raa\ and \rcp\ between light
quark and strange quark particles and among the strange particles
themselves.  It  seems that the Cronin effect scales with the hadronic
mass rather than the flavor content.  Thus the Cronin effect for
lambdas and protons is almost identical
\cite{STARCRONIN_ID,STARCRONIN_CH}, where the \raa\ values are very
different above \pt~=~2~\gevc.  

The momentum dependent strangeness enhancement is very similar in
central and peripheral collisions, while the protons are consistent
with \raa~=~1 in central collisions.  The extension of the measurement
to higher momenta might show if the protons and strange baryons show
the same \pt\ suppression.  It is notable that the \lam\ and $\Xi$
\raa\ values are in agreement with each other, which is not the case
for the strangeness enhancement factors shown in the previous section.
The \om\ might help further to separate the strangeness enhancement
from the high \pt\ suppression.


\section{Elliptic flow \vtwo\ of strange particles}

The elliptic flow \vtwo\ is the azimuthal momentum space anisotropy
of particle emission from non-central collisions in the transverse
direction with respect to the beam direction.  Elliptic flow is
described by the second harmonic coefficient of an azimuthal Fourier
decomposition of the momentum distribution.

Elliptic flow is an observable which is directly related to the
initial spatial anisotropy of the nuclear overlap region in the
transverse plane, translated into to the observed momentum
distribution of identified particles if quarks interact and thermalize
in the early stage.  It is important to understand the initial
geometry and how it varies with the collision centrality and system
size.  To compare the strength of \vtwo\ of different centralities one
has to correct for the initial eccentricity with respect to the
reaction-plane of the collisions.  A review of this experimental
measurements can be found in \cite{SORENSEN2009}.


\subsection{\vtwo\ scaling as a function of momentum}

The \vtwo\ results are not unique to strangeness and thus we will
only discuss them briefly in the context of mass and quark number
scaling.  Generally, any \vtwo\ discussion needs to distinguish between
the low \pt\ regime, where hydrodynamics holds, and the high \pt\
regime, where quark number scaling indicates non-bulk production of
the particles and their collective behavior.  Up to \pt~= 2~\gevc\ the
strange meson and baryon fit very well in the hydrodynamic systematics,
which assumes that the collective strength scales with the final hadron
mass.  Above 2~\gevc\ the strange and non-strange particles do not
scale with the hadron mass but rather with the number of valence
quarks.  This has been attributed to the fact that the partons in the
thermal system prior to hadronization are already flowing collectively
and then simply recombine into a hadron that flows with the added
collectivity of all valence quarks.  This can also be attributed to
hydrodynamics but in the partonic system rather than the hadronic
system, which is one of the signatures of the formation of the
deconfined collective partonic phase. 

\begin{figure}[ht!]
\begin{center}
\includegraphics[width=0.65\linewidth]{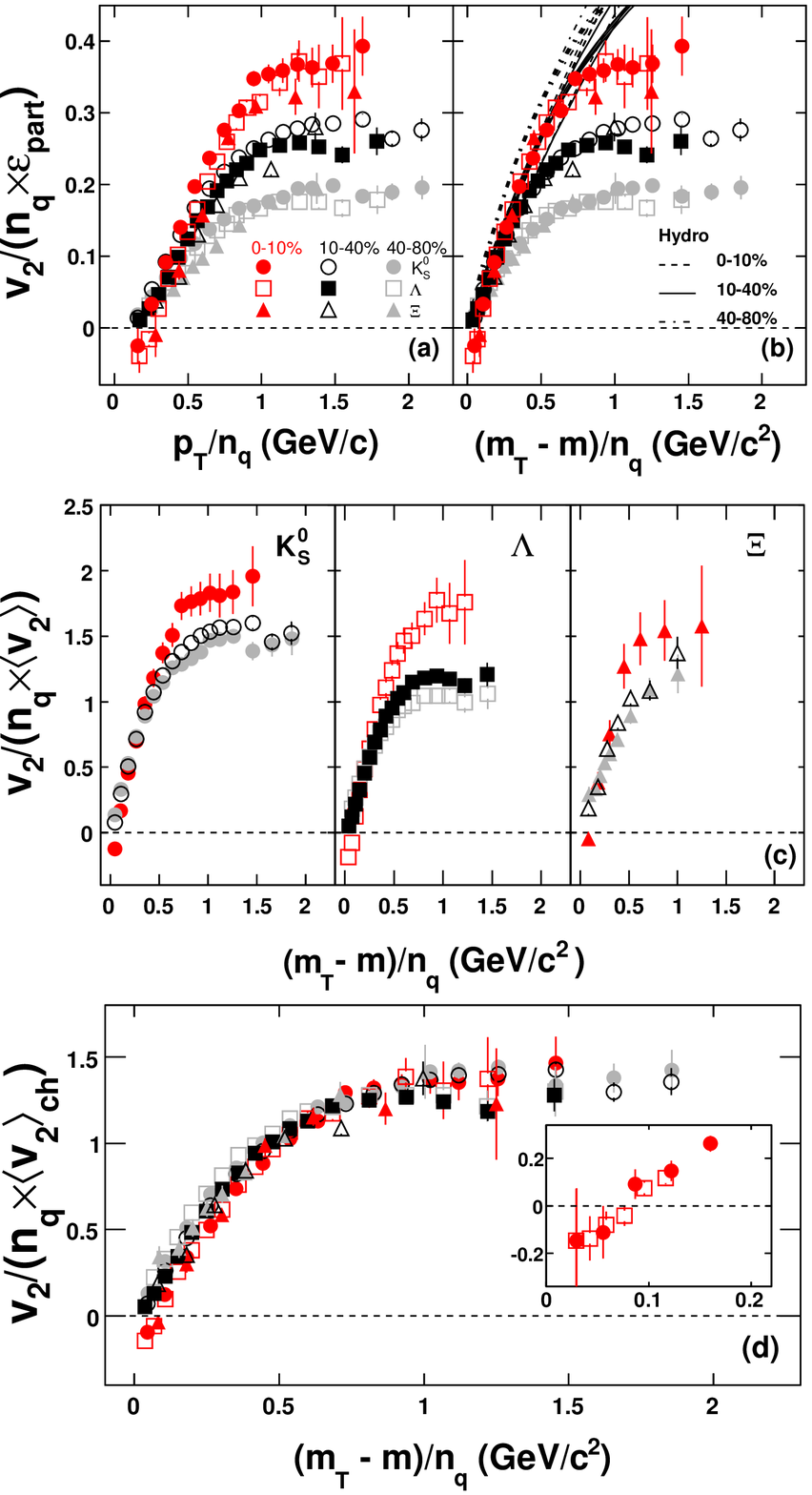}
\end{center}
\caption{Elliptic flow \vtwo\ scaled by the number of quarks (\nq)
and participant eccentricity (\epspart), (\vtwo\ / ($\nq \times
\epspart$)), of identified particles (particle + anti-particle) versus
(a) the scaled $\pt/\nq$ and (b) $(\mt\ - m)/\nq$ for three centrality
bins.  For comparison, ideal hydrodynamic model calculations
\cite{HUOVINENV22} are shown as lines in (b).  In (c) the same data
from (b) is shown, but scaled by the integrated \vtwo\ of each
particle, instead of \epspart.  In (d) the data from (b) scaled by the
integrated \vtwo\ of all charged hadrons is presented.  The inset in
(d) expands the low \mt\ region.  The error bars only represent the
statistical uncertainties.  All data are for \sqrts~= 200~GeV Au + Au
collisions
\cite{STARFLOW08}.}
\label{fig:flow_plot}
\end{figure}

Figure~\ref{fig:flow_plot} shows the experimental evidence for this
effect, and an attempt to address the universality of the quark number
scaling by analyzing the centrality dependence of the scaling properties.

\newpage

%
%

Here the scaled \vtwo\ is normalized first by the participant
eccentricity \epspart\ derived from a Monte Carlo Glauber
calculation.  The plots (a) and (b) show the doubly scaled quantities
from three centrality bins as a function of $\pt / \nq$ and ($\mt
- m) / \nq$ , respectively, which takes out the hadronic mass
dependence.  It is interesting to note that, in both cases, at a given
centrality the elliptic flow of all hadrons scales as observed for
minimum-bias events.  Upon dividing by the eccentricity the stronger
collective motion in more central collisions becomes apparent, which
is qualitatively consistent with ideal hydrodynamic model
calculations, shown as lines in (b).  However, there is no universal
scaling with eccentricity in the STAR data which is in disagreement
with the conclusions reached by PHENIX \cite{PHENIXFL}.

To further clarify the issue, instead of dividing the measured \vtwo\
by the corresponding eccentricity, in (c) the averaged \vtwo\ as a
function of transverse momentum weighted with the measured spectra is
shown.  This scaling seems to work better, however, different hadrons
seem to have different values of \vtwo, in particular for the top
10\% centrality bin at the higher \mt. 

In (d) the doubly scaled \vtwo\ is shown.  This time the integrated
values of \vtwo\ are extracted from the measurements of unidentified
charged hadrons in the corresponding centrality bins.  In this case the
scaling seems to work best.  It is interesting to point out that at the
most central bin (see inset in (d)) the values of \vtwo\ become
negative at low \pt\ for all hadrons.  This is most likely caused by
the strong radial flow developed in central Au+Au collisions, which
has also been observed at SPS energies. 


\subsection{Partonic versus hadronic flow}

\begin{figure}[h!]
\vspace{-1cm}

%
\begin{center}
\includegraphics[width=0.6\linewidth]{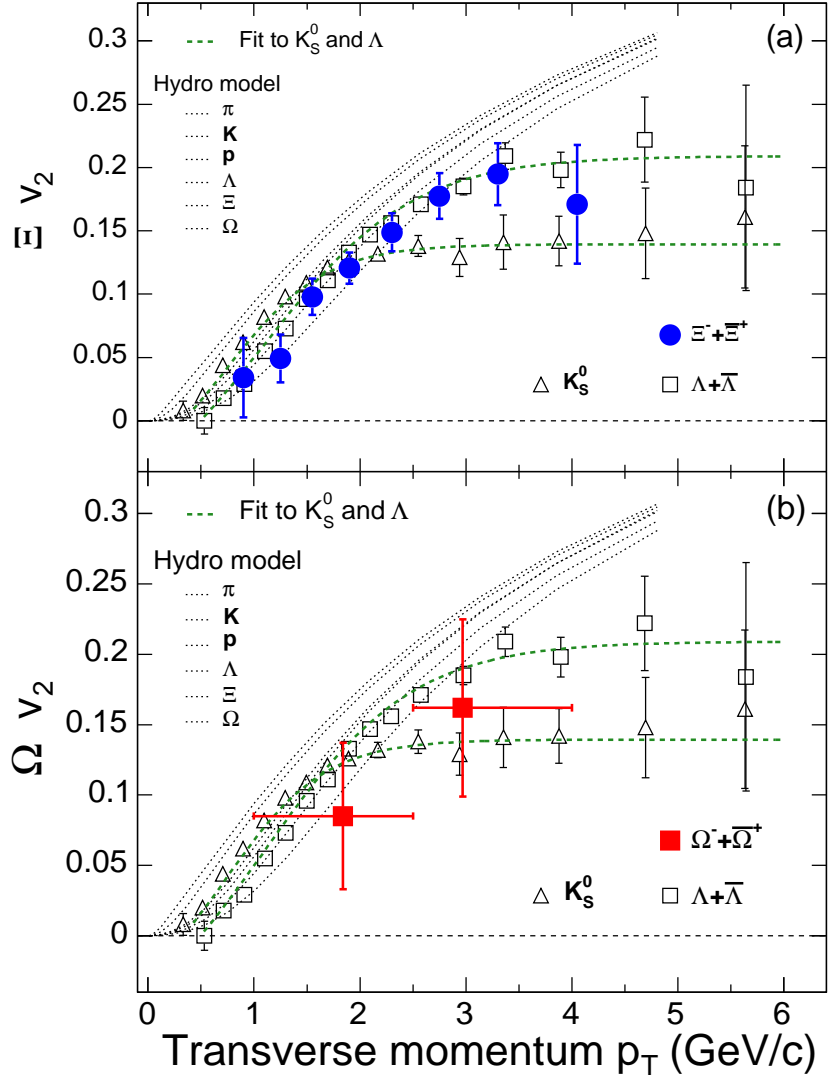}
\end{center}

%
\caption{Elliptic flow \vtwo\ of (a) \xim\ + \xip\ and (b) \omomp\
for \sqrts~= 200~GeV Au+Au minimum-bias collisions.  The  \vtwo\ of
\kzero\ and \lam\ \cite{STARPRLV2} are shown as open symbols, and the
results of the fits \cite{NQINSP} are shown as dashed lines.
Hydrodynamic model calculations \cite{HUOVINENV2} are shown as dotted
lines for K and \lam\ and as solid lines for \xim\ and \omm\ masses,
from top to bottom, respectively \cite{STARHYPFLOW}.}
\label{fig:flow_plot-strange_1}
\end{figure}

Further evidence for early partonic collectivity is given by
investigating in particular the flow of purely strange hadrons
i.e. the $\Omega$ and the $\phi$, both of which are expected to
exhibit a significantly reduced interaction cross section with the
surrounding hadronic medium.  Figure~\ref{fig:flow_plot-strange_1}
shows the \vtwo\ for multi-strange baryons in STAR.  From these early
measurements it could already be deduced that the $\Xi$ \vtwo\
follows the quark number scaling of the non-strange baryons.  Recent
preliminary high statistics measurements by STAR for vector mesons and
multi-strange baryons are shown in \Fi{fig:flow_plot-strange_2} 
\cite{SHI}.  It is evident that they follow in strength and scaling
exactly the same pattern than the light quark hadrons.  Thus a major
portion of the collective motion has to be developed in the early
partonic stage.

%
\begin{figure}[h!]
\begin{center}
\vspace{+6.4cm}
\includegraphics[width=0.35\linewidth, bb= 120 20 375 70]{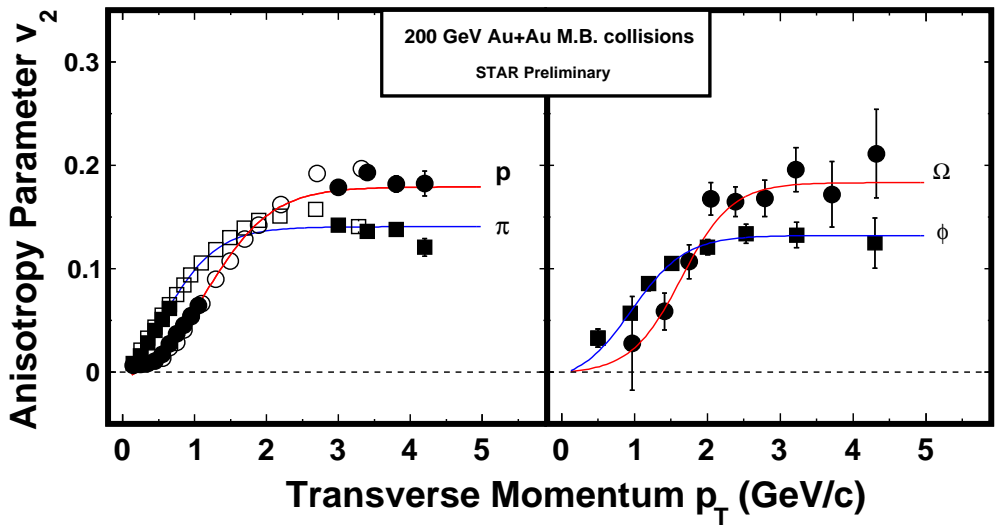}
\end{center}
\caption{Right: Elliptic flow \vtwo\ as a function of \pt\ for $\pi$, p (left)
and $\phi$, \om\ (right) in Au+Au minimum-bias collisions at \sqrts~=
200~GeV from STAR.  Open symbols represent results from PHENIX \cite{PHENIXFL}.
Lines represent a fit inspired by the number of quark scaling
\cite{NQINSP}.}
\label{fig:flow_plot-strange_2}

\end{figure}
%

%
%


\clearpage

\section{Conclusions}
\label{sec:conclusion}

The wealth of data on strange hadron production collected in the
recent years at the CERN-SPS and RHIC-BNL has allowed to establish a
long list of observations on the properties of hot and dense hadronic
matter produced in heavy ion reactions. The following list summarizes
the main findings as discussed in the previous sections:

\begin{itemize}

\item The enhancement of the production of strange hadrons per
  participating nucleon relative to elementary p+p(Be) collisions is a
  well established experimental fact.  The observed enhancement
  increases with the strangeness content of the hadron.

\item A clear onset of the strangeness enhancement with center-of-mass
  energy has not yet been observed.  In fact, the enhancement of
  particles with single strangeness increases toward lower energies
  in contrast to naive expectations.

\item Transport models involving only hadronic degrees of freedom
  are not able to describe the measured enhancement of multi-strange
  baryons.

\item Statistical models generally provide a reasonable description of
  strange particles yields at all energies.  The properties of the
  fireball are determined by the chemical freeze-out parameters \mub\
  and \tch.  These turn out to fall on a single freeze-out curve, if
  the center-of-mass energy is varied.

\item A pronounced maximum of the relative strangeness production is
  seen around \sqrts~= $7 - 8$~GeV and the energy evolution of the
  inverse slope parameters of kaon \pt~spectra exhibits a drastic
  change in the same region.  Both have been interpreted as an
  indication for a drastic change of the properties of the fireball
  due to the onset of deconfinement.

\item The scaling properties of the $\phi$ meson seem to rule out kaon
  coalescence as the dominant production mechanism.  It might be more
  likely that from SPS energies upwards its origin is to a large
  extend partonic.

\item The transverse mass spectra of $\Xi$ and $\Omega$ suggest that
  their kinetic freeze-out is happening in an earlier and hotter phase
  of the fireball evolution.  The radial flow observed in these
  particles would thus be mainly partonic by origin and starts to
  build up in the SPS energy regime already.  This is corroborated by
  the substantial elliptic flow of the $\Omega$ and $\phi$ measured at
  RHIC, which, due to their small hadronic cross section, can only
  develop in a partonic phase.

\item The measured system size dependence reveals a more complicated
  pattern as predicted by statistical models via the transition from a
  canonical to a grand-canonical ensemble.  The contribution of a
  deconfined phase might also alter the system size dependence in a
  significant way in addition to the volume effects in the statistical
  description.  However, the observed behavior can to a large extend
  be explained in a natural way by the core-corona approach.
  Alternatively, the assumption that the strange quark production, in
  contrast to the \npartch\ scaling of up and down quarks, scales with
  the number of binary collisions works with the notable exception of
  the $\phi$ meson.

\item Strange particles provide valuable information on the flavor
  dependence of intermediate and high \pt\ effects, such as high \pt\
  suppression.  The observed high baryon-meson ratio indicates that
  quark coalescence is dominating over fragmentation at intermediate
  \pt.  The observed quark number scaling of the elliptic flow at
  intermediate \pt\ points to a partonic origin of the early pressure
  in the fireball.

\end{itemize}

Starting from the original suggestion of using strange particles as a
probe for quark-gluon plasma formation, strangeness has turned out
to be a versatile tool to study the properties of the matter
produced in high energy nuclear reactions and has revealed many
unexpected effects.  Even though the interpretation of the observed
strangeness enhancement might not be as straight forward as originally
suggested, it seems there is no consistent explanation without
invoking partonic degrees of freedom.  Models that include only
hadronic degrees of freedom are not able to describe the enhancement
of multi-strange particles.  Instead one finds that statistical
approaches assuming full equilibration result in a much better
description.  Also the scaling properties of the $\phi$ meson yields
and spectra indicate that from SPS energies onwards its dominant
production mechanism is of partonic origin, i.e. quark coalescence,
while kaon coalescence cannot explain the observed features.  This
point of view is further strengthened by the phenomenology of radial
and elliptic flow of strange particles and follows from the fact that
particles with low hadronic cross section ($\phi$, $\Xi$, $\Omega$)
exhibit significant radial flow and that the elliptic flow of strange
particles scales with the number of constituent quarks.

Important questions that remain are for instance:  Is there a clear
onset of strangeness enhancement observable when increasing the
center-of-mass energy?  There are some features in the existing data
(the maximum of the \kplus/\pip~ratio and a sudden change in the
energy dependence of the inverse slopes parameters of kaon
\pt~spectra) that can be interpreted in this way, but additional
support for this findings is desirable.  Can a unique connection of
the enhancement of multi-strange particles to partonic equilibration
be established or is this a result of other properties of the
hadronization processes?  What is the excitation function of flow
phenomena (radial and elliptic) of strange particles and could this
help to uniquely define the onset of deconfinement?  Upcoming
measurements at high energies (LHC), but also at lower energies (FAIR,
RHIC, SPS) will provide important contributions to this strange and
exciting picture.


\section*{Acknowledgments}
The authors would like to thank A.~Andronic, J.~Aichelin,
F.~Becattini, R.~Bellwied, M.~Bleicher, E.~Bratkovskaya,
M.~Ga\'{z}dzicki, C.~Greiner, H.~Oeschler, K.~Redlich,
K.~\v{S}afa\v{r}ik, and H.~Str\"obele for many helpful and inspiring
discussions.  Special thanks go to A.~Andronic, R.~Bellwied,
K.~\v{S}afa\v{r}ik, H.~Str\"obele and T.~Schuster for reading the
document and providing valuable corrections and comments.

This work was supported by U.S. Department of Energy Office of Science
under contract numbers DE-FG02-94ER40845 and
DE-SC0003892/DE-PS02-09ER09-26.


\clearpage
\section{Data compilation}


\subsection{Energy dependence}
\label{sec:compedep}

\begin{figure}[h!]
\begin{center}
\begin{minipage}[b]{0.49\linewidth}
\begin{center}
\includegraphics[width=\linewidth]{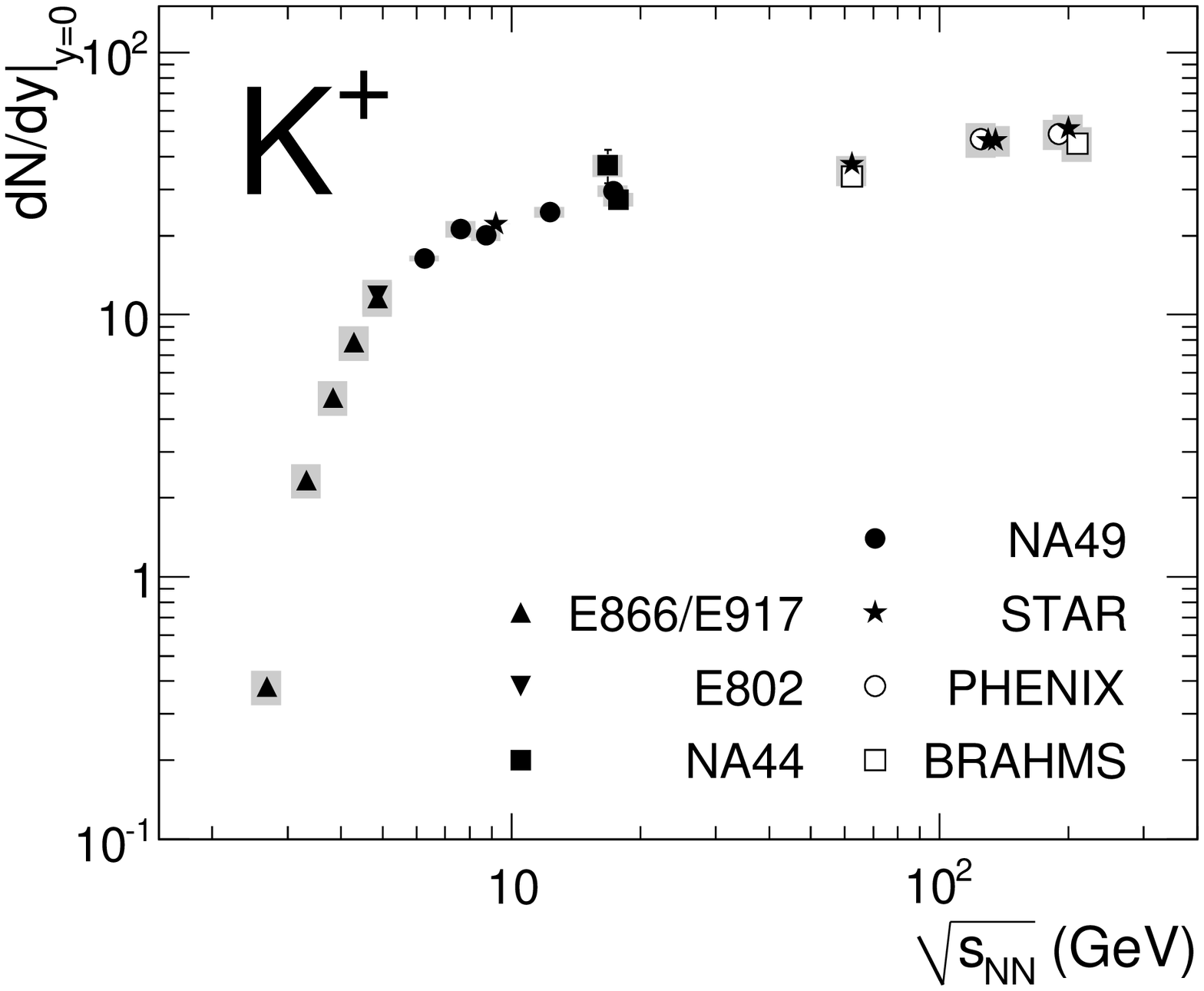}
\end{center}
\end{minipage}
\begin{minipage}[b]{0.49\linewidth}
\begin{center}
\includegraphics[width=\linewidth]{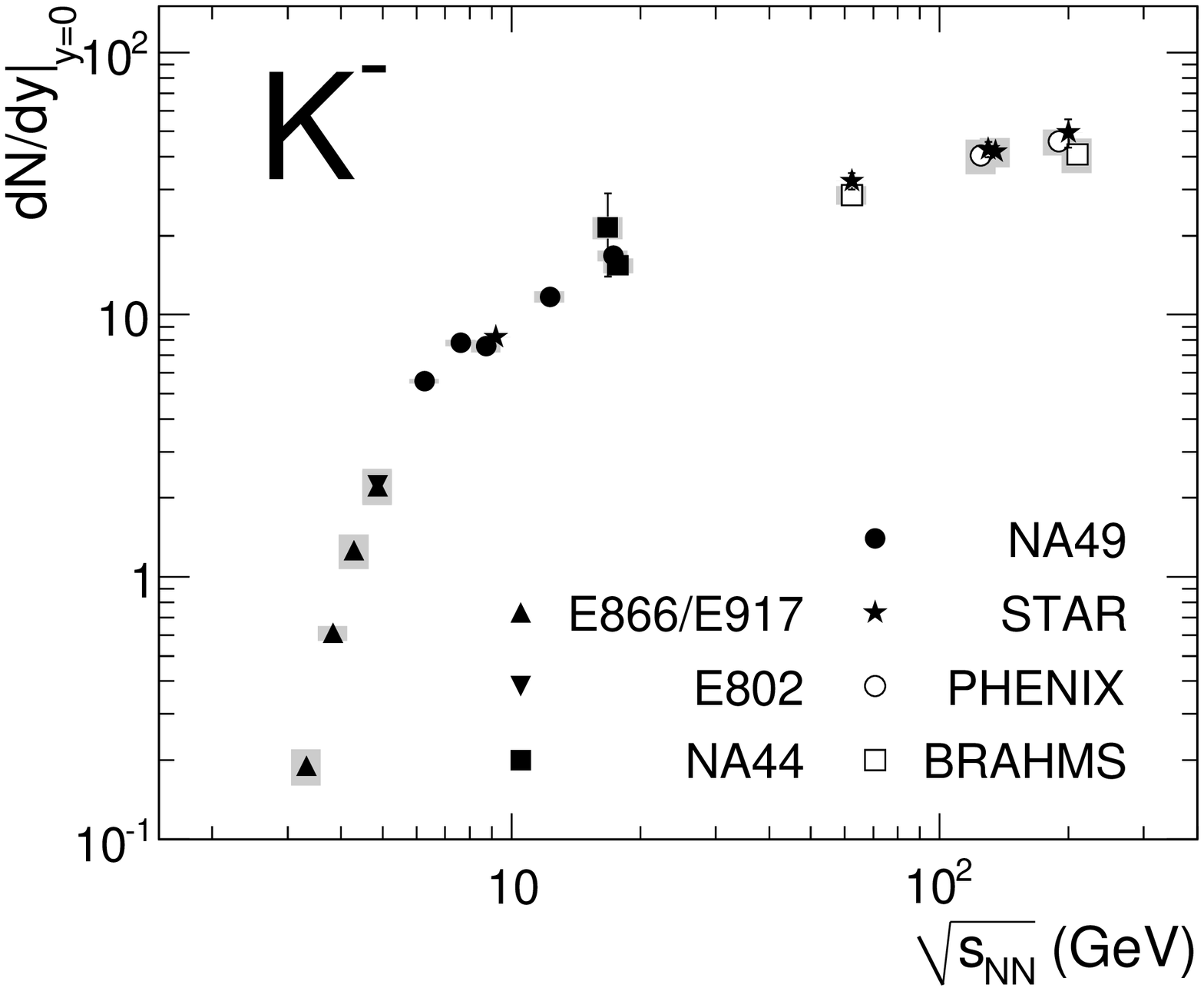}
\end{center}
\end{minipage}
\end{center}
\begin{center}
\begin{minipage}[b]{0.49\linewidth}
\begin{center}
\includegraphics[width=\linewidth]{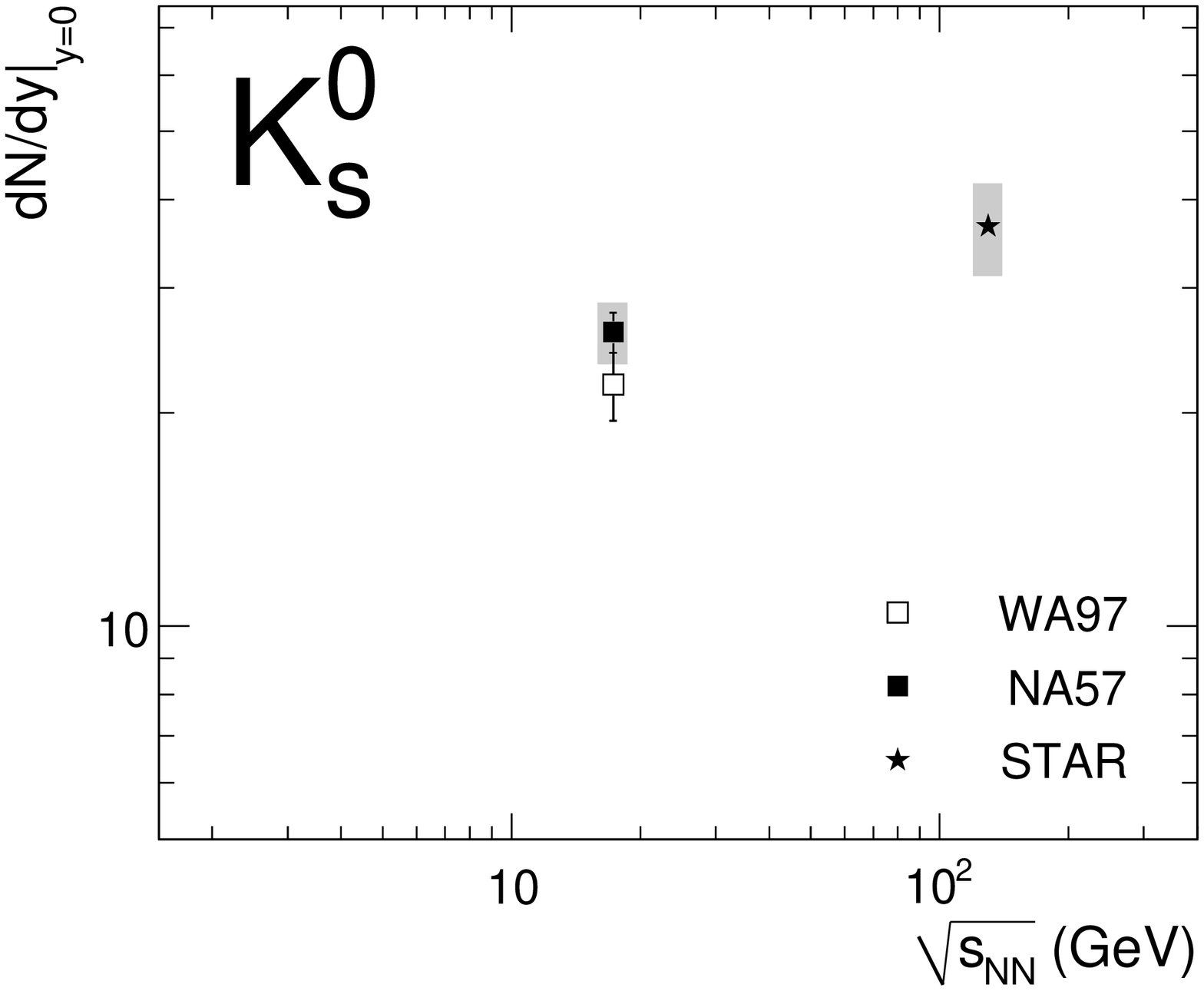}
\end{center}
\end{minipage}
\begin{minipage}[b]{0.49\linewidth}
\begin{center}
\includegraphics[width=\linewidth]{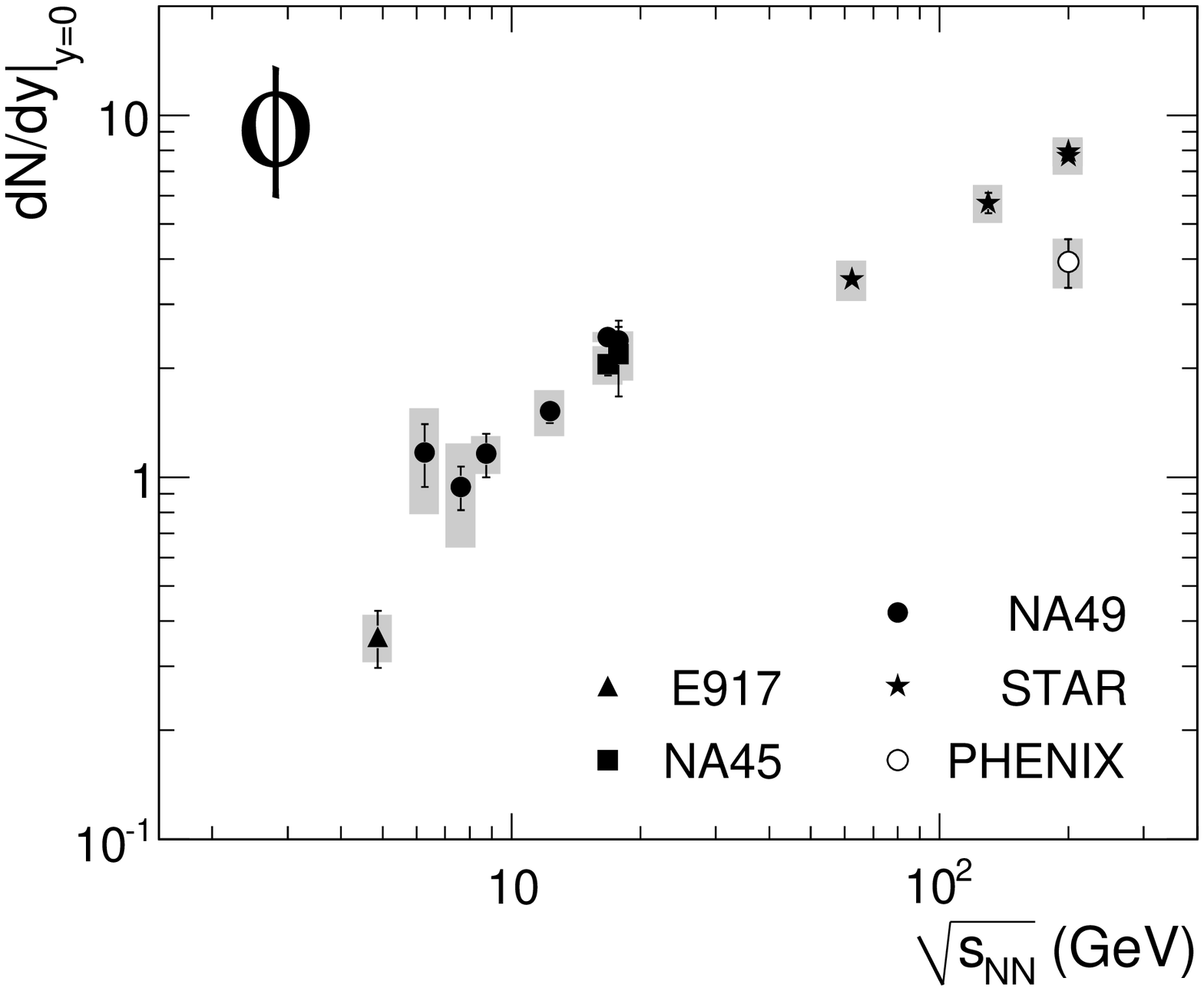}
\end{center}
\end{minipage}
\end{center}
\caption{The rapidity densities \dndy\ around midrapidity for K$^{+}$,
K$^{-}$, \kzero, and $\phi$ measured in central nucleus-nucleus
collisions as a function of \sqrts.  The systematic errors are
represented by the gray boxes.}
\label{fig:dndy_vs_sqrt_meson}
\end{figure}

\begin{figure}[t]
\begin{center}
\begin{minipage}[b]{0.49\linewidth}
\begin{center}
\includegraphics[width=\linewidth]{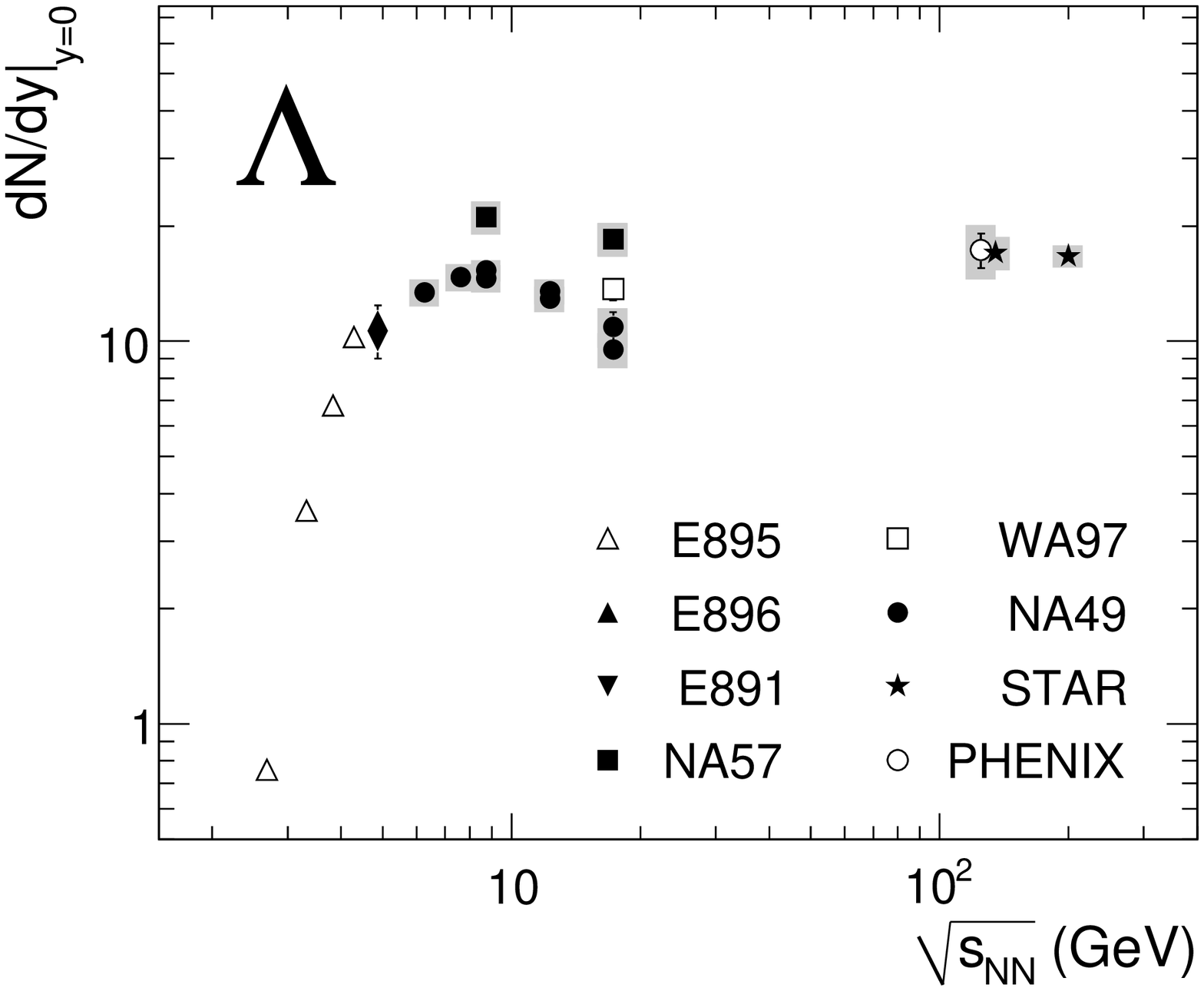}
\end{center}
\end{minipage}
\begin{minipage}[b]{0.49\linewidth}
\begin{center}
\includegraphics[width=\linewidth]{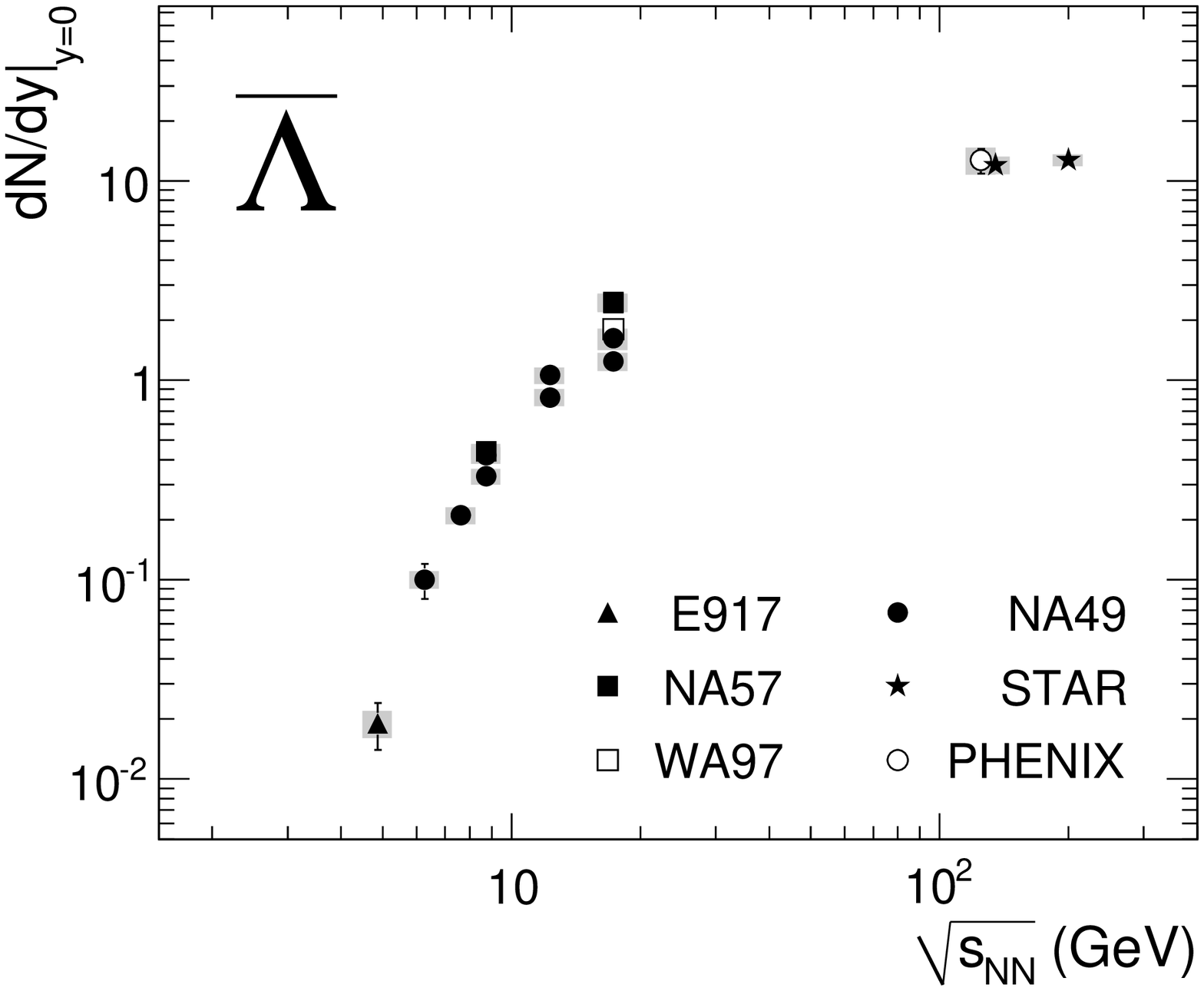}
\end{center}
\end{minipage}
\end{center}
\begin{center}
\begin{minipage}[b]{0.49\linewidth}
\begin{center}
\includegraphics[width=\linewidth]{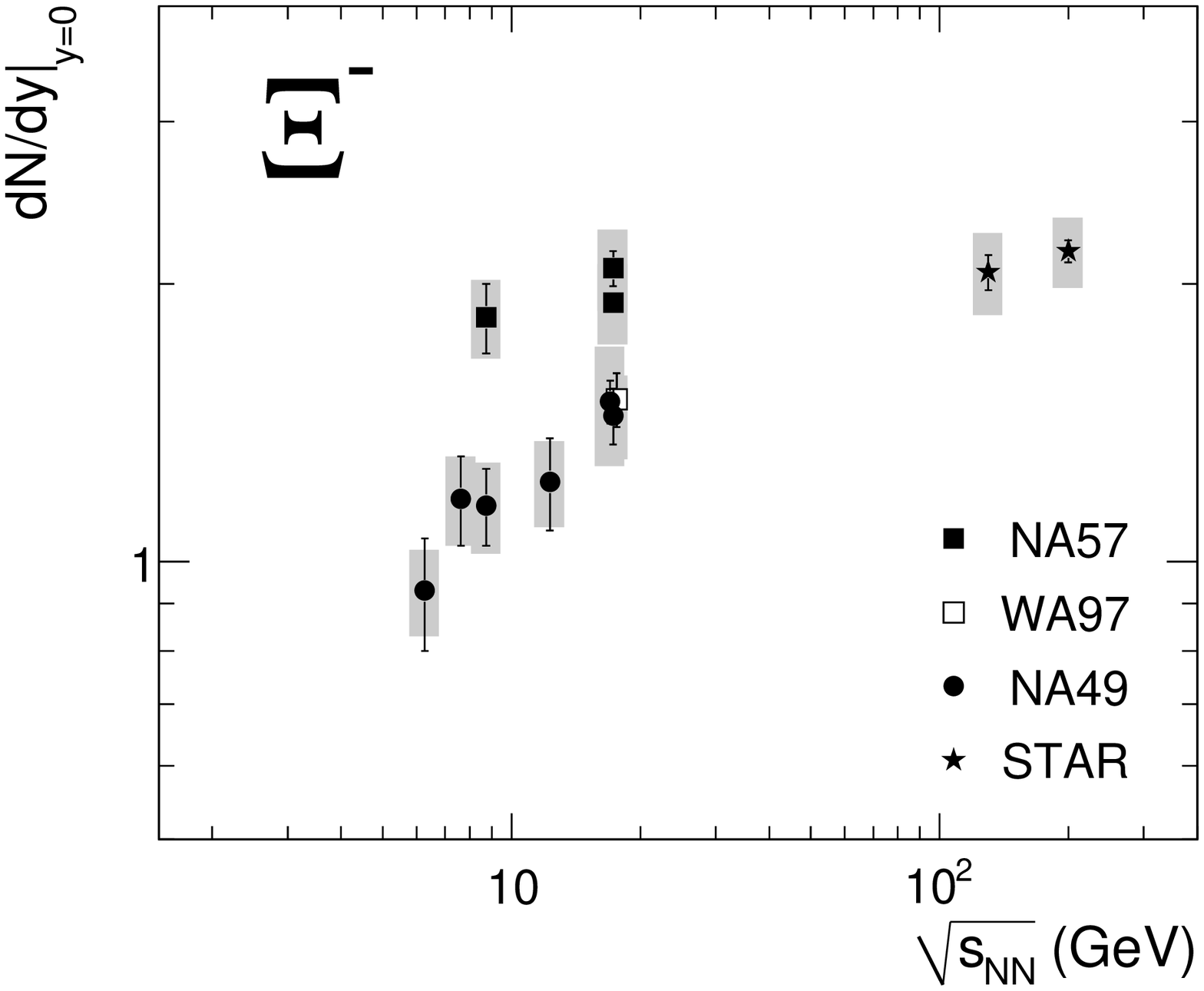}
\end{center}
\end{minipage}
\begin{minipage}[b]{0.49\linewidth}
\begin{center}
\includegraphics[width=\linewidth]{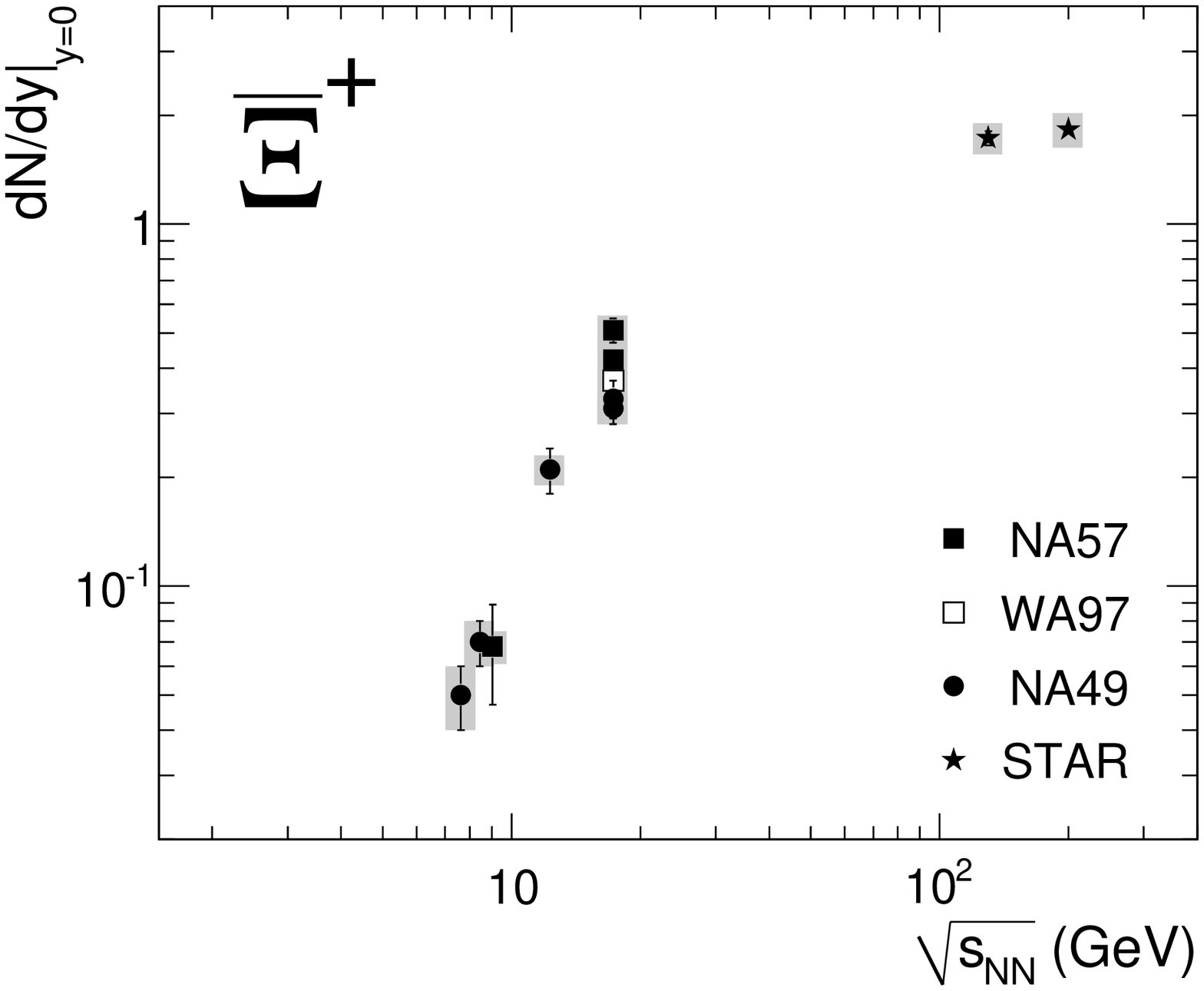}
\end{center}
\end{minipage}
\end{center}
\begin{center}
\begin{minipage}[b]{0.49\linewidth}
\begin{center}
\includegraphics[width=\linewidth]{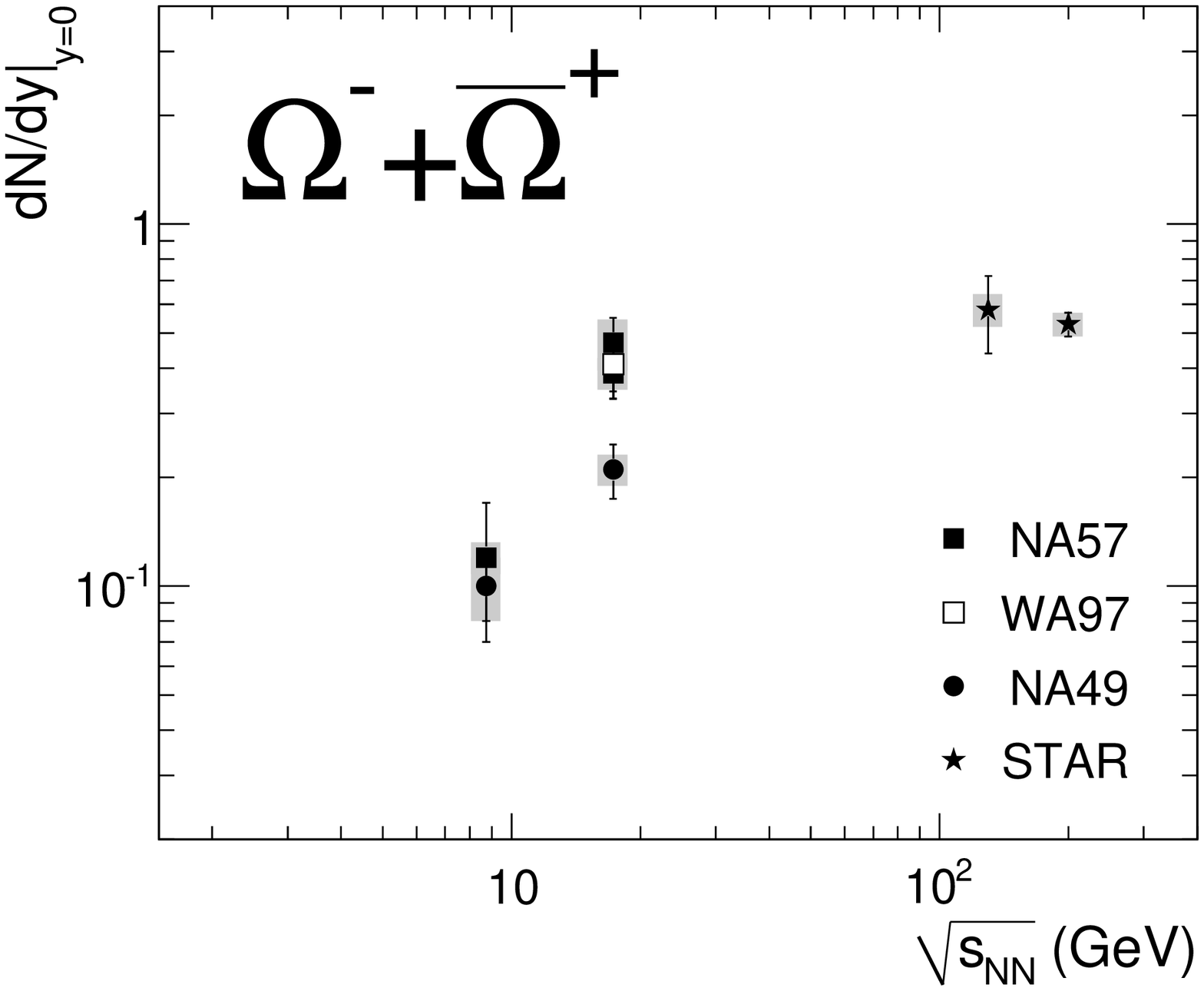}
\end{center}
\end{minipage}
\begin{minipage}[b]{0.49\linewidth}
\begin{center}
\end{center}
\end{minipage}
\end{center}
\caption{The rapidity densities \dndy\ around midrapidity for \lam,
\lab, \xim, \xip, and \omm+\omp\ measured in central nucleus-nucleus
collisions as a function of \sqrts.  The systematic errors are
represented by the gray boxes.}
\label{fig:dndy_vs_sqrt_baryon}
\end{figure}

Figures~\ref{fig:dndy_vs_sqrt_meson} and \ref{fig:dndy_vs_sqrt_baryon}
and the \Tar{tab:comp_kp}{tab:comp_om} summarize the current data
available on strange particle production in central Pb+Pb and Au+Au
collisions.

\vspace{40pt}

%
%

%
%

%
\begin{table*}[b]
\begin{center}
\caption{The rapidity densities \dndy\ around midrapidity and the
total multiplicities \navg\ of K$^{+}$ as measured in central
nucleus-nucleus collisions at different \sqrts.  Also listed are the
centrality selection (Cent.) and the averaged number of participants
(\npart).  In most cases the statistical (first) and systematic
(second) errors are given separately, while for a few measurements the
quadratic sum of both is quoted (denoted by 'sum').}
\vspace{5pt}
\begin{footnotesize}
\begin{tabular}{lllllllll} \hline\hline
\label{tab:comp_kp}
  \sqrts        &
  System        &
  Cent.         &
  \npart        &
  \dndy         &
  \navg         &
  Exp.          &
  Ref.          &
  Comnt.        \\
  (GeV) &       &         &     &                           &                        &           &                             &      \\ \hline\hline
   2.68 & Au+Au & 0-5\%   &     & 0.381$\pm$0.015$\pm$0.057 & ---                    & E866/E917 & \cite{E866E917KPM}          &      \\
   3.32 & Au+Au & 0-5\%   &     & 2.34$\pm$0.05$\pm$0.35    & ---                    & E866/E917 & \cite{E866E917KPM}          &      \\
   3.83 & Au+Au & 0-5\%   &     & 4.84$\pm$0.09$\pm$0.73    & ---                    & E866/E917 & \cite{E866E917KPM}          &      \\
   4.29 & Au+Au & 0-5\%   &     & 7.85$\pm$0.21$\pm$1.18    & ---                    & E866/E917 & \cite{E866E917KPM}          &      \\
   4.87 & Au+Au & 0-5\%   &     & 11.55$\pm$0.24$\pm$1.73   & ---                    & E866/E917 & \cite{E866E917KPM}          &      \\ \hline
   4.87 & Au+Au & 0-5\%   & 354 & 11.79$\pm$0.37$\pm$1.77   & 24.2$\pm$0.9$\pm$3.6   & E802      & \cite{E802KPM116}           &      \\
   4.96 & Au+Au & 0-4\%   & 363 & ---                       & 23.7$\pm$1.6$\pm$2.3   & E802      & \cite{E802SDEP111}          &      \\ \hline
   6.27 & Pb+Pb & 0-7\%   & 349 & 16.4$\pm$0.6$\pm$0.4      & 40.7$\pm$0.7$\pm$2.2   & NA49      & \cite{NA49KPI2030}          &      \\
   7.62 & Pb+Pb & 0-7\%   & 349 & 21.2$\pm$0.8$^{+1.5}_{-0.9}$ & 52.9$\pm$0.9$^{+3.0}_{-3.5}$ & NA49   & \cite{NA49KPI2030}          &      \\
   8.73 & Pb+Pb & 0-7\%   & 349 & 20.1$\pm$0.3$\pm$1.0      & 59.1$\pm$1.9$\pm$3.0   & NA49      & \cite{NA49KPI40158}         &      \\ \hline
   9.2  & Au+Au & 0-10\%  & 317 & 23.0$\pm$4.4(sum)         & ---                    & STAR      & \cite{STAR92}               &      \\ \hline
  12.3  & Pb+Pb & 0-7\%   & 349 & 24.6$\pm$0.2$\pm$1.2      & 76.9$\pm$2.0$\pm$4.0   & NA49      & \cite{NA49KPI40158}         &      \\
  17.3  & Pb+Pb & 0-5\%   & 362 & 29.6$\pm$0.3$\pm$1.5      & 103.0$\pm$5.0$\pm$5.0  & NA49      & \cite{NA49KPI40158}         &      \\ \hline
  17.3  & Pb+Pb & 0-4\%   &     & 37.1$\pm$5.4(sum)         & ---                    & NA44      & \cite{NA44A}                & a)   \\
  17.3  & Pb+Pb & 0-3.7\% &     & 27.5$\pm$0.2$\pm$1.6      & ---                    & NA44      & \cite{NA44B}                & b)   \\ \hline
  62.4  & Au+Au & 0-10\%  &     & 33.6$\pm$0.8$\pm$2.7      & ---                    & BRAHMS    & \cite{BRMSRAP62}            & c)   \\ \hline
  62.4  & Au+Au & 0-5\%   & 347 & 37.6$\pm$2.7(sum)         & ---                    & STAR      & \cite{STARSYST}             &      \\
 130.0  & Au+Au & 0-6\%   & 344 & 46.3$\pm$3.0(sum)         & ---                    & STAR      & \cite{STARSYST}             &      \\
 130.0  & Au+Au & 0-6\%   &     & 46.2$\pm$0.6$\pm$6.0      & ---                    & STAR      & \cite{STARKPI130}           &      \\ \hline
 130.0  & Au+Au & 0-5\%   & 348 & 46.7$\pm$1.5$\pm$7.0      & ---                    & PHENIX    & \cite{PHNXKPI130}           &      \\ \hline
 200.0  & Au+Au & 0-5\%   & 351 & 51.3$\pm$6.5(sum)         & ---                    & STAR      & \cite{STARSYST}             &      \\ \hline
 200.0  & Au+Au & 0-5\%   & 351 & 48.9$\pm$6.3              & ---                    & PHENIX    & \cite{PHNXKPI200}           &      \\ \hline
 200.0  & Au+Au & 0-10\%  & 328 & 45.0$\pm$0.67$\pm$6.75    & ---                    & BRAHMS    & \cite{BRMSCENT200}          &      \\
 200.0  & Au+Au & 0-5\%   &     & ---                       & 286.0$\pm$5.0$\pm$23.0 & BRAHMS    & \cite{BRMSRAP200}           & d)   \\ \hline\hline
\end{tabular}
\end{footnotesize}
\end{center}
\end{table*}
%

%
\begin{table*}[b]
\begin{center}
\caption{The rapidity densities \dndy\ around midrapidity and the
total multiplicities \navg\ of K$^{-}$ as measured in central
nucleus-nucleus collisions at different \sqrts.  Also listed are the
centrality selection (Cent.) and the averaged number of participants
(\npart).  In most cases the statistical (first) and systematic
(second) errors are given separately, while for a few measurements the
quadratic sum of both is quoted (denoted by 'sum').}
\vspace{5pt}
\begin{footnotesize}
\begin{tabular}{lllllllll} \hline\hline
\label{tab:comp_km}
  \sqrts        &
  System        &
  Cent.         &
  \npart        &
  \dndy         &
  \navg         &
  Exp.          &
  Ref.          &
  Comnt.        \\
  (GeV) &       &         &     &                           &                        &           &                             &      \\ \hline\hline
   3.32 & Au+Au & 0-5\%   &     & 0.19$\pm$0.01$\pm$0.03    & ---                    & E866/E917 & \cite{E866E917KPM}          &      \\
   3.83 & Au+Au & 0-5\%   &     & 0.61$\pm$0.02$\pm$0.09    & ---                    & E866/E917 & \cite{E866E917KPM}          &      \\
   4.29 & Au+Au & 0-5\%   &     & 1.26$\pm$0.04$\pm$0.19    & ---                    & E866/E917 & \cite{E866E917KPM}          &      \\
   4.87 & Au+Au & 0-5\%   &     & 2.21$\pm$0.03$\pm$0.33    & ---                    & E866/E917 & \cite{E866E917KPM}          &      \\ \hline
   4.87 & Au+Au & 0-5\%   & 354 & 2.24$\pm$0.05$\pm$0.34    & 4.14$\pm$0.09$\pm$0.62 & E802      & \cite{E802KPM116}           &      \\
   4.96 & Au+Au & 0-4\%   & 363 & ---                       & 3.76$\pm$0.28$\pm$0.38 & E802      & \cite{E802SDEP111}          &      \\ \hline
   6.27 & Pb+Pb & 0-7\%   & 349 & 5.58$\pm$0.07$\pm$0.11    & 10.3$\pm$0.1$\pm$0.2   & NA49      & \cite{NA49KPI2030}          &      \\
   7.62 & Pb+Pb & 0-7\%   & 349 & 7.8$\pm$0.1$\pm$0.2       & 16.0$\pm$0.2$\pm$0.4   & NA49      & \cite{NA49KPI2030}          &      \\
   8.73 & Pb+Pb & 0-7\%   & 349 & 7.58$\pm$0.12$\pm$0.4     & 19.2$\pm$0.5$\pm$1.0   & NA49      & \cite{NA49KPI40158}         &      \\ \hline
   9.2  & Au+Au & 0-10\%  & 317 & 8.7$\pm$2.0(sum)          & ---                    & STAR      & \cite{STAR92}               &      \\ \hline
  12.3  & Pb+Pb & 0-7\%   & 349 & 11.7$\pm$0.1$\pm$0.6      & 32.4$\pm$0.6$\pm$1.6   & NA49      & \cite{NA49KPI40158}         &      \\
  17.3  & Pb+Pb & 0-5\%   & 362 & 16.8$\pm$0.2$\pm$0.8      & 51.9$\pm$1.9$\pm$3.0   & NA49      & \cite{NA49KPI40158}         &      \\ \hline
  17.3  & Pb+Pb & 0-4\%   &     & 21.5$\pm$7.5(sum)         & ---                    & NA44      & \cite{NA44A}                & a)   \\
  17.3  & Pb+Pb & 0-3.7\% &     & 15.4$\pm$0.5$\pm$1.0      & ---                    & NA44      & \cite{NA44B}                & b)   \\ \hline
  62.4  & Au+Au & 0-10\%  &     & 28.6$\pm$0.8$\pm$2.3      & ---                    & BRAHMS    & \cite{BRMSRAP62}            & c)   \\ \hline
  62.4  & Au+Au & 0-5\%   & 347 & 32.4$\pm$2.3(sum)         & ---                    & STAR      & \cite{STARSYST}             &      \\
 130.0  & Au+Au & 0-6\%   & 344 & 42.7$\pm$2.8(sum)         & ---                    & STAR      & \cite{STARSYST}             &      \\
 130.0  & Au+Au & 0-6\%   &     & 41.9$\pm$0.6$\pm$5.4      & ---                    & STAR      & \cite{STARKPI130}           &      \\ \hline
 130.0  & Au+Au & 0-5\%   & 348 & 40.5$\pm$2.3$\pm$6.1      & ---                    & PHENIX    & \cite{PHNXKPI130}           &      \\ \hline
 200.0  & Au+Au & 0-5\%   & 351 & 49.5$\pm$6.2(sum)         & ---                    & STAR      & \cite{STARSYST}             &      \\ \hline
 200.0  & Au+Au & 0-5\%   & 351 & 45.7$\pm$5.2              & ---                    & PHENIX    & \cite{PHNXKPI200}           &      \\ \hline
 200.0  & Au+Au & 0-10\%  & 328 & 40.9$\pm$0.63$\pm$6.13    & ---                    & BRAHMS    & \cite{BRMSCENT200}          &      \\
 200.0  & Au+Au & 0-5\%   &     & ---                       & 242.0$\pm$4.0$\pm$19.0 & BRAHMS    & \cite{BRMSRAP200}           & d)   \\ \hline\hline
\end{tabular}
\end{footnotesize}
\end{center}
\end{table*}
%

%
\begin{table*}[b]
\begin{center}
\caption{The rapidity densities \dndy\ around midrapidity and the
total multiplicities \navg\ of \kzero\ as measured in central
nucleus-nucleus collisions at different \sqrts.  Also listed are the
centrality selection (Cent.) and the averaged number of participants
(\npart).  In most cases the statistical (first) and systematic
(second) errors are given separately, while for a few measurements the
quadratic sum of both is quoted (denoted by 'sum').}
\vspace{5pt}
\begin{footnotesize}
\begin{tabular}{lllllllll} \hline\hline
\label{tab:comp_k0s}
  \sqrts        &
  System        &
  Cent.         &
  \npart        &
  \dndy         &
  \navg         &
  Exp.          &
  Ref.          &
  Comnt.        \\
  (GeV) &       &         &     &                      &                      &        &                             &      \\ \hline\hline
  17.3  & Pb+Pb & 0-4.5\% &     & 26.0$\pm$1.7$\pm$2.6 & ---                  & NA57   & \cite{NA57RAP158}           &      \\ \hline
  17.3  & Pb+Pb &         & 351 & 21.9$\pm$2.4         & ---                  & WA97   & \cite{WA97ENHANCE,WA97QM99} &      \\ \hline
 130.0  & Au+Au & 0-6\%   &     & 33.9$\pm$1.1$\pm$5.1 & ---                  & STAR   & \cite{STARKPI130}           &      \\ \hline\hline
\end{tabular}
\end{footnotesize}
\end{center}
\end{table*}
%

%
\begin{table*}[b]
\begin{center}
\caption{The rapidity densities \dndy\ around midrapidity and the
total multiplicities \navg\ of $\phi$ as measured in central
nucleus-nucleus collisions at different \sqrts.  Also listed are the
centrality selection (Cent.) and the averaged number of participants
(\npart).  In most cases the statistical (first) and systematic
(second) errors are given separately, while for a few measurements the
quadratic sum of both is quoted (denoted by 'sum').}
\vspace{5pt}
\begin{footnotesize}
\begin{tabular}{lllllllll} \hline\hline
\label{tab:comp_phi}
  \sqrts        &
  System        &
  Cent.         &
  \npart        &
  \dndy         &
  \navg         &
  Exp.          &
  Ref.          &
  Comnt.        \\
  (GeV) &       &         &     &                           &                        &        &                             &      \\ \hline\hline
   4.87 & Au+Au & 0-5\%   & 340 & 0.362$\pm$0.065$\pm$0.054 & ---                    & E917   & \cite{E917PHI}              &      \\ \hline
   6.27 & Pb+Pb & 0-7\%   & 349 & 1.17$\pm$0.23$\pm$0.38    & 1.89$\pm$0.31$\pm$0.22 & NA49   & \cite{NA49EDEPPHI}          &      \\
   7.62 & Pb+Pb & 0-7\%   & 349 & 0.94$\pm$0.13$\pm$0.30    & 1.84$\pm$0.22$\pm$0.29 & NA49   & \cite{NA49EDEPPHI}          &      \\
   8.73 & Pb+Pb & 0-7\%   & 349 & 1.16$\pm$0.16$\pm$0.14    & 2.55$\pm$0.17$\pm$0.19 & NA49   & \cite{NA49EDEPPHI}          &      \\
  12.3  & Pb+Pb & 0-7\%   & 349 & 1.52$\pm$0.11$\pm$0.22    & 4.04$\pm$0.19$\pm$0.31 & NA49   & \cite{NA49EDEPPHI}          &      \\
  17.3  & Pb+Pb & 0-5\%   & 362 & 2.44$\pm$0.10$\pm$0.08    & 8.46$\pm$0.38$\pm$0.33 & NA49   & \cite{NA49EDEPPHI}          &      \\
  17.3  & Pb+Pb & 0-4\%   &     & 2.39$\pm$0.21(sum)        & 7.6$\pm$1.1(sum)       & NA49   & \cite{NA49PHI158}           &      \\ \hline
  17.3  & Pb+Au & 0-7\%   &     & 2.05$\pm$0.14$\pm$0.25    & ---                    & NA45   & \cite{NA45PHI158}           & e)   \\
  17.3  & Pb+Au & 0-7\%   &     & 2.04$\pm$0.49$\pm$0.32    & ---                    & NA45   & \cite{NA45PHI158}           & f)   \\ \hline
  62.4  & Au+Au & 0-20\%  &     & 3.52$\pm$0.08$\pm$0.45    & ---                    & STAR   & \cite{STAREDEPPHI}          &      \\
 130.0  & Au+Au & 0-11\%  &     & 5.73$\pm$0.37$\pm$0.69    & ---                    & STAR   & \cite{STARPHI130}           &      \\
 200.0  & Au+Au & 0-5\%   &     & 7.70$\pm$0.30$\pm$0.85    & ---                    & STAR   & \cite{STARPHI200}           &      \\
 200.0  & Au+Au & 0-5\%   &     & 7.95$\pm$0.11$\pm$0.73    & ---                    & STAR   & \cite{STAREDEPPHI}          &      \\ \hline
 200.0  & Au+Au & 0-10\%  & 325 & 3.94$\pm$0.60$\pm$0.62    & ---                    & PHENIX & \cite{PHNXPHI200}           &      \\ \hline\hline
\end{tabular}
\end{footnotesize}
\end{center}
\end{table*}
%

%
\begin{table*}[b]
\begin{center}
\caption{The rapidity densities \dndy\ around midrapidity and the
total multiplicities \navg\ of \lam\ as measured in central
nucleus-nucleus collisions at different \sqrts.  Also listed are the
centrality selection (Cent.) and the averaged number of participants
(\npart).  In most cases the statistical (first) and systematic
(second) errors are given separately, while for a few measurements the
quadratic sum of both is quoted (denoted by 'sum').}
\vspace{5pt}
\begin{footnotesize}
\begin{tabular}{lllllllll} \hline\hline
\label{tab:comp_lam}
  \sqrts        &
  System        &
  Cent.         &
  \npart        &
  \dndy         &
  \navg         &
  Exp.          &
  Ref.          &
  Comnt.        \\
  (GeV) &       &         &     &                      &                      &        &                             &    \\ \hline\hline
   2.68 & Au+Au & 0-5\%   &     & 0.76$\pm$0.03        & 0.58$\pm$0.04        & E895   & \cite{E895LAM}              & g) \\
   3.32 & Au+Au & 0-5\%   &     & 3.6$\pm$0.1          & 5.5$\pm$0.3          & E895   & \cite{E895LAM}              & g) \\
   3.83 & Au+Au & 0-5\%   &     & 6.8$\pm$0.2          & 11.6$\pm$0.9         & E895   & \cite{E895LAM}              & g) \\
   3.83 & Au+Au &         & 342 & ---                  & 12.0$\pm$1.0         & E895   & \cite{E895LAMXI}            &    \\
   4.29 & Au+Au & 0-5\%   &     & 10.25$\pm$0.3        & 16.0$\pm$1.0         & E895   & \cite{E895LAM}              & g) \\ \hline
   4.87 & Au+Au & 0-5\%   &     & 10.0$\pm$1.0         & ---                  & E891   & \cite{E891LAM}              &    \\ \hline
   4.87 & Au+Au & 0-5\%   &     &                      & 16.7$\pm$0.5$\pm$1.7 & E896   & \cite{E896LAM}              &    \\ \hline
   8.73 & Pb+Pb & 0-5\%   &     & 21.1$\pm$0.8$\pm$2.1 & ---                  & NA57   & \cite{NA57EDEPHYP}          &    \\
  17.3  & Pb+Pb & 0-5\%   &     & 18.5$\pm$1.1$\pm$1.9 & ---                  & NA57   & \cite{NA57EDEPHYP}          &    \\
  17.3  & Pb+Pb & 0-4.5\% & 349 & 18.5$\pm$1.1$\pm$1.8 & ---                  & NA57   & \cite{NA57ENHANCE}          &    \\ \hline
  17.3  & Pb+Pb &         & 351 & 13.7$\pm$0.9         & ---                  & WA97   & \cite{WA97ENHANCE,WA97QM99} &    \\ \hline
   6.27 & Pb+Pb & 0-7\%   & 349 & 13.4$\pm$0.1$\pm$1.1 & 27.1$\pm$0.2$\pm$2.2 & NA49   & \cite{NA49EDEPHYP}          &    \\
   7.62 & Pb+Pb & 0-7\%   & 349 & 14.7$\pm$0.2$\pm$1.2 & 36.9$\pm$0.3$\pm$3.3 & NA49   & \cite{NA49EDEPHYP}          &    \\
   8.73 & Pb+Pb & 0-7\%   & 349 & 14.6$\pm$0.2$\pm$1.2 & 43.1$\pm$0.4$\pm$4.3 & NA49   & \cite{NA49EDEPHYP}          &    \\
   8.73 & Pb+Pb & 0-7\%   & 349 & 15.3$\pm$0.6$\pm$1.0 & 45.6$\pm$1.9$\pm$3.4 & NA49   & \cite{NA49EDEPLAM}          & h) \\
  12.3  & Pb+Pb & 0-7\%   & 349 & 12.9$\pm$0.2$\pm$1.0 & 50.1$\pm$0.6$\pm$5.5 & NA49   & \cite{NA49EDEPHYP}          &    \\
  12.3  & Pb+Pb & 0-7\%   & 349 & 13.5$\pm$0.7$\pm$1.0 & 47.4$\pm$2.8$\pm$3.5 & NA49   & \cite{NA49EDEPLAM}          & h) \\
  17.3  & Pb+Pb & 0-10\%  & 335 &  9.5$\pm$0.1$\pm$1.0 & 44.9$\pm$0.6$\pm$8.0 & NA49   & \cite{NA49EDEPHYP}          &    \\
  17.3  & Pb+Pb & 0-10\%  & 335 & 10.9$\pm$1.0$\pm$1.3 & 44.1$\pm$3.2$\pm$5.0 & NA49   & \cite{NA49EDEPLAM}          & h) \\ \hline
 130.0  & Au+Au & 0-5\%   &     & 17.3$\pm$1.8$\pm$2.8 & ---                  & PHENIX & \cite{PHNXLAM130}           &    \\ \hline
 130.0  & Au+Au & 0-5\%   &     & 17.0$\pm$0.4$\pm$1.7 & ---                  & STAR   & \cite{STARLAM130}           & c) \\
 200.0  & Au+Au & 0-5\%   & 352 & 16.7$\pm$0.2$\pm$1.1 & ---                  & STAR   & \cite{STARHYP200}           & c) \\ \hline\hline
\end{tabular}
\end{footnotesize}
\end{center}
\end{table*}
%

%
\begin{table*}[b]
\begin{center}
\caption{The rapidity densities \dndy\ around midrapidity and the
total multiplicities \navg\ of \lab\ as measured in central
nucleus-nucleus collisions at different \sqrts.  Also listed are the
centrality selection (Cent.) and the averaged number of participants
(\npart).  In most cases the statistical (first) and systematic
(second) errors are given separately, while for a few measurements the
quadratic sum of both is quoted (denoted by 'sum').}
\vspace{5pt}
\begin{footnotesize}
\begin{tabular}{lllllllll} \hline\hline
\label{tab:comp_lab}
  \sqrts        &
  System        &
  Cent.         &
  \npart        &
  \dndy         &
  \navg         &
  Exp.          &
  Ref.          &
  Comnt.        \\
  (GeV) &       &         &     &                           &                        &        &                             &    \\ \hline\hline
   4.87 & Au+Au & 0-12\%  &     & 0.019$^{+0.004 \: +0.003}_{-0.005 \: -0.002}$ & ---         & E917   & \cite{E917LAB}              &    \\ \hline
   8.73 & Pb+Pb & 0-5\%   &     & 0.44$\pm$0.03$\pm$0.04    & ---                    & NA57   & \cite{NA57EDEPHYP}          &    \\
  17.3  & Pb+Pb & 0-5\%   &     & 2.47$\pm$0.14$\pm$0.25    & ---                    & NA57   & \cite{NA57EDEPHYP}          &    \\
  17.3  & Pb+Pb & 0-4.5\% & 349 & 2.44$\pm$0.14$\pm$0.24    & ---                    & NA57   & \cite{NA57ENHANCE}          &    \\ \hline
  17.3  & Pb+Pb &         & 351 & 1.8$\pm$0.2               & ---                    & WA97   & \cite{WA97ENHANCE,WA97QM99} &    \\ \hline
   6.27 & Pb+Pb & 0-7\%   & 349 & 0.10$\pm$0.02$\pm$0.01    & 0.16$\pm$0.02$\pm$0.03 & NA49   & \cite{NA49EDEPHYP}          &    \\
   7.62 & Pb+Pb & 0-7\%   & 349 & 0.21$\pm$0.02$\pm$0.02    & 0.39$\pm$0.02$\pm$0.04 & NA49   & \cite{NA49EDEPHYP}          &    \\
   8.73 & Pb+Pb & 0-7\%   & 349 & 0.33$\pm$0.02$\pm$0.03    & 0.68$\pm$0.03$\pm$0.07 & NA49   & \cite{NA49EDEPHYP}          &    \\
   8.73 & Pb+Pb & 0-7\%   & 349 & 0.42$\pm$0.04$\pm$0.04    & 0.74$\pm$0.04$\pm$0.06 & NA49   & \cite{NA49EDEPLAM}          & h) \\
  12.3  & Pb+Pb & 0-7\%   & 349 & 0.82$\pm$0.03$\pm$0.08    & 1.82$\pm$0.06$\pm$0.19 & NA49   & \cite{NA49EDEPHYP}          &    \\
  12.3  & Pb+Pb & 0-7\%   & 349 & 1.06$\pm$0.08$\pm$0.10    & 2.26$\pm$0.25$\pm$0.20 & NA49   & \cite{NA49EDEPLAM}          & h) \\
  17.3  & Pb+Pb & 0-10\%  & 335 & 1.24$\pm$0.03$\pm$0.13    & 3.07$\pm$0.06$\pm$0.31 & NA49   & \cite{NA49EDEPHYP}          &    \\
  17.3  & Pb+Pb & 0-10\%  & 335 & 1.62$\pm$0.16$\pm$0.2     & 3.87$\pm$0.18$\pm$0.40 & NA49   & \cite{NA49EDEPLAM}          & h) \\ \hline
 130.0  & Au+Au & 0-5\%   &     & 12.7$\pm$1.8$\pm$2.0      & ---                    & PHENIX & \cite{PHNXLAM130}           &    \\ \hline
 130.0  & Au+Au & 0-5\%   &     & 12.3$\pm$0.3$\pm$1.2      & ---                    & STAR   & \cite{STARLAM130}           &    \\
 200.0  & Au+Au & 0-5\%   & 352 & 12.7$\pm$0.2$\pm$0.9      & ---                    & STAR   & \cite{STARHYP200}           &    \\ \hline\hline
\end{tabular}
\end{footnotesize}
\end{center}
\end{table*}
%

%
\begin{table*}[b]
\begin{center}
\caption{The rapidity densities \dndy\ around midrapidity and the
total multiplicities \navg\ of \xim\ as measured in central
nucleus-nucleus collisions at different \sqrts.  Also listed are the
centrality selection (Cent.) and the averaged number of participants
(\npart).  In most cases the statistical (first) and systematic
(second) errors are given separately, while for a few measurements the
quadratic sum of both is quoted (denoted by 'sum').}
\vspace{5pt}
\begin{footnotesize}
\begin{tabular}{lllllllll} \hline\hline
\label{tab:comp_xim}
  \sqrts        &
  System        &
  Cent.         &
  \npart        &
  \dndy         &
  \navg         &
  Exp.          &
  Ref.          &
  Comnt.        \\
  (GeV) &       &         &     &                           &                        &        &                             &    \\ \hline\hline
   3.83 & Au+Au &         & 342 & ---                       & 0.17$\pm$0.05          & E895   & \cite{E895LAMXI}            &    \\ \hline
   8.73 & Pb+Pb & 0-10\%  &     & 1.84$\pm$0.16$\pm$0.18    & ---                    & NA57   & \cite{NA57EDEPHYP}          &    \\
  17.3  & Pb+Pb & 0-10\%  &     & 1.91$\pm$0.05$\pm$0.19    & ---                    & NA57   & \cite{NA57EDEPHYP}          &    \\
  17.3  & Pb+Pb & 0-4.5\% & 349 & 2.08$\pm$0.09$\pm$0.21    & ---                    & NA57   & \cite{NA57ENHANCE}          &    \\ \hline
  17.3  & Pb+Pb &         & 351 & 1.5$\pm$0.1               & ---                    & WA97   & \cite{WA97ENHANCE,WA97QM99} &    \\ \hline
   6.27 & Pb+Pb & 0-7\%   & 349 & 0.93$\pm$0.13$\pm$0.10    & 1.50$\pm$0.13$\pm$0.17 & NA49   & \cite{NA49EDEPHYP}          &    \\
   7.62 & Pb+Pb & 0-7\%   & 349 & 1.17$\pm$0.13$\pm$0.13    & 2.42$\pm$0.19$\pm$0.29 & NA49   & \cite{NA49EDEPHYP}          &    \\
   8.73 & Pb+Pb & 0-7\%   & 349 & 1.15$\pm$0.11$\pm$0.13    & 2.96$\pm$0.20$\pm$0.36 & NA49   & \cite{NA49EDEPHYP}          &    \\
  12.3  & Pb+Pb & 0-7\%   & 349 & 1.22$\pm$0.14$\pm$0.13    & 3.80$\pm$0.26$\pm$0.61 & NA49   & \cite{NA49EDEPHYP}          &    \\
  17.3  & Pb+Pb & 0-10\%  & 335 & 1.44$\pm$0.10$\pm$0.15    & 4.04$\pm$0.16$\pm$0.57 & NA49   & \cite{NA49EDEPHYP}          &    \\
  17.3  & Pb+Pb & 0-10\%  & 335 & 1.49$\pm$0.08$\pm$0.22    & 4.12$\pm$0.20$\pm$0.62 & NA49   & \cite{NA49XI158}            &    \\ \hline
 130.0  & Au+Au & 0-10\%  &     & 2.00$\pm$0.14$\pm$0.20    & ---                    & STAR   & \cite{STARHYP130}           & c) \\
 200.0  & Au+Au & 0-5\%   & 352 & 2.17$\pm$0.06$\pm$0.19    & ---                    & STAR   & \cite{STARHYP200}           &    \\ \hline\hline
\end{tabular}
\end{footnotesize}
\end{center}
\end{table*}
%

%
\begin{table*}[b]
\begin{center}
\caption{The rapidity densities \dndy\ around midrapidity and the
total multiplicities \navg\ of \xip\ as measured in central
nucleus-nucleus collisions at different \sqrts.  Also listed are the
centrality selection (Cent.) and the averaged number of participants
(\npart).  In most cases the statistical (first) and systematic
(second) errors are given separately, while for a few measurements the
quadratic sum of both is quoted (denoted by 'sum').}
\vspace{5pt}
\begin{footnotesize}
\begin{tabular}{lllllllll} \hline\hline
\label{tab:comp_xip}
  \sqrts        &
  System        &
  Cent.         &
  \npart        &
  \dndy         &
  \navg         &
  Exp.          &
  Ref.          &
  Comnt.        \\
  (GeV) &       &         &     &                            &                        &        &                             &    \\ \hline\hline
   8.73 & Pb+Pb & 0-10\%  &     & 0.068$\pm$0.021$\pm$0.007  & ---                    & NA57   & \cite{NA57EDEPHYP}          &    \\
  17.3  & Pb+Pb & 0-10\%  &     & 0.422$\pm$0.023$\pm$0.042  & ---                    & NA57   & \cite{NA57EDEPHYP}          &    \\
  17.3  & Pb+Pb & 0-4.5\% & 349 & 0.51$\pm$0.04$\pm$0.05     & ---                    & NA57   & \cite{NA57ENHANCE}          &    \\ \hline
  17.3  & Pb+Pb &         & 351 & 0.37$\pm$0.06              & ---                    & WA97   & \cite{WA97ENHANCE,WA97QM99} &    \\ \hline
   7.62 & Pb+Pb & 0-7\%   & 349 & 0.05$\pm$0.01$\pm$0.01     & 0.12$\pm$0.02$\pm$0.03 & NA49   & \cite{NA49EDEPHYP}          &    \\
   8.73 & Pb+Pb & 0-7\%   & 349 & 0.07$\pm$0.01$\pm$0.01     & 0.13$\pm$0.01$\pm$0.02 & NA49   & \cite{NA49EDEPHYP}          &    \\
  12.3  & Pb+Pb & 0-7\%   & 349 & 0.21$\pm$0.03$\pm$0.02     & 0.58$\pm$0.06$\pm$0.13 & NA49   & \cite{NA49EDEPHYP}          &    \\
  17.3  & Pb+Pb & 0-10\%  & 335 & 0.31$\pm$0.03$\pm$0.03     & 0.66$\pm$0.04$\pm$0.08 & NA49   & \cite{NA49EDEPHYP}          &    \\
  17.3  & Pb+Pb & 0-10\%  & 335 & 0.33$\pm$0.04$\pm$0.05     & 0.77$\pm$0.04$\pm$0.12 & NA49   & \cite{NA49XI158}            &    \\ \hline
 130.0  & Au+Au & 0-10\%  &     & 1.70$\pm$0.12$\pm$0.17     & ---                    & STAR   & \cite{STARHYP130}           & c) \\
 200.0  & Au+Au & 0-5\%   & 352 & 1.83$\pm$0.05$\pm$0.20     & ---                    & STAR   & \cite{STARHYP200}           &    \\ \hline\hline
\end{tabular}
\end{footnotesize}
\end{center}
\end{table*}
%

%
\begin{table*}[b]
\begin{center}
\caption{The rapidity densities \dndy\ around midrapidity and the
total multiplicities \navg\ of \omm+\omp\ as measured in central
nucleus-nucleus collisions at different \sqrts.  Also listed are the
centrality selection (Cent.) and the averaged number of participants
(\npart). In most cases the statistical (first) and systematic
(second) errors are given separately, while for a few measurements the
quadratic sum of both is quoted (denoted by 'sum').}
\vspace{5pt}
\begin{footnotesize}
\begin{tabular}{llllllllll} \hline\hline
\label{tab:comp_om}
  \sqrts        &
  System        &
  Cent.         &
  \npart        &
  Particle      &
  \dndy         &
  \navg         &
  Exp.          &
  Ref.          &
  Comnt.        \\
  (GeV) &       &          &     &           &                           &                        &        &                             &    \\ \hline\hline
   8.73 & Pb+Pb & 0-11\%   &     & \omm      & 0.085$\pm$0.046$\pm$0.009 & ---                    & NA57   & \cite{NA57EDEPHYP}          &    \\
   8.73 & Pb+Pb & 0-11\%   &     & \omp      & 0.035$\pm$0.020$\pm$0.004 & ---                    & NA57   & \cite{NA57EDEPHYP}          &    \\
  17.3  & Pb+Pb & 0-11\%   &     & \omm      & 0.259$\pm$0.037$\pm$0.026 & ---                    & NA57   & \cite{NA57EDEPHYP}          &    \\
  17.3  & Pb+Pb & 0-11\%   &     & \omp      & 0.129$\pm$0.022$\pm$0.013 & ---                    & NA57   & \cite{NA57EDEPHYP}          &    \\
  17.3  & Pb+Pb & 0-4.5\%  & 349 & \omm      & 0.31$\pm$0.07$\pm$0.05    & ---                    & NA57   & \cite{NA57ENHANCE}          &    \\
  17.3  & Pb+Pb & 0-4.5\%  & 349 & \omp      & 0.16$\pm$0.04$\pm$0.02    & ---                    & NA57   & \cite{NA57ENHANCE}          &    \\ \hline
  17.3  & Pb+Pb &          & 351 & \omm+\omp & 0.41$\pm$0.08             & ---                    & WA97   & \cite{WA97ENHANCE,WA97QM99} &    \\ \hline
   8.73 & Pb+Pb & 0-7\%    & 349 & \omm+\omp & 0.10$\pm$0.02$\pm$0.02    & 0.14$\pm$0.03$\pm$0.04 & NA49   & \cite{NA49EDEPOM}           &    \\
  17.3  & Pb+Pb & 0-23.5\% & 262 & \omm      & 0.14$\pm$0.03$\pm$0.01    & 0.43$\pm$0.09$\pm$0.03 & NA49   & \cite{NA49EDEPOM}           &    \\
  17.3  & Pb+Pb & 0-23.5\% & 262 & \omp      & 0.07$\pm$0.02$\pm$0.01    & 0.19$\pm$0.04$\pm$0.02 & NA49   & \cite{NA49EDEPOM}           &    \\ \hline
 130.0  & Au+Au & 0-10\%   &     & \omm+\omp & 0.55$\pm$0.11$\pm$0.06    & ---                    & STAR   & \cite{STARHYP130}           & c) \\
 200.0  & Au+Au & 0-5\%    & 352 & \omm+\omp & 0.53$\pm$0.04$\pm$0.04    & ---                    & STAR   & \cite{STARHYP200}           &    \\ \hline\hline
\end{tabular}
\end{footnotesize}
\end{center}
\end{table*}
%

\clearpage
\subsection*{Comments}

\begin{description}
\itemsep -2pt

\item[a)] Measured in the rapidity range 2.7~$< y <$~2.9.

\item[b)] Measured in the rapidity range 2.4~$< y <$~3.5.

\item[c)] The \dndy\ value is based on the \pt~extrapolation using the
          Boltzmann function.

\item[d)] The value of the total multiplicity is based on the
          extrapolation using the single Gaussian.

\item[e)] Measured in the rapidity range 2.0~$< y <$~2.4.

\item[f)] Measured via the decay channel
          $\phi \rightarrow e^{+}+e^{-}$ in the rapidity range 2.1~$<
          y <$~2.65.

\item[g)] Preliminary data only, shown on Quark Matter 2001 conference
          \cite{E895LAM}.

\item[h)] Data in \cite{NA49EDEPLAM} are not corrected for feed-down
          from weak decays and are therefore superseded by the values
          given in \cite{NA49EDEPHYP}.

\end{description}


\clearpage
\subsection*{List of Variables}

\begin{tabular}{l@{\hspace{20pt}:\hspace{5pt}}l}
\label{tab:variables}
$A$		                        & Atomic weight \\
$b$		                        & Impact parameter \\
$c$		                        & Velocity of light \\
$\beta$		                        & $v/c$ \\
\betas\		                        & Surface velocity \\
$\beta_{\perp}$	                        & Transverse expansion velocity \\
$c\tau$		                        & Lifetime \\
\dedx\	                	        & Specific energy loss \\
\dndy\		                        & Rapidity density \\
$E_{\rb{S}}$	                        & Strangeness enhancement factor \\
$\langle E \rangle/\langle N \rangle$ 	& Energy density / particle density \\
$\gamma$		                & Lorentz factor \\
\gams\	                                & Strange quark fugacity \\
\gamq\	                                & Fugacity for quark flavor $q$ \\
\epspart\	                        & Participant eccentricity \\
$I$		                        & Isospin \\
$J$		                        & Spin \\
$\kappa$	 	                & String tension \\
\lams\	                                & Wroblewski factor \\
$m$		                        & Mass \\
$m_{\rb{e}}$		                & Electron mass \\
$m_{\rb{P}}$		                & Proton mass \\
$m_{\rb{s}}$		                & Strange quark mass \\
\mt\		                        & Transverse mass \\
\mub\		                        & Baryo-chemical potential \\
$N$		                        & Multiplicity \\
\nbin\		                        & Number of binary collisions \\
$N_{\rb{light}}$		                & Number of light ($u$, $d$) quarks in a hadron \\
\npartch, \np		                & Number of participants \\
\nq\		                        & Number of quarks of type $q$ in a hadron \\
\npc\		                        & Number of corona participants \\
$N_{\rb{s}}$		                & Number of strange quarks in a hadron \\
$n_{\rb{B}}$		                & Baryon density \\
$n_{\rb{s}}$		                & Number density of (anti-)strange quarks \\
$P$		                        & Parity \\
\pt, $p_{\perp}$	                        & Transverse momentum \\
$p$		                        & Total momentum \\
$Q$		                        & $Q$-value of a given reaction \\
\end{tabular}
\newpage
\begin{tabular}{l@{\hspace{20pt}:\hspace{5pt}}l}
$R$		                        & Anti-baryon-baryon ratio \\
\raa\		                        & Nuclear suppression factor \\
\rcp\		                        & Nuclear suppression factor \\
$R_{\rb{s}}$		                & Fireball radius \\
$RMS_{\rb{y}}$		                & RMS width of rapidity distribution \\
$r$		                        & Radius parameter \\
$\rho$		                        & Boost angle $\rho = \tanh^{-1} \betap$ \\
$s$		                        & Entropy density \\
$S$		                        & Strangeness \\
$\sigma_{i}$	                        & Width of rapidity distribution of particle type $i$ (Gauss) \\
$\sigma^{\rb{NN}}_{\rb{inel}}$	        & Inelastic nucleon-nucleon cross section \\
$\sqrt{s_{\rb{coll}}}$	                & Center-of-mass energy of a binary collision \\
\sqrts\	                                & Center-or-mass energy in nucleon-nucleon system \\
$T$	 	                        & Temperature \\
\tch\		                        & Chemical freeze-out temperature \\
\tth\		                        & Thermal freeze-out temperature \\
\tc\		                        & Critical temperature \\
$T_{\rb{max}}$		                & Maximum kinetic energy transfer \\
$T^{*}$		                        & Inverse slope parameter \\
$\tau^{\rb{eq}}$	                        & Equilibration time \\
$\tau_{\rb{fo}}$		                & Decoupling times in hydro freeze-out \\
$V$		                        & Volume \\
$v_{\rb{s}}$		                & Surface velocity \\
$v_{\perp}$		                & Transverse expansion velocity \\
\vtwo		                        & Elliptic flow coefficient \\
$Y$		                        & Yield \\
$Y_{\rb{core}}$		                & Yield in core \\
$Y_{\rb{corona}}$	                        & Yield in corona \\
$y$		                        & Rapidity \\
\ybeam\		                        & Beam rapidity \\
$Z$		                        & Atomic charge \\
$z$		                        & Particle charge \\
\end{tabular}



\end{document}